\documentclass[aps,prb,twocolumn,10pt,superscriptaddress,showpacs]{revtex4-2}
\usepackage{amsmath,amssymb,graphics,epsfig,epstopdf,color,verbatim,ulem,braket,tabularx}
\usepackage[colorlinks,linkcolor=blue,citecolor=blue,urlcolor=blue]{hyperref}
\usepackage{physics}
\usepackage{bm}
\usepackage{blindtext}
\graphicspath{{Fig_Long/}}

\begin{document}

\title{Magnetic, thermodynamic, and dynamical properties of the three-dimensional fermionic Hubbard model: A comprehensive Monte Carlo study}

\author{Yu-Feng Song}
\affiliation{Hefei National Laboratory for Physical Sciences at Microscale and Department of Modern Physics, University of Science and Technology of China, Hefei, Anhui 230026, China}
\affiliation{Institute of Modern Physics, Northwest University, Xi'an 710127, China}

\author{Youjin Deng}
\email{yjdeng@ustc.edu.cn}
\affiliation{Hefei National Laboratory for Physical Sciences at Microscale and Department of Modern Physics, University of Science and Technology of China, Hefei, Anhui 230026, China}
\affiliation{Hefei National Laboratory, University of Science and Technology of China, Hefei 230088, China}

\author{Yuan-Yao He}
\email{heyuanyao@nwu.edu.cn}
\affiliation{Institute of Modern Physics, Northwest University, Xi'an 710127, China}
\affiliation{Shaanxi Key Laboratory for Theoretical Physics Frontiers, Xi'an 710127, China}
\affiliation{Hefei National Laboratory, University of Science and Technology of China, Hefei 230088, China}
\affiliation{Peng Huanwu Center for Fundamental Theory, Xian 710127, China}

\begin{abstract}
The interplay between quantum and thermal fluctuations can induce rich phenomena at finite temperatures in strongly correlated fermion systems. Here we report a {\it numerically exact} auxiliary-field quantum Monte Carlo (AFQMC) study for the finite-temperature properties of three-dimensional repulsive Hubbard model at half-filling. We concentrate on the complete temperature-interaction strength phase diagram of the model, which contains the low-temperature antiferromagnetic (AFM) long-range ordered phase and metal-insulator crossover (MIC) in the paramagnetic phase. Enabling access to unprecedented system sizes up to $20^3$, we achieve highly accurate results of the N\'{e}el transition temperature for representative values of on-site interaction $U$ via finite-size analysis of AFM structure factor. To quantitatively characterize the MIC above the N\'{e}el transition, we have developed efficient schemes to compute the thermal entropy versus $U$ at fixed temperature and to calculate the $U$-derivative of double occupancy in AFQMC simulations. Then combining variously thermodynamic and dynamical observables, we establish an efficient scheme to precisely determine the boundaries for the MIC by cross-checking different observables. We also demonstrate the temperature dependence of many commonly used observables. Away from half-filling, we explore the behavior of the sign problem and AFM spin correlation versus hole doping, and demonstrate the persistence of N\'{e}el AFM ordered phase to finite doping with limited results.
\end{abstract}

\date{\today}

\maketitle

\section{Introduction}
\label{sec:intro}

The fermionic Hubbard model has become the paradigmatic model for strongly correlated fermion systems, for which its importance is comparable to the Ising model for statistical mechanics. It was initially proposed in the 1960s~\cite{Hubbard1963,Kanamori1963,Gutzwiller1963} to model the interacting electrons and to correspondingly understand the magnetism and metal-insulator transitions in three-dimensional (3D) transition metals and their oxides~\cite{Fujimori1992,Tokura1993,Inoue1995,Morikawa1995}. Over the past decades, countless studies of the Hubbard model have gone far beyond the original intention, and have revealed a fantastic wealth of phases, phase transitions, and exotic correlation phenomena~\cite{Arovas2022,Qin2022} in this simple model. As a representative example, the two-dimensional (2D) repulsive Hubbard model on square lattice plays a crucial role in the study of high-temperature superconductivity in the cuprates~\cite{Dagotto1994,Bulut2010}. For this specific model, the modern state-of-art precision many-body simulations have made overwhelming progress~\cite{Qin2022} in the last 10 years, and have identified the N\'{e}el antiferromagnetic (AFM) order~\cite{LeBlanc2015}, pseudogap phenomena~\cite{Wuwei2018}, stripe orders~\cite{Chang2010,Qin2016a,Boxiao2017,Xiao2023}, and the $d$-wave superconductivity~\cite{Qin2020,Haoxu2024}, which all appear in the typical phase diagram of cuprates. 

Most of the essential physics in the repulsive Hubbard model originates from the competition between the hopping and the local interaction. It can be further diversified by important ingredients including the dimensionality, lattice geometry, temperature, next-nearest-neighbor (NNN) hopping and fermion filling~\cite{Arovas2022}. Despite its simplicity, only the one-dimensional Hubbard chain can be exactly solved~\cite{Lieb1968,Luo2023a,Luo2023b}, while in higher dimensions the quantitatively accurate results of the model fully count on quantum many-body numerical methods~\cite{Qin2022}. Comparing to its 2D correspondence, the 3D repulsive Hubbard model as its original form has been much less studied, due to the limitations of the numerical algorithms as well as the driving force from the cuprates. As a result, a large portion of the phase diagram for the model remains unknown, such as the normal phase at half-filling and away from half-filling. 

Since the 1990s, there have been a certain amount of numerical simulations to study the finite-temperature properties of the 3D repulsive Hubbard model, applying methods including auxiliary-field quantum Monte Carlo (AFQMC)~\cite{Ulmke1996,Staudt2000,Paiva2011,Ibarra2020,Fanjie2024}, second-order perturbation theory~\cite{Tahvildar1997}, dynamical mean-field theory (DMFT) and its extensions~\cite{Werner2005,Kent2005,Rohringer2011,Iskakov2022,Fuchs2011L,Fuchs2011B,Sotnikov2012,Gorelik2012,Mauro2014,Katanin2017}, numerical linked-cluster expansion (NLCE)~\cite{Khatami2016}, and diagrammatic Monte Carlo (DiagMC)~\cite{Kozik2013,Lenihan2022,Garioud2024}. Almost all of these studies stayed at half-filling. Among them, the majority focused on the magnetic phase transition from the paramagnetic (PM) phase (normal phase) to the N\'{e}el AFM ordered phase upon cooling~\cite{Staudt2000,Iskakov2022,Werner2005,Kent2005,Rohringer2011,Kozik2013,Lenihan2022,Garioud2024}. The rest studied the single-particle spectral properties~\cite{Ulmke1996,Fuchs2011B} and the thermodynamic properties~\cite{Fuchs2011L,Sotnikov2012,Gorelik2012,Mauro2014,Khatami2016,Ibarra2020} of the model. The further effects of NNN hopping~\cite{Fuchs2011B}, the mass imbalance~\cite{Sotnikov2012} and the hopping anisotropy~\cite{Ibarra2020,Mauro2014} on the half-filled system were also investigated. Away from half-filling, the magnetic phase diagram including both the N\'{e}el AFM order and an incommensurate spin-density wave (SDW) order has been revealed for specific parameter regions~\cite{Tahvildar1997,Katanin2017,Lenihan2022}. All these numerical results from different methods contribute significant pieces for the full scope of 3D repulsive Hubbard model. Nevertheless, there are still quite a lot of missing puzzles, such as the systematic and unbiased calculations of N\'{e}el transition temperatures and the normal phase. Besides, the previous AFQMC simulation by Staudt {\it et al.}~\cite{Staudt2000}, which has been used as the benchmark for many other studies, has apparent limitations in both the accuracy (see Sec.~\ref{sec:SignEleinAFQMC}) and the system size (up to $10^3$ lattice sites).

In alternative to the numerics, quantum simulation combining ultracold atoms and optical lattice potentials~\cite{Bloch2008} has become an important route to study the 3D repulsive Hubbard model~\cite{Kohl2005,Joerdens2008,Schneider2008,Duarte2015,Tarruell2010,Daniel2013,Hart2015,Shao2024}. The early stage of these optical lattice experiments reported the exploration of the single-particle physics~\cite{Kohl2005}, the realizations of band and Mott insulating phases~\cite{Joerdens2008,Schneider2008,Duarte2015}, and the achievement of short-range quantum magnetism~\cite{Daniel2013}. Then the AFM spin correlation was manifested via the measurement of AFM structure factor~\cite{Hart2015}, in an experimental setup at temperature around 1.4 times that of the N\'{e}el temperature. The impressive breakthrough appears most recently~\cite{Shao2024} that, in a nearly uniform optical lattice, the finite-temperature N\'{e}el transition of the model is clearly observed via the critical scaling behavior of AFM structure factor. With improving cooling and probing techniques, it is promising that the optical lattice experiments equipped with high-accuracy measurements can open a new avenue to explore the remaining open questions in 3D repulsive Hubbard model. In the meantime, the complementary high-precision numerical results are demanded, and can serve as crucial benchmarks or even guidelines for future experimental studies.

Based on the above discussions, we systematically study the finite-temperature properties of 3D repulsive Hubbard model applying the numerically unbiased AFQMC algorithm~\cite{Blankenbecler1981,Hirsch1983,White1989,Scalettar1991,McDaniel2017,Yuanyao2019,Yuanyao2019L}. Our motivations of this paper are two-fold. First, we expand on our companion paper~\cite{Yufeng2024} by presenting more detailed results for the full phase diagram of the model at half-filling. These include the algorithmic improvements and developments for AFQMC calculation, which are absolutely necessary to achieve the high-precision results. The rest are the more comprehensive results of the magnetic, thermodynamic and dynamical properties that we apply to compute the N\'{e}el transition temperatures and to characterize the metal-insulator crossover~\cite{Yufeng2024} in the normal phase. These results also offer more possibilities to benchmark for the ongoing optical lattice experiments. Second, we reach out to the preliminary attempts for the model with doping by AFQMC method, and study the sign problem and AFM spin correlation, to motivate the future precision many-body calculations for this more complicated and difficult problem. 

\begin{table}[h]
\label{Table:A3}
\centering
\caption{Contents reference table for all the involved physics in this work and our companion paper ~\cite{Yufeng2024}.}
\begin{tabular}{|c||c|}
\hline
Contents & Location in this work \\
\hline
Complete $U$-$T$ phase diagram & Sec.~\ref{sec:diagram}, Fig.~\ref{fig:PhaseDiagram} \\
N\'{e}el temperatures & Sec.~\ref{sec:diagram}, Fig.~\ref{fig:NeelTemp}, Fig.~\ref{fig:U4Neel} \\
Local spectrum $A_{\rm loc}(\omega)$ & Sec.~\ref{sec:FLtoBM}, Fig.~\ref{fig:Spectrum} \\
Determination of $U_{\rm BM}$ & Sec.~\ref{sec:FLtoBM}, Fig.~\ref{fig:CrossoverRatio} \\
Determination of $U_{\rm MI}$ & Sec.~\ref{sec:BMtoMI}, Fig.~\ref{fig:Quasiparwght} \\
AFM spin correlations & Sec.~\ref{sec:AFMSpinCrft}, Fig.~\ref{fig:AFMCorrelation} \\
Compute $S$ versus $U$ & Sec.~\ref{sec:ComputeEntropy}, Fig.~\ref{fig:CompareEntropy} \\
Entropy \& Double occupancy & Sec.~\ref{sec:EntDoucMax}, Fig.~\ref{fig:EntropyDouocc} \\
Specific heat & Sec.~\ref{sec:CvChie}, Fig.~\ref{fig:SpecificHeat}\\
Compressibility & Sec.~\ref{sec:CvChie}, Fig.~\ref{fig:Compressibility}\\
Directly compute $\partial D/\partial U$ & Sec.~\ref{sec:InflcDouOcc}, Appx.~\ref{sec:AppendixA}, Fig.~\ref{fig:DoubleOccInter}\\
Fidelity susceptibility& Sec.~\ref{sec:Chifidelity}, Fig.~\ref{fig:FidelitySuscep} \\
Self-energy & Sec.~\ref{sec:SelfEnergy}, Fig.~\ref{fig:SelfEnergy} \\
Away from half-filling & Sec.~\ref{sec:SignProblem}, Fig.~\ref{fig:SignProblem}, Fig.~\ref{fig:mSquVsfilling} \\
Finite-size effects & Appx.~\ref{sec:AppendixE} \\ 
\hline
\end{tabular}
\end{table}

The remainder of this paper is organized as follows. In Sec.~\ref{sec:modelmethod}, we describe the lattice model, AFQMC method and the physical quantities we compute. In Sec.~\ref{sec:SignEleinAFQMC}, we present important elements and improvements as applied in our AFQMC simulations. In Sec.~\ref{sec:diagram}, we concentrate on the complete phase diagram for 3D repulsive Hubbard model at half-filling and the corresponding numerical results. In Sec.~\ref{sec:ThermQuantity}, we show more results for the commonly used thermodynamic quantities for the phase diagram at half-filling, along with new methodological developments for their calculations. In Sec.~\ref{sec:SignProblem}, we study the sign problem and the AFM spin correlation for the doping case. Finally, we summarize our work and discuss the future opportunities for the 3D repulsive Hubbard model in Sec.~\ref{sec:Summary}. The Appendixes contain important derivations and formulas, Fermi surface properties and demonstrations of the finite-size effect. Considering the significant length of this work and its connection with our companion paper~\cite{Yufeng2024}, we present the contents reference for all the involved physics in Table.~\ref{Table:A3} to improve readability.

\section{Model, Method and physical observables}
\label{sec:modelmethod}

\subsection{Fermionic Hubbard model}
\label{sec:HubbardModel}

We study the single-band Hubbard model on simple cubic lattice described by the following Hamiltonian
\begin{equation}\begin{aligned}
\label{eq:Hamiltonian}
\hat{H}
=\sum_{\mathbf{k}\sigma}(\varepsilon_{\mathbf{k}}+\mu)c_{\mathbf{k}\sigma}^+c_{\mathbf{k}\sigma}
+ U\sum_{\mathbf{i}}\Big(\hat{n}_{\mathbf{i}\uparrow}\hat{n}_{\mathbf{i}\downarrow} - \frac{\hat{n}_{\mathbf{i}\uparrow} + \hat{n}_{\mathbf{i}\downarrow}}{2}\Big),
\end{aligned}\end{equation}
with $\sigma$ ($=\uparrow$ or $\downarrow$) denoting spin and $\hat{n}_{\mathbf{i}\sigma}=c_{\mathbf{i}\sigma}^+c_{\mathbf{i}\sigma}$ as the density operator. With the nearest-neighbor (NN) hopping strength $t$ and periodic boundary conditions (PBC), we have the kinetic energy dispersion $\varepsilon_{\mathbf{k}}=-2t(\cos k_x + \cos k_y + \cos k_z)$, where the momentum $k_x,k_y,k_z$ are defined in units of $2\pi/L$ with the system size $N_s=L^3$. The chemical potential term $\mu$ is a pure doping, and the system is half-filled with $n=1$ for $\mu=0$ due to the particle-hole symmetry. Throughout this work, we set $t$ as the energy unit, and focus on repulsive $(U>0)$ interaction.

Regarding the magnetic properties, the model Hamiltonian in Eq.~(\ref{eq:Hamiltonian}) has spin ${\rm SU}(2)$ symmetry which can be spontaneously broken with decreasing temperature. Thus at half-filling, the model hosts finite-temperature, continuous PM-AFM phase transition (N\'{e}el transition) belonging to the 3D Heisenberg university class~\cite{Campostrini2002}.

\subsection{Formalism of AFQMC method}
\label{sec:AFQMC}

We then employ the finite-temperature AFQMC algorithm~\cite{Blankenbecler1981,Hirsch1983,White1989,Scalettar1991,McDaniel2017,Yuanyao2019,Yuanyao2019L} to simulate the 3D Hubbard model in Eq.~(\ref{eq:Hamiltonian}). The model is free of the minus sign problem at half-filling due to the particle-hole symmetry~\cite{Wu2005}, while it generally suffers from the sign problem with doping.

The AFQMC algorithm deals with the partition function $Z={\rm Tr}(e^{-\beta\hat{H}})$ by discretizing the inverse temperature as $\beta = M\Delta\tau$ and applying the Trotter-Suzuki (TS) decomposition with either the asymmetric (second-order) formula as
\begin{equation}
\label{eq:SymTrot}
e^{-\Delta\tau\hat{H}}=e^{-\Delta\tau\hat{H}_0}e^{-\Delta\tau\hat{H}_I}+O[(\Delta\tau)^2],
\end{equation}
or the symmetric (third-order) formula
\begin{equation}
\label{eq:AsymTrot}
e^{-\Delta\tau\hat{H}}=e^{-\Delta\tau\hat{H}_0/2}e^{-\Delta\tau\hat{H}_I}e^{-\Delta\tau\hat{H}_0/2}+O[(\Delta\tau)^3],
\end{equation}
where $\hat{H}_0$ and $\hat{H}_I$ are free and interaction parts of the Hamiltonian, respectively. Practically, the systematic Trotter error in both decompositions can be removed by extrapolating multiple calculations with different $\Delta\tau$ values. Nevertheless, the above two formulas show different Trotter error scalings which indicates distinct convergence speed to the $\Delta\tau\to0$ limit. Comparing to the asymmetric formula, the symmetric TS decomposition can typically promise a faster convergence for general observables without increasing the computational complexity in practical calculations for the Hubbard model.

After the TS decomposition, the interaction term $e^{-\Delta\tau\hat{H}_I}$ remains to be handled. The AFQMC algorithm then decouples this two-body interaction term into free fermions coupled with auxiliary fields by the Hubbard-Stratonovich (HS) transformation. The widely used HS transformations for the Hubbard interaction are the formulas with two-component auxiliary fields~\cite{Hirsch1983} into the spin-$\hat{s}_z$ channel as
\begin{equation}\begin{aligned}
\label{eq:SzHS}
e^{-\Delta\tau U \big(\hat{n}_{\bm{\mathrm{i}}\uparrow} \hat{n}_{\bm{\mathrm{i}}\downarrow} - \frac{\hat{n}_{\bm{\mathrm{i}}\uparrow} + \hat{n}_{\bm{\mathrm{i}}\downarrow}}{2}\big) } = \frac{1}{2} \sum\limits_{x_{\bm{\mathrm{i}}} = \pm 1} e^{\gamma_{\mathrm{s}} x_{\bm{\mathrm{i}}} (\hat{n}_{\bm{\mathrm{i}}\uparrow} - \hat{n}_{\bm{\mathrm{i}}\downarrow})},
\end{aligned}\end{equation}
and into the charge channel as
\begin{equation}\begin{aligned}
\label{eq:DensityHS}
e^{-\Delta\tau U \big(\hat{n}_{\bm{\mathrm{i}}\uparrow} \hat{n}_{\bm{\mathrm{i}}\downarrow} - \frac{\hat{n}_{\bm{\mathrm{i}}\uparrow} + \hat{n}_{\bm{\mathrm{i}}\downarrow}}{2}\big) } = \frac{e^{\Delta\tau U/2}}{2} \sum\limits_{x_{\bm{\mathrm{i}}} = \pm 1} e^{\gamma_{\mathrm{c}} x_{\bm{\mathrm{i}}} (\hat{n}_{i\uparrow} + \hat{n}_{\bm{\mathrm{i}}\downarrow} - 1)},
\end{aligned}\end{equation}
where the coupling constants $\gamma_{\mathrm{s}}$ and $\gamma_{\mathrm{c}}$ satisfying
\begin{equation}
\left\{\begin{aligned}
\gamma_{\mathrm{s}} &= \cosh^{-1}(e^{+\Delta\tau U/2}) \\
\gamma_{\mathrm{c}} &= i\cos^{-1}(e^{-\Delta\tau U/2}).
\end{aligned}\right.
\end{equation}
Although both the transformations are exact, they can induce significant difference for physical quantities in practical calculations (with respect to finite-$\Delta\tau$ effects and Monte Carlo fluctuations). As an example, the HS transformation into the spin-$\hat{s}_z$ channel in Eq.~(\ref{eq:SzHS}) clearly breaks the spin ${\rm SU}(2)$ symmetry, and consequently results in considerable fluctuations for spin-related properties. It was also found that, at half-filling, this HS transformation can even produce wrong results for spin-spin correlations due to the ergodicity problem in the Monte carlo sampling with local update~\cite{Scalettar1991}. Alternatively, the transformation into charge channel in Eq.~(\ref{eq:DensityHS}) can almost get rid of these issues. Thus, for $U>0$, it is better to apply Eq.~(\ref{eq:DensityHS}) in AFQMC simulations to calculate the spin-related properties. On the other hand, the overall AFQMC computational effort using Eq.~(\ref{eq:DensityHS}) is about four times of that with the formula in Eq.~(\ref{eq:SzHS}), since the former deals with complex numbers.

Based on the above operations, the partition function can be evaluated as $Z={\rm Tr}(e^{-\beta\hat{H}})=\sum_{\mathbf{X}}W(\mathbf{X})$, where $\mathbf{X}$ is the auxiliary-field configuration and $W(\mathbf{X})$ as a matrix determinant is the corresponding weight. Then the fermionic observables are computed through importance sampling of field configurations via the Metropolis algorithm. Further details of the AFQMC algorithm can be found in the reviews in Refs.~\cite{Assaad2008,Chang2015}.

\subsection{Physical observables}
\label{sec:AFQMCObs}

It is straightforward to compute variously static and dynamical properties for correlated fermions in AFQMC simulations~\cite{White1989,Yuanyao2019}. In this work, we mainly concentrate on the magnetic, thermodynamic and dynamical properties of the 3D repulsive Hubbard model. We illustrate and discuss the corresponding physical observables in our calculations as follows. 

In order to study the magnetic properties, we measure the real-space spin-spin correlation function
\begin{equation}
\label{eq:RspSpinCrFt}
C(\bm{\mathrm{r}}) = \frac{1}{N_s} \sum\limits_{\bm{\mathrm{i}}} \langle \hat{s}_{\bm{\mathrm{i}}}^z \hat{s}_{\bm{\mathrm{i+r}}}^z \rangle,
\end{equation}
where $\hat{s}_{\bm{\mathrm{i}}}^z = (\hat{n}_{\bm{\mathrm{i}}\uparrow}-\hat{n}_{\bm{\mathrm{i}}\downarrow})/2$ is the $z$-component spin operator. Then the spin structure factor is defined from $C(\bm{\mathrm{r}})$ as 
\begin{equation}
\label{eq:Safmzz}
S(\mathbf{q}) = \sum_{\mathbf{r}} C(\mathbf{r}) e^{i\mathbf{q}\cdot\mathbf{r}}.
\end{equation}
Then the AFM structure factor is taken as $S_{\mathrm{AFM}}^{zz}=S(\bm{\pi})$ where $\bm{\mathrm{\pi}}=(\pi,\pi,\pi)$ is the AFM ordering vector. Then the mean-squared magnetization for the N\'{e}el AFM order can be expressed as $m^2=S_{\mathrm{AFM}}^{zz}/N_s$. Another observable to quantify the AFM spin correlation is the correlation length $\xi_\mathrm{AFM}$, which can be computed as~\cite{Sandvik2010,Hofmann2023}
\begin{equation}
\label{eq:CorrLength}
\xi_\mathrm{AFM} = \big[2\sin(\pi/L)\big]^{-1} \sqrt{\frac{S(\bm{\pi})}{S(\bm{\pi}+\delta\bm{\mathrm{q}})}-1},
\end{equation}
with $\delta \bm{\mathrm{q}}$ as the smallest momentum on the lattice, i.e., $\delta \bm{\mathrm{q}}=(2\pi/L,0,0)$ or $(0,2\pi/L,0)$ or $(0,0, 2\pi/L)$. 

For the thermodynamics, the thermal entropy $S$ is a very important quantity serving as the quantitative description of disorder in the system, and it also plays a crucial role in optical lattice experiments~\cite{Shao2024}. In previous studies~\cite{Ibarra2020,Dar2007,Paiva2010,Sushchyev2022}, it is usually evaluated by integrating energy across temperatures as
\begin{equation}
\label{eq:Entropy00}
\frac{S(T)}{N_s} = \ln4 + \frac{e(T)}{T}- \int_T^{\infty} \frac{e(T^{\prime})}{{T^{\prime}}^{2}} dT^{\prime},
\end{equation}
where $\ln4$ is the entropy density at infinite temperature for fermions, and $e(T)=\langle\hat{H}\rangle/N_s$ is the total energy density of the system. However, in this formula, the infinite temperature as the upper limit of the integral is an ambiguity for numerical calculations. We have solved this issue by applying the integration by substitution from $T$ to $\beta$ for the high temperature regime. Moreover, we have developed a very efficient scheme to compute the entropy versus increasing $U$ at fixed temperature. All the calculation details of both techniques are presented in Sec.~\ref{sec:ComputeEntropy}. 

Besides, we have also calculated the double occupancy $D=N_s^{-1}\sum_{\mathbf{i}}\langle \hat{n}_{\bm{\mathrm{i}}\uparrow} \hat{n}_{\bm{\mathrm{i}}\downarrow}\rangle$, and the charge compressibility
\begin{equation}\begin{aligned}
\label{eq:ChiCharge}
\chi_e = -\frac{dn}{d\mu} = \frac{\beta}{N_s}\sum_{\bm{\mathrm{ij}}}\big(\langle \hat{n}_{\bm{\mathrm{i}}} \hat{n}_{\bm{\mathrm{j}}} \rangle - \langle \hat{n}_{\bm{\mathrm{i}}} \rangle \langle \hat{n}_{\bm{\mathrm{j}}} \rangle\big),
\end{aligned}\end{equation}
with $\hat{n}_{\bm{\mathrm{i}}} = \hat{n}_{\bm{\mathrm{i}}\uparrow} + \hat{n}_{\bm{\mathrm{i}}\downarrow}$, and $n=N_s^{-1}\sum_{\mathbf{i}}\langle\hat{n}_{\mathbf{i}}\rangle$ as the fermion filling. The derivation for the second equality in Eq.~(\ref{eq:ChiCharge}) is packed in Appendix~\ref{sec:AppendixA}. Both $D$ and $\chi_e$ are closely related to the electron localization physics with increasing interaction strength. 

The dynamical quantities can offer more insights for the correlated systems beyond static observables, especially about the spectral and transport properties. In this work, we only pay attention to single-particle dynamical properties, revealed by the imaginary-time single-particle Green's function, the spectrum, and the self energy. Specifically, we obtain the local spectrum $A_{\rm loc}(\omega)$ from the corresponding imaginary-time Green's function as
\begin{equation}
G_{\rm loc}(\tau) = \frac{1}{2N_s} \sum_{\mathbf{i},\sigma} \big\langle c_{\mathbf{i}\sigma}(\tau) c_{\mathbf{i}\sigma}^+ \big\rangle, 
\end{equation}
using the stochastic numerical continuation (SAC) method~\cite{Sandvik2016,Shao2023}. This frequency-resolved spectra $A_{\rm loc}(\omega)$ relates to the electrical conductivity and also contains information about Mott physics. Another quantity is the quasiparticle weight $Z_{k_F}$ at Fermi surface, which quantifies the correlation enhancement for the quasiparticle mass. We compute $Z_{k_F}$ by the approximated relation~\cite{Liebsch2003,Liu2015}
\begin{equation}
\label{eq:QuasiZkf}
Z_{k_F} \approx \big[ 1-\Im \Sigma_{\sigma}(\mathbf{k}_F, i\omega_0)/\omega_0 \big]^{-1},
\end{equation}
where $\Sigma_{\sigma}(\mathbf{k}_F, i\omega_0)$ is the self energy and the Fermi wave vector $\mathbf{k}_F$, with $\omega_n=(2n+1)\pi/\beta$ as the Matsubara frequency for fermions. The $Z_{k_F}$ results can be applied as a tool to detect the entrance into the Mott insulator~\cite{Liebsch2003}.

Other than the aforementioned quantities, there are some other thermodynamic observables whose calculations in AFQMC depend on the dynamical correlations. First, instead of the numerical differentiation, we apply a method to compute the derivative of double occupancy over $U$ as
\begin{equation}\begin{aligned}
\label{eq:DouOccDerive}
\frac{\partial D}{\partial U} =
-\frac{2}{N_s}\int_0^{\beta/2} C_{\hat{H}_I}(\tau, 0) d\tau,
\end{aligned}\end{equation}
where $C_{\hat{H}_I}(\tau, 0)=\langle\hat{H}_I(\tau)\hat{H}_I(0)\rangle - \langle\hat{H}_I(\tau)\rangle\langle\hat{H}_I(0)\rangle$ is an imaginary-time correlation function for the operator $\hat{H}_I=\sum_{\mathbf{i}}\big[\hat{n}_{\mathbf{i}\uparrow} \hat{n}_{\mathbf{i} \downarrow} - (\hat{n}_{\mathbf{i}\uparrow} + \hat{n}_{\mathbf{i} \downarrow})/2\big]$. This formula is only valid at half-filling with $\mu=0$ and $n=1$. It can be further generalized to the other calculations with fixed $\mu$ ($\ne 0$) or with fixed $n$ ($\ne 1$). The calculation details for Eq.~(\ref{eq:DouOccDerive}) and its generalizations are packed in Appendix~\ref{sec:AppendixA}. The second quantity is the fidelity susceptibility $\chi_{\rm F}$, which serves as very useful probe of the phase transitions~\cite{You2007,Venuti2007,Gu2009,Schwandt2009,Albuquerque2010,WangLei2015,HuangLi2016}. 
At finite temperature, this quantity can be computed also from $C_{\hat{H}_I}(\tau, 0)$~\cite{Schwandt2009,Albuquerque2010,WangLei2015} as
\begin{equation}\begin{aligned}
\label{eq:fidelitysusp}
\chi_{\rm F} = \int_{0}^{\beta/2} \tau C_{\hat{H}_I}(\tau, 0) d\tau.
\end{aligned}\end{equation}
In the AFQMC simulation, the additional computational effort for the measurement of $C_{\hat{H}_I}(\tau, 0)$ shares the same scaling as general two-body imaginary-time correlation functions. 

In Eqs.~(\ref{eq:Entropy00}), (\ref{eq:DouOccDerive}), and (\ref{eq:fidelitysusp}), we need to deal with the integral while our AFQMC data points are surely discrete. The subsequent discretization error can even exceeds the overall Trotter error of the AFQMC simulation. For example, if the the trapezoidal method is applied to evaluate the integral in Eqs.~(\ref{eq:DouOccDerive}) and (\ref{eq:fidelitysusp}), the systematic error is proportional to $\Delta\tau$. Thus, we implement an efficient scheme to avoid this error. We first perform fitting (such as cubic-spline) for the integrand [as $E(T^{\prime})$ in Eq.~(\ref{eq:Entropy00}), and $C_{\hat{H}_I}(\tau, 0)$ in Eqs.~(\ref{eq:DouOccDerive}) and (\ref{eq:fidelitysusp})], and then evaluate the integral analytically using the fitting curve. We further estimate the uncertainty by the bootstrapping technique (see Appendix C in Ref.~\onlinecite{Song2024}).

\section{Significant elements in AFQMC}
\label{sec:SignEleinAFQMC}

Based on the fundamental concepts of the AFQMC algorithm presented in Sec.~\ref{sec:AFQMC}, we discuss the important elements implemented in our AFQMC simulations in this section. We mainly demonstrate the Trotter error and the effect of different HS transformations of the AFQMC results for 3D repulsive Hubbard model. The benchmark with previous AFQMC results in Ref.~\onlinecite{Staudt2000} is also presented and discussed. 

\begin{figure}[b]
\centering
\includegraphics[width=1.00\columnwidth]{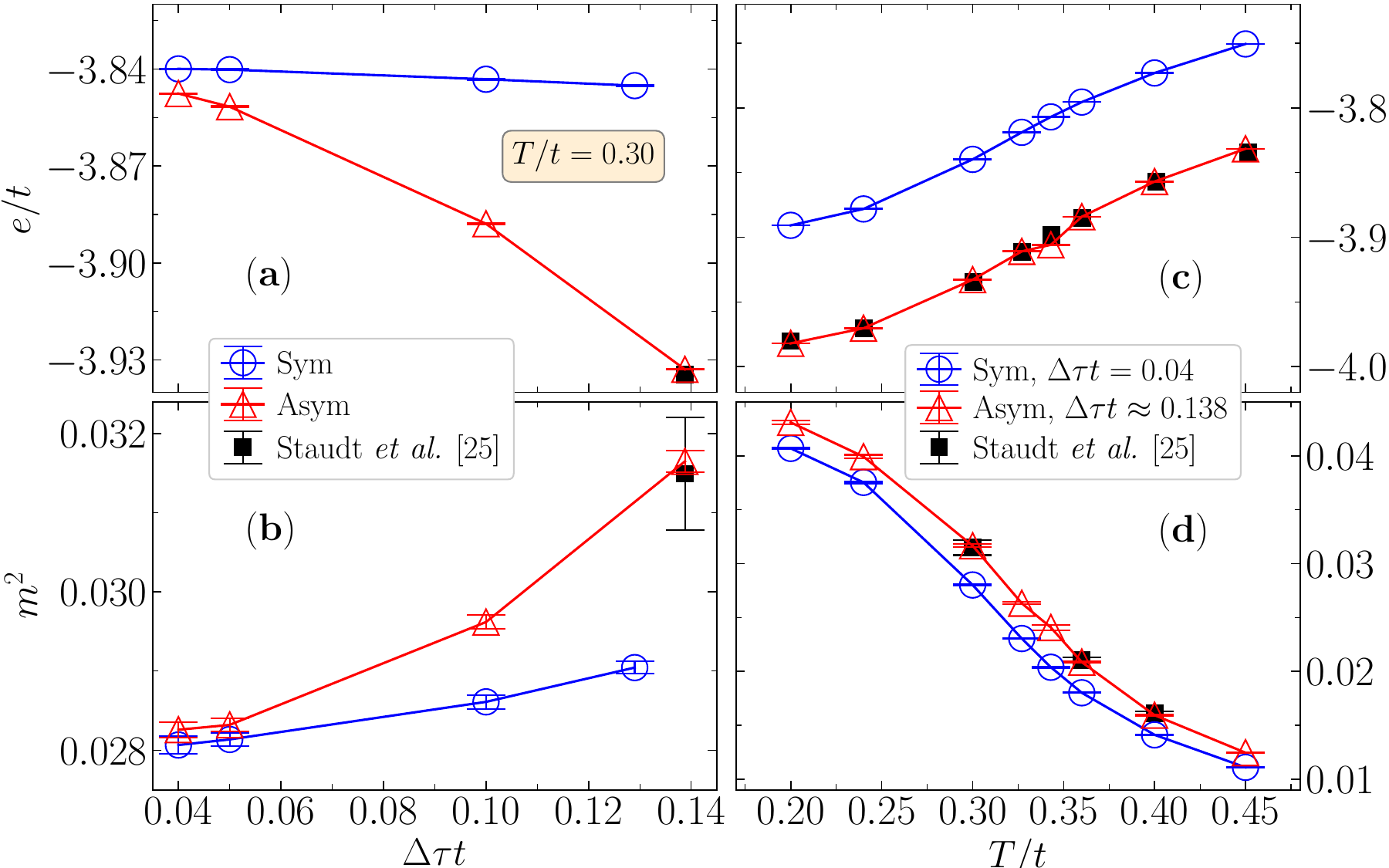}
\caption{Illustration of the Trotter error for total energy density $e/t$ and mean-squared magnetization $m^2$ for $L=4$ system with $U/t=6$ at half-filling, and the benchmark with results in Ref.~\onlinecite{Staudt2000}. (a), (b) Show the results of $e/t$ and $m^2$ versus $\Delta\tau t$ with both symmetric (denoted as ``Sym'') and asymmetric (as ``Asym'') TS decompositions for $T/t=0.30$, demonstrating the faster convergence as $\Delta \tau \to 0$ of the symmetric TS decomposition. (c), (d) Compare the results of $e/t$ and $m^2$ in a temperature range for the symmetric TS decomposition with $\Delta\tau t=0.04$ and the asymmetric decomposition with $\Delta\tau t\approx 0.138$, and the latter apparently recovers previous AFQMC results in Ref.~\onlinecite{Staudt2000}. This indicates that, in AFQMC calculations, a careful extrapolation for $\Delta \tau \to 0$ is needed to obtain reliable estimates of thermodynamic quantities. }
\label{fig:Trotter}
\end{figure}

The overall Trotter error of AFQMC results for general observables should be $O[(\Delta\tau)^2]$ and $O(\Delta\tau)$ within the symmetric TS decomposition in Eq.~(\ref{eq:SymTrot}) and the asymmetric one in Eq.~(\ref{eq:AsymTrot}), respectively. The difference between these two decompositions is clearly manifested in Fig.~\ref{fig:Trotter}(a) and \ref{fig:Trotter}(b), illustrating the Trotter error of total energy density $e/t$ and mean-squared magnetization $m^2$ for $L=4$ system. The Trotter errors from the symmetric decomposition are significantly smaller and the numerical results reach the convergence towards $\Delta\tau=0$ much faster than the asymmetric one. These results also elucidate that, for $\Delta\tau t=0.04$ with the symmetric decomposition, the Trotter error is already smaller than the statistical uncertainty and thus is negligible. Then panels (c) and (d) of Fig.~\ref{fig:Trotter} plot the results of $e/t$ and $m^2$ in a temperature range, from the simulations with (i) the symmetric decomposition with $\Delta\tau t=0.04$ and (ii) the asymmetric one with $\Delta\tau t\approx 0.138$. The simulation setup of (ii) is fairly close to the one used in Ref.~\onlinecite{Staudt2000}, whose results are apparently recovered by our simulations. Surprisingly, the corresponding relative errors induced by the Trotter error are about $3\%$ for $e/t$ and $14\%$ for $m^2$ for such a small system with the intermediate interaction $U/t=6$. These systematic bias is expected to even grow with the system size and interaction strength. Thus, the other relevant results obtained in Ref.~\onlinecite{Staudt2000} should also have the bias issue. For example, the authors computed the N\'{e}el transition temperatures via the extrapolation of the finite-size peak locations of the specific heat~\cite{Staudt2000}, which was estimated from the first-order derivative of total energy over temperature. 

\begin{figure}[h]
\centering
\includegraphics[width=1.00\columnwidth]{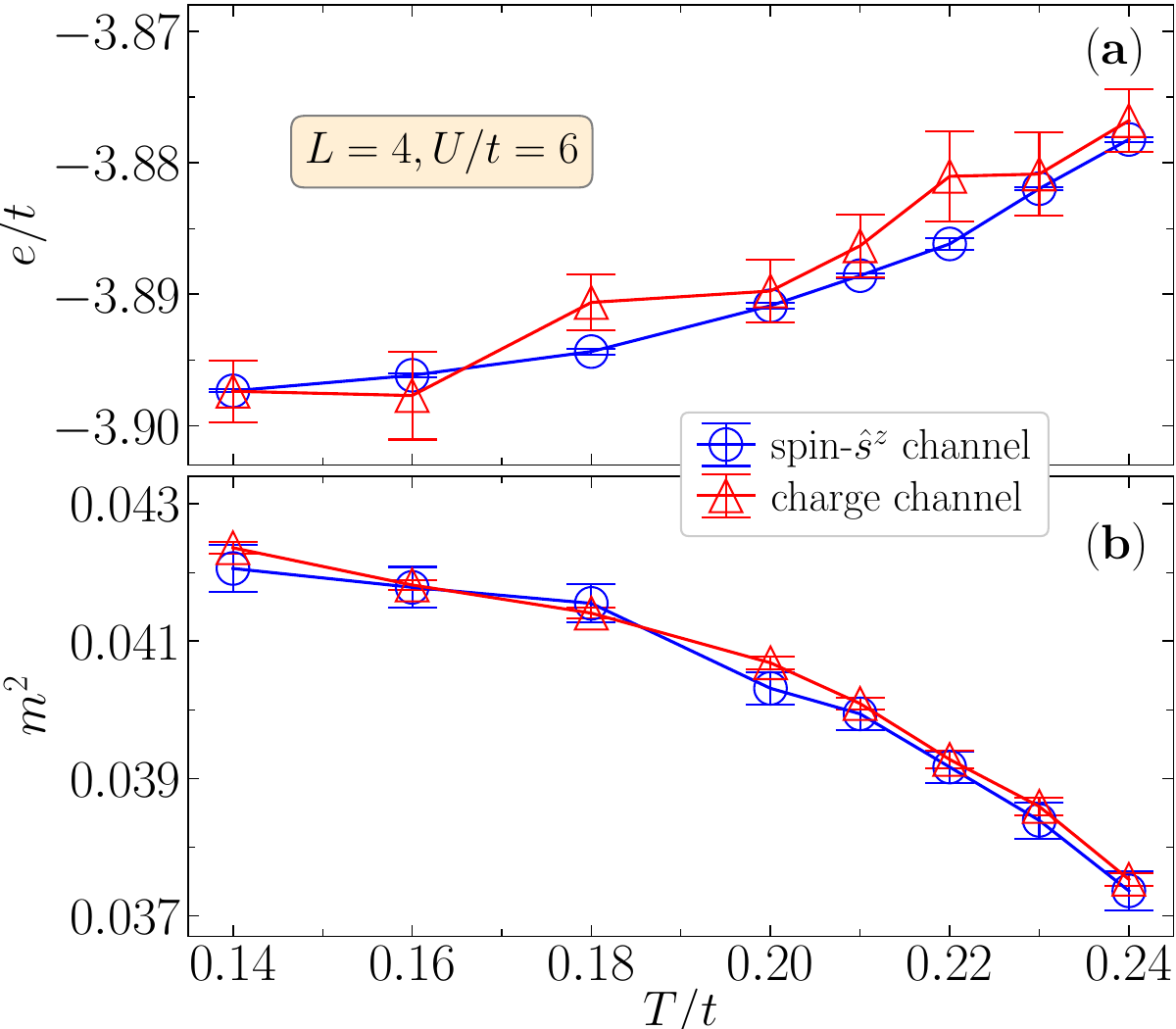}
\caption{Comparisons of AFQMC results of (a) total energy density $e/t$ and (b) mean-squared magnetization $m^2$ between the HS transformations into the spin-$\hat{s}^z$ channel and charge channel, for $L=4$ system with $U/t=6$ at half-filling. This suggests that the HS transformation in the charge channel is more suitable for computing magnetic observables and in the spin channel is more suitable for charge-related quantities. }
\label{fig:HSDecomp}
\end{figure}

Next we present the comparisons between the HS transformations into the spin-$\hat{s}^z$ channel in Eq.~(\ref{eq:SzHS}) and charge channel in Eq.~(\ref{eq:DensityHS}) in AFQMC simulations of 3D repulsive Hubbard model. The results of $e/t$ and $m^2$ are shown in Fig.~\ref{fig:HSDecomp}. First of all, the numerical results from different HS transformations are consistent within statistical uncertainties. In alignment with the discussion in Sec.~\ref{sec:AFQMC}, the HS transformation into spin-$\hat{s}^z$ channel indeed induces significantly bigger error bars than the charge channel transformation for the spin-related quantity $m^2$, due to the breaking of spin ${\rm SU}(2)$ symmetry. Oppositely, the energy results with the charge channel transformation has much larger statistical fluctuations than that with the formula into spin-$\hat{s}^z$ channel. This can be attributed to the breaking of the charge ${\rm SU}(2)$ symmetry~\cite{Cornelia2021} in the charge channel transformation in Eq.~(\ref{eq:DensityHS}), which similarly induces large fluctuations for density and pairing-related properties. The double occupancy belongs to density-related observables and thus suffers from the problem, and then contributes to the big error bars for both the interaction energy and the total energy. Practically, these symmetry-breaking effects tend to be more severe towards lower temperature, and with increasing system size and interaction strength~\cite{Scalettar1991}. All the $m^2$ results in Ref.~\onlinecite{Staudt2000} are rather noisy (see (b) and (d) panels of Fig.~\ref{fig:Trotter}), which indicates that only the HS transformation into spin-$\hat{s}^z$ channel was used. 

In our AFQMC simulations, we have performed systematic tests on the Trotter error with different parameters, and adopt gradually decreasing $\Delta\tau$ values for increasing $U/t$ to eliminate the Trotter error, i.e., from $\Delta\tau t=0.05$ for $U/t=4$ to $\Delta\tau t=0.02$ for $U/t=12$. Moreover, we combine the two different HS transformations to achieve high-precision results. We apply the HS transformation in the charge channel in Eq.~(\ref{eq:DensityHS}) to compute spin-related properties, and switch to the transformation in the spin-$\hat{s}^z$ channel in Eq.~(\ref{eq:SzHS}) for double occupancy and total energy. We also find dynamical single-particle properties are quite insensitive to the symmetry-breaking issue in HS transformations, and thus we simply use the spin-$\hat{s}^z$ transformation.

We have also implemented several efficient techniques in our AFQMC calculations. These include fast Fourier transform between the real and momentum spaces for the product of propagating matrices~\cite{Yuanyao2019,Yuanyao2019L} and the delayed version for the local update of auxiliary fields~\cite{McDaniel2017} to accelerate the simulation, and the $\tau$-line type of global update~\cite{Scalettar1991} to improve the efficiency of Monte Carlo sampling. These techniques allow us to perform the simulation for a large system with $20^3$ lattice sites using reasonable computational cost. Together with careful treatments of the Trotter error and HS transformations, our AFQMC simulations for 3D repulsive Hubbard model in this work simultaneously achieve the high-level realizations of the speed, precision, and efficiency.

\section{Phase diagram at half-filling}
\label{sec:diagram}

\begin{figure}[htp]
\centering
\includegraphics[width=0.99\linewidth]{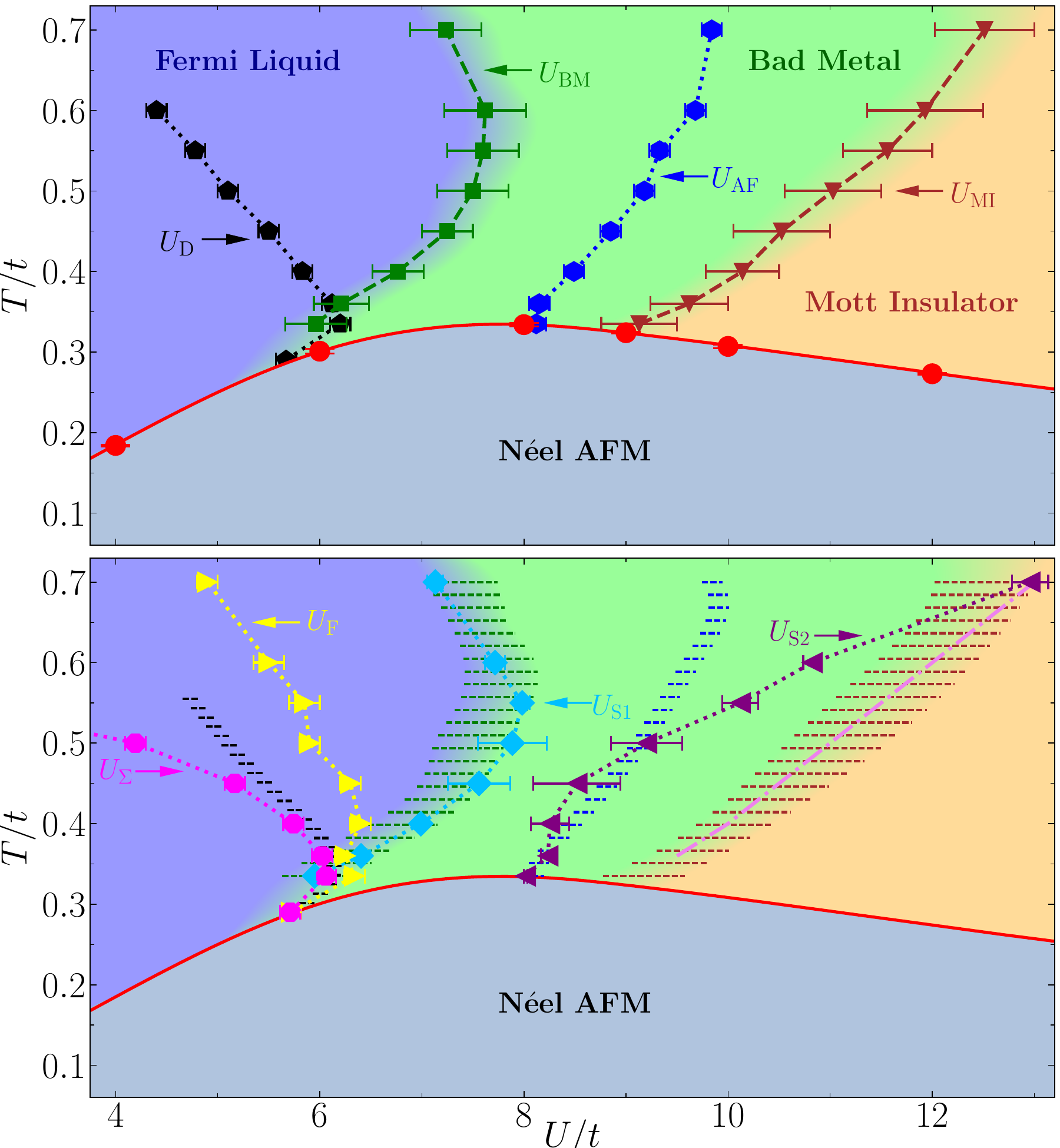}
\caption{Phase diagram of half-filled 3D repulsive Hubbard model, and the signatures of variously physical observables from our AFQMC calculations. (a) Shows the key ingredients of the phase diagram, including the N\'{e}el transitions and the metal-insulator crossover in the normal phase. Red circles and the corresponding interpolating connection (the solid red line) show the N\'{e}el transition temperature $T_{\rm N}$. The result of Heisenberg limit $T_{\rm H}=3.78t^2/U$~\cite{Sandvik1998} is also included (dashed red line). Above $T_{\rm N}$ and with increasing interaction, $U_{\rm BM}$ (green squares) as the onset of bad metal and $U_{\rm MI}$ (brown up triangles) as the entrance into Mott insulator, together characterize the crossover in the normal phase. Moreover, $U_{\rm AF}$ (blue hexagons) marks the peak locations of AFM structure factor, and $U_{\rm D}$ (black pentagons) tracks the most rapid suppression of double occupancy. In (b), the results of $U_{\rm BM}$, $U_{\rm MI}$, $U_{\rm AF}$ and $U_{\rm D}$ in (a) are plotted as horizontal dashed lines with the lengths indicating the error bars. Besides, the signatures from other physical observables are also presented, including the local minimum ($U_{\rm{S}1}$, light blue diamonds) and maximum ($U_{\rm{S}2}$, purple left triangles) of the thermal entropy, the peak location of fidelity susceptibility ($U_{\rm F}$, yellow right triangles), the crossing of the imaginary part of self-energy $\Im\Sigma(\mathbf{k}_F, i\omega_0)$ and $\Im\Sigma(\mathbf{k}_F, i\omega_1)$ ($U_{\mathrm{\Sigma}}$, magenta octagons), and the contour line of the charge compressibility with a very small threshold (pink dashed line, almost on top of $U_{\rm MI}$). More discussions about these results can be referred in the context of Sec.~\ref{sec:diagram}.}
\label{fig:PhaseDiagram}
\end{figure}

In this section, we concentrate on the complete $U$-$T$ phase diagram of 3D repulsive Hubbard model at half-filling from our numerical results, as shown in Fig.~\ref{fig:PhaseDiagram}. In AFQMC calculations, we cover a large range of the interaction strength with $0\le U/t\le 13$ and temperature with $T/t\le 0.70$. We mostly stay at far lower $T$ than the degeneracy temperature (namely, the Fermi temperature $T_F=6t$), above which the system can be taken as incoherent soup of fermions without long-lived quasiparticles. 

\begin{figure*}
\centering
\includegraphics[width=2.05\columnwidth]{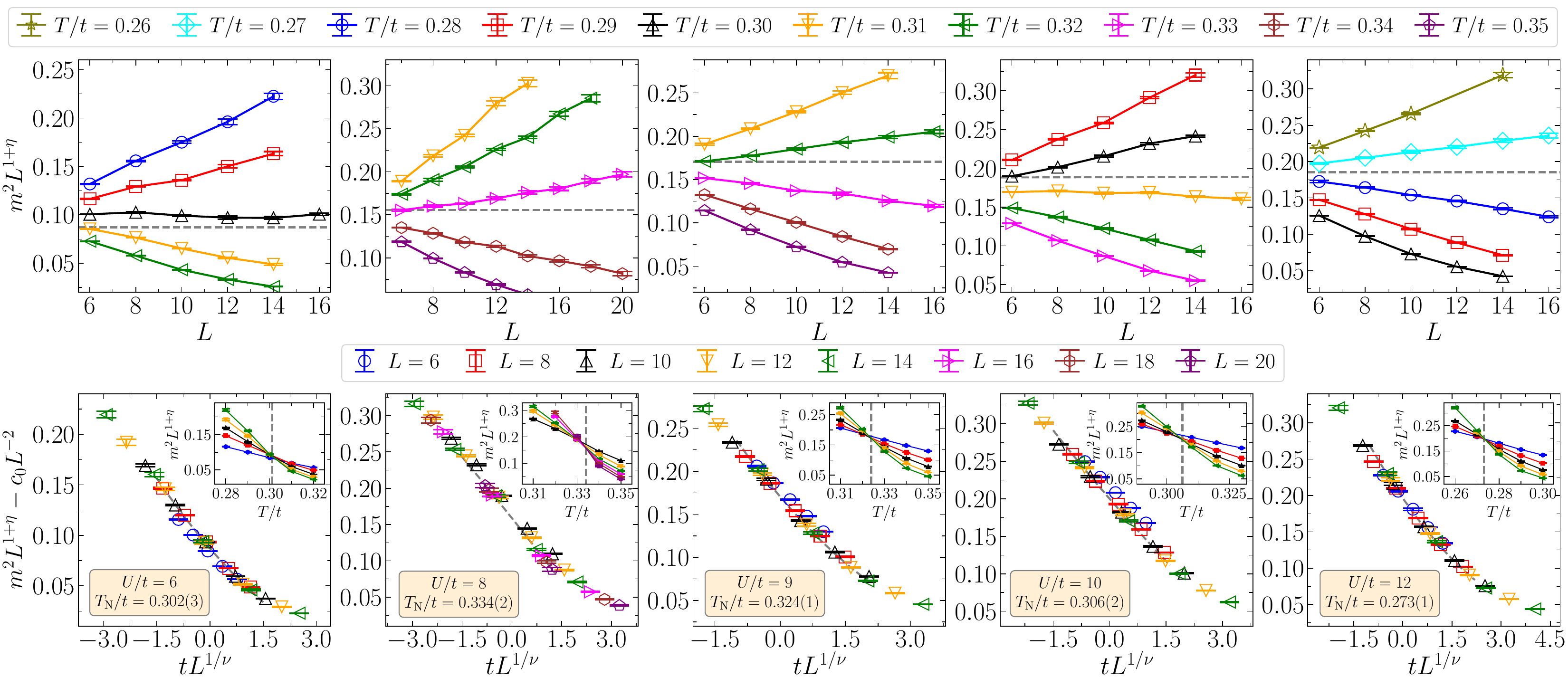}
\caption{Finite-size scalings for the mean-squared magnetization $m^2$ across the N\'{e}el transition in half-filled 3D repulsive Hubbard model, using the critical exponents $\eta=0.0375,\nu=0.7112$ from the 3D Heisenberg universality class~\cite{Campostrini2002}. The five panels in the upper row plot the rescaled quantity $m^2L^{1+\eta}$ versus linear system size $L$ at temperatures close to the transition, for the interaction strengths $U/t=6, 8, 9, 10$, and $12$. The corresponding panels in the lower row show the data collapse of $(m^2 L^{1+\eta}-c_0 L^{-2})$ versus $tL^{1/\nu}$ [with $t=(T-T_{\rm N})/T_{\rm N}$] based on the scaling formula in Eq.(\ref{eq:FSS}). The N\'{e}el temperatures $T_{\rm N}$ obtained from the fitting are also included. The insets replot $m^2 L^{1+\eta}$ versus temperature for different system sizes, with the crossings well consistent with the $T_{\rm N}$ results (vertical, gray dashed lines).}
\label{fig:NeelTemp}
\end{figure*}

Figure \ref{fig:PhaseDiagram}(a) illustrates the results of the N\'{e}el transition temperatures and the crossover boundaries in the normal phase. The latter is the focus of our companion paper~\cite{Yufeng2024}. We compute the N\'{e}el temperatures $T_{\rm N}$ for six representative interaction strengths ranging from $U/t=4$ to $U/t=12$, via the standard finite-size scaling for the mean-squared magnetization $m^2$. Above $T_{\rm N}$, the crossover from Fermi liquid to Mott insulator with increasing $U/t$ is prominent from the comparison of $U_{\rm BM}$ and $U_{\rm MI}$ as the onset of the bad metal and the entrance into Mott insulator, respectively. We note that some previous studies~\cite{Vu2015,Mousatov2019} use the term ``bad metal'' for the regime where the system exhibits linear resistivity versus temperature, which typically appears in higher temperature regime as $T/t>1.0$. Here, we adopt ``bad metal'' only to describe the metal-insulator crossover regime in which $A_{\rm loc}(\omega)$ has a dip around $\omega=0$ and $A_{\rm loc}(\omega=0)$ is finite~\cite{Deng2013,Ding2019}. We associate $U_{\rm BM}$ with the disappearance of the coherence peak of $A_{\rm loc}(\omega)$ around $\omega=0$, and determine $U_{\rm MI}$ from the quasiparticle weight accompanying the gap opening in $A_{\rm loc}(\omega)$. Close to the N\'{e}el transition around $(U/t=5.5,\ T/t=0.29)$, whether bad metal extends to even lower temperature or not is very hard to be resolved in AFQMC calculations and is also beyond the scope of our work. Thus, we only compute $U_{\rm BM}$ and $U_{\rm MI}$ for $T/t\ge 0.335$, which is close to the highest N\'{e}el temperature as $T_{\rm N}/t=0.334(2)$ achieved at $U/t=8$. 

Figure \ref{fig:PhaseDiagram}(b) together with the results of $U_{\rm D}$ and $U_{\rm AF}$ in Fig.~\ref{fig:PhaseDiagram}(a) summarizes the signatures of variously physical quantities from our numerical calculations with increasing $U/t$ at fixed temperatures. They include the maximum of AFM structure factor ($U_{\rm AF}$), the inflection point of double occupancy ($U_{\rm D}$), the local minimum and maximum of thermal entropy ($U_{\rm S1}$ and $U_{\rm S2}$), the peak location of fidelity susceptibility ($U_{\rm F}$), the crossing of the imaginary part of self-energy ($U_{\Sigma}$), and the vanishing charge compressibility. All these signal locations of $U$ are summarized in Appendix.~\ref{sec:AppendixB}. Some of these quantities have been commonly used to investigate the interaction-driven metal-insulator transition or crossover~\cite{Georges1996,Gebhard1997,Imada1998,Kotliar2004}. We pack all the important results of these quantities as well as the computational details in Sec.~\ref{sec:ThermQuantity}. 

In the following of this section, we show the calculations of the N\'{e}el transition temperatures in Sec.~\ref{sec:PmAfmTransition}, and offer more details on determining the crossover boundaries inside the normal phase in Sec.~\ref{sec:MICphysics}.

\subsection{Calculations of the N\'{e}el temperatures}
\label{sec:PmAfmTransition}

Considering the dimensionality and spin ${\rm SU}(2)$ symmetry, the N\'{e}el transition in half-filled 3D repulsive Hubbard model should belong to the 3D Heisenberg universality class~\cite{Campostrini2002,Staudt2000,Kozik2013}. We thus directly take the most accurate results to date of the critical exponents from Ref.~\onlinecite{Campostrini2002}, namely, the correlation length exponent $\nu=0.7112(5)$ and the anomalous dimension $\eta=0.0375(5)$, into the finite-size scaling analysis for the calculations of the N\'{e}el transition temperatures. 

We perform the standard finite-size scaling (FSS) calculations for the mean-squared magnetization $m^2$ based on the following formula~\cite{Binder1981}:
\begin{equation}\begin{aligned}
m^2 L^{1+\eta} = f(L/\xi) (1 + c_0 L^{-\omega} + \cdots),
\label{eq:FSS}
\end{aligned}\end{equation}
where the correlation length $\xi$ satisfies the scaling relation $\xi \propto |T-T_{\rm N}|^{-\nu}$ in the critical region, and $f(x)$ is the scaling invariant (universal) function with $f(x=0)$ as a finite constant. The correction terms [the the ellipsis in parentheses of Eq.~(\ref{eq:FSS})] like $c_0 L^{-\omega}$ (typically with $\omega>0$) arise from leading irrelevant operators, and account for the subleading finite-size effect around the critical point. Then Eq.~(\ref{eq:FSS}) indicates that $m^2 L^{1+\eta}$ should converge to the constant $f(x=0)$ with increasing $L$. Moreover, the temperature exhibiting monotonically increasing $m^2 L^{1+\eta}$ with increasing $L$ belongs to the AFM ordered phase, while $m^2 L^{1+\eta}$ decreases with growing $L$ in the normal phase. This can be used as an efficient way to pin down the transition temperature to a certain range. Correspondingly, we plot $m^2 L^{1+\eta}$ versus $L$ at temperatures close to the transition for $U/t=6,8,9,10,12$ in the upper row of Fig.~\ref{fig:NeelTemp}. These results clearly narrow the transition temperature $T_{\rm N}$ for all calculated $U/t$ down to a rather small temperature interval of $\Delta T=0.01t$.

To further achieve the high-precision results of $T_{\rm N}$ relies on the data collapse of $m^2 L^{1+\eta}$, for which we perform the least-square fitting by
\begin{equation}\begin{aligned}
\label{eq3:FSS}
m^2 L^{1+\eta} = \sum\limits_{k=0}^2 a_k \big[(T-T_{\rm N})L^{1/\nu}\big]^k + c_0 L^{-\omega},
\end{aligned}\end{equation}
The exponent $\omega$ is set to be 2 for simplicity (and the difference in $T_{\rm N}$ obtained with different $\omega$ is negligible considering the uncertainty). In the lower row of Fig.~\ref{fig:NeelTemp}, we demonstrate the data collapse results of $(m^2 L^{1+\eta}-c_0 L^{-2})$ versus $tL^{1/\nu}$ [with $t=(T-T_{\rm N})/T_{\rm N}$] for the corresponding $U/t$ in Figs.~\ref{fig:NeelTemp}(a)-\ref{fig:NeelTemp}(e). The results of $T_{\rm N}$ with the error bar obtained from the fitting are also presented in the plots, which are accurate to the third digits and all have relative errors smaller than $1\%$. These $T_{\rm N}$ results are furthermore verified by the finite-size crossings of $m^2 L^{1+\eta}$ versus temperature as shown in the insets. 

\begin{figure}[htp]
\centering
\includegraphics[width=0.95\columnwidth]{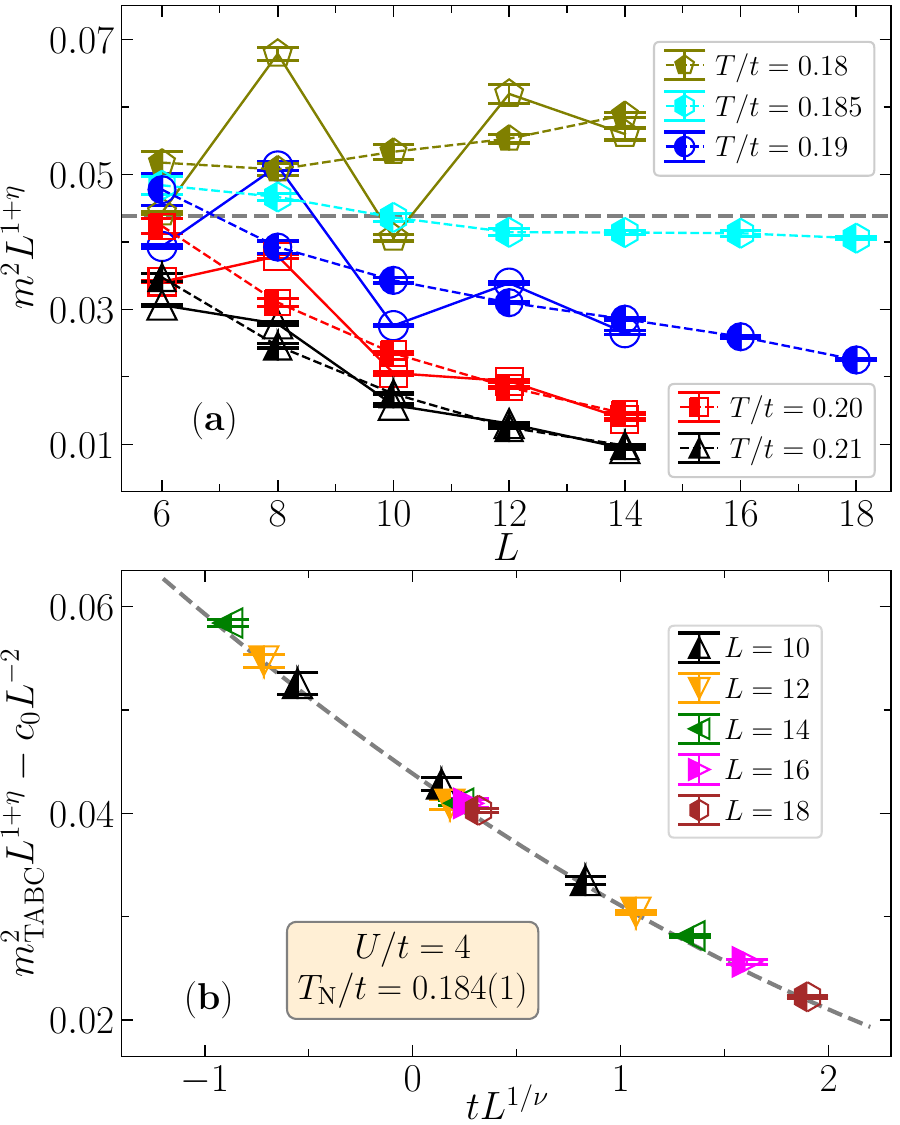}
\caption{Finite-size scaling for the mean-squared magnetization $m^2$ for $U/t=4$. (a) Plots the rescaled quantity $m^2 L^{1+\eta}$ versus $L$ at temperatures close to $T_{\rm N}$, with both PBC (open symbols) and TABC (left half-filled symbols) calculations. It is shown that, with PBC, it is difficult to analyze the data since the size dependence is oscillating. (b) Shows the data collapse of TABC results of $m^2$ as $(m_{\rm TABC}^2 L^{1+\eta}-c_0L^{-2})$ versus $tL^{1/\nu}$ with the dashed line as the universal function. The critical temperature from the fitting procedure is $T_{\rm N}/t=0.184(1)$.}
\label{fig:U4Neel}
\end{figure}

The above finite-size scaling procedure works well for $U/t\ge 6$. However, in the weakly interacting regime, we observe severe finite-size effect in $m^2$. A representative example with $U/t=4$ is shown in Fig.~\ref{fig:U4Neel}(a) for $m^2 L^{1+\eta}$ versus $L$. The AFQMC results from PBC calculations (the open symbols) clearly show strong oscillations even persisting to $L=14$. Such behavior originates from the single-particle finite-size effect of the noninteracting system, and it becomes invisible when the interaction term dominates. There are typically two techniques to deal with this issue. First, a finite-size correction with a reference system can be applied to remove the oscillations and accelerate the convergence towards the thermodynamic limit (TDL). However, we find that this method somehow fails for the system we study, with the $U=0$ or mean-field model (we find that it is hard to find an appropriate effective interaction) as the reference system. Therefore, we turn to the second technique applying twist averaged boundary conditions (TABC), which has been widely used in finite-size calculations~\cite{Lin2001,Qin2016b,Vitali2016,Haoxu2024}. 

Under the twisted boundary conditions, fermions pick up a phase when they wrap around the periodic boundaries as
\begin{equation}\begin{aligned}
\Psi(\cdots, \bm{\mathrm{r}}_j + L\hat{\mathbf{e}}_{\alpha},\cdots) = e^{i\Theta_{\alpha}}\Psi(\cdots, \bm{\mathrm{r}}_j,\cdots),
\end{aligned}\end{equation}
where $\hat{\mathbf{e}}_{\alpha}$ (with $\alpha=x,y,z$) is the unit vector along $\alpha$ direction and the twist angle $\Theta_{\alpha}$ satisfies $-\pi\le\Theta_{\alpha}<\pi$. To keep the translational symmetry in practical calculations, we adopt the gauge of evenly distributing the phase $e^{i\Theta_{\alpha}}$ into all the bonds along $\alpha$ direction and alter the hopping strength $t$ to $te^{i\Theta_{\alpha}/L}$. Note that the spin up and down sectors need to have the same phase to avoid the sign problem. To perform the TABC calculations for every finite-size system, we choose a group (with the number $N_{\Theta}$) of quasi-random $(\Theta_x,\Theta_y,\Theta_z)$~\cite{Qin2016b} and carry out the AFQMC simulation for each separately. We find that $N_{\Theta}$$\sim$$10$ is enough to achieve good statistics. Then the results of $m^2 L^{1+\eta}$ versus $L$ from the TABC calculations (the left half-filled symbols) for $U/t=4$ shown in Fig.~\ref{fig:U4Neel}(a) clearly exhibit monotonic dependence on $L$, indicating the removal of the single-particle finite-size effect. The further data collapse for the TABC results for $U/t=4$ is plotted in Fig.~\ref{fig:U4Neel}(b), presenting the critical temperature as $T_{\rm N}/t=0.184(1)$. 

The N\'{e}el temperatures from our AFQMC calculations are quantitatively comparable to the results presented in other studies~\cite{Staudt2000,Fanjie2024,Kozik2013,Garioud2024}. Especially, within error bars, our results are well consistent with another independent AFQMC study parallel to our work, which extracts $T_{\rm N}$ from the finite-size scaling of the correlation ratio~\cite{Fanjie2024}. However, the most recent DiagMC simulation~\cite{Garioud2024} has obtained slightly higher $T_{\rm N}$ than our results, i.e., $T_{\rm N}(U/t=4)/t=0.1925(25)$ and $T_{\rm N}(U/t=6)/t=0.32(1)$. This overestimate might be caused by the small but finite symmetry-breaking pinning field (which promotes the AFM ordering) applied in Ref.~\onlinecite{Garioud2024} in order to achieve precise results in DiagMC calculations for the magnetically ordered phase. With increasing $U/t$, our $T_{\rm N}$ results also evolve towards the critical temperature $T_{\rm H}/J=0.946(1)$~\cite{Sandvik1998} of effective spin-$1/2$ antiferromagnetic Heisenberg model with $J=4t^2/U$~\cite{MacDonald1988} in the strong interaction limit. The corresponding line of $T_{\rm H}=3.784(4)t^2/U$ in the $U$-$T$ plane is clearly traced by the trend of the interpolating connection [the solid red line in Fig.~\ref{fig:PhaseDiagram}(a)] of our $T_{\rm N}$ results approaching $U/t=\infty$. 

\begin{figure*}
\centering
\includegraphics[width=1.80\columnwidth]{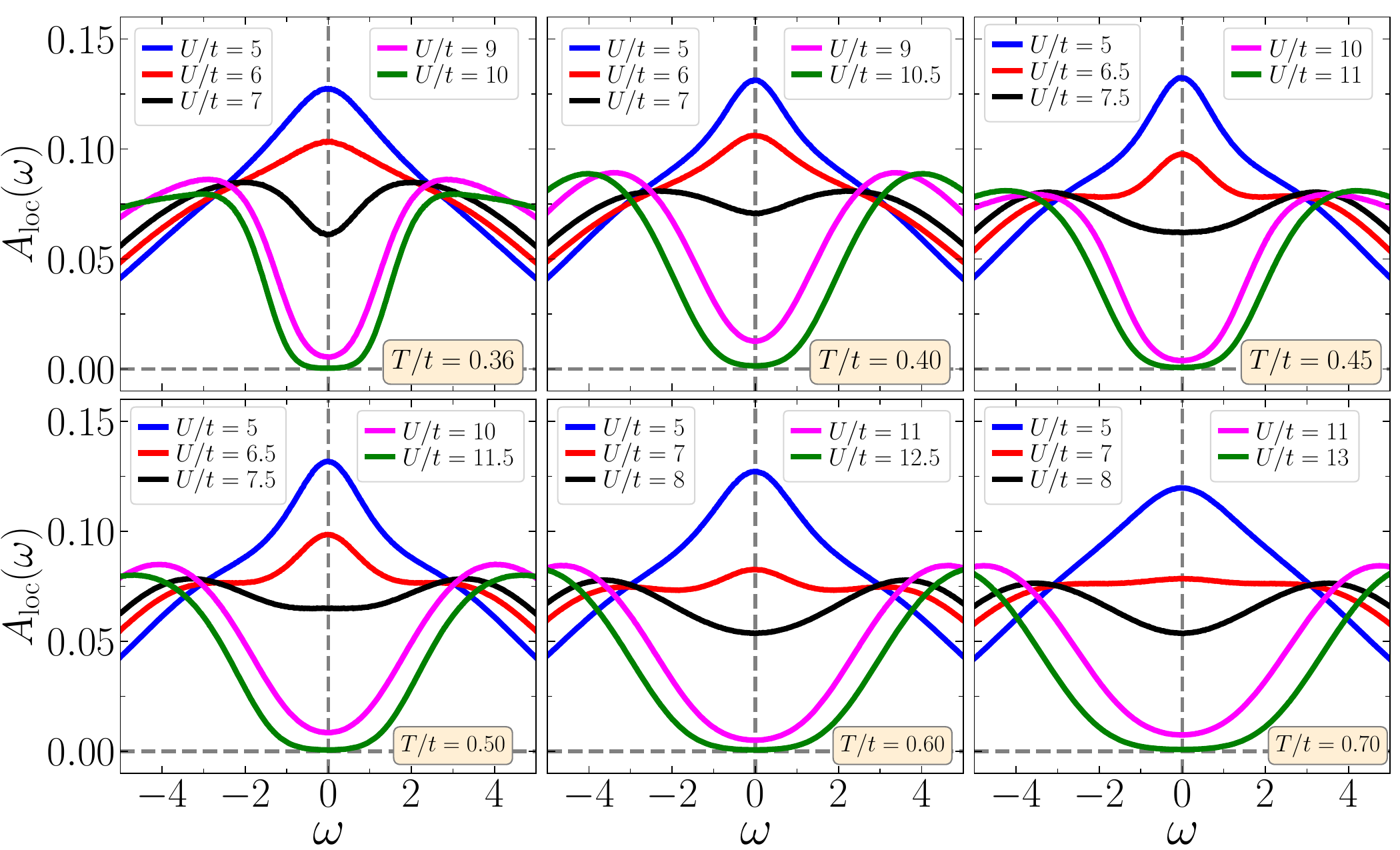}
\caption{Local single-particle spectrum $A_{\mathrm{loc}}(\omega)$ as a function of $\omega$ with different interaction strengths, at six temperatures. Note $A_{\mathrm{loc}}(\omega)$ is symmetric about $\omega=0$. For each temperature, the plotted interactions are specially chosen that the coherence peak disappears or the spectrum at $\omega=0$ approaches zero. These results are from $L=12$ system for $T/t=0.36$ and $L=8$ system for other temperatures, and the residual finite-size effects are negligible. }
\label{fig:Spectrum}
\end{figure*}

There are alternative ways to determine the transition temperatures. One example is extrapolating the peak location of the specific heat in finite-size systems to TDL, which was actually used by Staudt {\it et al.}~\cite{Staudt2000} for half-filled 3D repulsive Hubbard model. The foundation for this method to work is that the specific heat $C_v$ has the scaling form $C_v\propto|T-T_{\rm N}|^{-\alpha}$ in critical region, and at TDL it diverges at exactly the transition temperature $T_{\rm N}$. This requires that the critical exponent $\alpha$ must be non-negative. However, for the 3D Heisenberg universality class, $\alpha$ is negative as $\alpha=-0.1336(15)$~\cite{Campostrini2002}, which means $C_v$ does not diverge approaching $L=\infty$. As a result, whether the peak location of $C_v$ coincides with the critical temperature or not is uncertain and might depends on the transition. A negative example is the Berezinskii-Kosterlitz-Thouless transition in 2D $^4$He~\cite{Ceperley1989} and 2D XY model~\cite{Nguyen2021}, for which unbiased Monte Carlo calculations of large scale revealed peak locations of the non-divergent specific heat at about $1.6$ and $1.17$ times of the transition temperature, respectively. Thus, whether this method to compute $T_{\rm N}$ is valid for 3D Hubbard model still needs to be clarified with more careful and thorough study. 

\subsection{Metal-insulator crossover in the normal phase}
\label{sec:MICphysics}

The existence of N\'{e}el AFM long-range order prohibits the possible Mott metal-insulator transition~\cite{Georges1996,Gebhard1997,Imada1998} at low temperature in half-filled 3D Hubbard model. Alternatively, with increasing $U/t$ in the normal phase above $T_{\rm N}$, the system should evolve from Fermi liquid state in the weakly interacting regime to Mott insulator with strong interaction. As shown in Fig.~\ref{fig:PhaseDiagram}, our AFQMC calculations have identified a rather extended crossover regime (bad metal) in-between without any singularity in all physical observables~\cite{Yufeng2024}. 

Similar metal-insulator crossover (MIC) behaviors have been also revealed in previous studies for 2D Hubbard models~\cite{Ohashi2008,Park2008,Vuifmmode2013,Walsh2019,*Walsh2019L,Downey2023,Svistunov2020,Kim2021}. The ones implementing cluster DMFT (or dynamical cluster approximation, DCA)~\cite{Ohashi2008,Park2008,Vuifmmode2013,Walsh2019,*Walsh2019L,Downey2023} associate the crossover with the first-order transition ending at a critical point in low-temperature regime from their calculations. However, the existence of Mott transition in these 2D systems is still controversial, considering the systematic approximation in DMFT method and its major focus on paramagnetic solutions~\cite{Ohashi2008,Park2008,Vuifmmode2013,Walsh2019,*Walsh2019L,Downey2023}. Similarly, the results about the crossover behaviors in these DMFT calculations might be less reliable and demand careful verifications from unbiased calculations. As a comparison, for 2D half-filled Hubbard model (on square lattice), the DiagMC studies at finite temperatures~\cite{Svistunov2020,Kim2021} have calculated a bunch of different observables (without the fermion spectrum) and obtained quite diverse signals for the metal-insulator crossover. These results tend to complicate the crossover physics of the system with increasing interaction~\cite{Kim2021}. Thus, the quantitative characterization and precise determination of MIC still remain as significant challenges for numerics, though the system is a paramagnet in the crossover. 

We instead characterize the MIC in the normal phase of 3D half-filled Hubbard model using a self-contained scheme. We first resort to the fingerprint signatures for Fermi liquid and Mott insulator from local fermion spectrum $A_{\rm loc}(\omega)$, from which we determine the boundaries of the MIC and thus obtain the range of bad metal as the crossover regime. We further verify the boundaries from signatures of thermal entropy, the quasiparticle weight, and charge compressibility. We understand that the numerical analytical continuation used to compute $A_{\rm loc}(\omega)$ is an ill-posed problem, but the SAC method~\cite{Sandvik2016,Shao2023} we use has been proved in many other studies (see references in Ref.~\onlinecite{Shao2023}) to be one of the most trustworthy techniques to perform such calculations. This method allows us to carry out quantitative analysis for $A_{\rm loc}(\omega)$ results. We also need to clarify that, there is no phase transition in the MIC, and therefore we present the boundaries of the three regimes with reasonable uncertainties instead of sharp dividing lines. In the following subsections, we concentrate on the details in our AFQMC calculations to obtain the MIC boundaries. 

\subsubsection{Fermi liquid to bad metal}
\label{sec:FLtoBM}

The most prominent feature for Fermi liquid state is the coherence peak around Fermi energy in the local single-particle spectrum $A_{\rm loc}(\omega)$. Specifically for half-filled 3D Hubbard model, $A_{\rm loc}(\omega)$ is symmetric about $\omega=0$. Thus, at fixed temperature with increasing $U/t$, we take the disappearance of the peak and subsequent development of a dip at $\omega=0$ in $A_{\rm loc}(\omega)$ as stepping out of the Fermi liquid as well as the onset of bad metal state (as $U_{\rm BM}$), i.e., the boundary between these two regimes. Under this convention, $A_{\rm loc}(\omega)$ in bad metal should possess a local minimum at $\omega=0$ but with a finite value. 

In Fig.~\ref{fig:Spectrum}, we plot the $A_{\rm loc}(\omega)$ results for different temperatures from $T/t=0.36$ to $T/t=0.70$, each with carefully chosen interactions to emphasize the disappearance of the coherence peak and $A_{\rm loc}(\omega=0)$ approaching zero. We have verified that the $A_{\rm loc}(\omega)$ and $Z_{k_F}$ results already have negligible finite-size effect with $L=12$ for $T/t=0.36$ and $L=8$ for $T/t\ge 0.40$ (see more results in Appendix~\ref{sec:AppendixE}). The $A_{\rm loc}(\omega)$ results in Fig.~\ref{fig:Spectrum} actually contain quite a lot of information about the system. First, the monotonic suppression of $A_{\rm loc}$ around $\omega=0$ with increasing $U/t$ clearly signifies the electron localization tendency. Meanwhile, the spectral weight around $\omega=0$ is transferred to high energy part, which forms two shoulders at $\pm \omega_0$ in align with the appearance of the dip at $\omega=0$. These local maxima around $\pm \omega_0$ correspond to the upper and lower Hubbard bands~\cite{Hubbard1963}, which reside at $\omega=\pm U/2$ in the single-site limit. Our results apparently demonstrate $\omega_0 < U/2$, which is probably due to the quantum fluctuations induced by the hopping term. Second, we can observe a less obvious three-peak structure at $\omega=0$ and $\pm\omega_0$ in our results of $A_{\rm loc}(\omega)$, for $T/t\ge0.45$ with a specific range of $U/t$. For example, for $T/t=0.45$, $A_{\rm loc}(\omega)$ first develops shadow minimums between $\omega=0$ and $\pm\omega_0$ around $U/t=6.5$ and then the coherence peak disappears at around $U/t=7.5$. And we still take this intermediate region of $6.5\le U/t\le7.5$ as the Fermi liquid state. Such three-peak feature has been taken as a hallmark for the Mott metal-insulator transition especially from the aspect of DMFT calculations~\cite{Rozenberg1993,Georges1996,Gebhard1997,Imada1998,Kotliar2004}. 

The bad metal state in our work should also have differences with the well-known pseudogap phenomena in variously correlated fermion systems~\cite{Wuwei2018,Patrick2006,Fischer2007,Boschini2020,Bauer2014,Mueller2017,XiangLi2024}, whose definition also involves the appearance of a dip around Fermi energy in the fermionic spectrum. First, most of existing literatures define the pseudogap behavior from the momentum-space single-particle spectral function on Fermi surface~\cite{Patrick2006,Fischer2007,Mueller2017}, and it usually shows momentum anisotropy in the 2D Hubbard model and cuprates~\cite{Wuwei2018,Patrick2006,Fischer2007,Boschini2020}. Here we define bad metal from local fermion spectrum, and the Fermi surface of half-filled 3D Hubbard model is quite isotropic in momentum-resolved observables (see more details in Sec.~\ref{sec:BMtoMI} and Appendix~\ref{sec:AppendixD}). Second, the pseudogap behavior is mostly referred as a phenomenon appearing with lowering temperature~\cite{Wuwei2018,Patrick2006,Fischer2007,Boschini2020,Bauer2014,Mueller2017,XiangLi2024}. However, we probe the bad metal at fixed temperature with increasing $U$. Moreover, our results of the phase diagram in Fig.~\ref{fig:PhaseDiagram} explicitly shows that the bad metal regime should take a larger portion of the $U$-$T$ plane with $T/t>0.7$, and there is no well-defined pseudogap behavior along the temperature axis for $U/t\ge 8$. Thus, we prefer not to use the term ``pseudogap'' for the intermediate crossover regime. 

\begin{figure}[h]
\centering
\includegraphics[width=0.97\columnwidth]{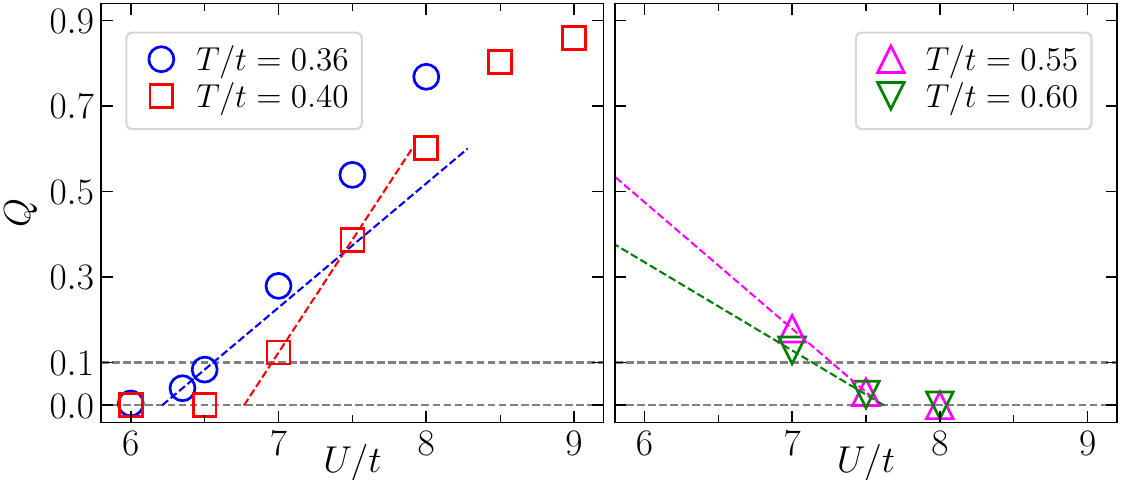}
\caption{The spectrum ratio $Q$ and its extrapolations versus $U/t$ for various temperatures. (a) and (b) represent the typical results for $T/t\le0.40$ and $T/t\ge0.45$. The extrapolations using linear fitting for $Q$ close to $Q=0$ are plotted as dashed lines. The corresponding extrapolated interaction for $Q=0$ is taken as $U_{\rm BM}$. The horizontal gray dashed lines marking $Q=0.1$ are used for estimating the uncertainty of $U_{\rm BM}$. The system sizes are the same as Fig.~\ref{fig:Spectrum}.}
\label{fig:CrossoverRatio}
\end{figure}

We then turn to the calculation of $U_{\rm BM}$ and its uncertainty based on the results of $A_{\rm loc}(\omega)$ presented in Fig.~\ref{fig:Spectrum}. To incorporate the two qualitatively different behaviors discussed above for $T/t\le0.40$ and $T/t\ge0.45$, we define the spectrum ratio $Q=|A_m-A_{\rm loc}(\omega=0)|/A_m$, where $A_m=\max_{0\le\omega<\infty}A_{\rm loc}(\omega)$ for $T/t\le0.40$ and $A_m=\min_{0\le\omega\le\omega_0}A_{\rm loc}(\omega)$ for $T/t\ge0.45$ (with $\pm\omega_0$ as the shoulder positions). Then the Fermi liquid regime for $T/t\le0.40$ has $A_m=A_{\rm loc}(\omega=0)$ and $Q=0$, while the bad metal regime acquires $Q>0$ with $A_m=A_{\rm loc}(\omega=\omega_0)$. Oppositely, for $T/t\ge0.45$, the bad metal regime has $Q=0$ while the Fermi liquid picks up a finite $Q$. Then we extrapolate $Q$ to zero, and take the corresponding $U$ for $Q=0$ as $U_{\rm BM}$. The examples of such extrapolation are shown in Fig.~\ref{fig:CrossoverRatio}. Then we further estimate the uncertainty as $\Delta U=|U^{\prime}-U_{\rm BM}|$ with $U^{\prime}$ as the interaction strength satisfying $Q=0.1$ (gray dashed lines in Fig.~\ref{fig:CrossoverRatio}), i.e., $10\%$ of the drop or rise of $A_{\rm loc}(\omega=0)$ at $U=U_{\rm BM}$. We take such criteria to account for the possible error in SAC calculations. A similar method was used in Ref.~\onlinecite{Bauer2014} to exact the pseudogap temperature for 2D interacting Fermi gas. The above procedure is repeated for every fixed temperature, and presents the final $U_{\rm BM}$ results (green squares) shown in Fig.~\ref{fig:PhaseDiagram}(a). In Sec.~\ref{sec:EntDoucMax}, we will further demonstrate that these $U_{\rm BM}$ results are well consistent with the positions of local maximum in thermal entropy.

\subsubsection{Bad metal to Mott insulator}
\label{sec:BMtoMI}

Continuing to increase the interaction, the system further evolves from the bad metal state into a Mott insulator. The characteristic feature is $A_{\rm loc}(\omega=0)$ decaying to zero, which also signifies the opening of a finite-temperature single-particle gap. Nevertheless, applying $A_{\rm loc}(\omega=0)=0$ as the criteria to determine the boundary $U_{\rm MI}$ between these two states can suffer the ambiguity of zero value for $A_{\rm loc}(\omega=0)$, which typically acquires a tiny but finite number induced by the accuracy of SAC calculations. Thus, we first compute $U_{\rm MI}$ from the quasiparticle weight $Z_{k_F}$, and then estimate its uncertainty combining the $A_{\rm loc}(\omega)$ results around $\omega=0$. 

According to Eq.~(\ref{eq:QuasiZkf}), the calculation of $Z_{k_F}$ needs the self-energy $\Sigma_{\sigma}(\mathbf{k},i\omega)$ ($\sigma$ as spin index), which is typically computed via the Dyson equation
\begin{equation}
\label{eq:EqSlfEng}
\Sigma_{\sigma}(\mathbf{k},i\omega_n) = G_{0,\sigma}^{-1}(\mathbf{k},i\omega_n) - G_{\sigma}^{-1}(\mathbf{k},i\omega_n),
\end{equation}
where $G_{0,\sigma}(\mathbf{k},i\omega_n)=[i\omega_n-(\varepsilon_{\mathbf{k}}+\mu)]^{-1}$ is the noninteracting single-particle Green's function, and its interacting correspondence is evaluated by the Fourier transform as $G_{\sigma}(\mathbf{k},i\omega_n)=\int_{0}^{\beta}G_{\sigma}(\mathbf{k},\tau)e^{i\omega_n\tau}d\tau$ with $G_{\sigma}(\mathbf{k},\tau)$ directly measured in AFQMC calculations. Regarding the choice of Fermi vector $\mathbf{k}_F$ in finite-size systems, we observe that the momentum-resolved properties in the half-filled 3D Hubbard model have rather small differences between independent $\mathbf{k}_F$ points (which can not be connected by symmetries) even for the weakly interacting regime. Such Fermi surface isotropy is more prominent with increasing interaction. This is quite different from the 2D case, for which the nodal and antinodal points can even show qualitatively different behaviors~\cite{Svistunov2020,Thomas2021}. The anisotropy originates from the fact that the antinodal points are the saddle points in kinetic energy dispersion and contributes to the van Hove singularity (vHs) with divergent density of states. Accordingly, such vHs does not exist in 3D and thus explains the Fermi surface isotropy. We have checked the AFQMC results of $G_{\sigma}(\mathbf{k}_F,\tau)$ and $Z_{k_F}$ for all the independent $\mathbf{k}_F$ points, and find that they are well consistent considering the error bars (see Appendix~\ref{sec:AppendixD}). Therefore, we take the averaged $Z_{k_F}$ over all $\mathbf{k}_F$ vectors as $\bar{Z}_{k_F}$, with an additional average of spin-up and -down sectors.

\begin{figure}[t]
\centering
\includegraphics[width=0.97\columnwidth]{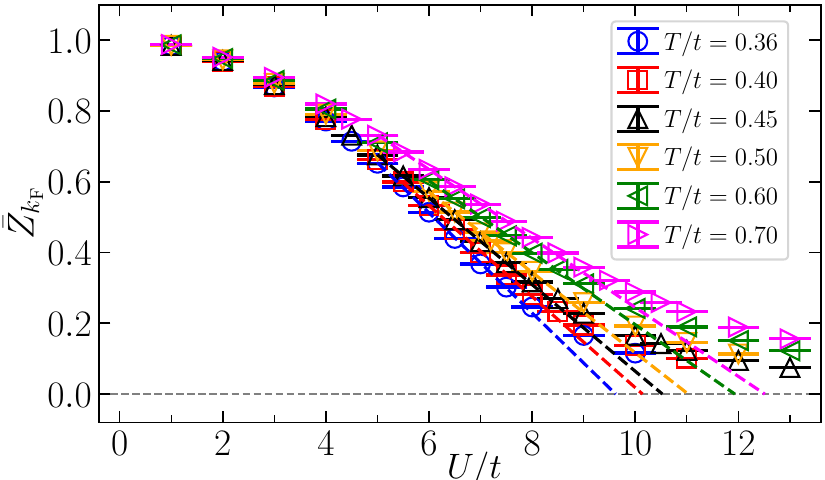}
\caption{The averaged quasiparticle weight $\bar{Z}_{k_F}$ versus $U/t$ at temperatures from $T/t=0.36$ to $T/t=0.70$. Linear fits (dashed lines with the same color as AFQMC results) are performed at intermediate interactions where $\bar{Z}_{k_F}$ is approximately linear. Then the extrapolated $U$ for $\bar{Z}_{k_F}=0$ is taken as $U_{\rm MI}$, the entrance into the Mott insulator. The system sizes are the same as Fig.~\ref{fig:Spectrum}.}
\label{fig:Quasiparwght}
\end{figure}

The results of $\bar{Z}_{k_F}$ versus interaction strength for different temperatures are shown in Fig.~\ref{fig:Quasiparwght}. The smooth suppression of $\bar{Z}_{k_F}$ with increasing $U$ reflects the MIC in the system. Along with the entrance into Mott insulator, $\bar{Z}_{k_F}$ should decay to zero at ground state. At finite temperatures, it is instead rounded off to a finite number~\cite{Liebsch2003}, as shown in Fig.~\ref{fig:Quasiparwght}. Hence we perform a linear fitting for $\bar{Z}_{k_F}$ with intermediate interactions, and take the extrapolated $U$ for $\bar{Z}_{k_F}=0$ as $U_{\rm MI}$. The same technique was used in Ref.~\onlinecite{Liebsch2003} to pinpoint the Mott transition. We then estimate the uncertainty of $U_{\rm MI}$ as $\Delta U=|U^{\prime\prime}-U_{\rm MI}|$ with $U^{\prime\prime}$ as the interaction strength rendering $A_{\rm loc}(\omega=0)<\epsilon$ (with $\epsilon\sim 10^{-3}$ as the threshold). This calculation procedure produces the final results of $U_{\rm MI}$ with error bars for $0.335\le T/t\le 0.70$ (brown up triangles) as shown in Fig.~\ref{fig:PhaseDiagram}(a). These $U_{\rm MI}$ results are further confirmed by the vanishing charge compressibility [the pink dashed line in Fig.~\ref{fig:PhaseDiagram}(b)], indicating Mott insulating state. 

The finite-temperature Mott insulator state we have identified in the phase diagram for $U/t>9$ can be taken as the result of the interplay between the temperature energy scale $\sim k_{\mathrm{B}} T$ and ground-state single-particle gap $\Delta_{sp}$, with the latter overtaking the former. The half-filled 3D Hubbard model should be fully gapped at $T=0$ for arbitrary $U$ due to the AFM long-range order. Then upon heating, the thermal fluctuation transfers the spectral weight above the gap in $A_{\rm loc}(\omega)$ to the in-gap region, and completely fill the gap as $A_{\rm loc}(\omega=0)>0$ at the corresponding temperature (as $T_{\rm MI}$) for every $U_{\rm MI}$ point. Moreover, $\Delta_{sp}$ should be asymptotically proportional to $U$ in strongly interacting regime, which suggests the linear dependence of $U_{\rm MI}$ on temperature. This conforms with our numerical results. A linear fitting for $T_{\rm MI}$ versus $U_{\rm MI}$ shows $T_{\rm MI}$$\sim$$0.11U$ for $U/t\ge10$ as the onset of Mott insulator state with lowering temperature. 

\section{Thermodynamic quantities at half-filling}
\label{sec:ThermQuantity}

Thermodynamic properties can offer more insights into the interplay between quantum and thermal fluctuations in the model, and contribute to more comprehensive understanding of its phase diagram with varying temperature and interaction strength. In this section, we continue to present detailed results with discussions of thermodynamic quantities for the half-filled 3D repulsive Hubbard model. 

Based on the underlying physics, we divide the numerical results into three parts. The first contains those quantities which further enrich and refine the phase diagram, involving the AFM spin correlations and thermal entropy. The second part deals with common thermodynamics along the temperature axis, such as double occupancy, charge compressibility, and specific heat. In the third part, we demonstrate that specific signatures applied in existing studies~\cite{Parcollet2004,WangLei2015,Svistunov2020} fail to characterize the MIC in the normal phase, including the inflection point of double occupancy, the peak of fidelity susceptibility, and the self-energy crossing. We also present important calculation details for these quantities in AFQMC simulations, especially including our method for computing the entropy versus $U$.

\subsection{AFM spin correlations}
\label{sec:AFMSpinCrft}

In the normal phase of the model, the thermal fluctuation destroys the AFM long-range order, and AFM spin correlation with a finite correlation length appears. Its magnetic properties can be characterized by AFM structure factor, short-range spin correlation, and the AFM correlation length. While the last two describe the short-range and long-range behaviors of spin correlations, respectively, in the system, AFM structure factor $S_{\rm AFM}^{zz}$ defined from Eq.~(\ref{eq:Safmzz}) is instead a balanced measurement combining correlations with all distances. Moreover, for 3D repulsive Hubbard model, $S_{\rm AFM}^{zz}$ can be directly measured in the optical lattice experiments~\cite{Hart2015,Shao2024}. For the short-range spin correlation, here we focus on the NN component computed as $|C_{\mathrm{NN}}|=|\sum_{\bm{\delta}} C(\bm{\delta})/N_{\bm{\delta}}|$ [with $C(\bm{\delta})$ defined in Eq.~(\ref{eq:RspSpinCrFt})], where $\bm{\delta}$ is the NN lattice vectors and $N_{\bm{\delta}}=6$ is coordination number. The AFM correlation length $\xi_{\rm AFM}$ is evaluated using Eq.~(\ref{eq:CorrLength}).

\begin{figure}[t]
\centering
\includegraphics[width=0.97\columnwidth]{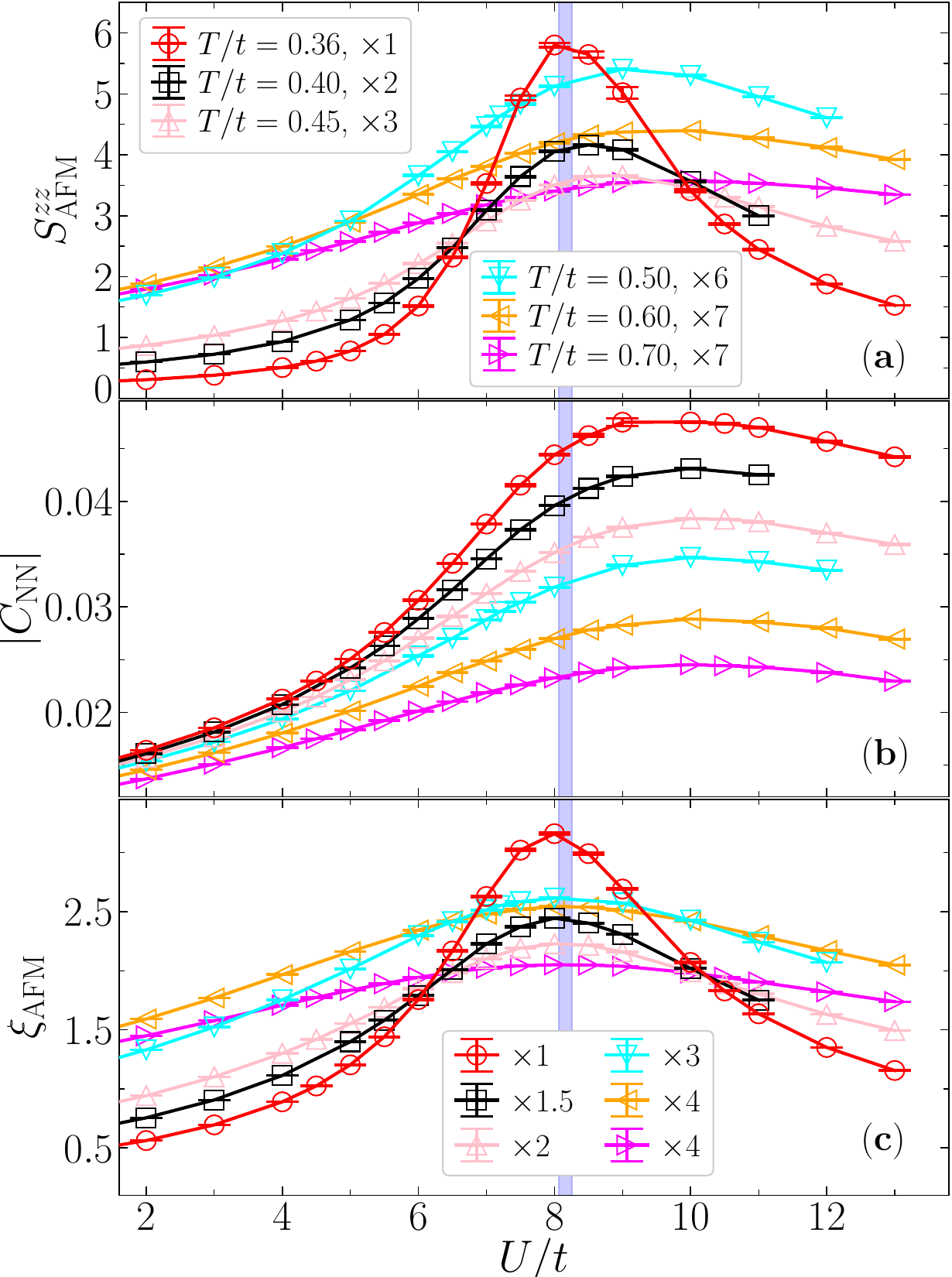}
\caption{AFM spin correlations in normal phase of half-filled 3D Hubbard model with (a) AFM structure factor $S_{\mathrm{AFM}}^{zz}$, (b) the NN spin-spin correlation function $|C_{\mathrm{NN}}|$, and (c) the correlation length $\xi_\mathrm{AFM}$, as a function of $U/t$ at temperatures from $T/t=0.36$ to $0.70$. In (a) and (c), the numerical results are rescaled with certain factors (see the legends) to fit into the plots. The peak location $U_{\rm AF}$ of $S_{\mathrm{AFM}}^{zz}$ for $T/t=0.36$ is shown as the vertical blue shading band. The results are from $L=12$ for $T/t=0.36$ and $L=8$ for other temperatures.}
\label{fig:AFMCorrelation}
\end{figure}

In Fig.~\ref{fig:AFMCorrelation}, we plot the numerical results of $S_{\rm AFM}^{zz}$, $|C_{\mathrm{NN}}|$ and $\chi_{\rm AFM}$ as a function of $U/t$ at various temperatures. With increasing interaction, all three quantities first increase, reach a maximum, and then decrease, resulting in peaks in the middle. The peaks tend to be more broadened at higher temperature due to stronger thermal fluctuations. The peak locations of $|C_{\mathrm{NN}}|$ [Fig.~\ref{fig:AFMCorrelation}(b)] are around $U/t=10$, accompanied by a slight shift towards larger $U$ with increasing temperature. For $\xi_{\rm AFM}$ [Fig.~\ref{fig:AFMCorrelation}(c)], the peak position instead stays almost unchanged around $U/t=8$. In this intermediate region, the AFM spin correlation in the system is dominated by the long-range part for $T/t=0.36$ as it's close to the N\'{e}el transitions. However, at high temperature as $T/t=0.70$, the correlation length is only $\sim$$0.5$ and thus the AFM spin correlation is primarily contributed by the short-range components. These together explain the feature of $S_{\rm AFM}^{zz}$ results [Fig.~\ref{fig:AFMCorrelation}(a)], for which the peak location $U_{\rm AF}$ moves from $U/t=8.3(1)$ at $T/t=0.36$ to $U/t=9.8(1)$ at $T/t=0.70$. Although the results presented in Fig.~\ref{fig:AFMCorrelation} might not converge, the peak locations are verified to have negligible finite-size effect (see Appendix~\ref{sec:AppendixE}). We then determine $U_{\rm AF}$ and its uncertainty by performing polynomial fitting for $S_{\rm AFM}^{zz}$ data around the peak, and reach the results (blue hexagon) in Fig.~\ref{fig:PhaseDiagram}(a). The $U_{\rm AF}$ curve resides almost in the center of the bad metal state, explicitly revealing that strong AFM spin correlation is another characteristic of this crossover regime. 

The competition between quantum and thermal fluctuations is responsible for the appearance of a maximum in AFM spin correlations at intermediate $U$ for finite temperature. Similar behavior was also observed in the 2D Hubbard model~\cite{Chiesa2011,Khatami2011}. While the interaction enhancement of the spin correlations in weakly interacting regime is apparent, the suppression towards the large $U$ side involves the effective AFM Heisenberg model for the Hubbard model. Approaching $U=\infty$, the Heisenberg AFM coupling $J=4t^2/U$~\cite{MacDonald1988} decreases with $U$. Then the fixed temperature $T$ in the Hubbard model becomes much larger than the coupling as $T/J\propto U \gg 1$, meaning the infinitely high temperature of the effective Heisenberg model. Thus, all three quantities $S_{\rm AFM}^{zz}$, $|C_{\mathrm{NN}}|$, and $\xi_{\rm AFM}$ characterizing the AFM spin correlations should decay to zero in the infinite-$U$ limit at fixed temperature. Then the peaks in AFM spin correlations naturally emerge in the middle, considering the enhancement and suppression in weakly and strongly interacting regimes, respectively. The situation is quite different at zero temperature, where there is no thermal fluctuation. At $T=0$, increasing $U$ constantly drives the system to the Heisenberg limit still staying at ground state. Thus, the AFM spin correlations should be monotonically enhanced with increasing $U$, which was confirmed in previous AFQMC study of the 2D Hubbard model~\cite{Mingpu2017}. Turning to the 3D case in our study, we have verified that, for $T/t<0.335$, the peak location of $S_{\rm AFM}^{zz}/N_s$ also moves towards larger $U$ with lower $T$ (not shown), which suggests $U_{\rm AF}=\infty$ as approaching $T=0$ and is consistent with the above discussion.

\subsection{Calculations of the thermal entropy}
\label{sec:ComputeEntropy}

As discussed in Sec.~\ref{sec:diagram}, the characterization of the MIC demands the calculations of physical observables versus $U$ at fixed temperatures. For the entropy, such calculations applying the conventional method presented in Sec.~\ref{sec:AFQMCObs} require numerical simulations covering a large temperature region for every $U$ point, which surely consumes large computational effort. In the following, we first present an important improvement for the conventional method to calculate entropy using Eq.~(\ref{eq:Entropy00}). And then we introduce a method to solve the issue and to compute the entropy versus interaction at a specific temperature. 

In previous studies~\cite{Ibarra2020,Dar2007,Paiva2010,Sushchyev2022}, the integral with infinity upper limit in Eq.~(\ref{eq:Entropy00}) was usually truncated using a very high temperature. However, the contribution of the residual tail to the entropy is hard to assess, and the numerical integration still demands numerous data points in the high-temperature region. These might cause quantitative deviation of the numerical results of entropy. A simple way to fix the high temperature issue is to rewrite the integral in $\beta=1/T$ axis as
\begin{equation}
\label{eq:Entropy01}
\frac{S(\beta)}{N_s} = \ln(4) + \beta e(\beta) - \int_0^{\beta} e(\beta^{\prime}) d \beta^{\prime}.
\end{equation}
In high temperature region, the total energy density $e(T)$ typically increases very slowly and smoothly to a constant with increasing temperature. The constant takes $e(T=\infty)=-U/4$ at half-filling. Thus, for the integral over $\beta^{\prime}$ in Eq.~(\ref{eq:Entropy01}), the high-temperature part can be evaluated very easily using a rather small number of data points. Nevertheless, the aforementioned issue in Eq.~(\ref{eq:Entropy00}) can appear if one wants to compute the entropy at low temperature using Eq.~(\ref{eq:Entropy01}). Evaluating the low-temperature part (with large $\beta^{\prime}$) of the integral needs a large amount of data points. Then it is straightforward to combine the advantages of Eqs.~(\ref{eq:Entropy00}) and (\ref{eq:Entropy01}), and divide the integral into two parts as
\begin{equation}\begin{aligned}
\label{eq:Entropy02}
\frac{S(T)}{N_s} 
= \ln(4) &+ \frac{e(T)}{T} - \int_T^{T_0}\frac{e(T^{\prime})}{{T^{\prime}}^2} dT^{\prime} \\
&- \int_0^{\beta_0} e(\beta^{\prime}) d\beta^{\prime},
\end{aligned}\end{equation}
where $T_0=1/\beta_0$ represents an intermediate temperature, and it can be tuned to double-check the final result of entropy. Practically, we first fit the energy results in the temperature region $T\le T^{\prime}\le T_0$ and the inverse temperature region $0\le\beta^{\prime}\le\beta_0$ individually [with $e(\beta=0)=-U/4$], and then compute the integrals using the fitting curve. We have also tested various choices of $T_0$, and reach well consistent results with different $T_0$ in the optimal range of $0.5\le T_0/t\le 1.5$ for the model.

\begin{figure}[h]
\centering
\includegraphics[width=0.99\columnwidth]{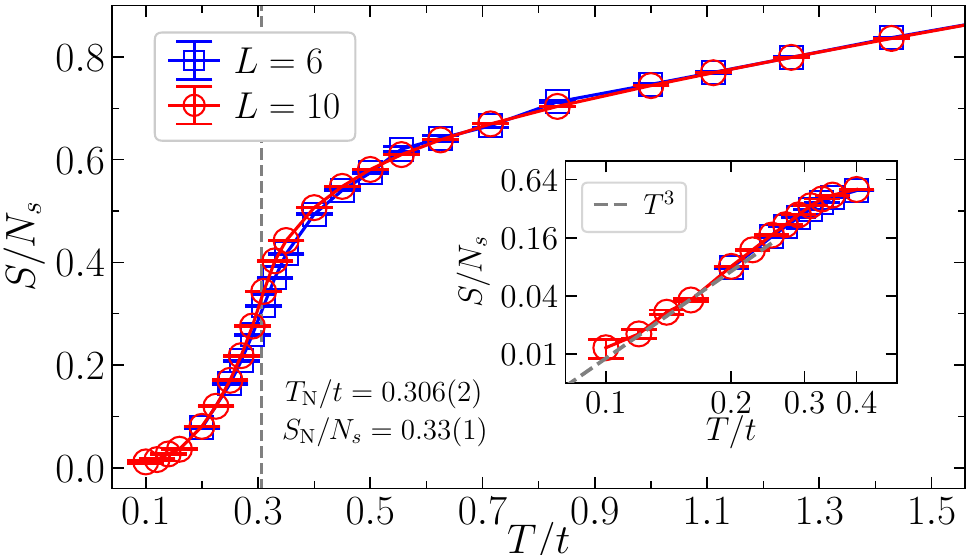}
\caption{The thermal entropy density $S/N_s$ for $U/t=10$ as a function of temperature in $L=6$ and $10$. The vertical gray dashed line marks the N\'{e}el temperature, and the critical entropy from $L=10$ is $S_{\rm N}/N_s=0.33(1)k_B$. The inset is a log-log plot for $S/N_s$ in the low-temperature region, and the limited results with $T/t<0.2$ conforms well with the $T^3$ scaling relation.}
\label{fig:EntropyU10}
\end{figure}

\begin{figure}[h]
\centering
\includegraphics[width=0.99\columnwidth]{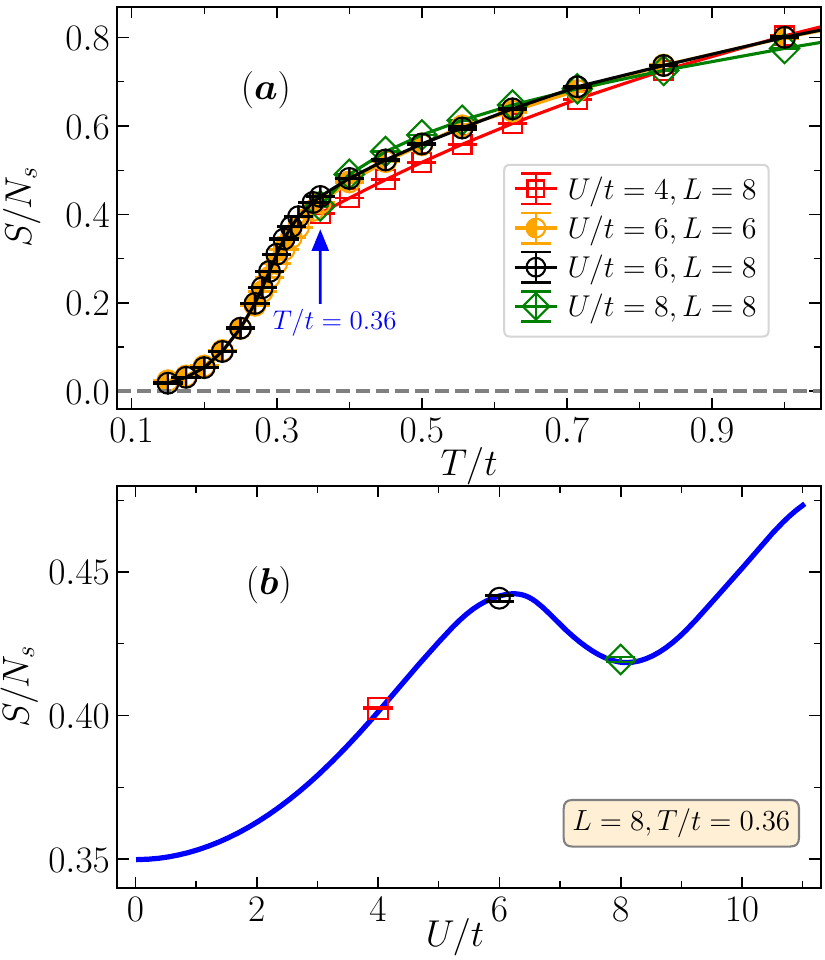}
\caption{(a) The thermal entropy density $S/N_s$ for $U/t=4,6,8$ as a function of temperature in $L=6$ and $8$. These results are computed from the improved formula Eq.~(\ref{eq:Entropy02}) of the conventional method. (b) $S/N_s$ results as a function of interaction strength at $T/t=0.36$ from $L=8$. The blue line is obtained from our new method, with the error bars smaller than the line width. The red square, black circle, and green diamond denote results from (a). The perfect consistency in this benchmark highlights the validity and efficiency of our method for calculating the entropy. }
\label{fig:CompareEntropy}
\end{figure}

With the above improvement using the hybrid formula in Eq.~(\ref{eq:Entropy02}), we can now access the high-precision results of thermal entropy along the temperature axis. In Figs.~\ref{fig:EntropyU10} and \ref{fig:CompareEntropy}(a), we plot the numerical results of $S/N_s$ versus $T/t$ for $U/t=10$ and $U/t=4,6,8$. The entropy simply decays to zero with lowering temperature, and an inflection point exists around the N\'{e}el transition. The $U/t=10$ and $U/t=6$ results reveal that the major finite-size effect of the entropy appears in a temperature region surrounding the N\'{e}el transition. For $U/t=10$, our results of $S/N_s$ at low temperatures (the inset of Fig.~\ref{fig:EntropyU10}) as $T/t\le 0.16$ conforms well with the $T^3$ scaling relation~\cite{Wessel2010}, which originates from the linear dispersion of spin-wave excitations in AFM Heisenberg model. We can also extract the critical entropy density as $S_N/N_s=0.33(1)k_B$ for $U/t=10$ with $L=10$ system. Even with finite-size effect, this number is rather close to the critical entropy in 3D AFM Heisenberg model as $S_N/N_s=0.341(5)k_B$~\cite{Wessel2010}, while it is almost half of the result from DMFT calculation~\cite{Werner2005}.

We then turn to our new scheme of directly computing the entropy as a function of $U$ at fixed temperature $T$. This method simply implements the definition of free energy as $F(U)=E(U)-TS(U)$ [with $E(U)$ as the total energy of the system], and expresses the entropy as $S(U)=[E(U)-F(U)]/T$. Then the only task is to compute the free energy, whose $U$ derivative actually involves the Hellmann-Feynman theorem at finite temperatures as
\begin{equation}
\frac{\dd F(U)}{\dd U} = \langle \hat{H}_I \rangle +\frac{\dd \mu(U)}{\dd U} \sum_{\bm{\mathrm{i}}} \langle \hat{n}_{\bm{\mathrm{i}}} \rangle, 
\label{eq:HFT}
\end{equation}
with $\hat{H}_I=\sum_{\bm{\mathrm{i}}} [\hat{n}_{\bm{\mathrm{i}}\uparrow} \hat{n}_{\bm{\mathrm{i}} \downarrow} - \frac{1}{2} (\hat{n}_{\bm{\mathrm{i}}\uparrow} + \hat{n}_{\bm{\mathrm{i}} \downarrow})]$. The proof of this formula is presented in Appendix~\ref{sec:AppendixA}. At half-filling, the chemical potential $\mu(U)$ is always zero and thus Eq.~(\ref{eq:HFT}) can be simplified as
\begin{equation}
\frac{1}{N_s}\frac{\dd F}{\dd U} 
= \frac{1}{N_s}\langle \hat{H}_I \rangle
= D(U) - \frac{1}{2}, 
\label{eq:HFT01}
\end{equation}
with $D(U)$ as the double occupancy. Then the free energy $F(U)$ can be evaluated via the integral of $D(U)$ as
\begin{equation}
\frac{F(U)}{N_s} = \frac{F(U=0)}{N_s} + \int_{0}^{U}D(U^{\prime})dU^{\prime} - \frac{U}{2},
\end{equation}
where $F(U=0)=F_0=-2T\sum_{\mathbf{k}}\ln[1+e^{-\beta(\varepsilon_{\mathbf{k}}+\mu)}]$ is the free energy of noninteracting system. Consequently, we arrive at the expression of the entropy density as
\begin{equation}
\label{eq:EntropyVsU}
\frac{S(U)}{N_s} = \frac{1}{T}\Big[\frac{E(U)-F_0}{N_s} - \int_{0}^{U}D(U^{\prime})dU^{\prime} + \frac{U}{2}\Big].
\end{equation}
This formula only needs the numerical results of $D(U)$ along the $U$ axis at fixed temperature $T$ to evaluate the integral and thus compute the entropy. We first obtain the high-precision results of $D(U)$ from AFQMC simulations, and then compute the integral using the fitting curve for $D(U)$. This method can be easily generated to the doping case. For example, for fixed filling away from $n=1$, the second term in Eq.~(\ref{eq:HFT}) is no longer zero as $\mu(U)$ takes $U$-dependent values. Thus, an additional integral of $\dd\mu(U)/\dd U$ over $U$ needs to be calculated to obtain the free energy $F(U)$, and the rest are the same as discussed above. This method should generally work for various Hubbard models regardless of the dimension and lattice geometry. 

In Fig.~\ref{fig:CompareEntropy}(b), we show the results of the entropy density versus $U/t$ at $T/t=0.36$ in the $L=8$ system. Our method is clearly validated, as the results computed with Eq.~(\ref{eq:EntropyVsU}) show perfect consistency with those from the conventional method at representative interaction strengths. Comparing to the conventional method, our method costs significantly less computational effort, and it can usually achieve a higher precision for the entropy due to the self-contained fixed-temperature calculations. These together demonstrate that our method is a generally valid and highly efficient scheme to compute the thermal entropy for Hubbard models. 
	
\subsection{Thermal entropy, double occupancy, and the Maxwell relation}
\label{sec:EntDoucMax}

Enabling the direct calculation of the thermal entropy along the interaction axis contributes a lot for studying fundamental properties in Hubbard models. First, the entropy results versus $U$ at fixed temperatures bring more insights into the MIC physics in the normal phase, and further enrich and refine the $U$-$T$ phase diagram. Second, combining with the fermion spectrum and AFM spin correlations, the entropy results offer a quantitative way to validate the effective Heisenberg picture in the strongly interacting regime. Third, via directly comparing the entropy density $S/N_s$ versus $U$ and the double occupancy $D$ versus $T$, we unambiguously demonstrate the Maxwell relation between these two quantities~\cite{Werner2005}
\begin{equation}
\label{eq:Maxwell}
\frac{1}{N_s}\Big(\frac{\partial S}{\partial U}\Big)_{(U_i,T_i)} 
= -\Big(\frac{\partial D}{\partial T}\Big)_{(U_i,T_i)}.
\end{equation}
This formula explicitly connects the $U$ derivative of the entropy density and the $T$-derivative of double occupancy at arbitrary point $(U_i,T_i)$ on the $U$-$T$ plane (as that in Fig.~\ref{fig:PhaseDiagram}). Incorporating this relation with the phase diagram, different behaviors in double occupancy with lowering temperature can be clearly understood, especially its anomalous decrease upon heating in a specific temperature range. Based on these points, we present the numerical results for the entropy and double occupancy, and discuss the underlying physics of the results. 

\begin{figure}[h]
\centering
\includegraphics[width=0.99\columnwidth]{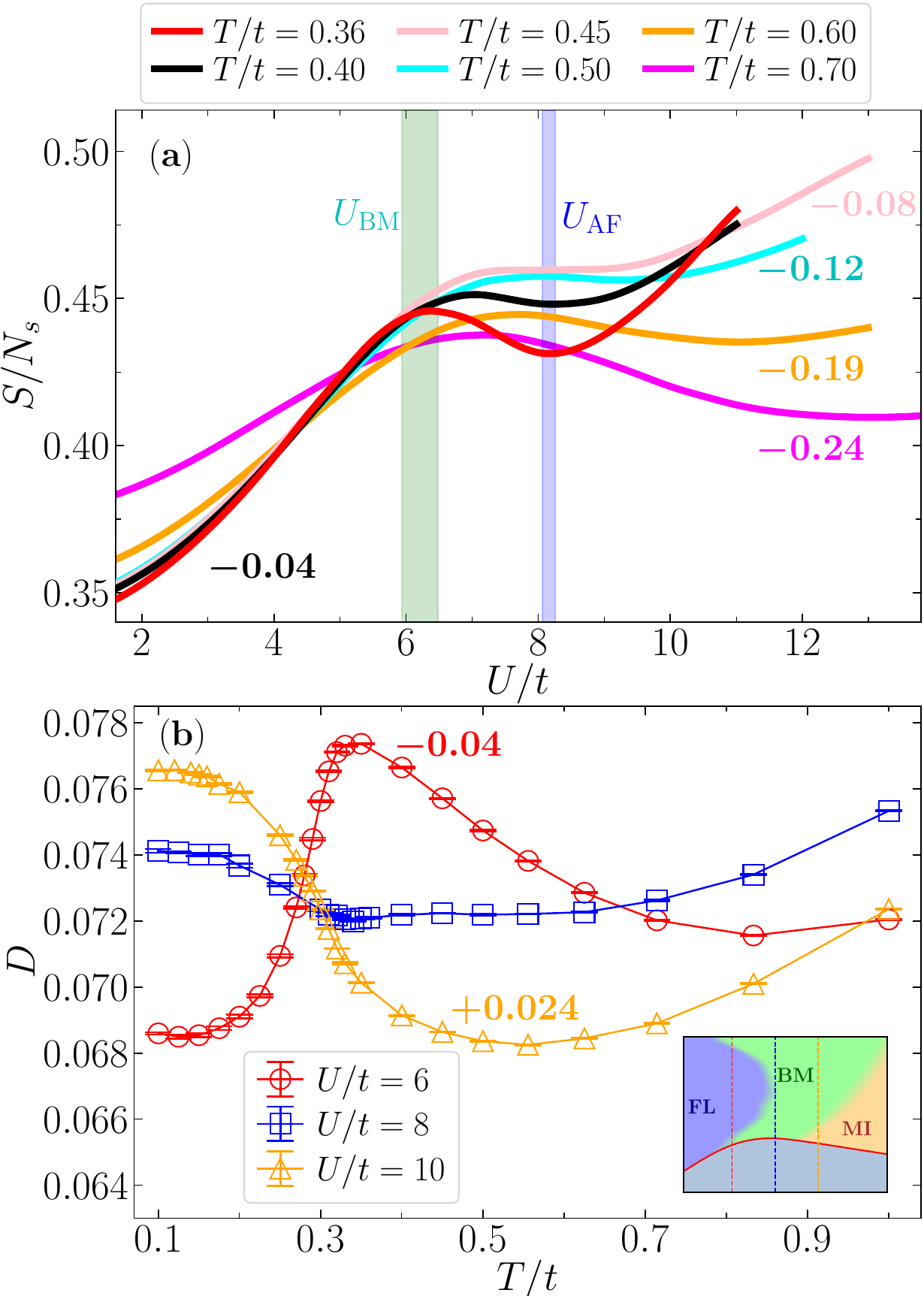}
\caption{(a) The thermal entropy density $S/N_s$ as a function of interaction strength at temperatures from $T/t=0.36$ to $T/t=0.70$. For $T/t=0.45,0.50,0.60,0.70$, the results are shifted by $-0.08,-0.12,-0.19,-0.24$, to fit into the plots. For $T/t=0.36$, the $U_{\rm BM}$ determined from $A_{\rm loc}(\omega)$ and the $U_{\rm AF}$ as the peak location of $S^{zz}_{\rm AFM}$ are plotted as the green and blue shading bands, respectively. The error bars are smaller than the linewidth and thus are neglected. The results are from $L=12$ for $T/t=0.36$ and $L=8$ for other temperatures. (b) Double occupancy $D$ as a function of temperature for $U/t=6,8,10$ from $L=10$. For $U/t=6$ and $10$, the results are shifted by $-0.04$ and $+0.024$ for the plot. The inset is the schematic phase diagram for Fig.~\ref{fig:PhaseDiagram}(a), with three vertical lines denoting $U/t=6,8,10$.}
\label{fig:EntropyDouocc}
\end{figure}

In Fig.~\ref{fig:EntropyDouocc}(a), we plot the entropy density versus interaction strength at temperatures from $T/t=0.36$ to $T/t=0.70$. The increase of the entropy with $U$ in weakly interacting regime characterizes the correlated Fermi liquid state~\cite{Walsh2019,*Walsh2019L,Downey2023}. Thus, the local maximum of $S/N_s$ [denoted as $U_{\rm S1}$, light blue diamonds in Fig.~\ref{fig:PhaseDiagram}(b)] also indicates stepping out of Fermi liquid and entering bad metal. This serves as a fully independent definition of the crossover boundary, in alternative to the scheme (and the $U_{\rm BM}$ results) via the fermion spectrum $A_{\rm loc}(\omega)$ discussed in Sec.~\ref{sec:FLtoBM}. Nevertheless, the separate results of $U_{\rm S1}$ and $U_{\rm BM}$ [see Fig.~\ref{fig:PhaseDiagram}(b)] are surprisingly consistent within the uncertainties, especially considering the lack of sharp signature pinpointing the boundary of the smooth crossover. This consistency illustrates the reliability of our characterizations for the MIC physics.

Once entering the bad metal state, the charge and spin contributions to the entropy exhibit totally different behaviors. The charge channel should follow the opposite trend of the electron localization described by $A_{\rm loc}(\omega)$, while the spin channel tracks the inverse of the AFM ordering tendency related to AFM structure factor $S_{\rm AFM}^{zz}$. Accordingly, in Fig.~\ref{fig:EntropyDouocc}(a), the decrease and local minimum of $S/N_s$ [denoted as $U_{\rm S2}$, purple left triangles in Fig.~\ref{fig:PhaseDiagram}(b)] with $U>U_{\rm S1}$ originates from the superposition of monotonically decreasing entropy in charge channel and the valley-like entropy in spin channel. For $T/t\le 0.50$, the $U_{\rm S2}$ almost coincides with the peak locations of $S_{\rm AFM}^{zz}$ [as $U_{\rm AF}$ in Fig.~\ref{fig:PhaseDiagram}(a)], which indicates the tiny charge contribution to the entropy with $U>U_{\rm AF}$. As a comparison, $U_{\rm S2}$ moves significantly to the right side of $U_{\rm AF}$ for $T/t>0.50$, due to the increased charge excitations and also the charge entropy as revealed by the $A_{\rm loc}(\omega)$ results. As a result, for $U>U_{\rm S2}$ at fixed temperature, the spin channel dominates the total entropy of the system, with vanishing contribution from the charge channel. And the most important low energy scale in this region corresponds to the NN spin-exchange coupling with energy $J\propto t^2/U$. Thus, the $U_{\rm S2}$ curve in the $U$-$T$ plane as shown in Fig.~\ref{fig:PhaseDiagram}(b) can be taken as the border, beyond which the half-filled 3D Hubbard model can be reasonably described by the spin-$1/2$ AFM Heisenberg model. 

In Fig.~\ref{fig:EntropyDouocc}(b), we present the numerical results of double occupancy $D$ as a function of temperature for $U/t=6,8,10$. These results apparently exhibit three different temperature dependencies of $D$. First of all, the infinite temperature limit has $D=0.25$ for arbitrary $U$, and this number begins to be suppressed when the decreasing temperature reaches the energy scale of the gap between the upper and lower Hubbard bands as $T$$\sim$$U$. Then the main features in these results of $D$ can be understood via the Maxwell relation in Eq.~(\ref{eq:Maxwell}) combining with the entropy results. For $U/t=6$ as a representative of weak interaction, the double occupancy has a a local minimum at $T/t$$\sim$$0.8$ and a local maximum at $T/t$$\sim$$0.34$, both with $\partial D/\partial T=0$. As shown in the inset of Fig.~\ref{fig:EntropyDouocc}(b), the vertical line of $U/t=6$ crosses the $U_{\rm S1}$ curve [the local maximum of the entropy versus $U$, see Fig.~\ref{fig:PhaseDiagram}(b)] for twice, indicating $\partial S/\partial U=0$ at the crossing temperatures which are actually the local minimum and maximum positions of $D$ in Fig.~\ref{fig:EntropyDouocc}(b). For $U/t=10$ standing for strong interaction, the double occupancy only has a broadened minima at $T/t$$\sim$$0.55$. Correspondingly, the vertical line of $U/t=10$ only encounters the $U_{\rm S2}$ curve [the local minimum of the entropy versus $U$, see Fig.~\ref{fig:PhaseDiagram}(b)] for once at $T/t$$\sim$$0.55$. For $U/t=8$, the shadow minimum at $T/t$$\sim$$0.33$ corresponds to crossing between the $U/t=8$ vertical line with the $U_{\rm S2}$ curve at the same temperature. Besides, the other features of double occupancy can also be understood from the entropy results. For example, Fig.~\ref{fig:EntropyDouocc}(a) shows that, for $T/t\ge 0.60$, the entropy decreases versus $U$ around $U/t=10$ indicating $\partial S/\partial U < 0$, which transfers to $\partial D/\partial T > 0$ according to Eq.~(\ref{eq:Maxwell}) and naturally explains the suppression of $D$ upon cooling in the same temperature range of $T/t\ge 0.60$ as shown in Fig.~\ref{fig:EntropyDouocc}(b). The entropy results for $T/t=0.45$ and $0.50$ are quite flat around $U/t=8$ as shown in Fig.~\ref{fig:EntropyDouocc}(a) rendering $\partial S/\partial U$$\sim$$0$, which then corresponds to the weak temperature dependence of $D$ for $U/t=8$ around $T/t=0.50$ as $\partial D/\partial T$$\sim$$0$ as shown in Fig.~\ref{fig:EntropyDouocc}(b). Other correspondences between the entropy and double occupancy can be identified similarly, which all together unambiguously verify the Maxwell's relation in a self-contained manner within AFQMC simulation results. 

We then pay more attention to the anomalous decrease of $D$ upon heating, which exists in specific temperature range as shown in Fig.~\ref{fig:EntropyDouocc}(b) for all three interaction strengths. This phenomenon resembles the Pomeranchuk effect in liquid $^3$He~\cite{Richardson1997}, and it has been predicted in numerical simulations of various Hubbard models~\cite{Gorelik2012,Kozik2013,Kim2021,Thomas2021,Georges1992,GangLi2014,Wietek2021,Qiaoyi2023} and experimentally observed in magic-angle graphene~\cite{Rozen2021}. Moreover, it has been theoretically proposed~\cite{Werner2005,Dar2007} and experimentally realized~\cite{Taie2012,Shao2024} as an interaction-induced adiabatic cooling scheme for cold atoms in an optical lattice. Nevertheless, in previous numerical studies, simple arguments based on the Maxwell's relation are usually used to explain this anomalous behavior~\cite{GangLi2014,Wietek2021,Qiaoyi2023}, which is indeed straightforward from our results but fails to reveal the underlying physics. For example, at $U/t=10$, the entropy in Fig.~\ref{fig:EntropyDouocc}(a) clearly has $\partial S/\partial U>0$ for $T/t\le 0.50$, indicating $\partial D/\partial T < 0$ as the heating induced decrease of $D$. Here, based on the phase diagram in Fig.~\ref{fig:PhaseDiagram}, we can now achieve a clear and complete physical understanding (especially for different $U$) for this phenomenon. For $U/t=6$ with lowering $T$, the anomalous increase of $D$ starting at $T/t$$\sim$$0.8$ can be attributed to the entrance into the Fermi liquid from the bad metal, during which the electrons become more delocalized and thus the double occupancy is promoted. After reaching the local maximum at $T/t$$\sim$$0.34$, the system re-enters into the bad metal state and continuously evolves into the N\'{e}el ordered phase and the fully gapped ground state. This evolution constantly increases the electron localization in the low $T$ region, and thus results in the monotonic suppression of $D$. Comparing to $U/t=6$, the interpretation for $U/t=10$ results is completely different. For $U/t=10$ with lowering $T$, the system first resides in the bad metal state, and gradually approaching the Mott insulator can explain the decrease of $D$ in the range of $0.55\le T/t\le 1.00$. Once crossing the $U_{\rm S2}$ curve where $D$ reaches the minimum at $T/t$$\sim$$0.55$, the system can be described by the effective Heisenberg model, and the related spin-exchange coupling and its origin as the virtual hopping of electrons appear, which slightly delocalizes the electrons and thus promotes the double occupancy. This process belongs to the quantum fluctuation which is further enhanced towards lowering $T$, and thus contributes to the monotonic increase of $D$ versus lowering $T$ for $T/t\le 0.55$. These analyses for $U/t=10$ also fit the results of $U/t=8$.

The residual finite-size effect of the entropy density ($L=12$ and $8$) and double occupancy ($L=10$) presented in Fig.~\ref{fig:EntropyDouocc} does not affect the understanding of our numerical results. For $S/N_s$, the finite-size effect indeed exists for $T<0.45$, but we have verified that the local maximum and minimum features keep unchanged with both $U_{\rm S1}$ and $U_{\rm S2}$ showing convergence for $L=8$ and $12$ (see Appendix~\ref{sec:AppendixE}). For $T\ge 0.45$, the $S/N_s$ data from the $L=8$ system can be safely taken as the TDL results [see $L=6$ and $8$ results for $U/t=6$ presented in Fig.~\ref{fig:CompareEntropy}(a)]. As for $D$, it only shows slight size dependence for $T/t<0.50$ (see Appendix~\ref{sec:AppendixE}) and the qualitative behaviors versus temperature discussed above persist to TDL.

\subsection{Specific heat and charge compressibility}
\label{sec:CvChie}

Specific heat is an important thermodynamic observable for condensed matter systems, for which it can reveal the fundamental excitations and detect the phase transitions. Charge compressibility is an alternative quantity to characterize the electron localization. In the following, we present the numerical results for these two quantities, which not only contribute additional verification for the phase diagram but also elucidate universal properties in Hubbard models. 

\begin{figure} [t]
\centering
\includegraphics[width=0.99\columnwidth]{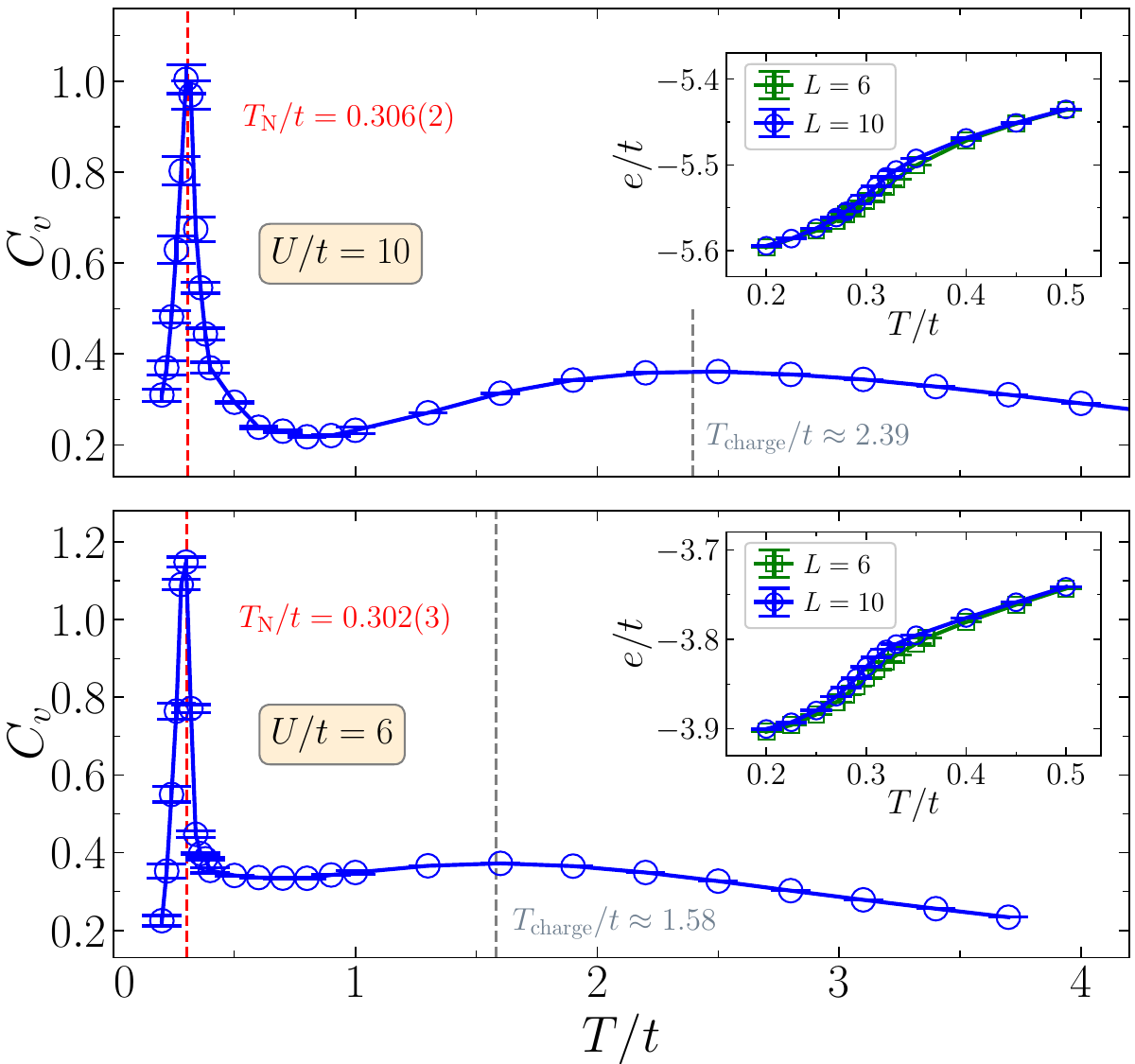}
\caption{The specific heat $C_v$ as a function of temperature for (a) $U/t=10$ and (b) $U/t=6$ from $L=10$. The N\'{e}el temperatures are presented with red vertical dashed lines. The high-temperature charge peak in $C_v$ appears at $T_{\rm charge}/t\simeq 2.39$ for $U/t=10$ and $1.58$ for $U/t=6$, as the gray vertical dashed lines. For each panel, the inset plots the total energy density $e/t$ from $L=6$ and $10$ in the temperature range of $0.2\le T/t\le 0.5$. }
\label{fig:SpecificHeat}
\end{figure}

We compute the specific heat using $C_v=\dd e(T)/\dd T$. Instead of numerical derivative, we first perform polynomial fitting for $e(T)$ versus $T$ and then compute its first-order derivative as $C_v$ with its uncertainty estimated by the bootstrapping technique. In Fig.~\ref{fig:SpecificHeat}, we present the results of $C_v$ as a function of temperature for $U/t=10$ and $U=6$. A double peak structure can be clearly observed, which was also verified to exist in 2D Hubbard model~\cite{Qiaoyi2023,Duffy1997,Paiva2001}. The sharp peak at low temperature and the broadened one at high temperature are called spin peak and charge peak, respectively~\cite{Duffy1997}. The spin peak is related to the N\'{e}el transition and the maximum is associated with spin excitations. However, as discussed in Sec.~\ref{sec:PmAfmTransition}, this spin peak in $C_v$ does not diverge and it is not clear whether its position coincides with N\'{e}el temperature $T_{\rm N}$ in TDL. The charge peak (at $T_{\rm charge}$) is instead contributed by the charge (fermionic) excitations across the gap relating to upper and lower Hubbard bands. It was found that $T_{\rm charge}\simeq0.24U$ within $7\le U/t\le 12$ in half-filled 2D Hubbard model~\cite{Duffy1997}. For our case of 3D, we obtain $T_{\rm charge}\simeq 2.39t=0.239U$ for $U/t=10$, $T_{\rm charge}\simeq 1.58t=0.263U$ for $U/t=6$, and $T_{\rm charge}\simeq 2.02t=0.253U$ for $U/t=8$ (not shown), which are very close to the 2D result. This charge peak feature in $C_v$ is actually inherited from the atomic (single site) limit, to which the Hubbard model degenerates at very high temperature as $T\gg U$. The $C_v$ result in atomic limit possesses the peak position at $T\simeq 0.208U$ (see Appendix~\ref{sec:AppendixC}). Our results of $T_{\rm charge}$ for $U/t=6,8,10$ clearly show drafting towards the atomic limit. The comparison of total energy density from $L=6$ and $10$ shown in insets of Fig.~\ref{fig:SpecificHeat} demonstrate that the finite-size effect only matters around $T_{\rm N}$ and it is negligible for $T/t\ge 0.5$ even with $L=6$. As a result, the $C_v$ results around the spin peak still have slight size dependence while the charge peak results safely reach the TDL. 

\begin{figure}[t]
\centering
\includegraphics[width=0.99\columnwidth]{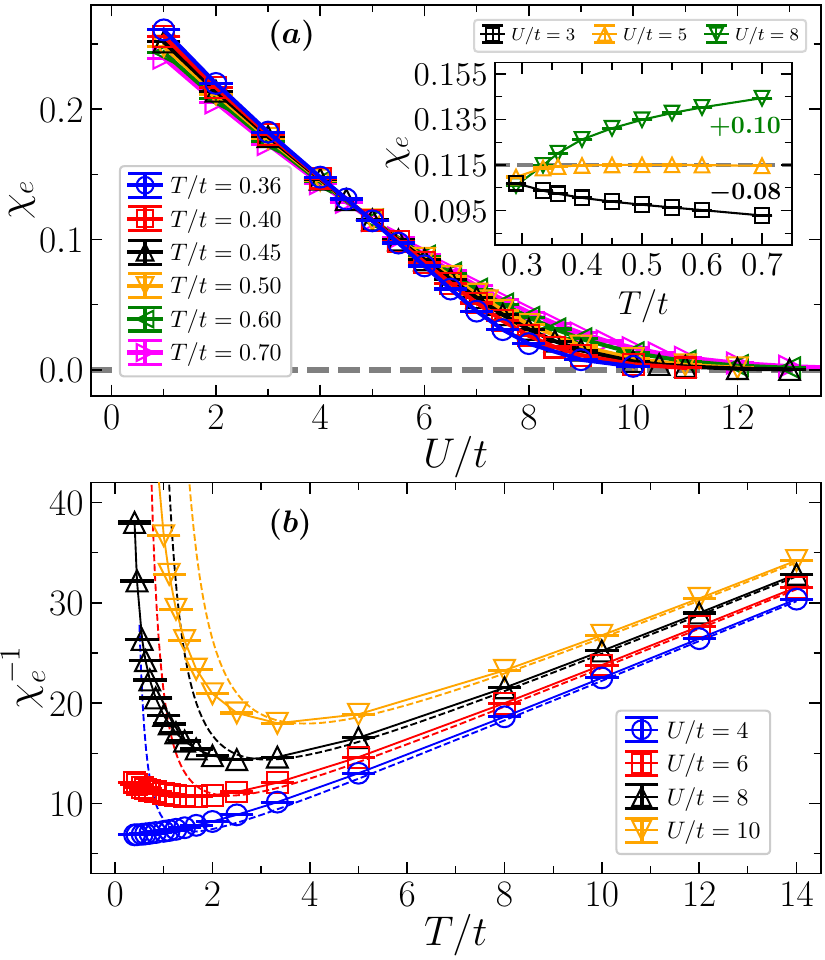}
\caption{(a) Charge compressibility $\chi_e$ as a function of interaction strength at temperatures from $T/t=0.36$ to $0.70$. All results are from $L=8$. The inset plots $\chi_e$ versus temperature in the range $0.3\le T/t\le 0.7$ for $U/t=3,5,8$. (b) Inverse charge compressibility $\chi_e^{-1}$ as a function of temperature for $U/t=4,6,8,10$. These results are from $L=10$. The dashed lines plot the $\chi_e^{-1}$ results of the atomic limit.}
\label{fig:Compressibility}
\end{figure}

According to Eq.~(\ref{eq:ChiCharge}), the calculation of charge compressibility $\chi_e$ only involves the static density-density correlation function. In Fig.~\ref{fig:Compressibility}, we present the results of $\chi_e$ as functions of interaction strength and temperature. As shown in Fig.~\ref{fig:Compressibility}(a), the MIC is again manifested by the smooth suppression of $\chi_e$ versus $U$ at fixed temperatures. Once entering the Mott insulator state, $\chi_e$ becomes tiny and finally vanishes, indicating the complete electron localization. With a small threshold $\epsilon$$\sim$$10^{-3}$, the criteria $\chi_e=\epsilon$ produces the pink dashed line in the phase diagram in Fig.~\ref{fig:PhaseDiagram}(b), which is well consistent with the $U_{\rm MI}$ results extracted from the spectrum. Besides, we also observe an interesting cross of the curves for different temperatures around $U/t=5$, which is apparent from the inset of Fig.~\ref{fig:Compressibility}(a) showing $\chi_e$ versus temperature for $U/t=3,5,8$. These results indicate the opposite sign of $\dd \chi_e/\dd T$ for $U/t<5$ and $U/t>5$, which are typical characterizations of metallic and insulating states. In Ref.~\onlinecite{Kim2021}, such a crossing of $\chi_e$ results was also observed in half-filled 2D Hubbard model, and it was taken as a signal of the crossover. A clear difference in 3D is that the crossing point almost stays unchanged at $U/t=5$ for $0.35\le T/t\le 0.70$, which fully resides in the Fermi liquid regime in the phase diagram. This feature of $\chi_e$ already converges regarding the system size reported in Fig.~\ref{fig:Compressibility}(a). Our understanding for this point is that the temperature dependence of $\chi_e$ might fail to characterize the Fermi liquid and bad metal states, which should vanish in the low-temperature region considering N\'{e}el AFM ordered phase. A related fact is that, even for $U/t=3$, $\chi_e$ should finally become zero approaching $T=0$ due to the gapped ground state. 

The $\chi_e^{-1}$ results, covering a large temperature range $0.36\le T/t\le 14$, are presented in Fig.~\ref{fig:Compressibility}(b) for $U/t=4,6,8,10$. The almost linear dependence on $T$ at high temperature region is prominent. Such behavior was also found in the 2D Hubbard model~\cite{Edwin2019,Qiaoyi2023}. Similar to specific heat, this feature at very high temperature can be explained by the results of the atomic limit, which are plotted in Fig.~\ref{fig:Compressibility}(b) as dashed lines. 
Combining the atomic limit within $T\gg U$, we can obtain $\chi_e^{-1}=T/(n-n^2/2)+U/2+O(U/T)$ with $n$ as the fermion filling (see Appendix~\ref{sec:AppendixC}), which clearly exhibits the linear dependence on $T$. At half-filling ($n=1$), the slope $(n-n^2/2)^{-1}$ is equal to $2$ and it increases with doping, which is consistent with the AFQMC results in Ref.~\onlinecite{Edwin2019}. Moreover, the DC resistivity $\rho$ can be computed via the Nernst-Einstein relation as $\rho=\chi_e^{-1}/D_{\rm diff}$ with $D_{\rm diff}$ denoting the diffusivity. Thus, the linear-$T$ dependence of $\chi_e^{-1}$ from our AFQMC calculations might serve as a qualitative explanation for the linear resistivity as observed in the Hubbard model~\cite{Edwin2019,Peter2019}, considering that $D_{\rm diff}$ only weakly depends on $T$ at very high temperatures. With decreasing temperature, $\chi_e^{-1}$ gradually deviates from the atomic limit due to the intervention of quantum fluctuations. For $U/t=6,8,10$, it further bends up around $T_{\rm charge}$ from $C_v$, and then rapidly increases indicating $\chi_e\to0$ towards $T=0$, which is an incipient signature of the gapped ground state. For $U/t=4$, this bending up of $\chi_e^{-1}$ should happen at a temperature even lower than the plotted data. 

As discussed above, while the low-temperature behaviors of specific heat and charge compressibility mainly depends on the quantum properties (ordering, metallic, or insulating), their high-temperature features including charge peak in $C_v$ and linear temperature dependence of $\chi_e$ are extensions of the atomic limit along the temperature axis. Consequently, these behaviors should generally exist in Hubbard models despite the dimension, lattice geometry, and additional hopping terms.

\subsection{Inflection point of double occupancy}
\label{sec:InflcDouOcc}

Based on Eq.~(\ref{eq:DouOccDerive}), we can directly measure the $U$-derivative of double occupancy $\partial D/\partial U$ at fixed temperature. With $\partial\langle\hat{N}\rangle/\partial U=0$ at half-filling, the calculation only requires additional measurement of the imaginary-time correlation $C_{\hat{H}_I}(\tau, 0)$, which is straightforward and similar to the computation of two-body correlation functions in AFQMC simulations. 

\begin{figure}[b]
\centering
\includegraphics[width=0.99\columnwidth]{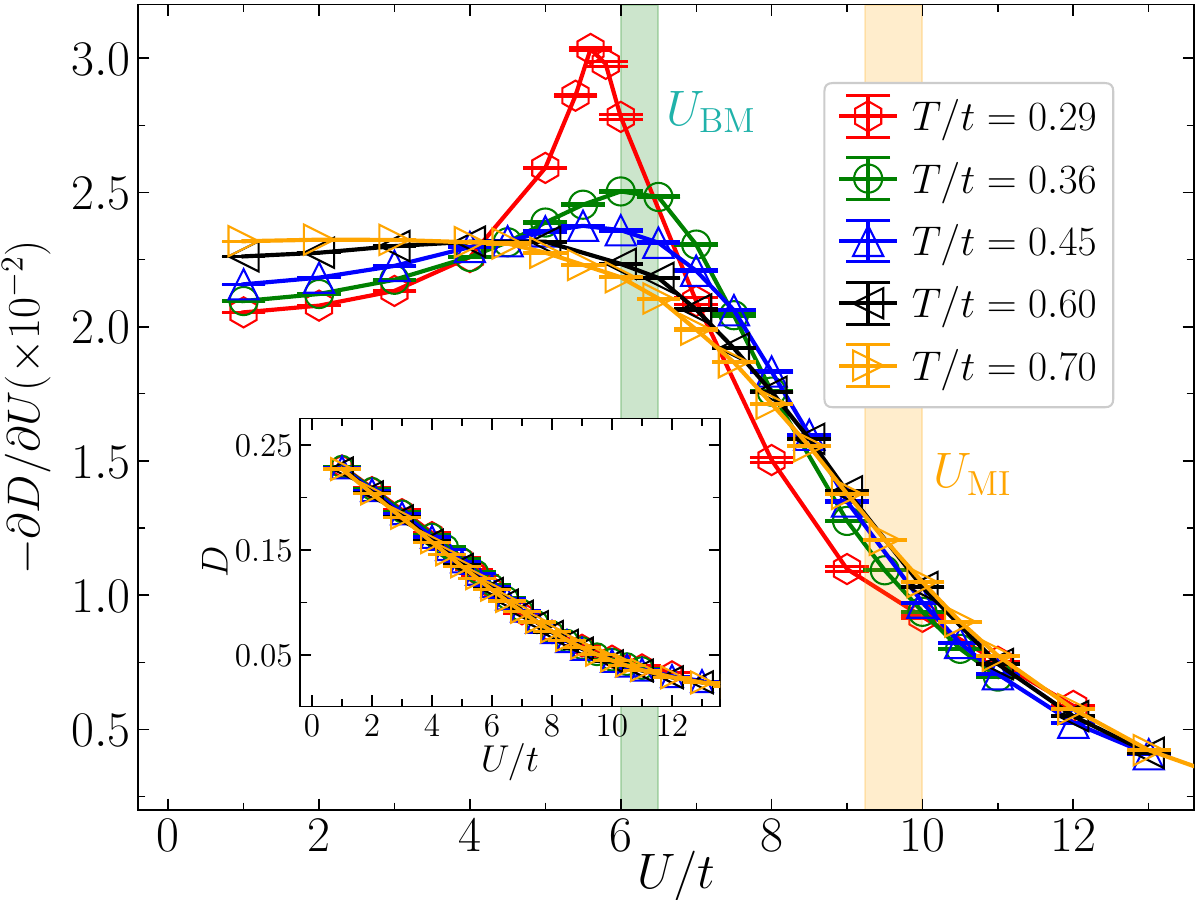}
\caption{Numerical results of $-\partial D/\partial U$ versus $U/t$ at temperatures from $T/t=0.29$ to $0.70$. The $U_{\rm BM}$ and $U_{\rm MI}$ results for $T/t=0.36$ are plotted as the green and yellow shading bands, respectively. The inset plots the raw data of $D$ versus $U/t$. For $T/t=0.29$, the peak location corresponds to the Neel transition. The results are from $L=12$ for $T/t=0.36$ and $L=8$ for other temperatures, and the finite-size effect is negligible for $T/t\ge0.36$. }
\label{fig:DoubleOccInter}
\end{figure}

In Fig.~\ref{fig:DoubleOccInter}, we show the numerical results of $-\partial D/\partial U$ as a function of $U/t$ at temperatures from $T/t=0.29$ to $T/t=0.70$. Broadened peaks in the intermediate interaction region can be clearly observed in this quantity, which reveals the inflection point of double occupancy $D$ as the most rapid suppression by interaction. The inflection point is less obvious regarding the raw data of $D$ plotted in the inset of Fig.~\ref{fig:DoubleOccInter}. The rather weak temperature dependence of $D$ is prominent, as it only changes in an interval of $\Delta D<0.01$ within $0\le T/t\le 1$ as shown in Fig.~\ref{fig:EntropyDouocc}(b). We take the peak location of $-\partial D/\partial U$ as the position of the inflection point of $D$, which accounts for the $U_{\rm D}$ results (black pentagons) in Fig.~\ref{fig:PhaseDiagram}(a). 

At $T/t=0.29$, the derivative $-\partial D/\partial U$ exhibits a sharper peak than other temperatures at $U/t\simeq 5.7$ (with a slight size dependence), which actually corresponds to the N\'{e}el transition. At $T/t=0.36$, the broadened peak in $-\partial D/\partial U$ with $U_{\rm D}/t=6.1(1)$ is consistent with $U_{\rm BM}/t=6.21(27)$. At higher temperatures, the peak location moves oppositely towards weaker interactions such as $U_{\rm D}/t=5.5(1)$ at $T/t=0.45$, and the peak finally disappears at $T/t=0.7$. Combining these results, the $U_{\rm D}$ curve shows a strange shape as increasing versus $T$ in the temperature range of $0.29\le T/t\le 0.36$ and then decreasing for $T/t>0.36$. As a result, the $U_{\rm D}$ curve at $T/t>0.36$ fully resides in the Fermi liquid regime in the phase diagram as shown in Fig.~\ref{fig:PhaseDiagram}(a). Moreover, the derivative $-\partial D/\partial U$ has no feature around $U_{\rm MI}$ for all the fixed temperatures in our study. These results unambiguously demonstrate that the double occupancy completely fails to capture the MIC physics in the normal phase (especially at $T/t>0.36$). In the conventional intuition, the most rapid decrease of double occupancy should be a representative feature at least for approaching Mott insulator~\cite{Rozenberg1999,Parcollet2004}. Our understanding for this counterintuitive behavior is that, the thermal fluctuation at $T/t>0.36$ might overtakes the quantum characteristic of double occupancy and thus drags the inflection point to the weaker interaction where the thermal fluctuation becomes more significant.

Similar results of $U_{\rm D}$ residing in the Fermi liquid regime was also obtained in a recent DCA study of the half-filled triangular lattice Hubbard model~\cite{Downey2023}. As a comparison, our AFQMC results of $U_{\rm D}$ are unbiased with negligible finite-size effect, which also benefit from Eq.~(\ref{eq:DouOccDerive}) as avoiding possible errors using numerical derivative. Nevertheless, the numerical results in our study and in Ref.~\onlinecite{Downey2023} together indicate that such a behavior of double occupancy might generally exist in various Hubbard models in the intermediate temperature region.

\subsection{The peak of fidelity susceptibility}
\label{sec:Chifidelity}

\begin{figure}[h]
\centering
\includegraphics[width=0.99\columnwidth]{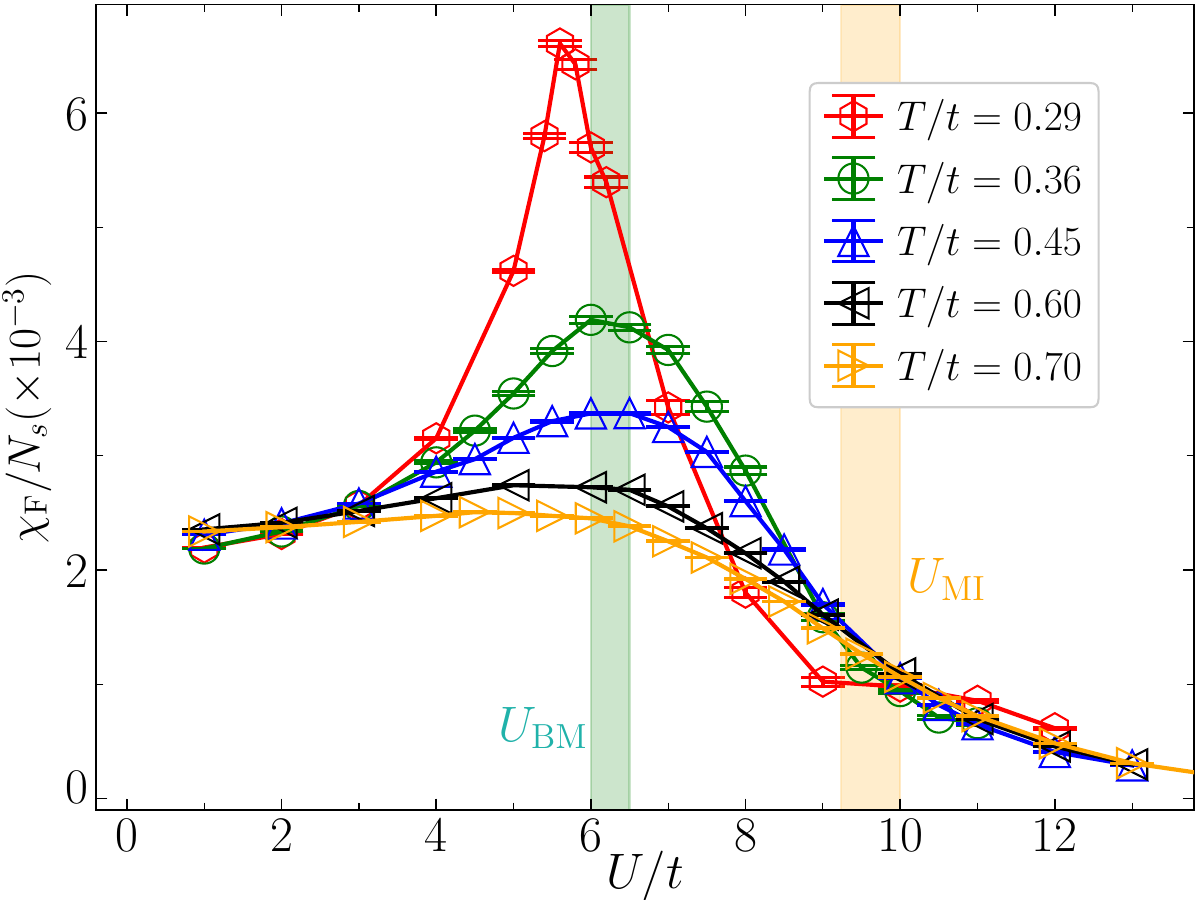}
\caption{Fidelity susceptibility per site $\chi_{\mathrm{F}}/N_{s}$versus $U/t$ at temperatures from $T/t=0.29$ to $T/t=0.70$. The $U_{\rm BM}$ and $U_{\rm MI}$ result for $T/t=0.36$ are plotted as the green and yellow shading bands respectively. For $T/t=0.29$, the peak location corresponds to the Neel transition. The results are from $L=12$ for $T/t=0.36$ and $L=8$ for other temperatures, and the finite-size effect is negligible for $T/t\ge0.36$. }
\label{fig:FidelitySuscep}
\end{figure}

In Fig.~\ref{fig:FidelitySuscep}, we present the results of the fidelity susceptibility per site $\chi_{\mathrm{F}}/N_s$ as a function of $U/t$ at temperatures from $T/t=0.29$ to $T/t=0.70$. All the peak locations of $\chi_{\mathrm{F}}/N_s$ constitutes the $U_{\rm F}$ results (yellow right triangles) in Fig.~\ref{fig:PhaseDiagram}(b).

At $T/t=0.29$, $\chi_{\mathrm{F}}/N_s$ shows a sharp peak at $U/t\simeq 5.7$ (with a slight size dependence) related to the N\'{e}el transition. This conforms with the results in Ref.~\onlinecite{WangLei2015} showing that, despite the original definition of the fidelity at $T=0$, the fidelity susceptibility generalized to $T>0$ can also be used as an efficient tool to probe the thermal phase transitions. The peak location of $\chi_{\mathrm{F}}/N_s$ acquires a consistent value of $U_{\rm F}/t=6.3(1)$ with $U_{\rm BM}/t=6.21(27)$ at $T/t=0.36$, reaches a maximum of $U_{\rm F}=6.4(1)$ at $T/t=0.40$, and then moves to the weaker interaction. Besides, no signals can be observed in $\chi_{\mathrm{F}}/N_s$ around $U_{\rm MI}$ for all the fixed temperatures. All these behaviors are very similar to those of $-\partial D/\partial U$, and the only difference lies in the specific values of $U_{\rm F}$ and $U_{\rm D}$. Thus, this quantity also fails to characterize the MIC physics in the normal phase. And the same interpretation as double occupancy should also fit to the fidelity susceptibility that the thermal fluctuation conquers its underlying physics which has the origin from the quantum aspect of the system~\cite{Albuquerque2010,WangLei2015}. 

In Ref.~\onlinecite{HuangLi2016}, it was found that the fidelity susceptibility is sensitive to the interaction strength and indeed show signals for Fermi-liquid to non-Fermi-liquid crossover in a half-filled two-band Hubbard model on Bethe lattice. However, the calculations in Ref.~\onlinecite{HuangLi2016} were performed at very low temperatures in order to reveal ground state properties, and there is no disturbance from long-range order. The situation in our study is quite different. We focus on the crossover physics at mediate to high temperatures, since the N\'{e}el AFM order occupies the low temperature region of the half-filled 3D repulsive Hubbard model. So it might be interesting to explore the possible MIC at very low (or zero) temperature in frustrated Hubbard models~\cite{Ohashi2008,Downey2023} with fidelity susceptibility (and double occupancy).

\subsection{The self-energy crossing}
\label{sec:SelfEnergy}

The self-energy contains important information about the quasiparticle properties of correlated fermion systems. For example, the imaginary part of momentum-resolved self-energy with real frequency as ${\rm Im}\Sigma(\mathbf{k},\omega)$ exhibits a zero (a pole) at $\omega=0$ for a metal at $\mathbf{k}=\mathbf{k}_F$ (an insulator at all $\mathbf{k}$). Correspondingly, at sufficiently low temperature when the several lowest $\omega_n$ are fairly close to zero ($i\omega_0\to0$), the imaginary part of $\Sigma(\mathbf{k},i\omega_n)$ defined in Eq.~(\ref{eq:EqSlfEng}) can be used as a metric to determine whether the system is metallic or insulating: if ${\rm Im}\Sigma(\mathbf{k}_F,i\omega_0)>{\rm Im}\Sigma(\mathbf{k}_F,i\omega_1)$ for all $\mathbf{k}_F$, the state is metallic; otherwise it is insulating. As a result, the crossing of ${\rm Im}\Sigma(\mathbf{k}_F,i\omega_0)$ and ${\rm Im}\Sigma(\mathbf{k}_F,i\omega_1)$ versus interaction indicates the change from metal to insulator. These simple criteria have been used in the study of half-filled 2D Hubbard model~\cite{Svistunov2020}, and contributed to the identification of an intermediate pseudogap regime at finite temperatures. Here we perform similar calculations for the 3D Hubbard model.

\begin{figure}[t]
\centering
\includegraphics[width=0.99\columnwidth]{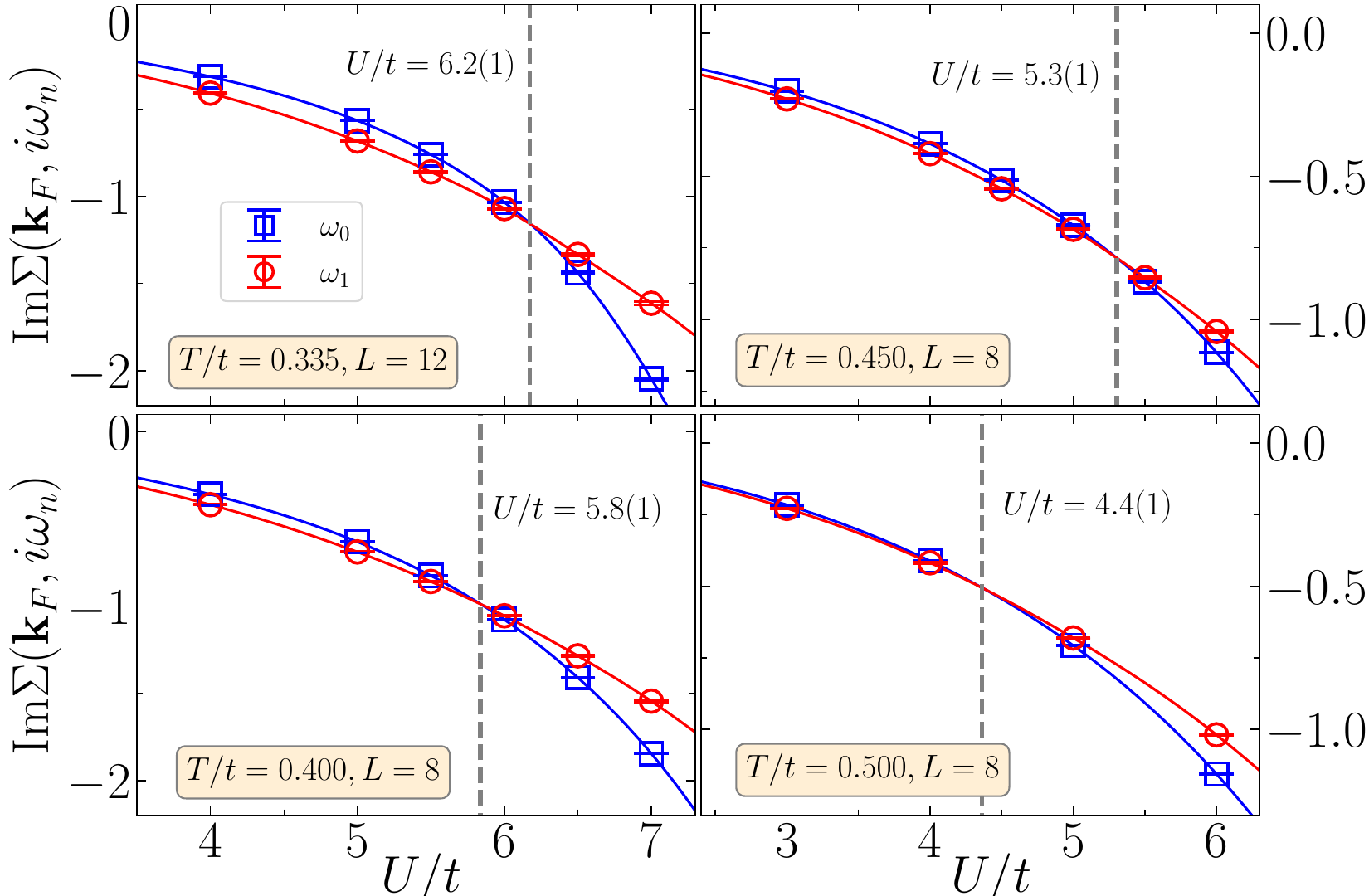}
\caption{The imaginary part of self-energy ${\rm Im}\Sigma(\mathbf{k}_F,i\omega_0)$ and ${\rm Im}\Sigma(\mathbf{k}_F,i\omega_1)$ with $\mathbf{k}_F = (\pi/2,\pi/2,\pi/2)$ as a function of $U/t$. The solid lines connecting the data points are from the polynomial fitting. The corresponding temperature, linear system size, and the crossing point (gray dashed lines) are presented in each panel.}
\label{fig:SelfEnergy}
\end{figure}

The AFQMC calculation of the self-energy $\Sigma(\mathbf{k},i\omega_n)$ based on Eq.~(\ref{eq:EqSlfEng}) has been discussed in Sec.~\ref{sec:BMtoMI}. In Fig.~\ref{fig:SelfEnergy}, we present the results of ${\rm Im}\Sigma(\mathbf{k}_F,i\omega_0)$ and ${\rm Im}\Sigma(\mathbf{k}_F,i\omega_1)$ with $\mathbf{k}_F = (\pi/2,\pi/2,\pi/2)$ versus interaction at four temperatures. The crossing (denoted as $U_{\Sigma}$) between these two quantities is apparent. The uncertainty of $U_{\Sigma}$ is estimated by the bootstrapping technique. Distinguished from the 2D case that $U_{\Sigma}$ take significantly different values at the nodal and antinodal points~\cite{Svistunov2020}, we find that, for the half-filled 3D Hubbard model, all the independent $\mathbf{k}_F$ points produce well consistent results for $U_{\Sigma}$ due to the nearly isotropic property of the Fermi surface. We have also verified that, within the systems reported in Fig.~\ref{fig:SelfEnergy}, the $U_{\Sigma}$ results have no finite-size effect. These results are summarized (magenta octagons) in Fig.~\ref{fig:PhaseDiagram}(b). 

While $U_{\Sigma}$ at $T/t=0.335$ and $0.36$ show consistency with the corresponding $U_{\rm BM}$, it decreases and moves into the Fermi liquid regime in the phase diagram at higher temperatures, resembling the results of $U_{\rm D}$ and $U_{\rm F}$. This suggests that the self-energy crossing metric also fails to characterize the MIC (especially for $T/t>0.36$) in the system. The reason for this failure is probably that the temperature region we study is too high and the above method stops working as the frequencies $\omega_0$ and $\omega_1$ are far from zero. For example, the temperature $T/t=0.40$ possesses $\omega_0=\pi T\simeq 1.257t$ and $\omega_1=3\pi T\simeq 3.770t$, and thus speculating the $\omega\to0$ behavior of ${\rm Im}\Sigma(\mathbf{k}_F,\omega)$ using ${\rm Im}\Sigma(\mathbf{k}_F,i\omega_0)$ and ${\rm Im}\Sigma(\mathbf{k}_F,i\omega_1)$ is obviously less reliable.

\section{Away from half-filling}
\label{sec:SignProblem}

The fermion filling spans a new dimension for correlated fermion systems, and the doping as deviation from half-filling can induce many unconventional phenomena, such as the high-temperature superconductivity in cuprates~\cite{Dagotto1994,Bulut2010} and the stripe orders in 2D repulsive Hubbard model~\cite{Chang2010,Qin2016a,Boxiao2017,Xiao2023}. As for the 3D repulsive Hubbard model, there are very few studies investigating its properties with doping~\cite{Tahvildar1997,Katanin2017,Lenihan2022,Shao2024}. The early work with the second-order perturbation theory~\cite{Tahvildar1997} and a subsequent one using dynamical vertex approximation~\cite{Katanin2017} both established the dome-shaped AFM ordered phase on the doping-temperature plane with fixed interaction. Moreover, both studies showed that, inside the AFM dome, the N\'{e}el AFM order gradually evolves into an incommensurate SDW order with increased doping (and lowering temperature)~\cite{Tahvildar1997,Katanin2017}. A more recent DiagMC study~\cite{Lenihan2022} also found such a N\'{e}el AFM-SDW crossover within limited accuracy. Partially aligned with the numerics, the optical lattice experiment~\cite{Shao2024} also observed the N\'{e}el AFM ordered phase up to $n\simeq 0.95$ for $U/t\simeq 11.75$ along a path with specific values of thermal entropy. 

For the 3D repulsive Hubbard model in Eq.~(\ref{eq:Hamiltonian}), the hole doping with $\mu>0$ and the electron doping with $\mu<0$ can be simply connected via a particle-hole transformation, i.e., $c_{\mathbf{i}\sigma}^+\to(-1)^{i_x+i_y+i_z}c_{\mathbf{i}\sigma}$ and $c_{\mathbf{i}\sigma}\to(-1)^{i_x+i_y+i_z}c_{\mathbf{i}\sigma}^+$. Thus we only concentrate on the hole doping case as follows. Away from half-filling, the minus sign problem~\cite{Loh1990} appears in AFQMC simulations, which here decays exponentially with $\beta$ and $N_s$ and thus prevents the access of high-precision results. In this section, we present limited AFQMC results demonstrating the behaviors of the sign problem and AFM spin correlation versus doping.

\begin{figure*}
\centering
\includegraphics[width=2.05\columnwidth]{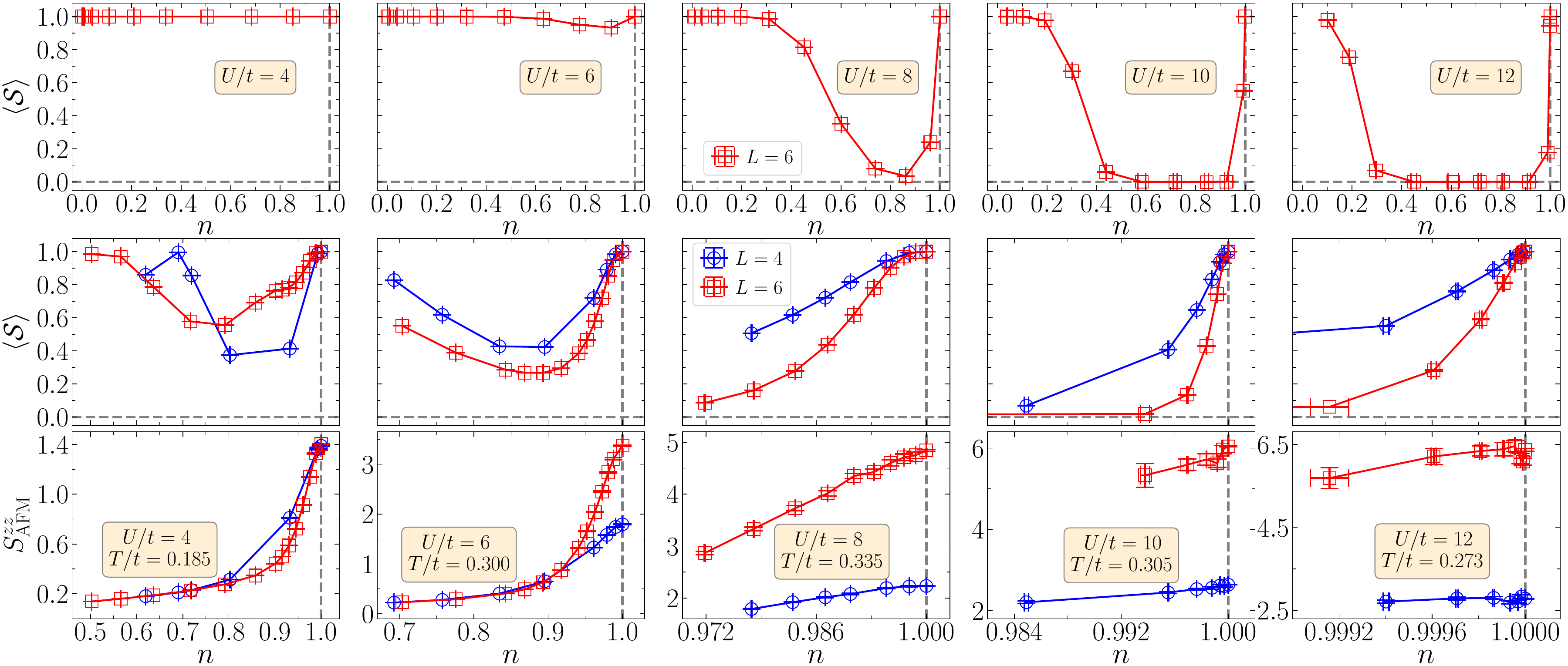}
\caption{(Top row) The sign average $\langle\mathcal{S}\rangle$ as a function of the fermion filling $n$ for $U/t=4,6,8,10,12$ from $L=6$ system at $T/t=0.36$, which is slightly higher than the highest N\'{e}el temperature $T_{\rm N}/t=0.334(2)$ for $U/t=8$ at half-filling. (Middle row) The sign average $\langle\mathcal{S}\rangle$, and (bottom row) AFM structure factor $S_{\rm AFM}^{zz}$ for $U/t=4,6,8,10,12$ as a function of $n$ at the temperature close or equal to the corresponding $T_{\rm N}$ at half-filling. Results from both $L=4$ and $L=6$ systems are presented. }
\label{fig:SignProblem}
\end{figure*}

In Fig.~\ref{fig:SignProblem}, we show the sign average $\langle\mathcal{S}\rangle$ and AFM structure factor $S_{\rm AFM}^{zz}$ as a function of fermion filling $n$, with different sets of interactions and temperatures. The top row plots the $L=6$ results of $\langle\mathcal{S}\rangle$ versus $n$ at $T/t=0.36$, slightly exceeding the highest $T_{\rm N}$ at half-filling. The severe sign problem is manifested, as $\langle\mathcal{S}\rangle$ for $U/t=8$ starts to approach zero around $n=0.86$ and it evolves into a zero zone of $n=0.4$$\sim$$0.9$ for $U/t=12$. The middle and bottom rows present the $L=4$ and $L=6$ results of $\langle\mathcal{S}\rangle$ and $S_{\rm AFM}^{zz}$ versus $n$ for $U/t=4,6,8,10,12$ at the temperature close or equal to the corresponding $T_{\rm N}$ at half-filling. Although all these parameter sets away from half-filling should fall into the normal phase, we can already observe the great challenge posed by the sign problem in these AFQMC calculations. For $U/t=4$ and $6$, the $\langle\mathcal{S}\rangle$ results are still not bad (as $\langle\mathcal{S}\rangle>0.2$ for $L=6$), and $S_{\rm AFM}^{zz}$ shows a quick suppression versus doping (as $\delta=1-n$), indicating going further away from N\'{e}el AFM ordered phase. For $U/t\ge8$, $\langle\mathcal{S}\rangle$ decays very quickly with doping. It is less than $0.1$ at $\delta\simeq0.028$ for $U/t=8$, and almost reaches zero at $\delta\simeq0.008$ for $U/t=10$ and at $\delta\simeq0.0008$ for $U/t=12$. The $S_{\rm AFM}^{zz}$ results correspondingly become noisy and are quickly suppressed by doping. These results clearly demonstrate that it is extremely hard to obtain meaningful conclusions about the AFM ordered phase from the present AFQMC simulations, especially for $U/t\ge8$.

\begin{figure}[b]
\centering
\includegraphics[width=0.99\columnwidth]{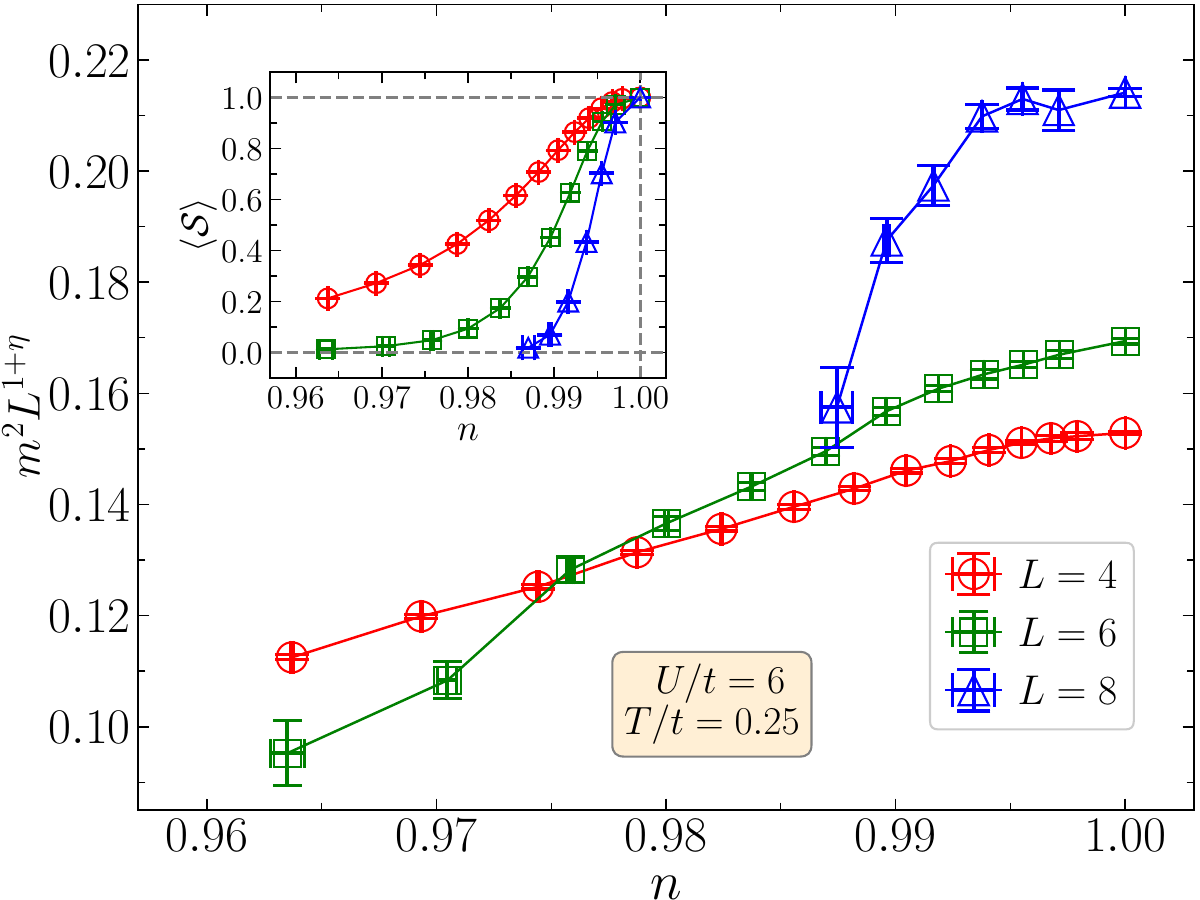}
\caption{The finite-size scaling of the mean-squared magnetization as $m^2 L^{1+\eta}$ as a function of fermion filling $n$ with $L=4,6,8$ for $U/t=6$ at $T/t=0.25$. The critical exponent $\eta=0.0375$ is used. The inset plots the corresponding results of the sign average $\langle\mathcal{S}\rangle$. }
\label{fig:mSquVsfilling}
\end{figure}

Considering the significant finite-size effect for $U/t=4$ (as discussed in Sec.~\ref{sec:PmAfmTransition}), we choose $U/t=6$ and $T/t=0.25$ to push the limit for our AFQMC calculations. For this case, a doping-driven N\'{e}el transition should exist at some specific filling, which also belongs to 3D Heisenberg universality class~\cite{Katanin2017}. In Fig.~\ref{fig:mSquVsfilling}, we present the finite-size scaling results of $m^2$ for N\'{e}el AFM order versus filling. The inset showing the sign average demonstrates the unavoidable inefficiency of these calculations, especially regarding $\langle\mathcal{S}\rangle=0.0181(6)$ at $n=0.9874(7)$ for $L=8$. The intersection of $m^2 L^{1+\eta}$ data from different system sizes signify the N\'{e}el transition, which appears at $n$$\sim$$0.975$ for $L=4$ and $L=6$. Unfortunately, the intersection between $L=6$ and $L=8$ is still not reached, which is indicated to be $n$$\sim$$0.985$ within the dropping tendency of $L=8$ results. The drifting of the intersection position to lower doping is due to the finite-size effect, which suggests the TDL transition point in the filling range of $0.985<n_c<1$. However, this is quite different from the DiagMC result in Ref.~\onlinecite{Lenihan2022} showing $n_c\simeq0.94$ for $U/t=5.8$ at $T/t=0.25$. Both the finite-size effect in our AFQMC calculations and the large uncertainties in the DiagMC result might be responsible for this inconsistency, which needs to be clarified with more precise results from alternative quantum many-body approaches.

Regarding the severe sign problem illustrated in Figs.~\ref{fig:SignProblem} and \ref{fig:mSquVsfilling}, the constrained-path (CP) AFQMC algorithm~\cite{Yuanyao2019,Zhang1999,Xiao2023} should be a promising way out to systematically study the AFM phase diagram with doping. This algorithm applies an appropriate constraint during sampling the configurations to control the sign problem, and thus to restore the polynomial scaling of computational effort. Although with the systematic bias, it was shown~\cite{Yuanyao2019,Xiao2023} that the CP-AFQMC algorithm can reach accurate results and has certain advantages in dealing with long-range orders including both the N\'{e}el AFM order and the incommensurate SDW order. We leave the CP-AFQMC study of the doped 3D repulsive Hubbard model to future work.

\section{summary and discussion}
\label{sec:Summary}

Understanding the complex interplay between thermal and quantum fluctuations in 3D repulsive Hubbard model as its original form is crucial, not only for studying fundamental properties of correlated fermions, but also for making connections with strongly correlated electron materials~\cite{Fujimori1992,Tokura1993,Inoue1995,Morikawa1995} and optical lattice experiments~\cite{Kohl2005,Joerdens2008,Schneider2008,Duarte2015,Tarruell2010,Daniel2013,Hart2015,Shao2024}. Associated with that aim, our numerical results in this work together with our companion paper~\cite{Yufeng2024} serve as a comprehensive study for 3D repulsive Hubbard model at nonzero temperature with the cutting-edge precision many-body simulations. 

In summary, we have applied the numerically exact AFQMC algorithm to clarify the finite-temperature properties of 3D repulsive Hubbard model on simple cubic lattice. Via significantly improving both the algorithmic implementations and the precise calculations of physical observables, we make important progress on both the phase diagram and thermodynamic properties of the model at half-filling, and also show the limitation of present AFQMC calculations with doping due to the sign problem. At half-filling, we have revealed the complete interaction-temperature phase diagram, which mainly consists of the N\'{e}el transitions and the MIC in the normal phase. For the former, we have presented highly accurate N\'{e}el transition temperatures to date via the finite-size scaling of the mean-squared magnetization results up to $L=20$. Especially, the elegant TABC technique is implemented to successfully overcome the strong finite-size effect for the weak interaction. As for the MIC physics, we have identified an extended crossover regime with strong AFM spin correlation between the Fermi liquid and Mott insulator in weakly and strongly interacting regimes, respectively. This is achieved by the combination analysis of AFQMC results for variously static and dynamic observables. All the related results and discussions in this paper can be taken as the supplementary expansion of our companion paper~\cite{Yufeng2024}. Besides, we have presented numerical results for the temperature dependence of double occupancy, thermal entropy, specific heat and charge compressibility, which show features generally existing in various Hubbard models. Away from half-filling, we have demonstrated the severe sign problem for the model, and have also provided limited results to show the N\'{e}el AFM ordered phase reaching out to finite doping. Our numerical results can surely provide more benchmark references for the ongoing optical lattice experiments as well as the future analytical and computational studies. 

Our work also sets up valuable foundations for the future research directions towards the missing puzzles of 3D repulsive Hubbard model. The first is the direct comparisons with results from optical lattice experiments. 
It requires the numerical simulations to involve more realistic effects encountered in experiments~\cite{Shao2024}, especially following the path of isentropic lines on the interaction-temperature plane. 
The second is to systematically investigate the phase diagram of the model away from half-filling. The most intriguing target is the precise characterization and determination of the magnetic phase diagram with doping. Other promising opportunities include the effect of NNN hopping, thermodynamics, possible pseudogap behavior and strange metallicity. 

\begin{acknowledgments}
Y.-Y.He acknowledges Yang Qi, Gang Li, Mingpu Qin, Hui Shao and Xiao Yan Xu for valuable discussions. This work was supported by the National Natural Science Foundation of China (under Grants No. 12247103, No. 12204377, and No. 12275263), the Innovation Program for Quantum Science and Technology (under Grant No. 2021ZD0301900), the Natural Science Foundation of Fujian province of China (under Grant No. 2023J02032), and the Youth Innovation Team of Shaanxi Universities.
\end{acknowledgments}

\appendix

\section{Proof of Eqs.~(\ref{eq:ChiCharge}), (\ref{eq:DouOccDerive}) and (\ref{eq:HFT})}
\label{sec:AppendixA}

In this appendix, we first present the general formulas for the derivative of observables over a specific model parameter, from which Eqs.~(\ref{eq:ChiCharge}) and (\ref{eq:DouOccDerive}) can be proved. Then we derive the finite-temperature generalization of the Hellmann-Feynman theorem as Eq.~(\ref{eq:HFT}). 

For a general observable $\langle \hat{O} \rangle=Z^{-1}{\Tr}(e^{-\beta\hat{H}}\hat{O})$, the first-order derivative over the model parameter $\alpha$ (which is not $T$ or $\beta$) can be written as 
\begin{equation}\begin{aligned}
\label{eq:derivative}
\frac{\partial \langle \hat{O} \rangle}{\partial \alpha} 
&= \frac{\partial }{\partial \alpha} {\frac{{\Tr}(e^{-\beta \hat{H}} \hat{O})}{{\Tr}(e^{-\beta \hat{H}})}} \\
&=\frac{1}{Z}\frac{\partial{\Tr}(e^{-\beta \hat{H}} \hat{O})}{\partial \alpha} + \beta \langle\hat{O}\rangle \Big\langle \frac{\partial \hat{H}}{\partial \alpha} \Big\rangle,
\end{aligned} \end{equation}
where the first term can be evaluated via the Taylor expansion for $e^{-\beta\hat{H}}$ as
\begin{equation}\begin{aligned}
\label{eq:A2}
&\frac{\partial{\Tr}(e^{-\beta \hat{H}} \hat{O})}{\partial \alpha} = {\Tr}\Big(e^{-\beta\hat{H}} \frac{\partial \hat{O}}{\partial \alpha}\Big) + \\
&{\Tr}\Big\{\Big[(-\beta)\Big(\frac{\partial \hat{H}}{\partial \alpha}\Big)  + \frac{(-\beta)^2}{2!} \Big(\hat{H} \frac{\partial \hat{H}}{\partial \alpha} + \frac{\partial \hat{H}}{\partial \alpha} \hat{H}\Big) + \cdots\Big]\hat{O}\Big\} \\
&\hspace{1.0cm}.
\end{aligned}\end{equation}
Then we can make further simplifications for several special cases. 

If the model parameter $\alpha$ satisfies $[\partial\hat{H}/\partial\alpha, \hat{H}]=0$ or the observable $\hat{O}$ is chosen with $[\hat{O}, \hat{H}]=0$, then Eq.~(\ref{eq:A2}) can be simplified as
\begin{equation}\begin{aligned}
\label{eq:A3}
\frac{\partial{\Tr}(e^{-\beta \hat{H}} \hat{O})}{\partial \alpha}
& =(-\beta){\Tr}\Big(e^{-\beta\hat{H}}\frac{\partial \hat{H}}{\partial \alpha}\hat{O}\Big) \\
& \hspace{2.0cm} + {\Tr}\Big(e^{-\beta\hat{H}}\frac{\partial \hat{O}}{\partial \alpha}\Big),
\end{aligned}\end{equation}
and thus $\partial\langle\hat{O}\rangle/\partial\alpha$ can be simplified as
\begin{equation}\begin{aligned}
\label{eq:A4}
\frac{\partial\langle\hat{O}\rangle}{\partial\alpha}
= (-\beta)\Big[\Big\langle\frac{\partial\hat{H}}{\partial\alpha} \hat{O}\Big\rangle - \Big\langle\frac{\partial \hat{H}}{\partial \alpha}\Big\rangle\langle\hat{O}\rangle\Big].
\end{aligned}\end{equation}
We have neglected the term $\langle\partial\hat{O}/\partial\alpha\rangle$, since it should typically vanish. For the Hubbard model in Eq.~(\ref{eq:Hamiltonian}), we can choose $\alpha=\mu$ with $\partial\hat{H}/\partial\mu=\hat{N}$ which satisfies $[\partial\hat{H}/\partial\alpha, \hat{H}]=0$. Then Eq.~(\ref{eq:ChiCharge}) corresponds $\hat{O}=\hat{N}=\sum_{\mathbf{i}}\hat{n}_{\mathbf{i}}$ with $\partial\hat{O}/\partial\mu=0$. Thus, we have
\begin{equation}\begin{aligned}
\label{eq:A5}
\frac{\partial\langle\hat{N}\rangle}{\partial\mu}
&= (-\beta)\big(\langle\hat{N}\hat{N}\rangle - \langle\hat{N}\rangle\langle\hat{N}\rangle\big) \\
&=(-\beta)\sum_{\mathbf{ij}}\big(\langle\hat{n}_{\mathbf{i}} \hat{n}_{\mathbf{j}}\rangle - \langle\hat{n}_{\mathbf{i}}\rangle\langle\hat{n}_{\mathbf{j}}\rangle\big).
\end{aligned}\end{equation}
With $\langle\hat{N}\rangle=N=N_s n$, we can reach the final expression of Eq.~(\ref{eq:ChiCharge}) as
\begin{equation}\begin{aligned}
\label{eq:A6}
\chi_e 
= -\frac{dn}{d\mu} 
&= -\frac{1}{N_s}\frac{\partial\langle\hat{N}\rangle}{\partial\mu} \\
&=\frac{\beta}{N_s}\sum_{\mathbf{ij}}\big(\langle\hat{n}_{\mathbf{i}} \hat{n}_{\mathbf{j}}\rangle - \langle\hat{n}_{\mathbf{i}}\rangle \langle\hat{n}_{\mathbf{j}}\rangle\big).
\end{aligned}\end{equation}
Similar formulas can be obtained for other derivatives such as $\partial E/\partial\mu=\partial\langle\hat{H}\rangle/\partial\mu$, $\partial E_K/\partial\mu$, and $\partial E_U/\partial\mu$, where $E_K$ and $E_U$ are the energies of the noninteracting and interaction terms respectively. Another choice of the parameter is $\alpha=U$ with $\hat{O}=\hat{N}$ or $\hat{H}$ satisfying $[\hat{O}, \hat{H}]=0$, and we can similarly reach the formulas for the derivatives $\partial n/\partial U$ and $\partial E/\partial U$ for fixed $\mu$ calculations (one needs to take care of the additional term of $\partial\mu/\partial U$ if $\mu$ is a function of $U$). The formula in Eq.~(\ref{eq:A4}) explicitly shows that the derivatives can be evaluated via the static correlation function $\langle(\partial\hat{H}/\partial\alpha)\hat{O}\rangle$, whose computation in AFQMC is rather straightforward. 

However, if neither $[\partial\hat{H}/\partial\alpha, \hat{H}]=0$ nor $[\hat{O}, \hat{H}]=0$ is satisfied, the above formula in Eq.~(\ref{eq:A4}) is no longer valid. Equation (\ref{eq:DouOccDerive}) actually belongs to this case, which corresponds to $\alpha=U$ and $\hat{O}=\hat{D}=\sum_{\mathbf{i}}\hat{n}_{\mathbf{i}\uparrow}\hat{n}_{\mathbf{i}\downarrow}$ (and $[\hat{D}, \hat{H}]\ne0$ is obvious). We can rewrite the Hamiltonian in Eq.~(\ref{eq:Hamiltonian}) as $\hat{H}=\hat{H}_0+U\hat{H}_I+\mu\hat{N}$ with $\hat{H}_I=\sum_{\mathbf{i}}\big[\hat{n}_{\mathbf{i}\uparrow} \hat{n}_{\mathbf{i} \downarrow} - (\hat{n}_{\mathbf{i}\uparrow} + \hat{n}_{\mathbf{i} \downarrow})/2\big]$, which results in $\partial\hat{H}/\partial U=\hat{H}_I + (\partial\mu/\partial U)\hat{N}$ (and thus $[\partial\hat{H}/\partial U, \hat{H}]
\ne 0$). Then for such general observable, we can derive the following formula as
\begin{equation}\begin{aligned}
\label{eq:A7}
\frac{\partial\langle\hat{O}\rangle}{\partial\alpha}
= -\int_0^{\beta} \Big(\Big\langle\frac{\partial\hat{H}}{\partial\alpha}(\tau)\hat{O}\Big\rangle - \Big\langle\frac{\partial\hat{H}}{\partial\alpha}(\tau)\Big\rangle\langle\hat{O}\rangle\Big) d\tau.
\end{aligned}\end{equation}
The calculation for this formula is based on the Lehmann representation, under which the observable can be expressed as
\begin{equation}\begin{aligned}
\label{eq:A8}
\langle\hat{O}\rangle 
= \frac{{\Tr}(e^{-\beta\hat{H}}\hat{O})}{Z}
= \frac{\sum_{m}e^{-\beta E_m}O_m}{\sum_{n}e^{-\beta E_n}},
\end{aligned}\end{equation}
with $O_m=\langle m|\hat{O}|m\rangle$ and $\hat{H}|m\rangle=E_m|m\rangle$. Thus, we can compute the derivative as
\begin{equation}\begin{aligned}
\label{eq:A9}
\frac{\partial\langle\hat{O}\rangle}{\partial\alpha}
&= \frac{\sum_{m} e^{-\beta E_{m}}\Big[(-\beta)\frac{\partial E_{m}}{\partial\alpha} O_{m} + \frac{\partial O_{m}}{\partial\alpha}\Big] }{\sum_{n} e^{-\beta E_{n}}} \\
&\hspace{0.4cm}
+\beta\frac{\Big(\sum_{m} e^{-\beta E_{m}}O_{m}\Big)\Big(\sum_{n} e^{-\beta E_{n}}\frac{\partial E_{n}}{\partial\alpha}\Big)}{ (\sum_{l} e^{-\beta E_{l}})^2 }.
\end{aligned}\end{equation}
The derivative $\partial E_{m}/\partial\alpha$ can be evaluated from the original version of Hellmann-Feynman theorem (at $T=0$) as $\partial E_{m}/\partial\alpha=\langle m|(\partial\hat{H}/\partial\alpha)|m\rangle$. Then based on similar derivations, we can obtain
\begin{equation}\begin{aligned}
\label{eq:A10}
&\frac{\sum_{m} e^{-\beta E_{m}}\Big[(-\beta)\frac{\partial E_{m}}{\partial\alpha} O_{m} + \frac{\partial O_{m}}{\partial\alpha}\Big] }{\sum_{n} e^{-\beta E_{n}}} \\
&\hspace{2.0cm} = -\int_0^{\beta} \Big\langle\frac{\partial\hat{H}}{\partial\alpha}(\tau)\hat{O}(0)\Big\rangle d\tau.
\end{aligned}\end{equation}
The proof for this equation is summarized in the end of this appendix. The second term in Eq.~(\ref{eq:A9}) can be computed as
\begin{equation}\begin{aligned}
\label{eq:A11}
\beta\langle\hat{O}\rangle\Big\langle\frac{\partial\hat{H}}{\partial\alpha}\Big\rangle
= \int_{0}^{\beta}\Big\langle\frac{\partial\hat{H}}{\partial\alpha}(\tau)\Big\rangle\langle\hat{O}\rangle d\tau,
\end{aligned}\end{equation}
which is written as an integral over $\tau$ intentionally. Combining Eqs.~(\ref{eq:A10}) and (\ref{eq:A11}), we can obtain the formula in Eq.~(\ref{eq:A7}). It is apparent that this formula degenerates to Eq.~(\ref{eq:A4}) under the condition of $[\partial\hat{H}/\partial\alpha, \hat{H}]=0$ or $[\hat{O}, \hat{H}]=0$. Based on Eq.~(\ref{eq:A7}), we can now compute $\partial D/\partial U$ explicitly. Since $\langle\hat{H}_I\rangle=N_sD - \langle\hat{N}\rangle/2$, we have
\begin{equation}\begin{aligned}
\label{eq:A12}
\frac{\partial D}{\partial U}
= \frac{1}{N_s}\frac{\partial\langle\hat{H}_I\rangle}{\partial U} + \frac{1}{2N_s}\frac{\partial\langle\hat{N}\rangle}{\partial U},
\end{aligned}\end{equation}
and the derivative $\partial\langle\hat{H}_I\rangle/\partial U$ can be evaluated with $\hat{O}=\hat{H}_I$ and $\alpha=U$ as
\begin{equation}\begin{aligned}
\label{eq:A13}
&\frac{\partial\langle\hat{H}_I\rangle}{\partial U}
= -\int_0^{\beta} \big(\big\langle\hat{H}_I(\tau)\hat{H}_I(0)\big\rangle - \big\langle\hat{H}_I(\tau)\big\rangle\langle\hat{H}_I\rangle\big) d\tau \\
&\hspace{0.7cm} 
-\frac{\partial\mu}{\partial U}\int_0^{\beta} \big(\big\langle\hat{N}(\tau)\hat{H}_I(0)\big\rangle - \big\langle\hat{N}(\tau)\big\rangle\langle\hat{H}_I\rangle\big) d\tau. 
\end{aligned}\end{equation}
Substituting this into Eq.~(\ref{eq:A12}), we now have
\begin{equation}\begin{aligned}
\label{eq:A14}
\frac{\partial D}{\partial U}
= &-\frac{1}{N_s}\int_0^{\beta} \big(\big\langle\hat{H}_I(\tau)\hat{H}_I(0)\big\rangle - \big\langle\hat{H}_I(\tau)\big\rangle\langle\hat{H}_I\rangle\big) d\tau \\
&-\frac{1}{N_s}\frac{\partial\mu}{\partial U}\int_0^{\beta} \big(\big\langle\hat{N}(\tau)\hat{H}_I(0)\big\rangle - \big\langle\hat{N}(\tau)\big\rangle\langle\hat{H}_I\rangle\big) d\tau.  \\
&+ \frac{1}{2N_s}\frac{\partial\langle\hat{N}\rangle}{\partial U},
\end{aligned}\end{equation}
and the second term is zero for a fixed-$\mu$ calculation, while the third term is zero for a fixed-$n$ calculation. At half-filling with $\mu=0$ and $n=1$, both second and third terms are zero. Besides, considering that the correlation function $C_{\hat{H}_I}(\tau, 0)=\langle\hat{H}_I(\tau)\hat{H}_I(0)\rangle - \langle\hat{H}_I(\tau)\rangle\langle\hat{H}_I(0)\rangle$ is symmetric about $\tau=\beta/2$, we can rewrite the integral $\int_{0}^{\beta}\langle\cdot\rangle d\tau$ in the first term as $2\int_{0}^{\beta/2}\langle\cdot\rangle d\tau$, which then arrives at the final expression in Eq.~(\ref{eq:DouOccDerive}). 

Then we focus on the finite-temperature generalization of the Hellmann-Feynman theorem. The derivation also involves the Lehmann representation. With the partition function $Z={\Tr}(e^{-\beta\hat{H}})=\sum_{m}e^{-\beta E_m}$, we can compute the free energy as
\begin{equation}
\label{eq:FreeEng}
F = -T\ln Z = -T\ln \Big({\sum_{m}} e^{-\beta E_{m}}\Big).
\end{equation}
Taking the derivative of free energy over $U$, we can obtain
\begin{equation}\begin{aligned}
\frac{\partial F}{\partial U} &= -T \frac{\sum_{m} e^{-\beta E_{m}} (-\beta)\frac{\partial E_{m}}{\partial U}}{\sum_{n} e^{-\beta E_{n}}}\\
&=\frac{\sum_{m} e^{-\beta E_{m}} \langle m| \frac{\partial \hat{H}}{\partial U}|m\rangle}{\sum_{n} e^{-\beta E_{n}}}\\
&=\frac{\sum_{m} \langle m| e^{-\beta \hat{H}} \frac{\partial \hat{H}}{\partial U}| m\rangle}{\sum_{n} e^{-\beta E_{n}}} = \Big\langle \frac{\partial \hat{H}}{\partial U} \Big\rangle.
\end{aligned}\end{equation}
This is the Hellmann-Feynman theorem at finite temperature. For the Hubbard model with $\hat{H}=\hat{H}_0+U\hat{H}_I+\mu\hat{N}$, the relation is now 
\begin{equation}
\frac{\partial F}{\partial U} = \langle \hat{H}_{I} \rangle
 + \frac{\partial\mu}{\partial U} \langle\hat{N}\rangle,
\end{equation}
which is exactly Eq.~(\ref{eq:HFT}). At the half-filling with $\mu=0$ and $n=1$, the second term is zero as $\partial\mu/\partial U=0$.

In the following, we present the proof of Eq.~(\ref{eq:A10}). For the left side of Eq.~(\ref{eq:A10}), we define 
\begin{equation}\begin{aligned}
I_1 &= -\beta \frac{\sum_{m} e^{-\beta E_{m}} \frac{\partial E_{m}}{\partial\alpha} O_{m}}{\sum_{n} e^{-\beta E_{n}}}, \\
I_2 &= \frac{\sum_{m} e^{-\beta E_{m}} \frac{\partial O_{m}}{\partial\alpha}}{\sum_{n} e^{-\beta E_{n}}}.
\end{aligned}\end{equation}
With the Hellmann-Feynman theorem at $T=0$, $I_1$ takes the form
\begin{equation}\begin{aligned}
\label{eq:DefineI1}
I_1 =-\beta \frac{\sum_{m} e^{-\beta E_{m}} \langle m|\frac{\partial \hat{H}}{\partial \alpha} |m \rangle O_{m}} {\sum_{n} e^{-\beta E_{n}}},
\end{aligned}\end{equation}
The $I_2$ can be evaluated by calculating $\partial O_{m}/\partial\alpha$ as
\begin{equation}\begin{aligned}
\label{eq:DefineI2}
I_2
&=\frac{\sum_{m} e^{-\beta E_{m}} \Big(\langle \frac{\partial}{\partial \alpha} m| \hat{O}|m\rangle
+\langle m|\hat{O}|\frac{\partial}{\partial \alpha} m \rangle \Big)  }{\sum_{n} e^{-\beta E_{n}}}\\
&=\frac{\sum_{m, n, m\neq n} e^{-\beta E_{m}} 
\langle \frac{\partial}{\partial \alpha} m|n\rangle \langle n|\hat{O}|m\rangle}{\sum_{l} e^{-\beta E_{l}}}\\
&+ \frac{\sum_{m, n, m\neq n} e^{-\beta E_{m}} \langle m|\hat{O}|n\rangle \langle n|\frac{\partial}{\partial \alpha} m \rangle}{\sum_{l} e^{-\beta E_{l}}}\\
&=\frac{\sum_{m, n,m\neq n} e^{-\beta E_{m}}\frac{\langle m|(\partial \hat{H}/\partial \alpha)|n\rangle}{E_{m} - E_{n}}
\langle n|\hat{O}|m\rangle}{\sum_{l} e^{-\beta E_{l}}}\\
&+ \frac{\sum_{m, n, m\neq n} e^{-\beta E_{m}} \langle m|\hat{O}|n\rangle 
\frac{\langle n|(\partial \hat{H}/\partial \alpha)|m\rangle}{E_{m} - E_{n}}}{\sum_{l} e^{-\beta E_{l}}},
\end{aligned}\end{equation}
where $m\neq n$ in the second equality comes from the fact that, if the state $m$ equals state $n$, the numerator can be simplified as $\sum_{m}e^{-\beta E_{m}} O_m \Big(\partial (\langle m|m\rangle)/\partial \alpha\Big)=0$. In the third equality in Eq.~(\ref{eq:DefineI2}), we have used the relations
\begin{equation}\begin{aligned}
\langle \frac{\partial}{\partial \alpha} m|n\rangle 
= \frac{\langle m|\frac{\partial \hat{H}}{\partial \alpha}|n \rangle}{E_{m}-E_{n}} \hspace{0.4cm}
\langle n|\frac{\partial }{\partial \alpha}m\rangle
= \frac{\langle n|\frac{\partial \hat{H}}{\partial \alpha}|m \rangle}{E_{m}-E_{n}},
\end{aligned}\end{equation}
which can be obtained from first calculating the $\alpha$ derivative of $\langle m|\hat{H} = \langle m |E_{m}$ and $\hat{H}|m\rangle = E_{m} |m\rangle$, and then apply $|n\rangle$ and $\langle n|$ to the right and left sides, respectively. 

For the right side of Eq.~(\ref{eq:A10}), we can compute it as 
\begin{equation}\begin{aligned}
&-\int_{0}^{\beta} \langle \frac{\partial \hat{H}}{\partial \alpha}(\tau) \hat{O}(0) \rangle d\tau\\
&=-\int_{0}^{\beta} \frac{\sum_{m} e^{-(\beta-\tau)E_{m}} \langle m|\frac{\partial \hat{H}}{\partial \alpha} e^{-\tau\hat{H}}\hat{O}|m\rangle}{\sum_{l}e^{-\beta E_{l}}}\\
&=-\int_{0}^{\beta} \frac{\sum_{m,n} e^{-(\beta-\tau)E_{m}-\tau E_{n}} \langle m|\frac{\partial \hat{H}}{\partial \alpha}
|n\rangle \langle n|\hat{O}|m\rangle}{\sum_{l}e^{-\beta E_{l}}}\\
&=-\int_{0}^{\beta} \frac{\sum_{m} e^{-\beta E_{m}} \langle m|\frac{\partial \hat{H}}{\partial \alpha}|m\rangle O_m}{\sum_{l}e^{-\beta E_{l}}}\\
&-\int_{0}^{\beta}\frac{\sum_{m,n,m\neq n} e^{-\beta E_{m}} e^{\tau(E_{m} -E_{n})} \langle m|\frac{\partial \hat{H}}{\partial \alpha}|n\rangle \langle n|\hat{O}|m\rangle}{\sum_{l}e^{-\beta E_{l}}},
\end{aligned}
\end{equation}
where the first term is exactly the $I_1$ in Eq.(\ref{eq:DefineI1}) (since the integrand has no $\tau$ dependence), and the second term can be further simplified by computing the integral as 
\begin{equation}\begin{aligned}
&-\frac{\sum_{m,n,m\neq n} e^{-\beta E_{m}} \langle m|\frac{\partial \hat{H}}{\partial \alpha}|n\rangle \langle n|\hat{O}|m\rangle \frac{e^{\beta(E_{m}-E_{n})} - 1}{E_{m} - E_{n}} }{\sum_{l}e^{-\beta E_{l}}}\\
&=\frac{\sum_{m,n,m\neq n} e^{-\beta E_{m}}  \langle n|\hat{O}|m\rangle \frac{\langle m|(\partial \hat{H}/\partial \alpha)|n\rangle}{E_{m} - E_{n}} }{\sum_{l}e^{-\beta E_{l}}}\\
&\hspace{1.0cm} -\frac{\sum_{m, n, m\neq n} e^{-\beta E_{n}} \langle n|\hat{O}|m\rangle 
\frac{\langle m|(\partial \hat{H}/\partial \alpha)|n\rangle}{E_{m} - E_{n}}}{\sum_{l} e^{-\beta E_{l}}}\\
&=\frac{\sum_{m,n,m\neq n} e^{-\beta E_{m}}  \langle n|\hat{O}|m\rangle \frac{\langle m|(\partial \hat{H}/\partial \alpha)|n\rangle}{E_{m} - E_{n}} }{\sum_{l}e^{-\beta E_{l}}}\\
&\hspace{1.0cm} +\frac{\sum_{m, n, m\neq n} e^{-\beta E_{m}} \langle m|\hat{O}|n\rangle 
\frac{\langle n|(\partial \hat{H}/\partial \alpha)|m\rangle}{E_{m} - E_{n}}}{\sum_{l} e^{-\beta E_{l}}},
\end{aligned}\end{equation}
where the final equality comes from the exchange of $m$ and $n$ labels in the second term, and the equation is exactly the $I_2$ in Eq.~(\ref{eq:DefineI2}). Therefore Eq.~(\ref{eq:A10}) is proved.

\begin{table}[h]
\centering
\caption{The signal locations in Fig.~\ref{fig:PhaseDiagram}(a).}
\begin{tabular}{|c|c|c|c|c|}
\hline
$T/t$&$U_{\mathrm{BM}}/t$ & $U_{\mathrm{MI}}/t$ & $U_{\mathrm{D}}/t$ & $U_{\mathrm{AF}}/t$\\
\hline
$0.29$& $\backslash$       & $\backslash$       & 5.7(1)	& $\backslash$        \\
\hline
$0.335$&6.0(3)  & 9.4(4)  & 6.2(1)  & 8.1(1)   \\
\hline
$0.36$&6.2(3)   & 9.6(4)  & 6.1(1)  & 8.15(10) \\
\hline
$0.40$&6.76(25) & 10.1(4) & 5.8(1)  & 8.5(1)   \\
\hline
$0.45$&7.25(25) & 10.5(5) & 5.5(1)  & 8.85(10) \\
\hline
$0.50$&7.50(35) & 11.0(5) & 5.1(1)  & 9.2(1)   \\
\hline
$0.55$&7.60(35) & 11.6(4) & 4.8(1)  & 9.3(1)   \\
\hline
$0.60$&7.6(4)   & 11.9(6) & 4.4(1)  & 9.7(1)   \\
\hline
$0.70$&7.24(35) & 12.5(5) & $\backslash$       & 9.8(1)   \\
\hline
\end{tabular}
\label{Table:A1}
\end{table}

\begin{table}[h]
\caption{The signal locations in Fig.~\ref{fig:PhaseDiagram}(b).}
\centering
\begin{tabular}{|c|c|c|c|c|}
\hline
$T/t$&$U_{\mathrm{S1}}/t$ & $U_{\mathrm{S2}}/t$ & $U_{\mathrm{F}}/t$ & $U_{\mathrm{\Sigma}}/t$\\
\hline
$0.29$& $\backslash$& $\backslash$  & 5.7(1) & 5.7(1) \\
\hline
$0.335$&5.9(1)  & 8.0(1)     & 6.3(1)        & 6.05(10)  \\
\hline
$0.36$&6.4(1)   & 8.2(1)     & 6.25(10)      & 6.0(1) \\
\hline
$0.40$&7.0(1)   & 8.25(15)   & 6.4(1)        & 5.7(1)   \\
\hline
$0.45$&7.56(25) & 8.52(35)   & 6.3(1)        & 5.2(1) \\
\hline
$0.50$&7.9(3)   & 9.2(3)     & 5.9(1)        & 4.2(1)  \\
\hline
$0.55$&8.0(1)   & 10.12(15)  & 5.85(15)      & $\backslash$  \\
\hline
$0.60$&7.7(1)   & 10.8(1)    & 5.50(15)      & $\backslash$   \\
\hline
$0.70$&7.1(1)   & 12.95(15)  & 4.9(1)        & $\backslash$   \\
\hline
\end{tabular}
\label{Table:A2}
\end{table}

\section{The signal locations in the phase diagram}
\label{sec:AppendixB}

In Tables ~\ref{Table:A1} and ~\ref{Table:A2} we list the signal locations in the phase diagram Fig.~\ref{fig:PhaseDiagram}, including the onset of bad metal $(U_{\rm BM})$ and Mott insulator ($U_{\rm MI}$), the peak locations of AFM structure factor ($U_{\rm AF}$), the inflection point of double occupancy ($U_{\rm D}$), the local maximum minimum and entropy ($U_{\rm S1}$ and $U_{\rm S2}$), the peak location of fidelity susceptibility ($U_{\rm F}$), and the crossing between $\Im\Sigma(\mathbf{k}_F, i\omega_0)$ and $\Im\Sigma(\mathbf{k}_F, i\omega_1)$ ($U_{\Sigma}$). 

\begin{figure}[h]
\centering
\includegraphics[width=0.99\columnwidth]{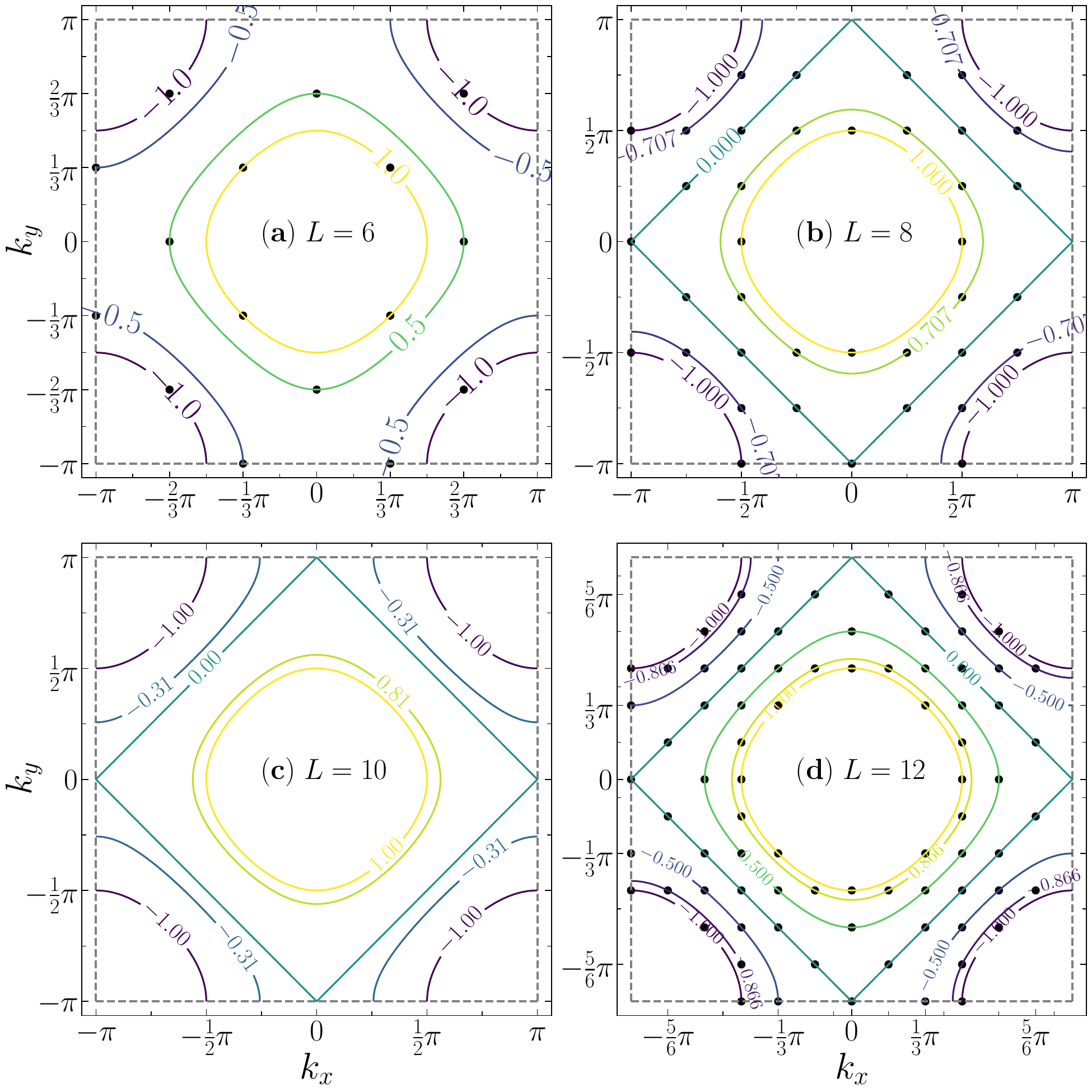}
\caption{Fermi vectors of finite-size systems with $L=6,8,10,12$. The contours represent curves of $k_x$ and $k_y$ with specific $k_z$ satisfying $-\cos(k_z)=c$, where $c$ is illustrated on the curve. Black circles mark the accessible Fermi vectors within the corresponding finite systems under periodic boundary conditions. Gray dashed lines denote the boundary of the first Brillouin zone.}
\label{figA1:FermiVector}
\end{figure}

\section{The specific heat and charge compressibility in the atomic limit}
\label{sec:AppendixC}

In this appendix, we present the calculation results in the atomic limit for Hubbard model. This limit is only valid under the condition $T\gg U$ (or $\beta U\ll 1$). Note that the atomic limit of Hubbard model is not simply the single atom case with the electron filling staying fixed. Instead, the Hubbard model in atomic limit degenerates to the single-site Hamiltonian in grand canonical ensemble [as Eq.~(\ref{eq:SingleSite})], and the hopping term between different sites is statistically transformed into the fermion filling fluctuation via the chemical potential term. Thus, the fermions of Hubbard model in the atomic limit can still move in the system via local fluctuation of the fermion density, which induces the resistivity $\rho=\chi_e^{-1}/D_{\rm diff}$ from the Ernst-Einstein relation.

The single-site Hamiltonian is given by
\begin{equation}
\label{eq:SingleSite}
\hat{h} = U\hat{n}_{\mathbf{i}\uparrow}\hat{n}_{\mathbf{i} \downarrow} +\mu(\hat{n}_{\mathbf{i}\uparrow}+ \hat{n}_{\mathbf{i}\downarrow}).
\end{equation}
The Hilbert space only has four states, and we can compute observables analytically. The fermion filling $n$ takes the equation as
\begin{equation}\begin{aligned}
\label{eq:particle}
n = \frac{2e^{-\beta\mu}+2e^{-\beta(U+2\mu)}}{2e^{-\beta\mu}+1+e^{-\beta(U+2\mu)}}
\end{aligned}\end{equation}
where $\mu=-U/2$ with $n=1$ corresponds to half-filling. For a general filling $n$, the chemical potential can be solved from the equation
\begin{equation}
e^{-\beta \mu} = \frac{(n-1) + \sqrt{(n-1)^2 - e^{-\beta U}n(n-2)}}{e^{-\beta U}(2-n)},
\end{equation}
which is further simplified within $\beta U\ll 1$ as
\begin{equation}\begin{aligned}
\label{eq2:mu}
-\beta \mu = \ln\frac{n}{2-n} + \frac{\beta U n}{2} + \mathcal{O}[(\beta U)^2]. 
\end{aligned}\end{equation}

The charge compressibility $\chi_{e}$ from Eq.~(\ref{eq:particle}) can be evaluated as
\begin{equation}\begin{aligned}
\chi_{e} 
= - \frac{\dd n}{\dd \mu}
= \beta \Big(2n-n^2- \frac{n}{e^{-\beta U}e^{-\beta\mu}+1}\Big).
\end{aligned}\end{equation}
Then its inverse $\chi_{e}^{-1}$ can be obtained within $\beta U\ll 1$ as
\begin{equation}\begin{aligned}
\label{eq:inversechi}
\chi_{e}^{-1} 
= T \Big( n-\frac{n^2}{2} \Big)^{-1} + \frac{U}{2} + \mathcal{O}(\beta U).
\end{aligned}\end{equation}

The specific heat $C_{v}$ is computed as the temperature derivative of the energy density $e$, which takes the form
\begin{equation}\begin{aligned}
\label{eq:energy}
e = \frac{2\mu e^{-\beta\mu} + (U+2\mu)e^{-\beta(U+2\mu)}}{2e^{-\beta\mu}+1+e^{-\beta(U+2\mu)}},
\end{aligned}\end{equation}
which can be simplified at half-filling ($\mu=-U/2$) as
\begin{equation}\begin{aligned}
\label{eq:energy1}
e = \frac{-Ue^{\beta U/2}}{2+2e^{\beta U/2}}.
\end{aligned}\end{equation}
Then the specific heat at half-filling is evaluated as 
\begin{equation}
C_{v} = \frac{\dd e}{\dd T} = \frac{U^2\beta^2}{4} \frac{e^{-\beta U/2}}{(e^{-\beta U/2}+1)^2}.
\end{equation}
The appearance of the charge peak corresponds to $\dd C_v/\dd T=0$, from which we can get the equation
\begin{equation}
\tanh{\frac{\beta_{\mathrm{charge}} U}{4}} = \frac{4}{\beta_{\mathrm{charge}} U}.
\end{equation}
Considering that the positive solution of $\tanh{x} = x^{-1}$ is about $1.19968$, the solution of $\dd C_v/\dd T=0$ is $\beta_{\mathrm{charge}}\approx 4.8/U$ which presents the corresponding temperature of the charge peak as $T_{\mathrm{charge}}\approx U/4.8=0.208U$.

\begin{figure}[h]
\centering
\includegraphics[width=0.95\columnwidth]{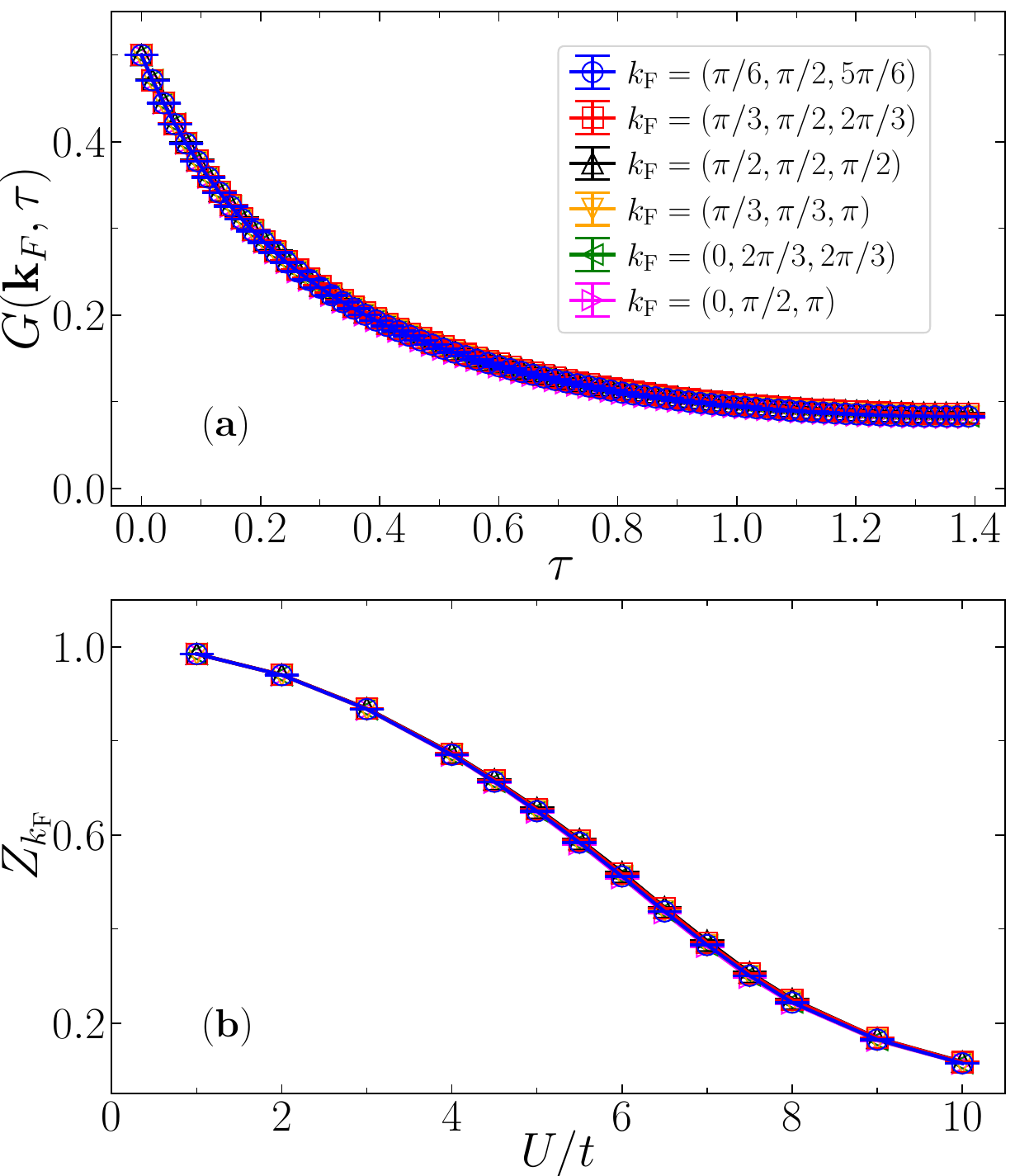}
\caption{(a) The dynamical single-particle Green's function $G(\mathbf{k}_F,\tau)$ for six independent Fermi vectors at $T/t=0.36$ for $U/t=8$ from $L=12$. The results almost coincide for all these Fermi vectors. (b) Quasiparticle weight $Z(k_F)$ versus $U/t$ at $T/t=0.36$ from $L=12$ for the six independent Fermi vectors shown in panel (a). For all interaction strengths plotted, the $Z(k_F)$ results are well consistent at different Fermi vectors showing the Fermi surface isotropy. }
\label{figA2:CompareDiffKF}
\end{figure}

\begin{figure}[h]
\centering
\includegraphics[width=0.99\columnwidth]{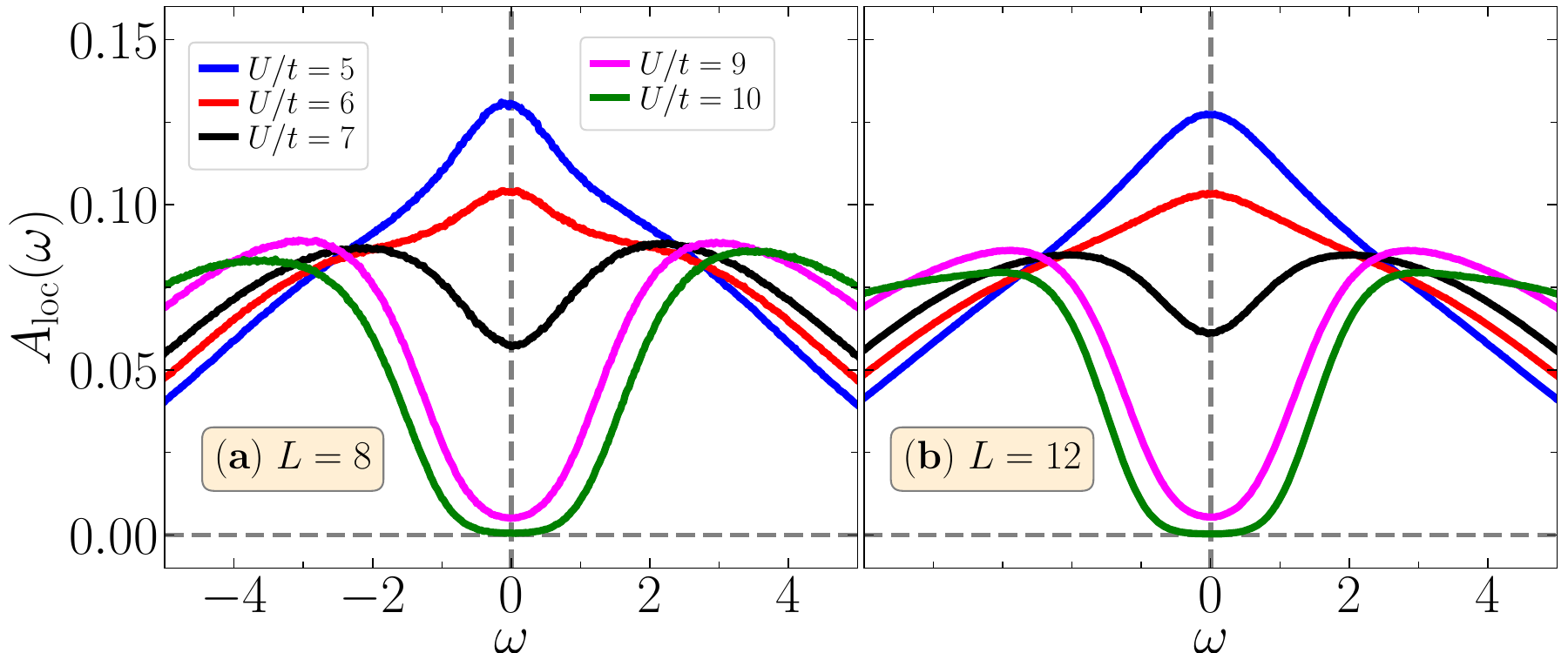}
\caption{Local fermion spectrum $A_{\mathrm{loc}}(\omega)$ at $T/t=0.36$ from $L=8$ (left) and $L=12$ (right). Only small quantitative difference can be observed, while they present consistent results of $U_{\rm BM}$. }
\label{figA3:LocalSpectraFSS}
\end{figure}

\section{The Fermi surface isotropy in 3D}
\label{sec:AppendixD}

For the kinetic energy dispersion on 3D simple cubic lattice, the Fermi vector $\mathbf{k}_F = (k_x, k_y, k_z)$ satisfies $\varepsilon_{\mathbf{k}}+\mu=0$, which at half-filling ($\mu=0$) transfers to $\cos(k_x) + \cos(k_y) + \cos(k_z) = 0$. In Fig.~\ref{figA1:FermiVector}, we demonstrate the accessible Fermi vectors for different finite-size systems of $L=6,8,10,12$, by plotting the curves of $k_x$ and $k_y$ for various $k_z$ values, with contour lines denoting $-\cos(k_z)$. The number of accessible Fermi vectors is non-monotonic versus $L$, for example, no Fermi vectors for $L=10$. This non-monotonic behavior indicates the strong finite-size effects, especially for weak interactions where the properties of the system is dominated by the Fermi surface structure. This also explains the oscillating effect of $m^2$ results for $U/t=4$ as discussed in Sec.~\ref{sec:PmAfmTransition}. 

In the main text, we determine the center value of the crossover boundary $U_{\mathrm{MI}}$ from the averaged quasiparticle weight over different Fermi vectors. Here we present the results of quasiparticle weights for independent Fermi vectors to validate the average process. As shown in Fig.~\ref{figA1:FermiVector}, for $L=12$, there are six independent Fermi vectors $(\pi/6, \pi/2, 5\pi/6)$, $(\pi/3, \pi/2, 2\pi/3)$, $(\pi/2, \pi/2, \pi/2)$, $(\pi/3, \pi/3, \pi)$, $(0, 2\pi/3, 2\pi/3)$, and $(0, \pi/2, \pi)$. Figure \ref{figA2:CompareDiffKF} plots the dynamical single-particle Green's function $G(\mathbf{k}_F,\tau)$ for $U/t=8$ and the quasiparticle weight $Z_{k_F}$ versus $U/t$ for the above six Fermi vectors, at temperature $T/t=0.36$. It is apparent that both $G(\mathbf{k}_F,\tau)$ and $Z_{k_F}$ show almost isotropic results across these different $\mathbf{k}_F$ points. Moreover, this behavior is insensitive to the temperature and system size. 

\section{The finite-size effect for various quantities}
\label{sec:AppendixE}

In this appendix, we present more results to demonstrate that the finite-size effect of the AFQMC results presented in the main text is negligible or does not affect the crossover boundaries determined from various quantities, including the local fermion spectrum $A_{\rm loc}(\omega)$, thermal entropy density $S/N_s$, AFM structure factor $S_{\rm AFM}^{zz}$, and the double occupancy $D$. 

\begin{figure}[h]
\centering
\includegraphics[width=0.95\columnwidth]{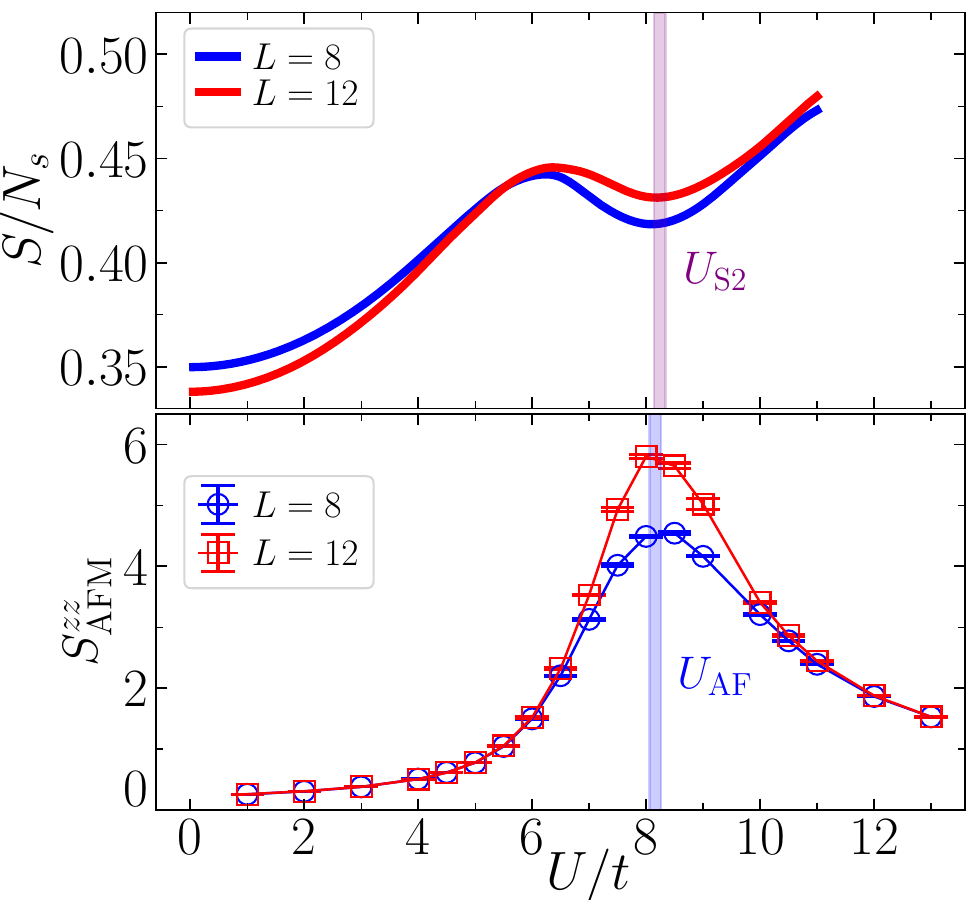}
\caption{The thermal entropy density $S/N_s$ (upper panel) and AFM structure factor $S_{\rm AFM}^{zz}$ (lower panel) versus interaction strength at $T/t=0.36$ from $L=8$ and $L=12$. The locations of the local minimum for $S/N_s$ and the peak for $S_{\rm AFM}^{zz}$ are consistent within $L=8$ and $12$. }
\label{figA4:ThermQuantFSS}
\end{figure}

Figure \ref{figA3:LocalSpectraFSS} shows the $A_{\rm loc}(\omega)$ results with different $U/t$ from $L=8$ and $L=12$ at $T/t=0.36$, where the finite-size effect is more pronounced comparing to higher temperatures. Only slight difference in $A_{\rm loc}(\omega)$ can be observed bewteen these two systems, which nevertheless presents consistent $U_{\rm BM}$ results within the uncertainties. This finite-size effect tends to vanish towards higher temperature. 

Figure \ref{figA4:ThermQuantFSS} displays the results of $S/N_s$ and $S_{\rm AFM}^{zz}$ from $L=8$ and $12$ at $T/t=0.36$. Although the absolute values of both quantities do not saturate (especially for $S_{\rm AFM}^{zz}$), both the local minimum location in $S/N_s$ (as $U_{\rm S2}$) and the peak location in $S_{\rm AFM}^{zz}$ (as $U_{\rm AF}$) already show convergence within $L=12$. For high temperatures, the $L=8$ results are already enough to present the converged results of $U_{\rm S1}$, $U_{\rm S2}$, and $U_{\rm AF}$. Moreover, we can clearly observe that, at $T/t=0.36$, the results of $U_{\rm S2}$ and $U_{\rm AF}$ coincide within the uncertainties. 

\begin{figure}[h]
\centering
\includegraphics[width=0.95\columnwidth]{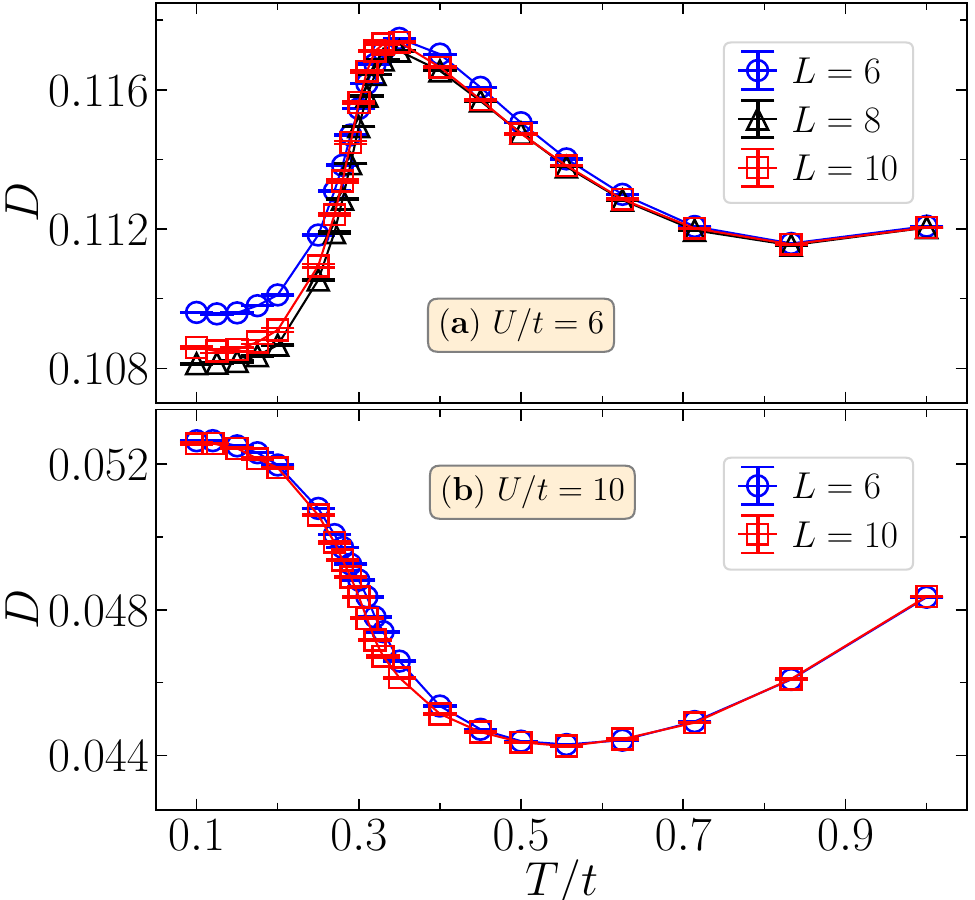}
\caption{Results of double occupancy for (a) $U/t=6$ from $L=6,8,10$ and (b) $U/t=10$ from $L=6,10$. For both interactions, the finite-size effect is only visible around the N\'{e}el transition, while it is negligible within $L=10$ for $T/t\ge 0.50$. }
\label{figA5:DouOcc}
\end{figure}

Figure \ref{figA5:DouOcc} plots the results of $D$ for $U/t=6$ from $L=6,8,10$ and for $U/t=10$ from $L=6,10$. The convergences for both interactions are apparent for $T/t\ge0.50$ within $L=10$. The only visible finite-size effect exists around the N\'{e}el transition (and also at low temperatures for $U/t=6$). Moreover, within $L=10$, the positions of the local minimum and maximum of $D$ already show convergence. In the main text, we explain the anomalous decrease of $D$ up heating from the view of lowering temperature. It is also clear from the opposite direction. For $U/t=6$ with increasing temperature, the anomalous decrease of $D$ is due to the incipient localization effects in a strongly correlated Fermi liquid~\cite{Georges1992} since further heating drives the system to the bad metal state. The increased localization also yields a gain in free energy~\cite{Georges1992}. So for weak interactions, this anomalous behavior is closely related to the correlated Fermi liquid state. For $U/t=10$ at low temperature, spin fluctuations dominates the system and the double occupancy is mainly contributed by the spin-exchange physics corresponding to the effective Heisenberg model, which are fully quantum fluctuations. Upon heating, thermal fluctuation sets in and it begins to suppress the above quantum fluctuations, resulting in the decrease of double occupancy. So for strong interaction, this anomalous behavior is instead overtaken by the spin-exchange physics.

\bibliography{3DHubbardRef_Long}

\begin{thebibliography}{132}%
\makeatletter
\providecommand \@ifxundefined [1]{%
 \@ifx{#1\undefined}
}%
\providecommand \@ifnum [1]{%
 \ifnum #1\expandafter \@firstoftwo
 \else \expandafter \@secondoftwo
 \fi
}%
\providecommand \@ifx [1]{%
 \ifx #1\expandafter \@firstoftwo
 \else \expandafter \@secondoftwo
 \fi
}%
\providecommand \natexlab [1]{#1}%
\providecommand \enquote  [1]{``#1''}%
\providecommand \bibnamefont  [1]{#1}%
\providecommand \bibfnamefont [1]{#1}%
\providecommand \citenamefont [1]{#1}%
\providecommand \href@noop [0]{\@secondoftwo}%
\providecommand \href [0]{\begingroup \@sanitize@url \@href}%
\providecommand \@href[1]{\@@startlink{#1}\@@href}%
\providecommand \@@href[1]{\endgroup#1\@@endlink}%
\providecommand \@sanitize@url [0]{\catcode `\\12\catcode `\$12\catcode
  `\&12\catcode `\#12\catcode `\^12\catcode `\_12\catcode `\%12\relax}%
\providecommand \@@startlink[1]{}%
\providecommand \@@endlink[0]{}%
\providecommand \url  [0]{\begingroup\@sanitize@url \@url }%
\providecommand \@url [1]{\endgroup\@href {#1}{\urlprefix }}%
\providecommand \urlprefix  [0]{URL }%
\providecommand \Eprint [0]{\href }%
\providecommand \doibase [0]{https://doi.org/}%
\providecommand \selectlanguage [0]{\@gobble}%
\providecommand \bibinfo  [0]{\@secondoftwo}%
\providecommand \bibfield  [0]{\@secondoftwo}%
\providecommand \translation [1]{[#1]}%
\providecommand \BibitemOpen [0]{}%
\providecommand \bibitemStop [0]{}%
\providecommand \bibitemNoStop [0]{.\EOS\space}%
\providecommand \EOS [0]{\spacefactor3000\relax}%
\providecommand \BibitemShut  [1]{\csname bibitem#1\endcsname}%
\let\auto@bib@innerbib\@empty
\bibitem [{\citenamefont {Hubbard}\ and\ \citenamefont
  {Flowers}(1963)}]{Hubbard1963}%
  \BibitemOpen
  \bibfield  {author} {\bibinfo {author} {\bibfnamefont {J.}~\bibnamefont
  {Hubbard}}\ and\ \bibinfo {author} {\bibfnamefont {B.~H.}\ \bibnamefont
  {Flowers}},\ }\bibfield  {title} {\bibinfo {title} {Electron correlations in
  narrow energy bands},\ }\href {https://doi.org/10.1098/rspa.1963.0204}
  {\bibfield  {journal} {\bibinfo  {journal} {Proc. R. Soc. Lond. Ser. A}\
  }\textbf {\bibinfo {volume} {276}},\ \bibinfo {pages} {238} (\bibinfo {year}
  {1963})}\BibitemShut {NoStop}%
\bibitem [{\citenamefont {Kanamori}(1963)}]{Kanamori1963}%
  \BibitemOpen
  \bibfield  {author} {\bibinfo {author} {\bibfnamefont {J.}~\bibnamefont
  {Kanamori}},\ }\bibfield  {title} {\bibinfo {title} {Electron correlation and
  ferromagnetism of transition metals},\ }\href
  {https://doi.org/10.1143/PTP.30.275} {\bibfield  {journal} {\bibinfo
  {journal} {Prog. Theor. Phys}\ }\textbf {\bibinfo {volume} {30}},\ \bibinfo
  {pages} {275} (\bibinfo {year} {1963})}\BibitemShut {NoStop}%
\bibitem [{\citenamefont {Gutzwiller}(1963)}]{Gutzwiller1963}%
  \BibitemOpen
  \bibfield  {author} {\bibinfo {author} {\bibfnamefont {M.~C.}\ \bibnamefont
  {Gutzwiller}},\ }\bibfield  {title} {\bibinfo {title} {Effect of correlation
  on the ferromagnetism of transition metals},\ }\href
  {https://doi.org/10.1103/PhysRevLett.10.159} {\bibfield  {journal} {\bibinfo
  {journal} {Phys. Rev. Lett.}\ }\textbf {\bibinfo {volume} {10}},\ \bibinfo
  {pages} {159} (\bibinfo {year} {1963})}\BibitemShut {NoStop}%
\bibitem [{\citenamefont {Fujimori}\ \emph {et~al.}(1992)\citenamefont
  {Fujimori}, \citenamefont {Hase}, \citenamefont {Namatame}, \citenamefont
  {Fujishima}, \citenamefont {Tokura}, \citenamefont {Eisaki}, \citenamefont
  {Uchida}, \citenamefont {Takegahara},\ and\ \citenamefont
  {de~Groot}}]{Fujimori1992}%
  \BibitemOpen
  \bibfield  {author} {\bibinfo {author} {\bibfnamefont {A.}~\bibnamefont
  {Fujimori}}, \bibinfo {author} {\bibfnamefont {I.}~\bibnamefont {Hase}},
  \bibinfo {author} {\bibfnamefont {H.}~\bibnamefont {Namatame}}, \bibinfo
  {author} {\bibfnamefont {Y.}~\bibnamefont {Fujishima}}, \bibinfo {author}
  {\bibfnamefont {Y.}~\bibnamefont {Tokura}}, \bibinfo {author} {\bibfnamefont
  {H.}~\bibnamefont {Eisaki}}, \bibinfo {author} {\bibfnamefont
  {S.}~\bibnamefont {Uchida}}, \bibinfo {author} {\bibfnamefont
  {K.}~\bibnamefont {Takegahara}},\ and\ \bibinfo {author} {\bibfnamefont
  {F.~M.~F.}\ \bibnamefont {de~Groot}},\ }\bibfield  {title} {\bibinfo {title}
  {Evolution of the spectral function in mott-hubbard systems with
  ${\mathit{d}}^{1}$ configuration},\ }\href
  {https://doi.org/10.1103/PhysRevLett.69.1796} {\bibfield  {journal} {\bibinfo
   {journal} {Phys. Rev. Lett.}\ }\textbf {\bibinfo {volume} {69}},\ \bibinfo
  {pages} {1796} (\bibinfo {year} {1992})}\BibitemShut {NoStop}%
\bibitem [{\citenamefont {Tokura}\ \emph {et~al.}(1993)\citenamefont {Tokura},
  \citenamefont {Taguchi}, \citenamefont {Okada}, \citenamefont {Fujishima},
  \citenamefont {Arima}, \citenamefont {Kumagai},\ and\ \citenamefont
  {Iye}}]{Tokura1993}%
  \BibitemOpen
  \bibfield  {author} {\bibinfo {author} {\bibfnamefont {Y.}~\bibnamefont
  {Tokura}}, \bibinfo {author} {\bibfnamefont {Y.}~\bibnamefont {Taguchi}},
  \bibinfo {author} {\bibfnamefont {Y.}~\bibnamefont {Okada}}, \bibinfo
  {author} {\bibfnamefont {Y.}~\bibnamefont {Fujishima}}, \bibinfo {author}
  {\bibfnamefont {T.}~\bibnamefont {Arima}}, \bibinfo {author} {\bibfnamefont
  {K.}~\bibnamefont {Kumagai}},\ and\ \bibinfo {author} {\bibfnamefont
  {Y.}~\bibnamefont {Iye}},\ }\bibfield  {title} {\bibinfo {title} {Filling
  dependence of electronic properties on the verge of metal--mott-insulator
  transition in
  ${\mathrm{sr}}_{1\mathrm{\ensuremath{-}}\mathit{x}}$${\mathrm{la}}_{\mathit{x}}$${\mathrm{tio}}_{3}$},\
  }\href {https://doi.org/10.1103/PhysRevLett.70.2126} {\bibfield  {journal}
  {\bibinfo  {journal} {Phys. Rev. Lett.}\ }\textbf {\bibinfo {volume} {70}},\
  \bibinfo {pages} {2126} (\bibinfo {year} {1993})}\BibitemShut {NoStop}%
\bibitem [{\citenamefont {Inoue}\ \emph {et~al.}(1995)\citenamefont {Inoue},
  \citenamefont {Hase}, \citenamefont {Aiura}, \citenamefont {Fujimori},
  \citenamefont {Haruyama}, \citenamefont {Maruyama},\ and\ \citenamefont
  {Nishihara}}]{Inoue1995}%
  \BibitemOpen
  \bibfield  {author} {\bibinfo {author} {\bibfnamefont {I.~H.}\ \bibnamefont
  {Inoue}}, \bibinfo {author} {\bibfnamefont {I.}~\bibnamefont {Hase}},
  \bibinfo {author} {\bibfnamefont {Y.}~\bibnamefont {Aiura}}, \bibinfo
  {author} {\bibfnamefont {A.}~\bibnamefont {Fujimori}}, \bibinfo {author}
  {\bibfnamefont {Y.}~\bibnamefont {Haruyama}}, \bibinfo {author}
  {\bibfnamefont {T.}~\bibnamefont {Maruyama}},\ and\ \bibinfo {author}
  {\bibfnamefont {Y.}~\bibnamefont {Nishihara}},\ }\bibfield  {title} {\bibinfo
  {title} {Systematic development of the spectral function in the
  $3{\mathit{d}}^{1}$ mott-hubbard system
  ${\mathrm{ca}}_{1\ensuremath{-}\mathit{x}}{\mathrm{sr}}_{\mathit{x}}{\mathrm{vo}}_{3}$},\
  }\href {https://doi.org/10.1103/PhysRevLett.74.2539} {\bibfield  {journal}
  {\bibinfo  {journal} {Phys. Rev. Lett.}\ }\textbf {\bibinfo {volume} {74}},\
  \bibinfo {pages} {2539} (\bibinfo {year} {1995})}\BibitemShut {NoStop}%
\bibitem [{\citenamefont {Morikawa}\ \emph {et~al.}(1995)\citenamefont
  {Morikawa}, \citenamefont {Mizokawa}, \citenamefont {Kobayashi},
  \citenamefont {Fujimori}, \citenamefont {Eisaki}, \citenamefont {Uchida},
  \citenamefont {Iga},\ and\ \citenamefont {Nishihara}}]{Morikawa1995}%
  \BibitemOpen
  \bibfield  {author} {\bibinfo {author} {\bibfnamefont {K.}~\bibnamefont
  {Morikawa}}, \bibinfo {author} {\bibfnamefont {T.}~\bibnamefont {Mizokawa}},
  \bibinfo {author} {\bibfnamefont {K.}~\bibnamefont {Kobayashi}}, \bibinfo
  {author} {\bibfnamefont {A.}~\bibnamefont {Fujimori}}, \bibinfo {author}
  {\bibfnamefont {H.}~\bibnamefont {Eisaki}}, \bibinfo {author} {\bibfnamefont
  {S.}~\bibnamefont {Uchida}}, \bibinfo {author} {\bibfnamefont
  {F.}~\bibnamefont {Iga}},\ and\ \bibinfo {author} {\bibfnamefont
  {Y.}~\bibnamefont {Nishihara}},\ }\bibfield  {title} {\bibinfo {title}
  {Spectral weight transfer and mass renormalization in mott-hubbard systems
  ${\mathrm{srvo}}_{3}$ and ${\mathrm{cavo}}_{3}$: Influence of long-range
  coulomb interaction},\ }\href {https://doi.org/10.1103/PhysRevB.52.13711}
  {\bibfield  {journal} {\bibinfo  {journal} {Phys. Rev. B}\ }\textbf {\bibinfo
  {volume} {52}},\ \bibinfo {pages} {13711} (\bibinfo {year}
  {1995})}\BibitemShut {NoStop}%
\bibitem [{\citenamefont {Arovas}\ \emph {et~al.}(2022)\citenamefont {Arovas},
  \citenamefont {Berg}, \citenamefont {Kivelson},\ and\ \citenamefont
  {Raghu}}]{Arovas2022}%
  \BibitemOpen
  \bibfield  {author} {\bibinfo {author} {\bibfnamefont {D.~P.}\ \bibnamefont
  {Arovas}}, \bibinfo {author} {\bibfnamefont {E.}~\bibnamefont {Berg}},
  \bibinfo {author} {\bibfnamefont {S.~A.}\ \bibnamefont {Kivelson}},\ and\
  \bibinfo {author} {\bibfnamefont {S.}~\bibnamefont {Raghu}},\ }\bibfield
  {title} {\bibinfo {title} {The hubbard model},\ }\href
  {https://doi.org/https://doi.org/10.1146/annurev-conmatphys-031620-102024}
  {\bibfield  {journal} {\bibinfo  {journal} {Annual Review of Condensed Matter
  Physics}\ }\textbf {\bibinfo {volume} {13}},\ \bibinfo {pages} {239}
  (\bibinfo {year} {2022})}\BibitemShut {NoStop}%
\bibitem [{\citenamefont {Qin}\ \emph {et~al.}(2022)\citenamefont {Qin},
  \citenamefont {Schäfer}, \citenamefont {Andergassen}, \citenamefont
  {Corboz},\ and\ \citenamefont {Gull}}]{Qin2022}%
  \BibitemOpen
  \bibfield  {author} {\bibinfo {author} {\bibfnamefont {M.}~\bibnamefont
  {Qin}}, \bibinfo {author} {\bibfnamefont {T.}~\bibnamefont {Schäfer}},
  \bibinfo {author} {\bibfnamefont {S.}~\bibnamefont {Andergassen}}, \bibinfo
  {author} {\bibfnamefont {P.}~\bibnamefont {Corboz}},\ and\ \bibinfo {author}
  {\bibfnamefont {E.}~\bibnamefont {Gull}},\ }\bibfield  {title} {\bibinfo
  {title} {The hubbard model: A computational perspective},\ }\href
  {https://doi.org/https://doi.org/10.1146/annurev-conmatphys-090921-033948}
  {\bibfield  {journal} {\bibinfo  {journal} {Annual Review of Condensed Matter
  Physics}\ }\textbf {\bibinfo {volume} {13}},\ \bibinfo {pages} {275}
  (\bibinfo {year} {2022})}\BibitemShut {NoStop}%
\bibitem [{\citenamefont {Dagotto}(1994)}]{Dagotto1994}%
  \BibitemOpen
  \bibfield  {author} {\bibinfo {author} {\bibfnamefont {E.}~\bibnamefont
  {Dagotto}},\ }\bibfield  {title} {\bibinfo {title} {Correlated electrons in
  high-temperature superconductors},\ }\href
  {https://doi.org/10.1103/RevModPhys.66.763} {\bibfield  {journal} {\bibinfo
  {journal} {Rev. Mod. Phys.}\ }\textbf {\bibinfo {volume} {66}},\ \bibinfo
  {pages} {763} (\bibinfo {year} {1994})}\BibitemShut {NoStop}%
\bibitem [{\citenamefont {Bulut}(2002)}]{Bulut2010}%
  \BibitemOpen
  \bibfield  {author} {\bibinfo {author} {\bibfnamefont {N.}~\bibnamefont
  {Bulut}},\ }\bibfield  {title} {\bibinfo {title} {$d_{x^2-y^2}$
  superconductivity and the hubbard model},\ }\href
  {https://doi.org/10.1080/00018730210155142} {\bibfield  {journal} {\bibinfo
  {journal} {Advances in Physics}\ }\textbf {\bibinfo {volume} {51}},\ \bibinfo
  {pages} {1587} (\bibinfo {year} {2002})}\BibitemShut {NoStop}%
\bibitem [{\citenamefont {LeBlanc}\ \emph {et~al.}(2015)\citenamefont
  {LeBlanc}, \citenamefont {Antipov}, \citenamefont {Becca}, \citenamefont
  {Bulik}, \citenamefont {Chan}, \citenamefont {Chung}, \citenamefont {Deng},
  \citenamefont {Ferrero}, \citenamefont {Henderson}, \citenamefont
  {Jim\'enez-Hoyos}, \citenamefont {Kozik}, \citenamefont {Liu}, \citenamefont
  {Millis}, \citenamefont {Prokof'ev}, \citenamefont {Qin}, \citenamefont
  {Scuseria}, \citenamefont {Shi}, \citenamefont {Svistunov}, \citenamefont
  {Tocchio}, \citenamefont {Tupitsyn}, \citenamefont {White}, \citenamefont
  {Zhang}, \citenamefont {Zheng}, \citenamefont {Zhu},\ and\ \citenamefont
  {Gull}}]{LeBlanc2015}%
  \BibitemOpen
  \bibfield  {author} {\bibinfo {author} {\bibfnamefont {J.~P.~F.}\
  \bibnamefont {LeBlanc}}, \bibinfo {author} {\bibfnamefont {A.~E.}\
  \bibnamefont {Antipov}}, \bibinfo {author} {\bibfnamefont {F.}~\bibnamefont
  {Becca}}, \bibinfo {author} {\bibfnamefont {I.~W.}\ \bibnamefont {Bulik}},
  \bibinfo {author} {\bibfnamefont {G.~K.-L.}\ \bibnamefont {Chan}}, \bibinfo
  {author} {\bibfnamefont {C.-M.}\ \bibnamefont {Chung}}, \bibinfo {author}
  {\bibfnamefont {Y.}~\bibnamefont {Deng}}, \bibinfo {author} {\bibfnamefont
  {M.}~\bibnamefont {Ferrero}}, \bibinfo {author} {\bibfnamefont {T.~M.}\
  \bibnamefont {Henderson}}, \bibinfo {author} {\bibfnamefont {C.~A.}\
  \bibnamefont {Jim\'enez-Hoyos}}, \bibinfo {author} {\bibfnamefont
  {E.}~\bibnamefont {Kozik}}, \bibinfo {author} {\bibfnamefont {X.-W.}\
  \bibnamefont {Liu}}, \bibinfo {author} {\bibfnamefont {A.~J.}\ \bibnamefont
  {Millis}}, \bibinfo {author} {\bibfnamefont {N.~V.}\ \bibnamefont
  {Prokof'ev}}, \bibinfo {author} {\bibfnamefont {M.}~\bibnamefont {Qin}},
  \bibinfo {author} {\bibfnamefont {G.~E.}\ \bibnamefont {Scuseria}}, \bibinfo
  {author} {\bibfnamefont {H.}~\bibnamefont {Shi}}, \bibinfo {author}
  {\bibfnamefont {B.~V.}\ \bibnamefont {Svistunov}}, \bibinfo {author}
  {\bibfnamefont {L.~F.}\ \bibnamefont {Tocchio}}, \bibinfo {author}
  {\bibfnamefont {I.~S.}\ \bibnamefont {Tupitsyn}}, \bibinfo {author}
  {\bibfnamefont {S.~R.}\ \bibnamefont {White}}, \bibinfo {author}
  {\bibfnamefont {S.}~\bibnamefont {Zhang}}, \bibinfo {author} {\bibfnamefont
  {B.-X.}\ \bibnamefont {Zheng}}, \bibinfo {author} {\bibfnamefont
  {Z.}~\bibnamefont {Zhu}},\ and\ \bibinfo {author} {\bibfnamefont
  {E.}~\bibnamefont {Gull}} (\bibinfo {collaboration} {Simons Collaboration on
  the Many-Electron Problem}),\ }\bibfield  {title} {\bibinfo {title}
  {Solutions of the two-dimensional hubbard model: Benchmarks and results from
  a wide range of numerical algorithms},\ }\href
  {https://doi.org/10.1103/PhysRevX.5.041041} {\bibfield  {journal} {\bibinfo
  {journal} {Phys. Rev. X}\ }\textbf {\bibinfo {volume} {5}},\ \bibinfo {pages}
  {041041} (\bibinfo {year} {2015})}\BibitemShut {NoStop}%
\bibitem [{\citenamefont {Wu}\ \emph {et~al.}(2018)\citenamefont {Wu},
  \citenamefont {Scheurer}, \citenamefont {Chatterjee}, \citenamefont
  {Sachdev}, \citenamefont {Georges},\ and\ \citenamefont
  {Ferrero}}]{Wuwei2018}%
  \BibitemOpen
  \bibfield  {author} {\bibinfo {author} {\bibfnamefont {W.}~\bibnamefont
  {Wu}}, \bibinfo {author} {\bibfnamefont {M.~S.}\ \bibnamefont {Scheurer}},
  \bibinfo {author} {\bibfnamefont {S.}~\bibnamefont {Chatterjee}}, \bibinfo
  {author} {\bibfnamefont {S.}~\bibnamefont {Sachdev}}, \bibinfo {author}
  {\bibfnamefont {A.}~\bibnamefont {Georges}},\ and\ \bibinfo {author}
  {\bibfnamefont {M.}~\bibnamefont {Ferrero}},\ }\bibfield  {title} {\bibinfo
  {title} {Pseudogap and fermi-surface topology in the two-dimensional hubbard
  model},\ }\href {https://doi.org/10.1103/PhysRevX.8.021048} {\bibfield
  {journal} {\bibinfo  {journal} {Phys. Rev. X}\ }\textbf {\bibinfo {volume}
  {8}},\ \bibinfo {pages} {021048} (\bibinfo {year} {2018})}\BibitemShut
  {NoStop}%
\bibitem [{\citenamefont {Chang}\ and\ \citenamefont
  {Zhang}(2010)}]{Chang2010}%
  \BibitemOpen
  \bibfield  {author} {\bibinfo {author} {\bibfnamefont {C.-C.}\ \bibnamefont
  {Chang}}\ and\ \bibinfo {author} {\bibfnamefont {S.}~\bibnamefont {Zhang}},\
  }\bibfield  {title} {\bibinfo {title} {Spin and charge order in the doped
  hubbard model: Long-wavelength collective modes},\ }\href
  {https://doi.org/10.1103/PhysRevLett.104.116402} {\bibfield  {journal}
  {\bibinfo  {journal} {Phys. Rev. Lett.}\ }\textbf {\bibinfo {volume} {104}},\
  \bibinfo {pages} {116402} (\bibinfo {year} {2010})}\BibitemShut {NoStop}%
\bibitem [{\citenamefont {Qin}\ \emph {et~al.}(2016{\natexlab{a}})\citenamefont
  {Qin}, \citenamefont {Shi},\ and\ \citenamefont {Zhang}}]{Qin2016a}%
  \BibitemOpen
  \bibfield  {author} {\bibinfo {author} {\bibfnamefont {M.}~\bibnamefont
  {Qin}}, \bibinfo {author} {\bibfnamefont {H.}~\bibnamefont {Shi}},\ and\
  \bibinfo {author} {\bibfnamefont {S.}~\bibnamefont {Zhang}},\ }\bibfield
  {title} {\bibinfo {title} {Coupling quantum monte carlo and
  independent-particle calculations: Self-consistent constraint for the sign
  problem based on the density or the density matrix},\ }\href
  {https://doi.org/10.1103/PhysRevB.94.235119} {\bibfield  {journal} {\bibinfo
  {journal} {Phys. Rev. B}\ }\textbf {\bibinfo {volume} {94}},\ \bibinfo
  {pages} {235119} (\bibinfo {year} {2016}{\natexlab{a}})}\BibitemShut
  {NoStop}%
\bibitem [{\citenamefont {Zheng}\ \emph {et~al.}(2017)\citenamefont {Zheng},
  \citenamefont {Chung}, \citenamefont {Corboz}, \citenamefont {Ehlers},
  \citenamefont {Qin}, \citenamefont {Noack}, \citenamefont {Shi},
  \citenamefont {White}, \citenamefont {Zhang},\ and\ \citenamefont
  {Chan}}]{Boxiao2017}%
  \BibitemOpen
  \bibfield  {author} {\bibinfo {author} {\bibfnamefont {B.-X.}\ \bibnamefont
  {Zheng}}, \bibinfo {author} {\bibfnamefont {C.-M.}\ \bibnamefont {Chung}},
  \bibinfo {author} {\bibfnamefont {P.}~\bibnamefont {Corboz}}, \bibinfo
  {author} {\bibfnamefont {G.}~\bibnamefont {Ehlers}}, \bibinfo {author}
  {\bibfnamefont {M.-P.}\ \bibnamefont {Qin}}, \bibinfo {author} {\bibfnamefont
  {R.~M.}\ \bibnamefont {Noack}}, \bibinfo {author} {\bibfnamefont
  {H.}~\bibnamefont {Shi}}, \bibinfo {author} {\bibfnamefont {S.~R.}\
  \bibnamefont {White}}, \bibinfo {author} {\bibfnamefont {S.}~\bibnamefont
  {Zhang}},\ and\ \bibinfo {author} {\bibfnamefont {G.~K.-L.}\ \bibnamefont
  {Chan}},\ }\bibfield  {title} {\bibinfo {title} {Stripe order in the
  underdoped region of the two-dimensional hubbard model},\ }\href
  {https://doi.org/10.1126/science.aam7127} {\bibfield  {journal} {\bibinfo
  {journal} {Science}\ }\textbf {\bibinfo {volume} {358}},\ \bibinfo {pages}
  {1155} (\bibinfo {year} {2017})}\BibitemShut {NoStop}%
\bibitem [{\citenamefont {Xiao}\ \emph {et~al.}(2023)\citenamefont {Xiao},
  \citenamefont {He}, \citenamefont {Georges},\ and\ \citenamefont
  {Zhang}}]{Xiao2023}%
  \BibitemOpen
  \bibfield  {author} {\bibinfo {author} {\bibfnamefont {B.}~\bibnamefont
  {Xiao}}, \bibinfo {author} {\bibfnamefont {Y.-Y.}\ \bibnamefont {He}},
  \bibinfo {author} {\bibfnamefont {A.}~\bibnamefont {Georges}},\ and\ \bibinfo
  {author} {\bibfnamefont {S.}~\bibnamefont {Zhang}},\ }\bibfield  {title}
  {\bibinfo {title} {Temperature dependence of spin and charge orders in the
  doped two-dimensional hubbard model},\ }\href
  {https://doi.org/10.1103/PhysRevX.13.011007} {\bibfield  {journal} {\bibinfo
  {journal} {Phys. Rev. X}\ }\textbf {\bibinfo {volume} {13}},\ \bibinfo
  {pages} {011007} (\bibinfo {year} {2023})}\BibitemShut {NoStop}%
\bibitem [{\citenamefont {Qin}\ \emph {et~al.}(2020)\citenamefont {Qin},
  \citenamefont {Chung}, \citenamefont {Shi}, \citenamefont {Vitali},
  \citenamefont {Hubig}, \citenamefont {Schollw\"ock}, \citenamefont {White},\
  and\ \citenamefont {Zhang}}]{Qin2020}%
  \BibitemOpen
  \bibfield  {author} {\bibinfo {author} {\bibfnamefont {M.}~\bibnamefont
  {Qin}}, \bibinfo {author} {\bibfnamefont {C.-M.}\ \bibnamefont {Chung}},
  \bibinfo {author} {\bibfnamefont {H.}~\bibnamefont {Shi}}, \bibinfo {author}
  {\bibfnamefont {E.}~\bibnamefont {Vitali}}, \bibinfo {author} {\bibfnamefont
  {C.}~\bibnamefont {Hubig}}, \bibinfo {author} {\bibfnamefont
  {U.}~\bibnamefont {Schollw\"ock}}, \bibinfo {author} {\bibfnamefont {S.~R.}\
  \bibnamefont {White}},\ and\ \bibinfo {author} {\bibfnamefont
  {S.}~\bibnamefont {Zhang}} (\bibinfo {collaboration} {Simons Collaboration on
  the Many-Electron Problem}),\ }\bibfield  {title} {\bibinfo {title} {Absence
  of superconductivity in the pure two-dimensional hubbard model},\ }\href
  {https://doi.org/10.1103/PhysRevX.10.031016} {\bibfield  {journal} {\bibinfo
  {journal} {Phys. Rev. X}\ }\textbf {\bibinfo {volume} {10}},\ \bibinfo
  {pages} {031016} (\bibinfo {year} {2020})}\BibitemShut {NoStop}%
\bibitem [{\citenamefont {Xu}\ \emph {et~al.}(2024)\citenamefont {Xu},
  \citenamefont {Chung}, \citenamefont {Qin}, \citenamefont {Schollwöck},
  \citenamefont {White},\ and\ \citenamefont {Zhang}}]{Haoxu2024}%
  \BibitemOpen
  \bibfield  {author} {\bibinfo {author} {\bibfnamefont {H.}~\bibnamefont
  {Xu}}, \bibinfo {author} {\bibfnamefont {C.-M.}\ \bibnamefont {Chung}},
  \bibinfo {author} {\bibfnamefont {M.}~\bibnamefont {Qin}}, \bibinfo {author}
  {\bibfnamefont {U.}~\bibnamefont {Schollwöck}}, \bibinfo {author}
  {\bibfnamefont {S.~R.}\ \bibnamefont {White}},\ and\ \bibinfo {author}
  {\bibfnamefont {S.}~\bibnamefont {Zhang}},\ }\bibfield  {title} {\bibinfo
  {title} {Coexistence of superconductivity with partially filled stripes in
  the hubbard model},\ }\href {https://doi.org/10.1126/science.adh7691}
  {\bibfield  {journal} {\bibinfo  {journal} {Science}\ }\textbf {\bibinfo
  {volume} {384}},\ \bibinfo {pages} {eadh7691} (\bibinfo {year}
  {2024})}\BibitemShut {NoStop}%
\bibitem [{\citenamefont {Lieb}\ and\ \citenamefont {Wu}(1968)}]{Lieb1968}%
  \BibitemOpen
  \bibfield  {author} {\bibinfo {author} {\bibfnamefont {E.~H.}\ \bibnamefont
  {Lieb}}\ and\ \bibinfo {author} {\bibfnamefont {F.~Y.}\ \bibnamefont {Wu}},\
  }\bibfield  {title} {\bibinfo {title} {Absence of mott transition in an exact
  solution of the short-range, one-band model in one dimension},\ }\href
  {https://doi.org/10.1103/PhysRevLett.20.1445} {\bibfield  {journal} {\bibinfo
   {journal} {Phys. Rev. Lett.}\ }\textbf {\bibinfo {volume} {20}},\ \bibinfo
  {pages} {1445} (\bibinfo {year} {1968})}\BibitemShut {NoStop}%
\bibitem [{\citenamefont {Luo}\ \emph {et~al.}(2023{\natexlab{a}})\citenamefont
  {Luo}, \citenamefont {Pu},\ and\ \citenamefont {Guan}}]{Luo2023a}%
  \BibitemOpen
  \bibfield  {author} {\bibinfo {author} {\bibfnamefont {J.-J.}\ \bibnamefont
  {Luo}}, \bibinfo {author} {\bibfnamefont {H.}~\bibnamefont {Pu}},\ and\
  \bibinfo {author} {\bibfnamefont {X.-W.}\ \bibnamefont {Guan}},\ }\bibfield
  {title} {\bibinfo {title} {Spin-incoherent liquid and interaction-driven
  criticality in the one-dimensional hubbard model},\ }\href
  {https://doi.org/10.1103/PhysRevB.107.L201103} {\bibfield  {journal}
  {\bibinfo  {journal} {Phys. Rev. B}\ }\textbf {\bibinfo {volume} {107}},\
  \bibinfo {pages} {L201103} (\bibinfo {year}
  {2023}{\natexlab{a}})}\BibitemShut {NoStop}%
\bibitem [{\citenamefont {Luo}\ \emph {et~al.}(2023{\natexlab{b}})\citenamefont
  {Luo}, \citenamefont {Pu},\ and\ \citenamefont {Guan}}]{Luo2023b}%
  \BibitemOpen
  \bibfield  {author} {\bibinfo {author} {\bibfnamefont {J.-J.}\ \bibnamefont
  {Luo}}, \bibinfo {author} {\bibfnamefont {H.}~\bibnamefont {Pu}},\ and\
  \bibinfo {author} {\bibfnamefont {X.-W.}\ \bibnamefont {Guan}},\ }\bibfield
  {title} {\bibinfo {title} {Exact results of one-dimensional repulsive hubbard
  model},\ }\href {https://arxiv.org/abs/2307.00890} {\bibfield  {journal}
  {\bibinfo  {journal} {arXiv:}\ }\textbf {\bibinfo {volume} {2307.00890}}
  (\bibinfo {year} {2023}{\natexlab{b}})}\BibitemShut {NoStop}%
\bibitem [{\citenamefont {Ulmke}\ \emph {et~al.}(1996)\citenamefont {Ulmke},
  \citenamefont {Scalettar}, \citenamefont {Nazarenko},\ and\ \citenamefont
  {Dagotto}}]{Ulmke1996}%
  \BibitemOpen
  \bibfield  {author} {\bibinfo {author} {\bibfnamefont {M.}~\bibnamefont
  {Ulmke}}, \bibinfo {author} {\bibfnamefont {R.~T.}\ \bibnamefont
  {Scalettar}}, \bibinfo {author} {\bibfnamefont {A.}~\bibnamefont
  {Nazarenko}},\ and\ \bibinfo {author} {\bibfnamefont {E.}~\bibnamefont
  {Dagotto}},\ }\bibfield  {title} {\bibinfo {title} {One-particle spectral
  weight of the three-dimensional single-band hubbard model},\ }\href
  {https://doi.org/10.1103/PhysRevB.54.16523} {\bibfield  {journal} {\bibinfo
  {journal} {Phys. Rev. B}\ }\textbf {\bibinfo {volume} {54}},\ \bibinfo
  {pages} {16523} (\bibinfo {year} {1996})}\BibitemShut {NoStop}%
\bibitem [{\citenamefont {Staudt}\ \emph {et~al.}(2000)\citenamefont {Staudt},
  \citenamefont {Dzierzawa},\ and\ \citenamefont {Muramatsu}}]{Staudt2000}%
  \BibitemOpen
  \bibfield  {author} {\bibinfo {author} {\bibfnamefont {R.}~\bibnamefont
  {Staudt}}, \bibinfo {author} {\bibfnamefont {M.}~\bibnamefont {Dzierzawa}},\
  and\ \bibinfo {author} {\bibfnamefont {A.}~\bibnamefont {Muramatsu}},\
  }\bibfield  {title} {\bibinfo {title} {Phase diagram of the three-dimensional
  hubbard model at half filling},\ }\href
  {https://doi.org/10.1007/s100510070120} {\bibfield  {journal} {\bibinfo
  {journal} {The European Physical Journal B}\ }\textbf {\bibinfo {volume}
  {17}},\ \bibinfo {pages} {411} (\bibinfo {year} {2000})}\BibitemShut
  {NoStop}%
\bibitem [{\citenamefont {Paiva}\ \emph {et~al.}(2011)\citenamefont {Paiva},
  \citenamefont {Loh}, \citenamefont {Randeria}, \citenamefont {Scalettar},\
  and\ \citenamefont {Trivedi}}]{Paiva2011}%
  \BibitemOpen
  \bibfield  {author} {\bibinfo {author} {\bibfnamefont {T.}~\bibnamefont
  {Paiva}}, \bibinfo {author} {\bibfnamefont {Y.~L.}\ \bibnamefont {Loh}},
  \bibinfo {author} {\bibfnamefont {M.}~\bibnamefont {Randeria}}, \bibinfo
  {author} {\bibfnamefont {R.~T.}\ \bibnamefont {Scalettar}},\ and\ \bibinfo
  {author} {\bibfnamefont {N.}~\bibnamefont {Trivedi}},\ }\bibfield  {title}
  {\bibinfo {title} {Fermions in 3d optical lattices: Cooling protocol to
  obtain antiferromagnetism},\ }\href
  {https://doi.org/10.1103/PhysRevLett.107.086401} {\bibfield  {journal}
  {\bibinfo  {journal} {Phys. Rev. Lett.}\ }\textbf {\bibinfo {volume} {107}},\
  \bibinfo {pages} {086401} (\bibinfo {year} {2011})}\BibitemShut {NoStop}%
\bibitem [{\citenamefont {Ibarra-Garc\'{\i}a-Padilla}\ \emph
  {et~al.}(2020)\citenamefont {Ibarra-Garc\'{\i}a-Padilla}, \citenamefont
  {Mukherjee}, \citenamefont {Hulet}, \citenamefont {Hazzard}, \citenamefont
  {Paiva},\ and\ \citenamefont {Scalettar}}]{Ibarra2020}%
  \BibitemOpen
  \bibfield  {author} {\bibinfo {author} {\bibfnamefont {E.}~\bibnamefont
  {Ibarra-Garc\'{\i}a-Padilla}}, \bibinfo {author} {\bibfnamefont
  {R.}~\bibnamefont {Mukherjee}}, \bibinfo {author} {\bibfnamefont {R.~G.}\
  \bibnamefont {Hulet}}, \bibinfo {author} {\bibfnamefont {K.~R.~A.}\
  \bibnamefont {Hazzard}}, \bibinfo {author} {\bibfnamefont {T.}~\bibnamefont
  {Paiva}},\ and\ \bibinfo {author} {\bibfnamefont {R.~T.}\ \bibnamefont
  {Scalettar}},\ }\bibfield  {title} {\bibinfo {title} {Thermodynamics and
  magnetism in the two-dimensional to three-dimensional crossover of the
  hubbard model},\ }\href {https://doi.org/10.1103/PhysRevA.102.033340}
  {\bibfield  {journal} {\bibinfo  {journal} {Phys. Rev. A}\ }\textbf {\bibinfo
  {volume} {102}},\ \bibinfo {pages} {033340} (\bibinfo {year}
  {2020})}\BibitemShut {NoStop}%
\bibitem [{\citenamefont {Sun}\ and\ \citenamefont {Xu}(2024)}]{Fanjie2024}%
  \BibitemOpen
  \bibfield  {author} {\bibinfo {author} {\bibfnamefont {F.}~\bibnamefont
  {Sun}}\ and\ \bibinfo {author} {\bibfnamefont {X.~Y.}\ \bibnamefont {Xu}},\
  }\bibfield  {title} {\bibinfo {title} {Boosting determinant quantum monte
  carlo with submatrix updates: Unveiling the phase diagram of the 3d hubbard
  model},\ }\href {https://arxiv.org/abs/2404.09989} {\bibfield  {journal}
  {\bibinfo  {journal} {arXiv:}\ }\textbf {\bibinfo {volume} {2404.09989}}
  (\bibinfo {year} {2024})}\BibitemShut {NoStop}%
\bibitem [{\citenamefont {Tahvildar-Zadeh}\ \emph {et~al.}(1997)\citenamefont
  {Tahvildar-Zadeh}, \citenamefont {Freericks},\ and\ \citenamefont
  {Jarrell}}]{Tahvildar1997}%
  \BibitemOpen
  \bibfield  {author} {\bibinfo {author} {\bibfnamefont {A.~N.}\ \bibnamefont
  {Tahvildar-Zadeh}}, \bibinfo {author} {\bibfnamefont {J.~K.}\ \bibnamefont
  {Freericks}},\ and\ \bibinfo {author} {\bibfnamefont {M.}~\bibnamefont
  {Jarrell}},\ }\bibfield  {title} {\bibinfo {title} {Magnetic phase diagram of
  the hubbard model in three dimensions:the second-order local approximation},\
  }\href {https://doi.org/10.1103/PhysRevB.55.942} {\bibfield  {journal}
  {\bibinfo  {journal} {Phys. Rev. B}\ }\textbf {\bibinfo {volume} {55}},\
  \bibinfo {pages} {942} (\bibinfo {year} {1997})}\BibitemShut {NoStop}%
\bibitem [{\citenamefont {Werner}\ \emph {et~al.}(2005)\citenamefont {Werner},
  \citenamefont {Parcollet}, \citenamefont {Georges},\ and\ \citenamefont
  {Hassan}}]{Werner2005}%
  \BibitemOpen
  \bibfield  {author} {\bibinfo {author} {\bibfnamefont {F.}~\bibnamefont
  {Werner}}, \bibinfo {author} {\bibfnamefont {O.}~\bibnamefont {Parcollet}},
  \bibinfo {author} {\bibfnamefont {A.}~\bibnamefont {Georges}},\ and\ \bibinfo
  {author} {\bibfnamefont {S.~R.}\ \bibnamefont {Hassan}},\ }\bibfield  {title}
  {\bibinfo {title} {Interaction-induced adiabatic cooling and
  antiferromagnetism of cold fermions in optical lattices},\ }\href
  {https://doi.org/10.1103/PhysRevLett.95.056401} {\bibfield  {journal}
  {\bibinfo  {journal} {Phys. Rev. Lett.}\ }\textbf {\bibinfo {volume} {95}},\
  \bibinfo {pages} {056401} (\bibinfo {year} {2005})}\BibitemShut {NoStop}%
\bibitem [{\citenamefont {Kent}\ \emph {et~al.}(2005)\citenamefont {Kent},
  \citenamefont {Jarrell}, \citenamefont {Maier},\ and\ \citenamefont
  {Pruschke}}]{Kent2005}%
  \BibitemOpen
  \bibfield  {author} {\bibinfo {author} {\bibfnamefont {P.~R.~C.}\
  \bibnamefont {Kent}}, \bibinfo {author} {\bibfnamefont {M.}~\bibnamefont
  {Jarrell}}, \bibinfo {author} {\bibfnamefont {T.~A.}\ \bibnamefont {Maier}},\
  and\ \bibinfo {author} {\bibfnamefont {T.}~\bibnamefont {Pruschke}},\
  }\bibfield  {title} {\bibinfo {title} {Efficient calculation of the
  antiferromagnetic phase diagram of the three-dimensional hubbard model},\
  }\href {https://doi.org/10.1103/PhysRevB.72.060411} {\bibfield  {journal}
  {\bibinfo  {journal} {Phys. Rev. B}\ }\textbf {\bibinfo {volume} {72}},\
  \bibinfo {pages} {060411} (\bibinfo {year} {2005})}\BibitemShut {NoStop}%
\bibitem [{\citenamefont {Rohringer}\ \emph {et~al.}(2011)\citenamefont
  {Rohringer}, \citenamefont {Toschi}, \citenamefont {Katanin},\ and\
  \citenamefont {Held}}]{Rohringer2011}%
  \BibitemOpen
  \bibfield  {author} {\bibinfo {author} {\bibfnamefont {G.}~\bibnamefont
  {Rohringer}}, \bibinfo {author} {\bibfnamefont {A.}~\bibnamefont {Toschi}},
  \bibinfo {author} {\bibfnamefont {A.}~\bibnamefont {Katanin}},\ and\ \bibinfo
  {author} {\bibfnamefont {K.}~\bibnamefont {Held}},\ }\bibfield  {title}
  {\bibinfo {title} {Critical properties of the half-filled hubbard model in
  three dimensions},\ }\href {https://doi.org/10.1103/PhysRevLett.107.256402}
  {\bibfield  {journal} {\bibinfo  {journal} {Phys. Rev. Lett.}\ }\textbf
  {\bibinfo {volume} {107}},\ \bibinfo {pages} {256402} (\bibinfo {year}
  {2011})}\BibitemShut {NoStop}%
\bibitem [{\citenamefont {Iskakov}\ and\ \citenamefont
  {Gull}(2022)}]{Iskakov2022}%
  \BibitemOpen
  \bibfield  {author} {\bibinfo {author} {\bibfnamefont {S.}~\bibnamefont
  {Iskakov}}\ and\ \bibinfo {author} {\bibfnamefont {E.}~\bibnamefont {Gull}},\
  }\bibfield  {title} {\bibinfo {title} {Phase transitions in partial summation
  methods: Results from the three-dimensional hubbard model},\ }\href
  {https://doi.org/10.1103/PhysRevB.105.045109} {\bibfield  {journal} {\bibinfo
   {journal} {Phys. Rev. B}\ }\textbf {\bibinfo {volume} {105}},\ \bibinfo
  {pages} {045109} (\bibinfo {year} {2022})}\BibitemShut {NoStop}%
\bibitem [{\citenamefont {Fuchs}\ \emph
  {et~al.}(2011{\natexlab{a}})\citenamefont {Fuchs}, \citenamefont {Gull},
  \citenamefont {Pollet}, \citenamefont {Burovski}, \citenamefont {Kozik},
  \citenamefont {Pruschke},\ and\ \citenamefont {Troyer}}]{Fuchs2011L}%
  \BibitemOpen
  \bibfield  {author} {\bibinfo {author} {\bibfnamefont {S.}~\bibnamefont
  {Fuchs}}, \bibinfo {author} {\bibfnamefont {E.}~\bibnamefont {Gull}},
  \bibinfo {author} {\bibfnamefont {L.}~\bibnamefont {Pollet}}, \bibinfo
  {author} {\bibfnamefont {E.}~\bibnamefont {Burovski}}, \bibinfo {author}
  {\bibfnamefont {E.}~\bibnamefont {Kozik}}, \bibinfo {author} {\bibfnamefont
  {T.}~\bibnamefont {Pruschke}},\ and\ \bibinfo {author} {\bibfnamefont
  {M.}~\bibnamefont {Troyer}},\ }\bibfield  {title} {\bibinfo {title}
  {Thermodynamics of the 3d hubbard model on approaching the n\'eel
  transition},\ }\href {https://doi.org/10.1103/PhysRevLett.106.030401}
  {\bibfield  {journal} {\bibinfo  {journal} {Phys. Rev. Lett.}\ }\textbf
  {\bibinfo {volume} {106}},\ \bibinfo {pages} {030401} (\bibinfo {year}
  {2011}{\natexlab{a}})}\BibitemShut {NoStop}%
\bibitem [{\citenamefont {Fuchs}\ \emph
  {et~al.}(2011{\natexlab{b}})\citenamefont {Fuchs}, \citenamefont {Gull},
  \citenamefont {Troyer}, \citenamefont {Jarrell},\ and\ \citenamefont
  {Pruschke}}]{Fuchs2011B}%
  \BibitemOpen
  \bibfield  {author} {\bibinfo {author} {\bibfnamefont {S.}~\bibnamefont
  {Fuchs}}, \bibinfo {author} {\bibfnamefont {E.}~\bibnamefont {Gull}},
  \bibinfo {author} {\bibfnamefont {M.}~\bibnamefont {Troyer}}, \bibinfo
  {author} {\bibfnamefont {M.}~\bibnamefont {Jarrell}},\ and\ \bibinfo {author}
  {\bibfnamefont {T.}~\bibnamefont {Pruschke}},\ }\bibfield  {title} {\bibinfo
  {title} {Spectral properties of the three-dimensional hubbard model},\ }\href
  {https://doi.org/10.1103/PhysRevB.83.235113} {\bibfield  {journal} {\bibinfo
  {journal} {Phys. Rev. B}\ }\textbf {\bibinfo {volume} {83}},\ \bibinfo
  {pages} {235113} (\bibinfo {year} {2011}{\natexlab{b}})}\BibitemShut
  {NoStop}%
\bibitem [{\citenamefont {Sotnikov}\ \emph {et~al.}(2012)\citenamefont
  {Sotnikov}, \citenamefont {Cocks},\ and\ \citenamefont
  {Hofstetter}}]{Sotnikov2012}%
  \BibitemOpen
  \bibfield  {author} {\bibinfo {author} {\bibfnamefont {A.}~\bibnamefont
  {Sotnikov}}, \bibinfo {author} {\bibfnamefont {D.}~\bibnamefont {Cocks}},\
  and\ \bibinfo {author} {\bibfnamefont {W.}~\bibnamefont {Hofstetter}},\
  }\bibfield  {title} {\bibinfo {title} {Advantages of mass-imbalanced
  ultracold fermionic mixtures for approaching quantum magnetism in optical
  lattices},\ }\href {https://doi.org/10.1103/PhysRevLett.109.065301}
  {\bibfield  {journal} {\bibinfo  {journal} {Phys. Rev. Lett.}\ }\textbf
  {\bibinfo {volume} {109}},\ \bibinfo {pages} {065301} (\bibinfo {year}
  {2012})}\BibitemShut {NoStop}%
\bibitem [{\citenamefont {Gorelik}\ \emph {et~al.}(2012)\citenamefont
  {Gorelik}, \citenamefont {Rost}, \citenamefont {Paiva}, \citenamefont
  {Scalettar}, \citenamefont {Kl\"umper},\ and\ \citenamefont
  {Bl\"umer}}]{Gorelik2012}%
  \BibitemOpen
  \bibfield  {author} {\bibinfo {author} {\bibfnamefont {E.~V.}\ \bibnamefont
  {Gorelik}}, \bibinfo {author} {\bibfnamefont {D.}~\bibnamefont {Rost}},
  \bibinfo {author} {\bibfnamefont {T.}~\bibnamefont {Paiva}}, \bibinfo
  {author} {\bibfnamefont {R.}~\bibnamefont {Scalettar}}, \bibinfo {author}
  {\bibfnamefont {A.}~\bibnamefont {Kl\"umper}},\ and\ \bibinfo {author}
  {\bibfnamefont {N.}~\bibnamefont {Bl\"umer}},\ }\bibfield  {title} {\bibinfo
  {title} {Universal probes for antiferromagnetic correlations and entropy in
  cold fermions on optical lattices},\ }\href
  {https://doi.org/10.1103/PhysRevA.85.061602} {\bibfield  {journal} {\bibinfo
  {journal} {Phys. Rev. A}\ }\textbf {\bibinfo {volume} {85}},\ \bibinfo
  {pages} {061602} (\bibinfo {year} {2012})}\BibitemShut {NoStop}%
\bibitem [{\citenamefont {Imri\ifmmode~\check{s}\else \v{s}\fi{}ka}\ \emph
  {et~al.}(2014)\citenamefont {Imri\ifmmode~\check{s}\else \v{s}\fi{}ka},
  \citenamefont {Iazzi}, \citenamefont {Wang}, \citenamefont {Gull},
  \citenamefont {Greif}, \citenamefont {Uehlinger}, \citenamefont {Jotzu},
  \citenamefont {Tarruell}, \citenamefont {Esslinger},\ and\ \citenamefont
  {Troyer}}]{Mauro2014}%
  \BibitemOpen
  \bibfield  {author} {\bibinfo {author} {\bibfnamefont {J.}~\bibnamefont
  {Imri\ifmmode~\check{s}\else \v{s}\fi{}ka}}, \bibinfo {author} {\bibfnamefont
  {M.}~\bibnamefont {Iazzi}}, \bibinfo {author} {\bibfnamefont
  {L.}~\bibnamefont {Wang}}, \bibinfo {author} {\bibfnamefont {E.}~\bibnamefont
  {Gull}}, \bibinfo {author} {\bibfnamefont {D.}~\bibnamefont {Greif}},
  \bibinfo {author} {\bibfnamefont {T.}~\bibnamefont {Uehlinger}}, \bibinfo
  {author} {\bibfnamefont {G.}~\bibnamefont {Jotzu}}, \bibinfo {author}
  {\bibfnamefont {L.}~\bibnamefont {Tarruell}}, \bibinfo {author}
  {\bibfnamefont {T.}~\bibnamefont {Esslinger}},\ and\ \bibinfo {author}
  {\bibfnamefont {M.}~\bibnamefont {Troyer}},\ }\bibfield  {title} {\bibinfo
  {title} {Thermodynamics and magnetic properties of the anisotropic 3d hubbard
  model},\ }\href {https://doi.org/10.1103/PhysRevLett.112.115301} {\bibfield
  {journal} {\bibinfo  {journal} {Phys. Rev. Lett.}\ }\textbf {\bibinfo
  {volume} {112}},\ \bibinfo {pages} {115301} (\bibinfo {year}
  {2014})}\BibitemShut {NoStop}%
\bibitem [{\citenamefont {Sch\"afer}\ \emph {et~al.}(2017)\citenamefont
  {Sch\"afer}, \citenamefont {Katanin}, \citenamefont {Held},\ and\
  \citenamefont {Toschi}}]{Katanin2017}%
  \BibitemOpen
  \bibfield  {author} {\bibinfo {author} {\bibfnamefont {T.}~\bibnamefont
  {Sch\"afer}}, \bibinfo {author} {\bibfnamefont {A.~A.}\ \bibnamefont
  {Katanin}}, \bibinfo {author} {\bibfnamefont {K.}~\bibnamefont {Held}},\ and\
  \bibinfo {author} {\bibfnamefont {A.}~\bibnamefont {Toschi}},\ }\bibfield
  {title} {\bibinfo {title} {Interplay of correlations and kohn anomalies in
  three dimensions: Quantum criticality with a twist},\ }\href
  {https://doi.org/10.1103/PhysRevLett.119.046402} {\bibfield  {journal}
  {\bibinfo  {journal} {Phys. Rev. Lett.}\ }\textbf {\bibinfo {volume} {119}},\
  \bibinfo {pages} {046402} (\bibinfo {year} {2017})}\BibitemShut {NoStop}%
\bibitem [{\citenamefont {Khatami}(2016)}]{Khatami2016}%
  \BibitemOpen
  \bibfield  {author} {\bibinfo {author} {\bibfnamefont {E.}~\bibnamefont
  {Khatami}},\ }\bibfield  {title} {\bibinfo {title} {Three-dimensional hubbard
  model in the thermodynamic limit},\ }\href
  {https://doi.org/10.1103/PhysRevB.94.125114} {\bibfield  {journal} {\bibinfo
  {journal} {Phys. Rev. B}\ }\textbf {\bibinfo {volume} {94}},\ \bibinfo
  {pages} {125114} (\bibinfo {year} {2016})}\BibitemShut {NoStop}%
\bibitem [{\citenamefont {Kozik}\ \emph {et~al.}(2013)\citenamefont {Kozik},
  \citenamefont {Burovski}, \citenamefont {Scarola},\ and\ \citenamefont
  {Troyer}}]{Kozik2013}%
  \BibitemOpen
  \bibfield  {author} {\bibinfo {author} {\bibfnamefont {E.}~\bibnamefont
  {Kozik}}, \bibinfo {author} {\bibfnamefont {E.}~\bibnamefont {Burovski}},
  \bibinfo {author} {\bibfnamefont {V.~W.}\ \bibnamefont {Scarola}},\ and\
  \bibinfo {author} {\bibfnamefont {M.}~\bibnamefont {Troyer}},\ }\bibfield
  {title} {\bibinfo {title} {N\'eel temperature and thermodynamics of the
  half-filled three-dimensional hubbard model by diagrammatic determinant monte
  carlo},\ }\href {https://doi.org/10.1103/PhysRevB.87.205102} {\bibfield
  {journal} {\bibinfo  {journal} {Phys. Rev. B}\ }\textbf {\bibinfo {volume}
  {87}},\ \bibinfo {pages} {205102} (\bibinfo {year} {2013})}\BibitemShut
  {NoStop}%
\bibitem [{\citenamefont {Lenihan}\ \emph {et~al.}(2022)\citenamefont
  {Lenihan}, \citenamefont {Kim}, \citenamefont {\ifmmode~\check{S}\else
  \v{S}\fi{}imkovic},\ and\ \citenamefont {Kozik}}]{Lenihan2022}%
  \BibitemOpen
  \bibfield  {author} {\bibinfo {author} {\bibfnamefont {C.}~\bibnamefont
  {Lenihan}}, \bibinfo {author} {\bibfnamefont {A.~J.}\ \bibnamefont {Kim}},
  \bibinfo {author} {\bibfnamefont {F.}~\bibnamefont {\ifmmode~\check{S}\else
  \v{S}\fi{}imkovic}},\ and\ \bibinfo {author} {\bibfnamefont {E.}~\bibnamefont
  {Kozik}},\ }\bibfield  {title} {\bibinfo {title} {Evaluating second-order
  phase transitions with diagrammatic monte carlo: N\'eel transition in the
  doped three-dimensional hubbard model},\ }\href
  {https://doi.org/10.1103/PhysRevLett.129.107202} {\bibfield  {journal}
  {\bibinfo  {journal} {Phys. Rev. Lett.}\ }\textbf {\bibinfo {volume} {129}},\
  \bibinfo {pages} {107202} (\bibinfo {year} {2022})}\BibitemShut {NoStop}%
\bibitem [{\citenamefont {Garioud}\ \emph {et~al.}(2024)\citenamefont
  {Garioud}, \citenamefont {\ifmmode~\check{S}\else \v{S}\fi{}imkovic},
  \citenamefont {Rossi}, \citenamefont {Spada}, \citenamefont {Sch\"afer},
  \citenamefont {Werner},\ and\ \citenamefont {Ferrero}}]{Garioud2024}%
  \BibitemOpen
  \bibfield  {author} {\bibinfo {author} {\bibfnamefont {R.}~\bibnamefont
  {Garioud}}, \bibinfo {author} {\bibfnamefont {F.}~\bibnamefont
  {\ifmmode~\check{S}\else \v{S}\fi{}imkovic}}, \bibinfo {author}
  {\bibfnamefont {R.}~\bibnamefont {Rossi}}, \bibinfo {author} {\bibfnamefont
  {G.}~\bibnamefont {Spada}}, \bibinfo {author} {\bibfnamefont
  {T.}~\bibnamefont {Sch\"afer}}, \bibinfo {author} {\bibfnamefont
  {F.}~\bibnamefont {Werner}},\ and\ \bibinfo {author} {\bibfnamefont
  {M.}~\bibnamefont {Ferrero}},\ }\bibfield  {title} {\bibinfo {title}
  {Symmetry-broken perturbation theory to large orders in antiferromagnetic
  phases},\ }\href {https://doi.org/10.1103/PhysRevLett.132.246505} {\bibfield
  {journal} {\bibinfo  {journal} {Phys. Rev. Lett.}\ }\textbf {\bibinfo
  {volume} {132}},\ \bibinfo {pages} {246505} (\bibinfo {year}
  {2024})}\BibitemShut {NoStop}%
\bibitem [{\citenamefont {Bloch}\ \emph {et~al.}(2008)\citenamefont {Bloch},
  \citenamefont {Dalibard},\ and\ \citenamefont {Zwerger}}]{Bloch2008}%
  \BibitemOpen
  \bibfield  {author} {\bibinfo {author} {\bibfnamefont {I.}~\bibnamefont
  {Bloch}}, \bibinfo {author} {\bibfnamefont {J.}~\bibnamefont {Dalibard}},\
  and\ \bibinfo {author} {\bibfnamefont {W.}~\bibnamefont {Zwerger}},\
  }\bibfield  {title} {\bibinfo {title} {Many-body physics with ultracold
  gases},\ }\href {https://doi.org/10.1103/RevModPhys.80.885} {\bibfield
  {journal} {\bibinfo  {journal} {Rev. Mod. Phys.}\ }\textbf {\bibinfo {volume}
  {80}},\ \bibinfo {pages} {885} (\bibinfo {year} {2008})}\BibitemShut
  {NoStop}%
\bibitem [{\citenamefont {K\"ohl}\ \emph {et~al.}(2005)\citenamefont {K\"ohl},
  \citenamefont {Moritz}, \citenamefont {St\"oferle}, \citenamefont
  {G\"unter},\ and\ \citenamefont {Esslinger}}]{Kohl2005}%
  \BibitemOpen
  \bibfield  {author} {\bibinfo {author} {\bibfnamefont {M.}~\bibnamefont
  {K\"ohl}}, \bibinfo {author} {\bibfnamefont {H.}~\bibnamefont {Moritz}},
  \bibinfo {author} {\bibfnamefont {T.}~\bibnamefont {St\"oferle}}, \bibinfo
  {author} {\bibfnamefont {K.}~\bibnamefont {G\"unter}},\ and\ \bibinfo
  {author} {\bibfnamefont {T.}~\bibnamefont {Esslinger}},\ }\bibfield  {title}
  {\bibinfo {title} {Fermionic atoms in a three dimensional optical lattice:
  Observing fermi surfaces, dynamics, and interactions},\ }\href
  {https://doi.org/10.1103/PhysRevLett.94.080403} {\bibfield  {journal}
  {\bibinfo  {journal} {Phys. Rev. Lett.}\ }\textbf {\bibinfo {volume} {94}},\
  \bibinfo {pages} {080403} (\bibinfo {year} {2005})}\BibitemShut {NoStop}%
\bibitem [{\citenamefont {Joerdens}\ \emph {et~al.}(2008)\citenamefont
  {Joerdens}, \citenamefont {Strohmaier}, \citenamefont {Guenther},
  \citenamefont {Moritz},\ and\ \citenamefont {Esslinger}}]{Joerdens2008}%
  \BibitemOpen
  \bibfield  {author} {\bibinfo {author} {\bibfnamefont {R.}~\bibnamefont
  {Joerdens}}, \bibinfo {author} {\bibfnamefont {N.}~\bibnamefont
  {Strohmaier}}, \bibinfo {author} {\bibfnamefont {K.}~\bibnamefont
  {Guenther}}, \bibinfo {author} {\bibfnamefont {H.}~\bibnamefont {Moritz}},\
  and\ \bibinfo {author} {\bibfnamefont {T.}~\bibnamefont {Esslinger}},\
  }\bibfield  {title} {\bibinfo {title} {A mott insulator of fermionic atoms in
  an optical lattice},\ }\href {https://doi.org/10.1038/nature07244} {\bibfield
   {journal} {\bibinfo  {journal} {Nature}\ }\textbf {\bibinfo {volume}
  {455}},\ \bibinfo {pages} {204} (\bibinfo {year} {2008})}\BibitemShut
  {NoStop}%
\bibitem [{\citenamefont {Schneider}\ \emph {et~al.}(2008)\citenamefont
  {Schneider}, \citenamefont {Hackermueller}, \citenamefont {Will},
  \citenamefont {Best}, \citenamefont {Bloch}, \citenamefont {Costi},
  \citenamefont {Helmes}, \citenamefont {Rasch},\ and\ \citenamefont
  {Rosch}}]{Schneider2008}%
  \BibitemOpen
  \bibfield  {author} {\bibinfo {author} {\bibfnamefont {U.}~\bibnamefont
  {Schneider}}, \bibinfo {author} {\bibfnamefont {L.}~\bibnamefont
  {Hackermueller}}, \bibinfo {author} {\bibfnamefont {S.}~\bibnamefont {Will}},
  \bibinfo {author} {\bibfnamefont {T.}~\bibnamefont {Best}}, \bibinfo {author}
  {\bibfnamefont {I.}~\bibnamefont {Bloch}}, \bibinfo {author} {\bibfnamefont
  {T.~A.}\ \bibnamefont {Costi}}, \bibinfo {author} {\bibfnamefont {R.~W.}\
  \bibnamefont {Helmes}}, \bibinfo {author} {\bibfnamefont {D.}~\bibnamefont
  {Rasch}},\ and\ \bibinfo {author} {\bibfnamefont {A.}~\bibnamefont {Rosch}},\
  }\bibfield  {title} {\bibinfo {title} {Metallic and insulating phases of
  repulsively interacting fermions in a 3d optical lattice},\ }\href
  {https://doi.org/10.1126/science.1165449} {\bibfield  {journal} {\bibinfo
  {journal} {Science}\ }\textbf {\bibinfo {volume} {322}},\ \bibinfo {pages}
  {1520} (\bibinfo {year} {2008})}\BibitemShut {NoStop}%
\bibitem [{\citenamefont {Duarte}\ \emph {et~al.}(2015)\citenamefont {Duarte},
  \citenamefont {Hart}, \citenamefont {Yang}, \citenamefont {Liu},
  \citenamefont {Paiva}, \citenamefont {Khatami}, \citenamefont {Scalettar},
  \citenamefont {Trivedi},\ and\ \citenamefont {Hulet}}]{Duarte2015}%
  \BibitemOpen
  \bibfield  {author} {\bibinfo {author} {\bibfnamefont {P.~M.}\ \bibnamefont
  {Duarte}}, \bibinfo {author} {\bibfnamefont {R.~A.}\ \bibnamefont {Hart}},
  \bibinfo {author} {\bibfnamefont {T.-L.}\ \bibnamefont {Yang}}, \bibinfo
  {author} {\bibfnamefont {X.}~\bibnamefont {Liu}}, \bibinfo {author}
  {\bibfnamefont {T.}~\bibnamefont {Paiva}}, \bibinfo {author} {\bibfnamefont
  {E.}~\bibnamefont {Khatami}}, \bibinfo {author} {\bibfnamefont {R.~T.}\
  \bibnamefont {Scalettar}}, \bibinfo {author} {\bibfnamefont {N.}~\bibnamefont
  {Trivedi}},\ and\ \bibinfo {author} {\bibfnamefont {R.~G.}\ \bibnamefont
  {Hulet}},\ }\bibfield  {title} {\bibinfo {title} {Compressibility of a
  fermionic mott insulator of ultracold atoms},\ }\href
  {https://doi.org/10.1103/PhysRevLett.114.070403} {\bibfield  {journal}
  {\bibinfo  {journal} {Phys. Rev. Lett.}\ }\textbf {\bibinfo {volume} {114}},\
  \bibinfo {pages} {070403} (\bibinfo {year} {2015})}\BibitemShut {NoStop}%
\bibitem [{\citenamefont {J\"ordens}\ \emph {et~al.}(2010)\citenamefont
  {J\"ordens}, \citenamefont {Tarruell}, \citenamefont {Greif}, \citenamefont
  {Uehlinger}, \citenamefont {Strohmaier}, \citenamefont {Moritz},
  \citenamefont {Esslinger}, \citenamefont {De~Leo}, \citenamefont {Kollath},
  \citenamefont {Georges}, \citenamefont {Scarola}, \citenamefont {Pollet},
  \citenamefont {Burovski}, \citenamefont {Kozik},\ and\ \citenamefont
  {Troyer}}]{Tarruell2010}%
  \BibitemOpen
  \bibfield  {author} {\bibinfo {author} {\bibfnamefont {R.}~\bibnamefont
  {J\"ordens}}, \bibinfo {author} {\bibfnamefont {L.}~\bibnamefont {Tarruell}},
  \bibinfo {author} {\bibfnamefont {D.}~\bibnamefont {Greif}}, \bibinfo
  {author} {\bibfnamefont {T.}~\bibnamefont {Uehlinger}}, \bibinfo {author}
  {\bibfnamefont {N.}~\bibnamefont {Strohmaier}}, \bibinfo {author}
  {\bibfnamefont {H.}~\bibnamefont {Moritz}}, \bibinfo {author} {\bibfnamefont
  {T.}~\bibnamefont {Esslinger}}, \bibinfo {author} {\bibfnamefont
  {L.}~\bibnamefont {De~Leo}}, \bibinfo {author} {\bibfnamefont
  {C.}~\bibnamefont {Kollath}}, \bibinfo {author} {\bibfnamefont
  {A.}~\bibnamefont {Georges}}, \bibinfo {author} {\bibfnamefont
  {V.}~\bibnamefont {Scarola}}, \bibinfo {author} {\bibfnamefont
  {L.}~\bibnamefont {Pollet}}, \bibinfo {author} {\bibfnamefont
  {E.}~\bibnamefont {Burovski}}, \bibinfo {author} {\bibfnamefont
  {E.}~\bibnamefont {Kozik}},\ and\ \bibinfo {author} {\bibfnamefont
  {M.}~\bibnamefont {Troyer}},\ }\bibfield  {title} {\bibinfo {title}
  {Quantitative determination of temperature in the approach to magnetic order
  of ultracold fermions in an optical lattice},\ }\href
  {https://doi.org/10.1103/PhysRevLett.104.180401} {\bibfield  {journal}
  {\bibinfo  {journal} {Phys. Rev. Lett.}\ }\textbf {\bibinfo {volume} {104}},\
  \bibinfo {pages} {180401} (\bibinfo {year} {2010})}\BibitemShut {NoStop}%
\bibitem [{\citenamefont {Greif}\ \emph {et~al.}(2013)\citenamefont {Greif},
  \citenamefont {Uehlinger}, \citenamefont {Jotzu}, \citenamefont {Tarruell},\
  and\ \citenamefont {Esslinger}}]{Daniel2013}%
  \BibitemOpen
  \bibfield  {author} {\bibinfo {author} {\bibfnamefont {D.}~\bibnamefont
  {Greif}}, \bibinfo {author} {\bibfnamefont {T.}~\bibnamefont {Uehlinger}},
  \bibinfo {author} {\bibfnamefont {G.}~\bibnamefont {Jotzu}}, \bibinfo
  {author} {\bibfnamefont {L.}~\bibnamefont {Tarruell}},\ and\ \bibinfo
  {author} {\bibfnamefont {T.}~\bibnamefont {Esslinger}},\ }\bibfield  {title}
  {\bibinfo {title} {Short-range quantum magnetism of ultracold fermions in an
  optical lattice},\ }\href {https://doi.org/10.1126/science.1236362}
  {\bibfield  {journal} {\bibinfo  {journal} {Science}\ }\textbf {\bibinfo
  {volume} {340}},\ \bibinfo {pages} {1307} (\bibinfo {year}
  {2013})}\BibitemShut {NoStop}%
\bibitem [{\citenamefont {Hart}\ \emph {et~al.}(2015)\citenamefont {Hart},
  \citenamefont {Duarte}, \citenamefont {Yang}, \citenamefont {Liu},
  \citenamefont {Paiva}, \citenamefont {Khatami}, \citenamefont {Scalettar},
  \citenamefont {Trivedi}, \citenamefont {Huse},\ and\ \citenamefont
  {Hulet}}]{Hart2015}%
  \BibitemOpen
  \bibfield  {author} {\bibinfo {author} {\bibfnamefont {R.~A.}\ \bibnamefont
  {Hart}}, \bibinfo {author} {\bibfnamefont {P.~M.}\ \bibnamefont {Duarte}},
  \bibinfo {author} {\bibfnamefont {T.-L.}\ \bibnamefont {Yang}}, \bibinfo
  {author} {\bibfnamefont {X.}~\bibnamefont {Liu}}, \bibinfo {author}
  {\bibfnamefont {T.}~\bibnamefont {Paiva}}, \bibinfo {author} {\bibfnamefont
  {E.}~\bibnamefont {Khatami}}, \bibinfo {author} {\bibfnamefont {R.~T.}\
  \bibnamefont {Scalettar}}, \bibinfo {author} {\bibfnamefont {N.}~\bibnamefont
  {Trivedi}}, \bibinfo {author} {\bibfnamefont {D.~A.}\ \bibnamefont {Huse}},\
  and\ \bibinfo {author} {\bibfnamefont {R.~G.}\ \bibnamefont {Hulet}},\
  }\bibfield  {title} {\bibinfo {title} {Observation of antiferromagnetic
  correlations in the hubbard model with ultracold atoms},\ }\href
  {https://doi.org/10.1038/nature14223} {\bibfield  {journal} {\bibinfo
  {journal} {Nature}\ }\textbf {\bibinfo {volume} {519}},\ \bibinfo {pages}
  {211} (\bibinfo {year} {2015})}\BibitemShut {NoStop}%
\bibitem [{\citenamefont {Shao}\ \emph {et~al.}(2024)\citenamefont {Shao},
  \citenamefont {Wang}, \citenamefont {Zhu}, \citenamefont {Zhu}, \citenamefont
  {Sun}, \citenamefont {Chen}, \citenamefont {Zhang}, \citenamefont {Fan},
  \citenamefont {Deng}, \citenamefont {Yao}, \citenamefont {Chen},\ and\
  \citenamefont {Pan}}]{Shao2024}%
  \BibitemOpen
  \bibfield  {author} {\bibinfo {author} {\bibfnamefont {H.-J.}\ \bibnamefont
  {Shao}}, \bibinfo {author} {\bibfnamefont {Y.-X.}\ \bibnamefont {Wang}},
  \bibinfo {author} {\bibfnamefont {D.-Z.}\ \bibnamefont {Zhu}}, \bibinfo
  {author} {\bibfnamefont {Y.-S.}\ \bibnamefont {Zhu}}, \bibinfo {author}
  {\bibfnamefont {H.-N.}\ \bibnamefont {Sun}}, \bibinfo {author} {\bibfnamefont
  {S.-Y.}\ \bibnamefont {Chen}}, \bibinfo {author} {\bibfnamefont
  {C.}~\bibnamefont {Zhang}}, \bibinfo {author} {\bibfnamefont {Z.-J.}\
  \bibnamefont {Fan}}, \bibinfo {author} {\bibfnamefont {Y.}~\bibnamefont
  {Deng}}, \bibinfo {author} {\bibfnamefont {X.-C.}\ \bibnamefont {Yao}},
  \bibinfo {author} {\bibfnamefont {Y.-A.}\ \bibnamefont {Chen}},\ and\
  \bibinfo {author} {\bibfnamefont {J.-W.}\ \bibnamefont {Pan}},\ }\bibfield
  {title} {\bibinfo {title} {Antiferromagnetic phase transition in a 3d
  fermionic hubbard model},\ }\href
  {https://doi.org/10.1038/s41586-024-07689-2} {\bibfield  {journal} {\bibinfo
  {journal} {Nature}\ }\textbf {\bibinfo {volume} {632}},\ \bibinfo {pages}
  {267} (\bibinfo {year} {2024})}\BibitemShut {NoStop}%
\bibitem [{\citenamefont {Blankenbecler}\ \emph {et~al.}(1981)\citenamefont
  {Blankenbecler}, \citenamefont {Scalapino},\ and\ \citenamefont
  {Sugar}}]{Blankenbecler1981}%
  \BibitemOpen
  \bibfield  {author} {\bibinfo {author} {\bibfnamefont {R.}~\bibnamefont
  {Blankenbecler}}, \bibinfo {author} {\bibfnamefont {D.~J.}\ \bibnamefont
  {Scalapino}},\ and\ \bibinfo {author} {\bibfnamefont {R.~L.}\ \bibnamefont
  {Sugar}},\ }\bibfield  {title} {\bibinfo {title} {Monte carlo calculations of
  coupled boson-fermion systems. i},\ }\href
  {https://doi.org/10.1103/PhysRevD.24.2278} {\bibfield  {journal} {\bibinfo
  {journal} {Phys. Rev. D}\ }\textbf {\bibinfo {volume} {24}},\ \bibinfo
  {pages} {2278} (\bibinfo {year} {1981})}\BibitemShut {NoStop}%
\bibitem [{\citenamefont {Hirsch}(1983)}]{Hirsch1983}%
  \BibitemOpen
  \bibfield  {author} {\bibinfo {author} {\bibfnamefont {J.~E.}\ \bibnamefont
  {Hirsch}},\ }\bibfield  {title} {\bibinfo {title} {Discrete
  hubbard-stratonovich transformation for fermion lattice models},\ }\href
  {https://doi.org/10.1103/PhysRevB.28.4059} {\bibfield  {journal} {\bibinfo
  {journal} {Phys. Rev. B}\ }\textbf {\bibinfo {volume} {28}},\ \bibinfo
  {pages} {4059} (\bibinfo {year} {1983})}\BibitemShut {NoStop}%
\bibitem [{\citenamefont {White}\ \emph {et~al.}(1989)\citenamefont {White},
  \citenamefont {Scalapino}, \citenamefont {Sugar}, \citenamefont {Loh},
  \citenamefont {Gubernatis},\ and\ \citenamefont {Scalettar}}]{White1989}%
  \BibitemOpen
  \bibfield  {author} {\bibinfo {author} {\bibfnamefont {S.~R.}\ \bibnamefont
  {White}}, \bibinfo {author} {\bibfnamefont {D.~J.}\ \bibnamefont
  {Scalapino}}, \bibinfo {author} {\bibfnamefont {R.~L.}\ \bibnamefont
  {Sugar}}, \bibinfo {author} {\bibfnamefont {E.~Y.}\ \bibnamefont {Loh}},
  \bibinfo {author} {\bibfnamefont {J.~E.}\ \bibnamefont {Gubernatis}},\ and\
  \bibinfo {author} {\bibfnamefont {R.~T.}\ \bibnamefont {Scalettar}},\
  }\bibfield  {title} {\bibinfo {title} {Numerical study of the two-dimensional
  hubbard model},\ }\href {https://doi.org/10.1103/PhysRevB.40.506} {\bibfield
  {journal} {\bibinfo  {journal} {Phys. Rev. B}\ }\textbf {\bibinfo {volume}
  {40}},\ \bibinfo {pages} {506} (\bibinfo {year} {1989})}\BibitemShut
  {NoStop}%
\bibitem [{\citenamefont {Scalettar}\ \emph {et~al.}(1991)\citenamefont
  {Scalettar}, \citenamefont {Noack},\ and\ \citenamefont
  {Singh}}]{Scalettar1991}%
  \BibitemOpen
  \bibfield  {author} {\bibinfo {author} {\bibfnamefont {R.~T.}\ \bibnamefont
  {Scalettar}}, \bibinfo {author} {\bibfnamefont {R.~M.}\ \bibnamefont
  {Noack}},\ and\ \bibinfo {author} {\bibfnamefont {R.~R.~P.}\ \bibnamefont
  {Singh}},\ }\bibfield  {title} {\bibinfo {title} {Ergodicity at large
  couplings with the determinant monte carlo algorithm},\ }\href
  {https://doi.org/10.1103/PhysRevB.44.10502} {\bibfield  {journal} {\bibinfo
  {journal} {Phys. Rev. B}\ }\textbf {\bibinfo {volume} {44}},\ \bibinfo
  {pages} {10502} (\bibinfo {year} {1991})}\BibitemShut {NoStop}%
\bibitem [{\citenamefont {McDaniel}\ \emph {et~al.}(2017)\citenamefont
  {McDaniel}, \citenamefont {D’Azevedo}, \citenamefont {Li}, \citenamefont
  {Wong},\ and\ \citenamefont {Kent}}]{McDaniel2017}%
  \BibitemOpen
  \bibfield  {author} {\bibinfo {author} {\bibfnamefont {T.}~\bibnamefont
  {McDaniel}}, \bibinfo {author} {\bibfnamefont {E.~F.}\ \bibnamefont
  {D’Azevedo}}, \bibinfo {author} {\bibfnamefont {Y.~W.}\ \bibnamefont {Li}},
  \bibinfo {author} {\bibfnamefont {K.}~\bibnamefont {Wong}},\ and\ \bibinfo
  {author} {\bibfnamefont {P.~R.~C.}\ \bibnamefont {Kent}},\ }\bibfield
  {title} {\bibinfo {title} {{Delayed Slater determinant update algorithms for
  high efficiency quantum Monte Carlo}},\ }\href
  {https://doi.org/10.1063/1.4998616} {\bibfield  {journal} {\bibinfo
  {journal} {The Journal of Chemical Physics}\ }\textbf {\bibinfo {volume}
  {147}},\ \bibinfo {pages} {174107} (\bibinfo {year} {2017})}\BibitemShut
  {NoStop}%
\bibitem [{\citenamefont {He}\ \emph {et~al.}(2019{\natexlab{a}})\citenamefont
  {He}, \citenamefont {Qin}, \citenamefont {Shi}, \citenamefont {Lu},\ and\
  \citenamefont {Zhang}}]{Yuanyao2019}%
  \BibitemOpen
  \bibfield  {author} {\bibinfo {author} {\bibfnamefont {Y.-Y.}\ \bibnamefont
  {He}}, \bibinfo {author} {\bibfnamefont {M.}~\bibnamefont {Qin}}, \bibinfo
  {author} {\bibfnamefont {H.}~\bibnamefont {Shi}}, \bibinfo {author}
  {\bibfnamefont {Z.-Y.}\ \bibnamefont {Lu}},\ and\ \bibinfo {author}
  {\bibfnamefont {S.}~\bibnamefont {Zhang}},\ }\bibfield  {title} {\bibinfo
  {title} {Finite-temperature auxiliary-field quantum monte carlo:
  Self-consistent constraint and systematic approach to low temperatures},\
  }\href {https://doi.org/10.1103/PhysRevB.99.045108} {\bibfield  {journal}
  {\bibinfo  {journal} {Phys. Rev. B}\ }\textbf {\bibinfo {volume} {99}},\
  \bibinfo {pages} {045108} (\bibinfo {year} {2019}{\natexlab{a}})}\BibitemShut
  {NoStop}%
\bibitem [{\citenamefont {He}\ \emph {et~al.}(2019{\natexlab{b}})\citenamefont
  {He}, \citenamefont {Shi},\ and\ \citenamefont {Zhang}}]{Yuanyao2019L}%
  \BibitemOpen
  \bibfield  {author} {\bibinfo {author} {\bibfnamefont {Y.-Y.}\ \bibnamefont
  {He}}, \bibinfo {author} {\bibfnamefont {H.}~\bibnamefont {Shi}},\ and\
  \bibinfo {author} {\bibfnamefont {S.}~\bibnamefont {Zhang}},\ }\bibfield
  {title} {\bibinfo {title} {Reaching the continuum limit in finite-temperature
  ab initio field-theory computations in many-fermion systems},\ }\href
  {https://doi.org/10.1103/PhysRevLett.123.136402} {\bibfield  {journal}
  {\bibinfo  {journal} {Phys. Rev. Lett.}\ }\textbf {\bibinfo {volume} {123}},\
  \bibinfo {pages} {136402} (\bibinfo {year} {2019}{\natexlab{b}})}\BibitemShut
  {NoStop}%
\bibitem [{\citenamefont {Song}\ \emph
  {et~al.}(2024{\natexlab{a}})\citenamefont {Song}, \citenamefont {Deng},\ and\
  \citenamefont {He}}]{Yufeng2024}%
  \BibitemOpen
  \bibfield  {author} {\bibinfo {author} {\bibfnamefont {Y.-F.}\ \bibnamefont
  {Song}}, \bibinfo {author} {\bibfnamefont {Y.}~\bibnamefont {Deng}},\ and\
  \bibinfo {author} {\bibfnamefont {Y.-Y.}\ \bibnamefont {He}},\ }\bibfield
  {title} {\bibinfo {title} {Extended metal-insulator crossover with strong
  antiferromagnetic spin correlation in half-filled 3d hubbard model},\ }\href
  {https://arxiv.org/abs/2404.08745} {\bibfield  {journal} {\bibinfo  {journal}
  {arXiv:}\ }\textbf {\bibinfo {volume} {2404.08745}} (\bibinfo {year}
  {2024}{\natexlab{a}})}\BibitemShut {NoStop}%
\bibitem [{\citenamefont {Campostrini}\ \emph {et~al.}(2002)\citenamefont
  {Campostrini}, \citenamefont {Hasenbusch}, \citenamefont {Pelissetto},
  \citenamefont {Rossi},\ and\ \citenamefont {Vicari}}]{Campostrini2002}%
  \BibitemOpen
  \bibfield  {author} {\bibinfo {author} {\bibfnamefont {M.}~\bibnamefont
  {Campostrini}}, \bibinfo {author} {\bibfnamefont {M.}~\bibnamefont
  {Hasenbusch}}, \bibinfo {author} {\bibfnamefont {A.}~\bibnamefont
  {Pelissetto}}, \bibinfo {author} {\bibfnamefont {P.}~\bibnamefont {Rossi}},\
  and\ \bibinfo {author} {\bibfnamefont {E.}~\bibnamefont {Vicari}},\
  }\bibfield  {title} {\bibinfo {title} {Critical exponents and equation of
  state of the three-dimensional heisenberg universality class},\ }\href
  {https://doi.org/10.1103/PhysRevB.65.144520} {\bibfield  {journal} {\bibinfo
  {journal} {Phys. Rev. B}\ }\textbf {\bibinfo {volume} {65}},\ \bibinfo
  {pages} {144520} (\bibinfo {year} {2002})}\BibitemShut {NoStop}%
\bibitem [{\citenamefont {Wu}\ and\ \citenamefont {Zhang}(2005)}]{Wu2005}%
  \BibitemOpen
  \bibfield  {author} {\bibinfo {author} {\bibfnamefont {C.}~\bibnamefont
  {Wu}}\ and\ \bibinfo {author} {\bibfnamefont {S.-C.}\ \bibnamefont {Zhang}},\
  }\bibfield  {title} {\bibinfo {title} {Sufficient condition for absence of
  the sign problem in the fermionic quantum monte carlo algorithm},\ }\href
  {https://doi.org/10.1103/PhysRevB.71.155115} {\bibfield  {journal} {\bibinfo
  {journal} {Phys. Rev. B}\ }\textbf {\bibinfo {volume} {71}},\ \bibinfo
  {pages} {155115} (\bibinfo {year} {2005})}\BibitemShut {NoStop}%
\bibitem [{\citenamefont {Assaad}\ and\ \citenamefont
  {Evertz}(2008)}]{Assaad2008}%
  \BibitemOpen
  \bibfield  {author} {\bibinfo {author} {\bibfnamefont {F.}~\bibnamefont
  {Assaad}}\ and\ \bibinfo {author} {\bibfnamefont {H.}~\bibnamefont
  {Evertz}},\ }\bibfield  {title} {\bibinfo {title} {World-line and
  determinantal quantum monte carlo methods for spins, phonons and electrons},\
  }in\ \href {https://doi.org/10.1007/978-3-540-74686-7_10} {\emph {\bibinfo
  {booktitle} {Computational Many-Particle Physics}}},\ \bibinfo {series}
  {Lecture Notes in Physics}, Vol.\ \bibinfo {volume} {739},\ \bibinfo {editor}
  {edited by\ \bibinfo {editor} {\bibfnamefont {H.}~\bibnamefont {Fehske}},
  \bibinfo {editor} {\bibfnamefont {R.}~\bibnamefont {Schneider}},\ and\
  \bibinfo {editor} {\bibfnamefont {A.}~\bibnamefont {Wei{\ss}e}}}\ (\bibinfo
  {publisher} {Springer Berlin Heidelberg},\ \bibinfo {year} {2008})\ pp.\
  \bibinfo {pages} {277--356}\BibitemShut {NoStop}%
\bibitem [{\citenamefont {Chang}\ \emph {et~al.}(2015)\citenamefont {Chang},
  \citenamefont {Gogolenko}, \citenamefont {Perez}, \citenamefont {Bai},\ and\
  \citenamefont {Scalettar}}]{Chang2015}%
  \BibitemOpen
  \bibfield  {author} {\bibinfo {author} {\bibfnamefont {C.-C.}\ \bibnamefont
  {Chang}}, \bibinfo {author} {\bibfnamefont {S.}~\bibnamefont {Gogolenko}},
  \bibinfo {author} {\bibfnamefont {J.}~\bibnamefont {Perez}}, \bibinfo
  {author} {\bibfnamefont {Z.}~\bibnamefont {Bai}},\ and\ \bibinfo {author}
  {\bibfnamefont {R.~T.}\ \bibnamefont {Scalettar}},\ }\bibfield  {title}
  {\bibinfo {title} {Recent advances in determinant quantum monte carlo},\
  }\href {https://doi.org/10.1080/14786435.2013.845314} {\bibfield  {journal}
  {\bibinfo  {journal} {Philosophical Magazine}\ }\textbf {\bibinfo {volume}
  {95}},\ \bibinfo {pages} {1260} (\bibinfo {year} {2015})}\BibitemShut
  {NoStop}%
\bibitem [{\citenamefont {Sandvik}(2010)}]{Sandvik2010}%
  \BibitemOpen
  \bibfield  {author} {\bibinfo {author} {\bibfnamefont {A.~W.}\ \bibnamefont
  {Sandvik}},\ }\bibfield  {title} {\bibinfo {title} {{Computational Studies of
  Quantum Spin Systems}},\ }\href {https://doi.org/10.1063/1.3518900}
  {\bibfield  {journal} {\bibinfo  {journal} {AIP Conference Proceedings}\
  }\textbf {\bibinfo {volume} {1297}},\ \bibinfo {pages} {135} (\bibinfo {year}
  {2010})}\BibitemShut {NoStop}%
\bibitem [{\citenamefont {Hofmann}\ \emph {et~al.}(2023)\citenamefont
  {Hofmann}, \citenamefont {Berg},\ and\ \citenamefont
  {Chowdhury}}]{Hofmann2023}%
  \BibitemOpen
  \bibfield  {author} {\bibinfo {author} {\bibfnamefont {J.~S.}\ \bibnamefont
  {Hofmann}}, \bibinfo {author} {\bibfnamefont {E.}~\bibnamefont {Berg}},\ and\
  \bibinfo {author} {\bibfnamefont {D.}~\bibnamefont {Chowdhury}},\ }\bibfield
  {title} {\bibinfo {title} {Superconductivity, charge density wave, and
  supersolidity in flat bands with a tunable quantum metric},\ }\href
  {https://doi.org/10.1103/PhysRevLett.130.226001} {\bibfield  {journal}
  {\bibinfo  {journal} {Phys. Rev. Lett.}\ }\textbf {\bibinfo {volume} {130}},\
  \bibinfo {pages} {226001} (\bibinfo {year} {2023})}\BibitemShut {NoStop}%
\bibitem [{\citenamefont {Dar\'e}\ \emph {et~al.}(2007)\citenamefont {Dar\'e},
  \citenamefont {Raymond}, \citenamefont {Albinet},\ and\ \citenamefont
  {Tremblay}}]{Dar2007}%
  \BibitemOpen
  \bibfield  {author} {\bibinfo {author} {\bibfnamefont {A.-M.}\ \bibnamefont
  {Dar\'e}}, \bibinfo {author} {\bibfnamefont {L.}~\bibnamefont {Raymond}},
  \bibinfo {author} {\bibfnamefont {G.}~\bibnamefont {Albinet}},\ and\ \bibinfo
  {author} {\bibfnamefont {A.-M.~S.}\ \bibnamefont {Tremblay}},\ }\bibfield
  {title} {\bibinfo {title} {Interaction-induced adiabatic cooling for
  antiferromagnetism in optical lattices},\ }\href
  {https://doi.org/10.1103/PhysRevB.76.064402} {\bibfield  {journal} {\bibinfo
  {journal} {Phys. Rev. B}\ }\textbf {\bibinfo {volume} {76}},\ \bibinfo
  {pages} {064402} (\bibinfo {year} {2007})}\BibitemShut {NoStop}%
\bibitem [{\citenamefont {Paiva}\ \emph {et~al.}(2010)\citenamefont {Paiva},
  \citenamefont {Scalettar}, \citenamefont {Randeria},\ and\ \citenamefont
  {Trivedi}}]{Paiva2010}%
  \BibitemOpen
  \bibfield  {author} {\bibinfo {author} {\bibfnamefont {T.}~\bibnamefont
  {Paiva}}, \bibinfo {author} {\bibfnamefont {R.}~\bibnamefont {Scalettar}},
  \bibinfo {author} {\bibfnamefont {M.}~\bibnamefont {Randeria}},\ and\
  \bibinfo {author} {\bibfnamefont {N.}~\bibnamefont {Trivedi}},\ }\bibfield
  {title} {\bibinfo {title} {Fermions in 2d optical lattices: Temperature and
  entropy scales for observing antiferromagnetism and superfluidity},\ }\href
  {https://doi.org/10.1103/PhysRevLett.104.066406} {\bibfield  {journal}
  {\bibinfo  {journal} {Phys. Rev. Lett.}\ }\textbf {\bibinfo {volume} {104}},\
  \bibinfo {pages} {066406} (\bibinfo {year} {2010})}\BibitemShut {NoStop}%
\bibitem [{\citenamefont {Sushchyev}\ and\ \citenamefont
  {Wessel}(2022)}]{Sushchyev2022}%
  \BibitemOpen
  \bibfield  {author} {\bibinfo {author} {\bibfnamefont {A.}~\bibnamefont
  {Sushchyev}}\ and\ \bibinfo {author} {\bibfnamefont {S.}~\bibnamefont
  {Wessel}},\ }\bibfield  {title} {\bibinfo {title} {Thermodynamics of the
  metal-insulator transition in the extended hubbard model from determinantal
  quantum monte carlo},\ }\href {https://doi.org/10.1103/PhysRevB.106.155121}
  {\bibfield  {journal} {\bibinfo  {journal} {Phys. Rev. B}\ }\textbf {\bibinfo
  {volume} {106}},\ \bibinfo {pages} {155121} (\bibinfo {year}
  {2022})}\BibitemShut {NoStop}%
\bibitem [{\citenamefont {Sandvik}(2016)}]{Sandvik2016}%
  \BibitemOpen
  \bibfield  {author} {\bibinfo {author} {\bibfnamefont {A.~W.}\ \bibnamefont
  {Sandvik}},\ }\bibfield  {title} {\bibinfo {title} {Constrained sampling
  method for analytic continuation},\ }\href
  {https://doi.org/10.1103/PhysRevE.94.063308} {\bibfield  {journal} {\bibinfo
  {journal} {Phys. Rev. E}\ }\textbf {\bibinfo {volume} {94}},\ \bibinfo
  {pages} {063308} (\bibinfo {year} {2016})}\BibitemShut {NoStop}%
\bibitem [{\citenamefont {Shao}\ and\ \citenamefont
  {Sandvik}(2023)}]{Shao2023}%
  \BibitemOpen
  \bibfield  {author} {\bibinfo {author} {\bibfnamefont {H.}~\bibnamefont
  {Shao}}\ and\ \bibinfo {author} {\bibfnamefont {A.~W.}\ \bibnamefont
  {Sandvik}},\ }\bibfield  {title} {\bibinfo {title} {Progress on stochastic
  analytic continuation of quantum monte carlo data},\ }\href
  {https://doi.org/https://doi.org/10.1016/j.physrep.2022.11.002} {\bibfield
  {journal} {\bibinfo  {journal} {Physics Reports}\ }\textbf {\bibinfo {volume}
  {1003}},\ \bibinfo {pages} {1} (\bibinfo {year} {2023})},\ \bibinfo {note}
  {progress on stochastic analytic continuation of quantum Monte Carlo
  data}\BibitemShut {NoStop}%
\bibitem [{\citenamefont {Liebsch}(2003)}]{Liebsch2003}%
  \BibitemOpen
  \bibfield  {author} {\bibinfo {author} {\bibfnamefont {A.}~\bibnamefont
  {Liebsch}},\ }\bibfield  {title} {\bibinfo {title} {Mott transitions in
  multiorbital systems},\ }\href
  {https://doi.org/10.1103/PhysRevLett.91.226401} {\bibfield  {journal}
  {\bibinfo  {journal} {Phys. Rev. Lett.}\ }\textbf {\bibinfo {volume} {91}},\
  \bibinfo {pages} {226401} (\bibinfo {year} {2003})}\BibitemShut {NoStop}%
\bibitem [{\citenamefont {Liu}\ and\ \citenamefont {Wang}(2015)}]{Liu2015}%
  \BibitemOpen
  \bibfield  {author} {\bibinfo {author} {\bibfnamefont {Y.-H.}\ \bibnamefont
  {Liu}}\ and\ \bibinfo {author} {\bibfnamefont {L.}~\bibnamefont {Wang}},\
  }\bibfield  {title} {\bibinfo {title} {Quantum monte carlo study of
  mass-imbalanced hubbard models},\ }\href
  {https://doi.org/10.1103/PhysRevB.92.235129} {\bibfield  {journal} {\bibinfo
  {journal} {Phys. Rev. B}\ }\textbf {\bibinfo {volume} {92}},\ \bibinfo
  {pages} {235129} (\bibinfo {year} {2015})}\BibitemShut {NoStop}%
\bibitem [{\citenamefont {You}\ \emph {et~al.}(2007)\citenamefont {You},
  \citenamefont {Li},\ and\ \citenamefont {Gu}}]{You2007}%
  \BibitemOpen
  \bibfield  {author} {\bibinfo {author} {\bibfnamefont {W.-L.}\ \bibnamefont
  {You}}, \bibinfo {author} {\bibfnamefont {Y.-W.}\ \bibnamefont {Li}},\ and\
  \bibinfo {author} {\bibfnamefont {S.-J.}\ \bibnamefont {Gu}},\ }\bibfield
  {title} {\bibinfo {title} {Fidelity, dynamic structure factor, and
  susceptibility in critical phenomena},\ }\href
  {https://doi.org/10.1103/PhysRevE.76.022101} {\bibfield  {journal} {\bibinfo
  {journal} {Phys. Rev. E}\ }\textbf {\bibinfo {volume} {76}},\ \bibinfo
  {pages} {022101} (\bibinfo {year} {2007})}\BibitemShut {NoStop}%
\bibitem [{\citenamefont {Campos~Venuti}\ and\ \citenamefont
  {Zanardi}(2007)}]{Venuti2007}%
  \BibitemOpen
  \bibfield  {author} {\bibinfo {author} {\bibfnamefont {L.}~\bibnamefont
  {Campos~Venuti}}\ and\ \bibinfo {author} {\bibfnamefont {P.}~\bibnamefont
  {Zanardi}},\ }\bibfield  {title} {\bibinfo {title} {Quantum critical scaling
  of the geometric tensors},\ }\href
  {https://doi.org/10.1103/PhysRevLett.99.095701} {\bibfield  {journal}
  {\bibinfo  {journal} {Phys. Rev. Lett.}\ }\textbf {\bibinfo {volume} {99}},\
  \bibinfo {pages} {095701} (\bibinfo {year} {2007})}\BibitemShut {NoStop}%
\bibitem [{\citenamefont {Gu}\ and\ \citenamefont {Lin}(2009)}]{Gu2009}%
  \BibitemOpen
  \bibfield  {author} {\bibinfo {author} {\bibfnamefont {S.-J.}\ \bibnamefont
  {Gu}}\ and\ \bibinfo {author} {\bibfnamefont {H.-Q.}\ \bibnamefont {Lin}},\
  }\bibfield  {title} {\bibinfo {title} {Scaling dimension of fidelity
  susceptibility in quantum phase transitions},\ }\href
  {https://doi.org/10.1209/0295-5075/87/10003} {\bibfield  {journal} {\bibinfo
  {journal} {Europhysics Letters}\ }\textbf {\bibinfo {volume} {87}},\ \bibinfo
  {pages} {10003} (\bibinfo {year} {2009})}\BibitemShut {NoStop}%
\bibitem [{\citenamefont {Schwandt}\ \emph {et~al.}(2009)\citenamefont
  {Schwandt}, \citenamefont {Alet},\ and\ \citenamefont
  {Capponi}}]{Schwandt2009}%
  \BibitemOpen
  \bibfield  {author} {\bibinfo {author} {\bibfnamefont {D.}~\bibnamefont
  {Schwandt}}, \bibinfo {author} {\bibfnamefont {F.}~\bibnamefont {Alet}},\
  and\ \bibinfo {author} {\bibfnamefont {S.}~\bibnamefont {Capponi}},\
  }\bibfield  {title} {\bibinfo {title} {Quantum monte carlo simulations of
  fidelity at magnetic quantum phase transitions},\ }\href
  {https://doi.org/10.1103/PhysRevLett.103.170501} {\bibfield  {journal}
  {\bibinfo  {journal} {Phys. Rev. Lett.}\ }\textbf {\bibinfo {volume} {103}},\
  \bibinfo {pages} {170501} (\bibinfo {year} {2009})}\BibitemShut {NoStop}%
\bibitem [{\citenamefont {Albuquerque}\ \emph {et~al.}(2010)\citenamefont
  {Albuquerque}, \citenamefont {Alet}, \citenamefont {Sire},\ and\
  \citenamefont {Capponi}}]{Albuquerque2010}%
  \BibitemOpen
  \bibfield  {author} {\bibinfo {author} {\bibfnamefont {A.~F.}\ \bibnamefont
  {Albuquerque}}, \bibinfo {author} {\bibfnamefont {F.}~\bibnamefont {Alet}},
  \bibinfo {author} {\bibfnamefont {C.}~\bibnamefont {Sire}},\ and\ \bibinfo
  {author} {\bibfnamefont {S.}~\bibnamefont {Capponi}},\ }\bibfield  {title}
  {\bibinfo {title} {Quantum critical scaling of fidelity susceptibility},\
  }\href {https://doi.org/10.1103/PhysRevB.81.064418} {\bibfield  {journal}
  {\bibinfo  {journal} {Phys. Rev. B}\ }\textbf {\bibinfo {volume} {81}},\
  \bibinfo {pages} {064418} (\bibinfo {year} {2010})}\BibitemShut {NoStop}%
\bibitem [{\citenamefont {Wang}\ \emph {et~al.}(2015)\citenamefont {Wang},
  \citenamefont {Liu}, \citenamefont {Imri\ifmmode~\check{s}\else
  \v{s}\fi{}ka}, \citenamefont {Ma},\ and\ \citenamefont
  {Troyer}}]{WangLei2015}%
  \BibitemOpen
  \bibfield  {author} {\bibinfo {author} {\bibfnamefont {L.}~\bibnamefont
  {Wang}}, \bibinfo {author} {\bibfnamefont {Y.-H.}\ \bibnamefont {Liu}},
  \bibinfo {author} {\bibfnamefont {J.}~\bibnamefont
  {Imri\ifmmode~\check{s}\else \v{s}\fi{}ka}}, \bibinfo {author} {\bibfnamefont
  {P.~N.}\ \bibnamefont {Ma}},\ and\ \bibinfo {author} {\bibfnamefont
  {M.}~\bibnamefont {Troyer}},\ }\bibfield  {title} {\bibinfo {title} {Fidelity
  susceptibility made simple: A unified quantum monte carlo approach},\ }\href
  {https://doi.org/10.1103/PhysRevX.5.031007} {\bibfield  {journal} {\bibinfo
  {journal} {Phys. Rev. X}\ }\textbf {\bibinfo {volume} {5}},\ \bibinfo {pages}
  {031007} (\bibinfo {year} {2015})}\BibitemShut {NoStop}%
\bibitem [{\citenamefont {Huang}\ \emph {et~al.}(2016)\citenamefont {Huang},
  \citenamefont {Wang}, \citenamefont {Wang},\ and\ \citenamefont
  {Werner}}]{HuangLi2016}%
  \BibitemOpen
  \bibfield  {author} {\bibinfo {author} {\bibfnamefont {L.}~\bibnamefont
  {Huang}}, \bibinfo {author} {\bibfnamefont {Y.}~\bibnamefont {Wang}},
  \bibinfo {author} {\bibfnamefont {L.}~\bibnamefont {Wang}},\ and\ \bibinfo
  {author} {\bibfnamefont {P.}~\bibnamefont {Werner}},\ }\bibfield  {title}
  {\bibinfo {title} {Detecting phase transitions and crossovers in hubbard
  models using the fidelity susceptibility},\ }\href
  {https://doi.org/10.1103/PhysRevB.94.235110} {\bibfield  {journal} {\bibinfo
  {journal} {Phys. Rev. B}\ }\textbf {\bibinfo {volume} {94}},\ \bibinfo
  {pages} {235110} (\bibinfo {year} {2016})}\BibitemShut {NoStop}%
\bibitem [{\citenamefont {Song}\ \emph
  {et~al.}(2024{\natexlab{b}})\citenamefont {Song}, \citenamefont {Deng},\ and\
  \citenamefont {He}}]{Song2024}%
  \BibitemOpen
  \bibfield  {author} {\bibinfo {author} {\bibfnamefont {Y.-F.}\ \bibnamefont
  {Song}}, \bibinfo {author} {\bibfnamefont {Y.}~\bibnamefont {Deng}},\ and\
  \bibinfo {author} {\bibfnamefont {Y.-Y.}\ \bibnamefont {He}},\ }\bibfield
  {title} {\bibinfo {title} {Nature of the mixed-parity pairing of attractive
  fermions with spin-orbit coupling in an optical lattice},\ }\href
  {https://doi.org/10.1103/PhysRevB.109.094504} {\bibfield  {journal} {\bibinfo
   {journal} {Phys. Rev. B}\ }\textbf {\bibinfo {volume} {109}},\ \bibinfo
  {pages} {094504} (\bibinfo {year} {2024}{\natexlab{b}})}\BibitemShut
  {NoStop}%
\bibitem [{\citenamefont {Hille}\ \emph {et~al.}(2020)\citenamefont {Hille},
  \citenamefont {Kugler}, \citenamefont {Eckhardt}, \citenamefont {He},
  \citenamefont {Kauch}, \citenamefont {Honerkamp}, \citenamefont {Toschi},\
  and\ \citenamefont {Andergassen}}]{Cornelia2021}%
  \BibitemOpen
  \bibfield  {author} {\bibinfo {author} {\bibfnamefont {C.}~\bibnamefont
  {Hille}}, \bibinfo {author} {\bibfnamefont {F.~B.}\ \bibnamefont {Kugler}},
  \bibinfo {author} {\bibfnamefont {C.~J.}\ \bibnamefont {Eckhardt}}, \bibinfo
  {author} {\bibfnamefont {Y.-Y.}\ \bibnamefont {He}}, \bibinfo {author}
  {\bibfnamefont {A.}~\bibnamefont {Kauch}}, \bibinfo {author} {\bibfnamefont
  {C.}~\bibnamefont {Honerkamp}}, \bibinfo {author} {\bibfnamefont
  {A.}~\bibnamefont {Toschi}},\ and\ \bibinfo {author} {\bibfnamefont
  {S.}~\bibnamefont {Andergassen}},\ }\bibfield  {title} {\bibinfo {title}
  {Quantitative functional renormalization group description of the
  two-dimensional hubbard model},\ }\href
  {https://doi.org/10.1103/PhysRevResearch.2.033372} {\bibfield  {journal}
  {\bibinfo  {journal} {Phys. Rev. Res.}\ }\textbf {\bibinfo {volume} {2}},\
  \bibinfo {pages} {033372} (\bibinfo {year} {2020})}\BibitemShut {NoStop}%
\bibitem [{\citenamefont {Sandvik}(1998)}]{Sandvik1998}%
  \BibitemOpen
  \bibfield  {author} {\bibinfo {author} {\bibfnamefont {A.~W.}\ \bibnamefont
  {Sandvik}},\ }\bibfield  {title} {\bibinfo {title} {Critical temperature and
  the transition from quantum to classical order parameter fluctuations in the
  three-dimensional heisenberg antiferromagnet},\ }\href
  {https://doi.org/10.1103/PhysRevLett.80.5196} {\bibfield  {journal} {\bibinfo
   {journal} {Phys. Rev. Lett.}\ }\textbf {\bibinfo {volume} {80}},\ \bibinfo
  {pages} {5196} (\bibinfo {year} {1998})}\BibitemShut {NoStop}%
\bibitem [{\citenamefont {Vu\ifmmode \check{c}\else \v{c}\fi{}i\ifmmode
  \check{c}\else \v{c}\fi{}evi\ifmmode~\acute{c}\else \'{c}\fi{}}\ \emph
  {et~al.}(2015)\citenamefont {Vu\ifmmode \check{c}\else \v{c}\fi{}i\ifmmode
  \check{c}\else \v{c}\fi{}evi\ifmmode~\acute{c}\else \'{c}\fi{}},
  \citenamefont {Tanaskovi\ifmmode~\acute{c}\else \'{c}\fi{}}, \citenamefont
  {Rozenberg},\ and\ \citenamefont {Dobrosavljevi\ifmmode~\acute{c}\else
  \'{c}\fi{}}}]{Vu2015}%
  \BibitemOpen
  \bibfield  {author} {\bibinfo {author} {\bibfnamefont {J.}~\bibnamefont
  {Vu\ifmmode \check{c}\else \v{c}\fi{}i\ifmmode \check{c}\else
  \v{c}\fi{}evi\ifmmode~\acute{c}\else \'{c}\fi{}}}, \bibinfo {author}
  {\bibfnamefont {D.}~\bibnamefont {Tanaskovi\ifmmode~\acute{c}\else
  \'{c}\fi{}}}, \bibinfo {author} {\bibfnamefont {M.~J.}\ \bibnamefont
  {Rozenberg}},\ and\ \bibinfo {author} {\bibfnamefont {V.}~\bibnamefont
  {Dobrosavljevi\ifmmode~\acute{c}\else \'{c}\fi{}}},\ }\bibfield  {title}
  {\bibinfo {title} {Bad-metal behavior reveals mott quantum criticality in
  doped hubbard models},\ }\href
  {https://doi.org/10.1103/PhysRevLett.114.246402} {\bibfield  {journal}
  {\bibinfo  {journal} {Phys. Rev. Lett.}\ }\textbf {\bibinfo {volume} {114}},\
  \bibinfo {pages} {246402} (\bibinfo {year} {2015})}\BibitemShut {NoStop}%
\bibitem [{\citenamefont {Mousatov}\ \emph {et~al.}(2019)\citenamefont
  {Mousatov}, \citenamefont {Esterlis},\ and\ \citenamefont
  {Hartnoll}}]{Mousatov2019}%
  \BibitemOpen
  \bibfield  {author} {\bibinfo {author} {\bibfnamefont {C.~H.}\ \bibnamefont
  {Mousatov}}, \bibinfo {author} {\bibfnamefont {I.}~\bibnamefont {Esterlis}},\
  and\ \bibinfo {author} {\bibfnamefont {S.~A.}\ \bibnamefont {Hartnoll}},\
  }\bibfield  {title} {\bibinfo {title} {Bad metallic transport in a modified
  hubbard model},\ }\href {https://doi.org/10.1103/PhysRevLett.122.186601}
  {\bibfield  {journal} {\bibinfo  {journal} {Phys. Rev. Lett.}\ }\textbf
  {\bibinfo {volume} {122}},\ \bibinfo {pages} {186601} (\bibinfo {year}
  {2019})}\BibitemShut {NoStop}%
\bibitem [{\citenamefont {Deng}\ \emph {et~al.}(2013)\citenamefont {Deng},
  \citenamefont {Mravlje}, \citenamefont {\ifmmode~\check{Z}\else
  \v{Z}\fi{}itko}, \citenamefont {Ferrero}, \citenamefont {Kotliar},\ and\
  \citenamefont {Georges}}]{Deng2013}%
  \BibitemOpen
  \bibfield  {author} {\bibinfo {author} {\bibfnamefont {X.}~\bibnamefont
  {Deng}}, \bibinfo {author} {\bibfnamefont {J.}~\bibnamefont {Mravlje}},
  \bibinfo {author} {\bibfnamefont {R.}~\bibnamefont {\ifmmode~\check{Z}\else
  \v{Z}\fi{}itko}}, \bibinfo {author} {\bibfnamefont {M.}~\bibnamefont
  {Ferrero}}, \bibinfo {author} {\bibfnamefont {G.}~\bibnamefont {Kotliar}},\
  and\ \bibinfo {author} {\bibfnamefont {A.}~\bibnamefont {Georges}},\
  }\bibfield  {title} {\bibinfo {title} {How bad metals turn good:
  Spectroscopic signatures of resilient quasiparticles},\ }\href
  {https://doi.org/10.1103/PhysRevLett.110.086401} {\bibfield  {journal}
  {\bibinfo  {journal} {Phys. Rev. Lett.}\ }\textbf {\bibinfo {volume} {110}},\
  \bibinfo {pages} {086401} (\bibinfo {year} {2013})}\BibitemShut {NoStop}%
\bibitem [{\citenamefont {Ding}\ \emph {et~al.}(2019)\citenamefont {Ding},
  \citenamefont {Yu}, \citenamefont {Si},\ and\ \citenamefont
  {Abrahams}}]{Ding2019}%
  \BibitemOpen
  \bibfield  {author} {\bibinfo {author} {\bibfnamefont {W.}~\bibnamefont
  {Ding}}, \bibinfo {author} {\bibfnamefont {R.}~\bibnamefont {Yu}}, \bibinfo
  {author} {\bibfnamefont {Q.}~\bibnamefont {Si}},\ and\ \bibinfo {author}
  {\bibfnamefont {E.}~\bibnamefont {Abrahams}},\ }\bibfield  {title} {\bibinfo
  {title} {Effective exchange interactions for bad metals and implications for
  iron-based superconductors},\ }\href
  {https://doi.org/10.1103/PhysRevB.100.235113} {\bibfield  {journal} {\bibinfo
   {journal} {Phys. Rev. B}\ }\textbf {\bibinfo {volume} {100}},\ \bibinfo
  {pages} {235113} (\bibinfo {year} {2019})}\BibitemShut {NoStop}%
\bibitem [{\citenamefont {Georges}\ \emph {et~al.}(1996)\citenamefont
  {Georges}, \citenamefont {Kotliar}, \citenamefont {Krauth},\ and\
  \citenamefont {Rozenberg}}]{Georges1996}%
  \BibitemOpen
  \bibfield  {author} {\bibinfo {author} {\bibfnamefont {A.}~\bibnamefont
  {Georges}}, \bibinfo {author} {\bibfnamefont {G.}~\bibnamefont {Kotliar}},
  \bibinfo {author} {\bibfnamefont {W.}~\bibnamefont {Krauth}},\ and\ \bibinfo
  {author} {\bibfnamefont {M.~J.}\ \bibnamefont {Rozenberg}},\ }\bibfield
  {title} {\bibinfo {title} {Dynamical mean-field theory of strongly correlated
  fermion systems and the limit of infinite dimensions},\ }\href
  {https://doi.org/10.1103/RevModPhys.68.13} {\bibfield  {journal} {\bibinfo
  {journal} {Rev. Mod. Phys.}\ }\textbf {\bibinfo {volume} {68}},\ \bibinfo
  {pages} {13} (\bibinfo {year} {1996})}\BibitemShut {NoStop}%
\bibitem [{\citenamefont {Gebhard}(1997)}]{Gebhard1997}%
  \BibitemOpen
  \bibfield  {author} {\bibinfo {author} {\bibfnamefont {F.}~\bibnamefont
  {Gebhard}},\ }\bibfield  {title} {\bibinfo {title} {The mott metal-insulator
  transition: Models and methods}\ }(\bibinfo  {publisher} {Springer-Verlag,
  Berlin Heidelberg},\ \bibinfo {year} {1997})\BibitemShut {NoStop}%
\bibitem [{\citenamefont {Imada}\ \emph {et~al.}(1998)\citenamefont {Imada},
  \citenamefont {Fujimori},\ and\ \citenamefont {Tokura}}]{Imada1998}%
  \BibitemOpen
  \bibfield  {author} {\bibinfo {author} {\bibfnamefont {M.}~\bibnamefont
  {Imada}}, \bibinfo {author} {\bibfnamefont {A.}~\bibnamefont {Fujimori}},\
  and\ \bibinfo {author} {\bibfnamefont {Y.}~\bibnamefont {Tokura}},\
  }\bibfield  {title} {\bibinfo {title} {Metal-insulator transitions},\ }\href
  {https://doi.org/10.1103/RevModPhys.70.1039} {\bibfield  {journal} {\bibinfo
  {journal} {Rev. Mod. Phys.}\ }\textbf {\bibinfo {volume} {70}},\ \bibinfo
  {pages} {1039} (\bibinfo {year} {1998})}\BibitemShut {NoStop}%
\bibitem [{\citenamefont {Kotliar}\ and\ \citenamefont
  {Vollhardt}(2004)}]{Kotliar2004}%
  \BibitemOpen
  \bibfield  {author} {\bibinfo {author} {\bibfnamefont {G.}~\bibnamefont
  {Kotliar}}\ and\ \bibinfo {author} {\bibfnamefont {D.}~\bibnamefont
  {Vollhardt}},\ }\bibfield  {title} {\bibinfo {title} {{Strongly Correlated
  Materials: Insights From Dynamical Mean-Field Theory}},\ }\href
  {https://doi.org/10.1063/1.1712502} {\bibfield  {journal} {\bibinfo
  {journal} {Physics Today}\ }\textbf {\bibinfo {volume} {57}},\ \bibinfo
  {pages} {53} (\bibinfo {year} {2004})}\BibitemShut {NoStop}%
\bibitem [{\citenamefont {Binder}(1981)}]{Binder1981}%
  \BibitemOpen
  \bibfield  {author} {\bibinfo {author} {\bibfnamefont {K.}~\bibnamefont
  {Binder}},\ }\bibfield  {title} {\bibinfo {title} {Critical properties from
  monte carlo coarse graining and renormalization},\ }\href
  {https://doi.org/10.1103/PhysRevLett.47.693} {\bibfield  {journal} {\bibinfo
  {journal} {Phys. Rev. Lett.}\ }\textbf {\bibinfo {volume} {47}},\ \bibinfo
  {pages} {693} (\bibinfo {year} {1981})}\BibitemShut {NoStop}%
\bibitem [{\citenamefont {Lin}\ \emph {et~al.}(2001)\citenamefont {Lin},
  \citenamefont {Zong},\ and\ \citenamefont {Ceperley}}]{Lin2001}%
  \BibitemOpen
  \bibfield  {author} {\bibinfo {author} {\bibfnamefont {C.}~\bibnamefont
  {Lin}}, \bibinfo {author} {\bibfnamefont {F.~H.}\ \bibnamefont {Zong}},\ and\
  \bibinfo {author} {\bibfnamefont {D.~M.}\ \bibnamefont {Ceperley}},\
  }\bibfield  {title} {\bibinfo {title} {Twist-averaged boundary conditions in
  continuum quantum monte carlo algorithms},\ }\href
  {https://doi.org/10.1103/PhysRevE.64.016702} {\bibfield  {journal} {\bibinfo
  {journal} {Phys. Rev. E}\ }\textbf {\bibinfo {volume} {64}},\ \bibinfo
  {pages} {016702} (\bibinfo {year} {2001})}\BibitemShut {NoStop}%
\bibitem [{\citenamefont {Qin}\ \emph {et~al.}(2016{\natexlab{b}})\citenamefont
  {Qin}, \citenamefont {Shi},\ and\ \citenamefont {Zhang}}]{Qin2016b}%
  \BibitemOpen
  \bibfield  {author} {\bibinfo {author} {\bibfnamefont {M.}~\bibnamefont
  {Qin}}, \bibinfo {author} {\bibfnamefont {H.}~\bibnamefont {Shi}},\ and\
  \bibinfo {author} {\bibfnamefont {S.}~\bibnamefont {Zhang}},\ }\bibfield
  {title} {\bibinfo {title} {Benchmark study of the two-dimensional hubbard
  model with auxiliary-field quantum monte carlo method},\ }\href
  {https://doi.org/10.1103/PhysRevB.94.085103} {\bibfield  {journal} {\bibinfo
  {journal} {Phys. Rev. B}\ }\textbf {\bibinfo {volume} {94}},\ \bibinfo
  {pages} {085103} (\bibinfo {year} {2016}{\natexlab{b}})}\BibitemShut
  {NoStop}%
\bibitem [{\citenamefont {Vitali}\ \emph {et~al.}(2016)\citenamefont {Vitali},
  \citenamefont {Shi}, \citenamefont {Qin},\ and\ \citenamefont
  {Zhang}}]{Vitali2016}%
  \BibitemOpen
  \bibfield  {author} {\bibinfo {author} {\bibfnamefont {E.}~\bibnamefont
  {Vitali}}, \bibinfo {author} {\bibfnamefont {H.}~\bibnamefont {Shi}},
  \bibinfo {author} {\bibfnamefont {M.}~\bibnamefont {Qin}},\ and\ \bibinfo
  {author} {\bibfnamefont {S.}~\bibnamefont {Zhang}},\ }\bibfield  {title}
  {\bibinfo {title} {Computation of dynamical correlation functions for
  many-fermion systems with auxiliary-field quantum monte carlo},\ }\href
  {https://doi.org/10.1103/PhysRevB.94.085140} {\bibfield  {journal} {\bibinfo
  {journal} {Phys. Rev. B}\ }\textbf {\bibinfo {volume} {94}},\ \bibinfo
  {pages} {085140} (\bibinfo {year} {2016})}\BibitemShut {NoStop}%
\bibitem [{\citenamefont {MacDonald}\ \emph {et~al.}(1988)\citenamefont
  {MacDonald}, \citenamefont {Girvin},\ and\ \citenamefont
  {Yoshioka}}]{MacDonald1988}%
  \BibitemOpen
  \bibfield  {author} {\bibinfo {author} {\bibfnamefont {A.~H.}\ \bibnamefont
  {MacDonald}}, \bibinfo {author} {\bibfnamefont {S.~M.}\ \bibnamefont
  {Girvin}},\ and\ \bibinfo {author} {\bibfnamefont {D.}~\bibnamefont
  {Yoshioka}},\ }\bibfield  {title} {\bibinfo {title} {$\frac{t}{U}$ expansion
  for the hubbard model},\ }\href {https://doi.org/10.1103/PhysRevB.37.9753}
  {\bibfield  {journal} {\bibinfo  {journal} {Phys. Rev. B}\ }\textbf {\bibinfo
  {volume} {37}},\ \bibinfo {pages} {9753} (\bibinfo {year}
  {1988})}\BibitemShut {NoStop}%
\bibitem [{\citenamefont {Ceperley}\ and\ \citenamefont
  {Pollock}(1989)}]{Ceperley1989}%
  \BibitemOpen
  \bibfield  {author} {\bibinfo {author} {\bibfnamefont {D.~M.}\ \bibnamefont
  {Ceperley}}\ and\ \bibinfo {author} {\bibfnamefont {E.~L.}\ \bibnamefont
  {Pollock}},\ }\bibfield  {title} {\bibinfo {title} {Path-integral simulation
  of the superfluid transition in two-dimensional $^{4}\mathrm{He}$},\ }\href
  {https://doi.org/10.1103/PhysRevB.39.2084} {\bibfield  {journal} {\bibinfo
  {journal} {Phys. Rev. B}\ }\textbf {\bibinfo {volume} {39}},\ \bibinfo
  {pages} {2084} (\bibinfo {year} {1989})}\BibitemShut {NoStop}%
\bibitem [{\citenamefont {Nguyen}\ and\ \citenamefont
  {Boninsegni}(2021)}]{Nguyen2021}%
  \BibitemOpen
  \bibfield  {author} {\bibinfo {author} {\bibfnamefont {P.~H.}\ \bibnamefont
  {Nguyen}}\ and\ \bibinfo {author} {\bibfnamefont {M.}~\bibnamefont
  {Boninsegni}},\ }\bibfield  {title} {\bibinfo {title} {Superfluid transition
  and specific heat of the 2d x-y model: Monte carlo simulation},\ }\bibfield
  {journal} {\bibinfo  {journal} {Applied Sciences}\ }\textbf {\bibinfo
  {volume} {11}},\ \href {https://doi.org/10.3390/app11114931}
  {10.3390/app11114931} (\bibinfo {year} {2021})\BibitemShut {NoStop}%
\bibitem [{\citenamefont {Ohashi}\ \emph {et~al.}(2008)\citenamefont {Ohashi},
  \citenamefont {Momoi}, \citenamefont {Tsunetsugu},\ and\ \citenamefont
  {Kawakami}}]{Ohashi2008}%
  \BibitemOpen
  \bibfield  {author} {\bibinfo {author} {\bibfnamefont {T.}~\bibnamefont
  {Ohashi}}, \bibinfo {author} {\bibfnamefont {T.}~\bibnamefont {Momoi}},
  \bibinfo {author} {\bibfnamefont {H.}~\bibnamefont {Tsunetsugu}},\ and\
  \bibinfo {author} {\bibfnamefont {N.}~\bibnamefont {Kawakami}},\ }\bibfield
  {title} {\bibinfo {title} {Finite temperature mott transition in hubbard
  model on anisotropic triangular lattice},\ }\href
  {https://doi.org/10.1103/PhysRevLett.100.076402} {\bibfield  {journal}
  {\bibinfo  {journal} {Phys. Rev. Lett.}\ }\textbf {\bibinfo {volume} {100}},\
  \bibinfo {pages} {076402} (\bibinfo {year} {2008})}\BibitemShut {NoStop}%
\bibitem [{\citenamefont {Park}\ \emph {et~al.}(2008)\citenamefont {Park},
  \citenamefont {Haule},\ and\ \citenamefont {Kotliar}}]{Park2008}%
  \BibitemOpen
  \bibfield  {author} {\bibinfo {author} {\bibfnamefont {H.}~\bibnamefont
  {Park}}, \bibinfo {author} {\bibfnamefont {K.}~\bibnamefont {Haule}},\ and\
  \bibinfo {author} {\bibfnamefont {G.}~\bibnamefont {Kotliar}},\ }\bibfield
  {title} {\bibinfo {title} {Cluster dynamical mean field theory of the mott
  transition},\ }\href {https://doi.org/10.1103/PhysRevLett.101.186403}
  {\bibfield  {journal} {\bibinfo  {journal} {Phys. Rev. Lett.}\ }\textbf
  {\bibinfo {volume} {101}},\ \bibinfo {pages} {186403} (\bibinfo {year}
  {2008})}\BibitemShut {NoStop}%
\bibitem [{\citenamefont {Vu\ifmmode \check{c}\else \v{c}\fi{}i\ifmmode
  \check{c}\else \v{c}\fi{}evi\ifmmode~\acute{c}\else \'{c}\fi{}}\ \emph
  {et~al.}(2013)\citenamefont {Vu\ifmmode \check{c}\else \v{c}\fi{}i\ifmmode
  \check{c}\else \v{c}\fi{}evi\ifmmode~\acute{c}\else \'{c}\fi{}},
  \citenamefont {Terletska}, \citenamefont {Tanaskovi\ifmmode~\acute{c}\else
  \'{c}\fi{}},\ and\ \citenamefont {Dobrosavljevi\ifmmode~\acute{c}\else
  \'{c}\fi{}}}]{Vuifmmode2013}%
  \BibitemOpen
  \bibfield  {author} {\bibinfo {author} {\bibfnamefont {J.}~\bibnamefont
  {Vu\ifmmode \check{c}\else \v{c}\fi{}i\ifmmode \check{c}\else
  \v{c}\fi{}evi\ifmmode~\acute{c}\else \'{c}\fi{}}}, \bibinfo {author}
  {\bibfnamefont {H.}~\bibnamefont {Terletska}}, \bibinfo {author}
  {\bibfnamefont {D.}~\bibnamefont {Tanaskovi\ifmmode~\acute{c}\else
  \'{c}\fi{}}},\ and\ \bibinfo {author} {\bibfnamefont {V.}~\bibnamefont
  {Dobrosavljevi\ifmmode~\acute{c}\else \'{c}\fi{}}},\ }\bibfield  {title}
  {\bibinfo {title} {Finite-temperature crossover and the quantum widom line
  near the mott transition},\ }\href
  {https://doi.org/10.1103/PhysRevB.88.075143} {\bibfield  {journal} {\bibinfo
  {journal} {Phys. Rev. B}\ }\textbf {\bibinfo {volume} {88}},\ \bibinfo
  {pages} {075143} (\bibinfo {year} {2013})}\BibitemShut {NoStop}%
\bibitem [{\citenamefont {Walsh}\ \emph
  {et~al.}(2019{\natexlab{a}})\citenamefont {Walsh}, \citenamefont {S\'emon},
  \citenamefont {Poulin}, \citenamefont {Sordi},\ and\ \citenamefont
  {Tremblay}}]{Walsh2019}%
  \BibitemOpen
  \bibfield  {author} {\bibinfo {author} {\bibfnamefont {C.}~\bibnamefont
  {Walsh}}, \bibinfo {author} {\bibfnamefont {P.}~\bibnamefont {S\'emon}},
  \bibinfo {author} {\bibfnamefont {D.}~\bibnamefont {Poulin}}, \bibinfo
  {author} {\bibfnamefont {G.}~\bibnamefont {Sordi}},\ and\ \bibinfo {author}
  {\bibfnamefont {A.-M.~S.}\ \bibnamefont {Tremblay}},\ }\bibfield  {title}
  {\bibinfo {title} {Thermodynamic and information-theoretic description of the
  mott transition in the two-dimensional hubbard model},\ }\href
  {https://doi.org/10.1103/PhysRevB.99.075122} {\bibfield  {journal} {\bibinfo
  {journal} {Phys. Rev. B}\ }\textbf {\bibinfo {volume} {99}},\ \bibinfo
  {pages} {075122} (\bibinfo {year} {2019}{\natexlab{a}})}\BibitemShut
  {NoStop}%
\bibitem [{\citenamefont {Walsh}\ \emph
  {et~al.}(2019{\natexlab{b}})\citenamefont {Walsh}, \citenamefont {S\'emon},
  \citenamefont {Poulin}, \citenamefont {Sordi},\ and\ \citenamefont
  {Tremblay}}]{Walsh2019L}%
  \BibitemOpen
  \bibfield  {author} {\bibinfo {author} {\bibfnamefont {C.}~\bibnamefont
  {Walsh}}, \bibinfo {author} {\bibfnamefont {P.}~\bibnamefont {S\'emon}},
  \bibinfo {author} {\bibfnamefont {D.}~\bibnamefont {Poulin}}, \bibinfo
  {author} {\bibfnamefont {G.}~\bibnamefont {Sordi}},\ and\ \bibinfo {author}
  {\bibfnamefont {A.-M.~S.}\ \bibnamefont {Tremblay}},\ }\bibfield  {title}
  {\bibinfo {title} {Local entanglement entropy and mutual information across
  the mott transition in the two-dimensional hubbard model},\ }\href
  {https://doi.org/10.1103/PhysRevLett.122.067203} {\bibfield  {journal}
  {\bibinfo  {journal} {Phys. Rev. Lett.}\ }\textbf {\bibinfo {volume} {122}},\
  \bibinfo {pages} {067203} (\bibinfo {year} {2019}{\natexlab{b}})}\BibitemShut
  {NoStop}%
\bibitem [{\citenamefont {Downey}\ \emph {et~al.}(2023)\citenamefont {Downey},
  \citenamefont {Gingras}, \citenamefont {Fournier}, \citenamefont {H\'ebert},
  \citenamefont {Charlebois},\ and\ \citenamefont {Tremblay}}]{Downey2023}%
  \BibitemOpen
  \bibfield  {author} {\bibinfo {author} {\bibfnamefont {P.-O.}\ \bibnamefont
  {Downey}}, \bibinfo {author} {\bibfnamefont {O.}~\bibnamefont {Gingras}},
  \bibinfo {author} {\bibfnamefont {J.}~\bibnamefont {Fournier}}, \bibinfo
  {author} {\bibfnamefont {C.-D.}\ \bibnamefont {H\'ebert}}, \bibinfo {author}
  {\bibfnamefont {M.}~\bibnamefont {Charlebois}},\ and\ \bibinfo {author}
  {\bibfnamefont {A.-M.~S.}\ \bibnamefont {Tremblay}},\ }\bibfield  {title}
  {\bibinfo {title} {Mott transition, widom line, and pseudogap in the
  half-filled triangular lattice hubbard model},\ }\href
  {https://doi.org/10.1103/PhysRevB.107.125159} {\bibfield  {journal} {\bibinfo
   {journal} {Phys. Rev. B}\ }\textbf {\bibinfo {volume} {107}},\ \bibinfo
  {pages} {125159} (\bibinfo {year} {2023})}\BibitemShut {NoStop}%
\bibitem [{\citenamefont {\ifmmode~\check{S}\else \v{S}\fi{}imkovic}\ \emph
  {et~al.}(2020)\citenamefont {\ifmmode~\check{S}\else \v{S}\fi{}imkovic},
  \citenamefont {LeBlanc}, \citenamefont {Kim}, \citenamefont {Deng},
  \citenamefont {Prokof'ev}, \citenamefont {Svistunov},\ and\ \citenamefont
  {Kozik}}]{Svistunov2020}%
  \BibitemOpen
  \bibfield  {author} {\bibinfo {author} {\bibfnamefont {F.}~\bibnamefont
  {\ifmmode~\check{S}\else \v{S}\fi{}imkovic}}, \bibinfo {author}
  {\bibfnamefont {J.~P.~F.}\ \bibnamefont {LeBlanc}}, \bibinfo {author}
  {\bibfnamefont {A.~J.}\ \bibnamefont {Kim}}, \bibinfo {author} {\bibfnamefont
  {Y.}~\bibnamefont {Deng}}, \bibinfo {author} {\bibfnamefont {N.~V.}\
  \bibnamefont {Prokof'ev}}, \bibinfo {author} {\bibfnamefont {B.~V.}\
  \bibnamefont {Svistunov}},\ and\ \bibinfo {author} {\bibfnamefont
  {E.}~\bibnamefont {Kozik}},\ }\bibfield  {title} {\bibinfo {title} {Extended
  crossover from a fermi liquid to a quasiantiferromagnet in the half-filled 2d
  hubbard model},\ }\href {https://doi.org/10.1103/PhysRevLett.124.017003}
  {\bibfield  {journal} {\bibinfo  {journal} {Phys. Rev. Lett.}\ }\textbf
  {\bibinfo {volume} {124}},\ \bibinfo {pages} {017003} (\bibinfo {year}
  {2020})}\BibitemShut {NoStop}%
\bibitem [{\citenamefont {Kim}\ \emph {et~al.}(2020)\citenamefont {Kim},
  \citenamefont {Simkovic},\ and\ \citenamefont {Kozik}}]{Kim2021}%
  \BibitemOpen
  \bibfield  {author} {\bibinfo {author} {\bibfnamefont {A.~J.}\ \bibnamefont
  {Kim}}, \bibinfo {author} {\bibfnamefont {F.}~\bibnamefont {Simkovic}},\ and\
  \bibinfo {author} {\bibfnamefont {E.}~\bibnamefont {Kozik}},\ }\bibfield
  {title} {\bibinfo {title} {Spin and charge correlations across the
  metal-to-insulator crossover in the half-filled 2d hubbard model},\ }\href
  {https://doi.org/10.1103/PhysRevLett.124.117602} {\bibfield  {journal}
  {\bibinfo  {journal} {Phys. Rev. Lett.}\ }\textbf {\bibinfo {volume} {124}},\
  \bibinfo {pages} {117602} (\bibinfo {year} {2020})}\BibitemShut {NoStop}%
\bibitem [{\citenamefont {Zhang}\ \emph {et~al.}(1993)\citenamefont {Zhang},
  \citenamefont {Rozenberg},\ and\ \citenamefont {Kotliar}}]{Rozenberg1993}%
  \BibitemOpen
  \bibfield  {author} {\bibinfo {author} {\bibfnamefont {X.~Y.}\ \bibnamefont
  {Zhang}}, \bibinfo {author} {\bibfnamefont {M.~J.}\ \bibnamefont
  {Rozenberg}},\ and\ \bibinfo {author} {\bibfnamefont {G.}~\bibnamefont
  {Kotliar}},\ }\bibfield  {title} {\bibinfo {title} {Mott transition in the
  d=\ensuremath{\infty} hubbard model at zero temperature},\ }\href
  {https://doi.org/10.1103/PhysRevLett.70.1666} {\bibfield  {journal} {\bibinfo
   {journal} {Phys. Rev. Lett.}\ }\textbf {\bibinfo {volume} {70}},\ \bibinfo
  {pages} {1666} (\bibinfo {year} {1993})}\BibitemShut {NoStop}%
\bibitem [{\citenamefont {Lee}\ \emph {et~al.}(2006)\citenamefont {Lee},
  \citenamefont {Nagaosa},\ and\ \citenamefont {Wen}}]{Patrick2006}%
  \BibitemOpen
  \bibfield  {author} {\bibinfo {author} {\bibfnamefont {P.~A.}\ \bibnamefont
  {Lee}}, \bibinfo {author} {\bibfnamefont {N.}~\bibnamefont {Nagaosa}},\ and\
  \bibinfo {author} {\bibfnamefont {X.-G.}\ \bibnamefont {Wen}},\ }\bibfield
  {title} {\bibinfo {title} {Doping a mott insulator: Physics of
  high-temperature superconductivity},\ }\href
  {https://doi.org/10.1103/RevModPhys.78.17} {\bibfield  {journal} {\bibinfo
  {journal} {Rev. Mod. Phys.}\ }\textbf {\bibinfo {volume} {78}},\ \bibinfo
  {pages} {17} (\bibinfo {year} {2006})}\BibitemShut {NoStop}%
\bibitem [{\citenamefont {Fischer}\ \emph {et~al.}(2007)\citenamefont
  {Fischer}, \citenamefont {Kugler}, \citenamefont {Maggio-Aprile},
  \citenamefont {Berthod},\ and\ \citenamefont {Renner}}]{Fischer2007}%
  \BibitemOpen
  \bibfield  {author} {\bibinfo {author} {\bibfnamefont {O.}~\bibnamefont
  {Fischer}}, \bibinfo {author} {\bibfnamefont {M.}~\bibnamefont {Kugler}},
  \bibinfo {author} {\bibfnamefont {I.}~\bibnamefont {Maggio-Aprile}}, \bibinfo
  {author} {\bibfnamefont {C.}~\bibnamefont {Berthod}},\ and\ \bibinfo {author}
  {\bibfnamefont {C.}~\bibnamefont {Renner}},\ }\bibfield  {title} {\bibinfo
  {title} {Scanning tunneling spectroscopy of high-temperature
  superconductors},\ }\href {https://doi.org/10.1103/RevModPhys.79.353}
  {\bibfield  {journal} {\bibinfo  {journal} {Rev. Mod. Phys.}\ }\textbf
  {\bibinfo {volume} {79}},\ \bibinfo {pages} {353} (\bibinfo {year}
  {2007})}\BibitemShut {NoStop}%
\bibitem [{\citenamefont {Boschini}\ \emph {et~al.}(2020)\citenamefont
  {Boschini}, \citenamefont {Zonno}, \citenamefont {Razzoli}, \citenamefont
  {Day}, \citenamefont {Michiardi}, \citenamefont {Zwartsenberg}, \citenamefont
  {Nigge}, \citenamefont {Schneider}, \citenamefont {da~Silva~Neto},
  \citenamefont {Erb}, \citenamefont {Zhdanovich}, \citenamefont {Mills},
  \citenamefont {Levy}, \citenamefont {Giannetti}, \citenamefont {Jones},\ and\
  \citenamefont {Damascelli}}]{Boschini2020}%
  \BibitemOpen
  \bibfield  {author} {\bibinfo {author} {\bibfnamefont {F.}~\bibnamefont
  {Boschini}}, \bibinfo {author} {\bibfnamefont {M.}~\bibnamefont {Zonno}},
  \bibinfo {author} {\bibfnamefont {E.}~\bibnamefont {Razzoli}}, \bibinfo
  {author} {\bibfnamefont {R.~P.}\ \bibnamefont {Day}}, \bibinfo {author}
  {\bibfnamefont {M.}~\bibnamefont {Michiardi}}, \bibinfo {author}
  {\bibfnamefont {B.}~\bibnamefont {Zwartsenberg}}, \bibinfo {author}
  {\bibfnamefont {P.}~\bibnamefont {Nigge}}, \bibinfo {author} {\bibfnamefont
  {M.}~\bibnamefont {Schneider}}, \bibinfo {author} {\bibfnamefont {E.~H.}\
  \bibnamefont {da~Silva~Neto}}, \bibinfo {author} {\bibfnamefont
  {A.}~\bibnamefont {Erb}}, \bibinfo {author} {\bibfnamefont {S.}~\bibnamefont
  {Zhdanovich}}, \bibinfo {author} {\bibfnamefont {A.~K.}\ \bibnamefont
  {Mills}}, \bibinfo {author} {\bibfnamefont {G.}~\bibnamefont {Levy}},
  \bibinfo {author} {\bibfnamefont {C.}~\bibnamefont {Giannetti}}, \bibinfo
  {author} {\bibfnamefont {D.~J.}\ \bibnamefont {Jones}},\ and\ \bibinfo
  {author} {\bibfnamefont {A.}~\bibnamefont {Damascelli}},\ }\bibfield  {title}
  {\bibinfo {title} {Emergence of pseudogap from short-range spin-correlations
  in electron-doped cuprates},\ }\href
  {https://doi.org/10.1038/s41535-020-0208-6} {\bibfield  {journal} {\bibinfo
  {journal} {npj Quantum Materials}\ }\textbf {\bibinfo {volume} {5}},\
  \bibinfo {pages} {6} (\bibinfo {year} {2020})}\BibitemShut {NoStop}%
\bibitem [{\citenamefont {Bauer}\ \emph {et~al.}(2014)\citenamefont {Bauer},
  \citenamefont {Parish},\ and\ \citenamefont {Enss}}]{Bauer2014}%
  \BibitemOpen
  \bibfield  {author} {\bibinfo {author} {\bibfnamefont {M.}~\bibnamefont
  {Bauer}}, \bibinfo {author} {\bibfnamefont {M.~M.}\ \bibnamefont {Parish}},\
  and\ \bibinfo {author} {\bibfnamefont {T.}~\bibnamefont {Enss}},\ }\bibfield
  {title} {\bibinfo {title} {Universal equation of state and pseudogap in the
  two-dimensional fermi gas},\ }\href
  {https://doi.org/10.1103/PhysRevLett.112.135302} {\bibfield  {journal}
  {\bibinfo  {journal} {Phys. Rev. Lett.}\ }\textbf {\bibinfo {volume} {112}},\
  \bibinfo {pages} {135302} (\bibinfo {year} {2014})}\BibitemShut {NoStop}%
\bibitem [{\citenamefont {Mueller}(2017)}]{Mueller2017}%
  \BibitemOpen
  \bibfield  {author} {\bibinfo {author} {\bibfnamefont {E.~J.}\ \bibnamefont
  {Mueller}},\ }\bibfield  {title} {\bibinfo {title} {Review of pseudogaps in
  strongly interacting fermi gases},\ }\href
  {https://doi.org/10.1088/1361-6633/aa7e53} {\bibfield  {journal} {\bibinfo
  {journal} {Reports on Progress in Physics}\ }\textbf {\bibinfo {volume}
  {80}},\ \bibinfo {pages} {104401} (\bibinfo {year} {2017})}\BibitemShut
  {NoStop}%
\bibitem [{\citenamefont {Li}\ \emph {et~al.}(2024)\citenamefont {Li},
  \citenamefont {Wang}, \citenamefont {Luo}, \citenamefont {Zhou},
  \citenamefont {Xie}, \citenamefont {Shen}, \citenamefont {Nie}, \citenamefont
  {Chen}, \citenamefont {Hu}, \citenamefont {Chen}, \citenamefont {Yao},\ and\
  \citenamefont {Pan}}]{XiangLi2024}%
  \BibitemOpen
  \bibfield  {author} {\bibinfo {author} {\bibfnamefont {X.}~\bibnamefont
  {Li}}, \bibinfo {author} {\bibfnamefont {S.}~\bibnamefont {Wang}}, \bibinfo
  {author} {\bibfnamefont {X.}~\bibnamefont {Luo}}, \bibinfo {author}
  {\bibfnamefont {Y.-Y.}\ \bibnamefont {Zhou}}, \bibinfo {author}
  {\bibfnamefont {K.}~\bibnamefont {Xie}}, \bibinfo {author} {\bibfnamefont
  {H.-C.}\ \bibnamefont {Shen}}, \bibinfo {author} {\bibfnamefont {Y.-Z.}\
  \bibnamefont {Nie}}, \bibinfo {author} {\bibfnamefont {Q.}~\bibnamefont
  {Chen}}, \bibinfo {author} {\bibfnamefont {H.}~\bibnamefont {Hu}}, \bibinfo
  {author} {\bibfnamefont {Y.-A.}\ \bibnamefont {Chen}}, \bibinfo {author}
  {\bibfnamefont {X.-C.}\ \bibnamefont {Yao}},\ and\ \bibinfo {author}
  {\bibfnamefont {J.-W.}\ \bibnamefont {Pan}},\ }\bibfield  {title} {\bibinfo
  {title} {Observation and quantification of the pseudogap in unitary fermi
  gases},\ }\href {https://doi.org/10.1038/s41586-023-06964-y} {\bibfield
  {journal} {\bibinfo  {journal} {Nature}\ }\textbf {\bibinfo {volume} {626}},\
  \bibinfo {pages} {288} (\bibinfo {year} {2024})}\BibitemShut {NoStop}%
\bibitem [{\citenamefont {Sch\"afer}\ \emph {et~al.}(2021)\citenamefont
  {Sch\"afer}, \citenamefont {Wentzell}, \citenamefont {\ifmmode~\check{S}\else
  \v{S}\fi{}imkovic}, \citenamefont {He}, \citenamefont {Hille}, \citenamefont
  {Klett}, \citenamefont {Eckhardt}, \citenamefont {Arzhang}, \citenamefont
  {Harkov}, \citenamefont {Le~R\'egent}, \citenamefont {Kirsch}, \citenamefont
  {Wang}, \citenamefont {Kim}, \citenamefont {Kozik}, \citenamefont {Stepanov},
  \citenamefont {Kauch}, \citenamefont {Andergassen}, \citenamefont {Hansmann},
  \citenamefont {Rohe}, \citenamefont {Vilk}, \citenamefont {LeBlanc},
  \citenamefont {Zhang}, \citenamefont {Tremblay}, \citenamefont {Ferrero},
  \citenamefont {Parcollet},\ and\ \citenamefont {Georges}}]{Thomas2021}%
  \BibitemOpen
  \bibfield  {author} {\bibinfo {author} {\bibfnamefont {T.}~\bibnamefont
  {Sch\"afer}}, \bibinfo {author} {\bibfnamefont {N.}~\bibnamefont {Wentzell}},
  \bibinfo {author} {\bibfnamefont {F.}~\bibnamefont {\ifmmode~\check{S}\else
  \v{S}\fi{}imkovic}}, \bibinfo {author} {\bibfnamefont {Y.-Y.}\ \bibnamefont
  {He}}, \bibinfo {author} {\bibfnamefont {C.}~\bibnamefont {Hille}}, \bibinfo
  {author} {\bibfnamefont {M.}~\bibnamefont {Klett}}, \bibinfo {author}
  {\bibfnamefont {C.~J.}\ \bibnamefont {Eckhardt}}, \bibinfo {author}
  {\bibfnamefont {B.}~\bibnamefont {Arzhang}}, \bibinfo {author} {\bibfnamefont
  {V.}~\bibnamefont {Harkov}}, \bibinfo {author} {\bibfnamefont {F.~m. c.-M.}\
  \bibnamefont {Le~R\'egent}}, \bibinfo {author} {\bibfnamefont
  {A.}~\bibnamefont {Kirsch}}, \bibinfo {author} {\bibfnamefont
  {Y.}~\bibnamefont {Wang}}, \bibinfo {author} {\bibfnamefont {A.~J.}\
  \bibnamefont {Kim}}, \bibinfo {author} {\bibfnamefont {E.}~\bibnamefont
  {Kozik}}, \bibinfo {author} {\bibfnamefont {E.~A.}\ \bibnamefont {Stepanov}},
  \bibinfo {author} {\bibfnamefont {A.}~\bibnamefont {Kauch}}, \bibinfo
  {author} {\bibfnamefont {S.}~\bibnamefont {Andergassen}}, \bibinfo {author}
  {\bibfnamefont {P.}~\bibnamefont {Hansmann}}, \bibinfo {author}
  {\bibfnamefont {D.}~\bibnamefont {Rohe}}, \bibinfo {author} {\bibfnamefont
  {Y.~M.}\ \bibnamefont {Vilk}}, \bibinfo {author} {\bibfnamefont {J.~P.~F.}\
  \bibnamefont {LeBlanc}}, \bibinfo {author} {\bibfnamefont {S.}~\bibnamefont
  {Zhang}}, \bibinfo {author} {\bibfnamefont {A.-M.~S.}\ \bibnamefont
  {Tremblay}}, \bibinfo {author} {\bibfnamefont {M.}~\bibnamefont {Ferrero}},
  \bibinfo {author} {\bibfnamefont {O.}~\bibnamefont {Parcollet}},\ and\
  \bibinfo {author} {\bibfnamefont {A.}~\bibnamefont {Georges}},\ }\bibfield
  {title} {\bibinfo {title} {Tracking the footprints of spin fluctuations: A
  multimethod, multimessenger study of the two-dimensional hubbard model},\
  }\href {https://doi.org/10.1103/PhysRevX.11.011058} {\bibfield  {journal}
  {\bibinfo  {journal} {Phys. Rev. X}\ }\textbf {\bibinfo {volume} {11}},\
  \bibinfo {pages} {011058} (\bibinfo {year} {2021})}\BibitemShut {NoStop}%
\bibitem [{\citenamefont {Parcollet}\ \emph {et~al.}(2004)\citenamefont
  {Parcollet}, \citenamefont {Biroli},\ and\ \citenamefont
  {Kotliar}}]{Parcollet2004}%
  \BibitemOpen
  \bibfield  {author} {\bibinfo {author} {\bibfnamefont {O.}~\bibnamefont
  {Parcollet}}, \bibinfo {author} {\bibfnamefont {G.}~\bibnamefont {Biroli}},\
  and\ \bibinfo {author} {\bibfnamefont {G.}~\bibnamefont {Kotliar}},\
  }\bibfield  {title} {\bibinfo {title} {Cluster dynamical mean field analysis
  of the mott transition},\ }\href
  {https://doi.org/10.1103/PhysRevLett.92.226402} {\bibfield  {journal}
  {\bibinfo  {journal} {Phys. Rev. Lett.}\ }\textbf {\bibinfo {volume} {92}},\
  \bibinfo {pages} {226402} (\bibinfo {year} {2004})}\BibitemShut {NoStop}%
\bibitem [{\citenamefont {Chiesa}\ \emph {et~al.}(2011)\citenamefont {Chiesa},
  \citenamefont {Varney}, \citenamefont {Rigol},\ and\ \citenamefont
  {Scalettar}}]{Chiesa2011}%
  \BibitemOpen
  \bibfield  {author} {\bibinfo {author} {\bibfnamefont {S.}~\bibnamefont
  {Chiesa}}, \bibinfo {author} {\bibfnamefont {C.~N.}\ \bibnamefont {Varney}},
  \bibinfo {author} {\bibfnamefont {M.}~\bibnamefont {Rigol}},\ and\ \bibinfo
  {author} {\bibfnamefont {R.~T.}\ \bibnamefont {Scalettar}},\ }\bibfield
  {title} {\bibinfo {title} {Magnetism and pairing of two-dimensional trapped
  fermions},\ }\href {https://doi.org/10.1103/PhysRevLett.106.035301}
  {\bibfield  {journal} {\bibinfo  {journal} {Phys. Rev. Lett.}\ }\textbf
  {\bibinfo {volume} {106}},\ \bibinfo {pages} {035301} (\bibinfo {year}
  {2011})}\BibitemShut {NoStop}%
\bibitem [{\citenamefont {Khatami}\ and\ \citenamefont
  {Rigol}(2011)}]{Khatami2011}%
  \BibitemOpen
  \bibfield  {author} {\bibinfo {author} {\bibfnamefont {E.}~\bibnamefont
  {Khatami}}\ and\ \bibinfo {author} {\bibfnamefont {M.}~\bibnamefont
  {Rigol}},\ }\bibfield  {title} {\bibinfo {title} {Thermodynamics of strongly
  interacting fermions in two-dimensional optical lattices},\ }\href
  {https://doi.org/10.1103/PhysRevA.84.053611} {\bibfield  {journal} {\bibinfo
  {journal} {Phys. Rev. A}\ }\textbf {\bibinfo {volume} {84}},\ \bibinfo
  {pages} {053611} (\bibinfo {year} {2011})}\BibitemShut {NoStop}%
\bibitem [{\citenamefont {Qin}\ \emph {et~al.}(2017)\citenamefont {Qin},
  \citenamefont {Shi},\ and\ \citenamefont {Zhang}}]{Mingpu2017}%
  \BibitemOpen
  \bibfield  {author} {\bibinfo {author} {\bibfnamefont {M.}~\bibnamefont
  {Qin}}, \bibinfo {author} {\bibfnamefont {H.}~\bibnamefont {Shi}},\ and\
  \bibinfo {author} {\bibfnamefont {S.}~\bibnamefont {Zhang}},\ }\bibfield
  {title} {\bibinfo {title} {Numerical results on the short-range spin
  correlation functions in the ground state of the two-dimensional hubbard
  model},\ }\href {https://doi.org/10.1103/PhysRevB.96.075156} {\bibfield
  {journal} {\bibinfo  {journal} {Phys. Rev. B}\ }\textbf {\bibinfo {volume}
  {96}},\ \bibinfo {pages} {075156} (\bibinfo {year} {2017})}\BibitemShut
  {NoStop}%
\bibitem [{\citenamefont {Wessel}(2010)}]{Wessel2010}%
  \BibitemOpen
  \bibfield  {author} {\bibinfo {author} {\bibfnamefont {S.}~\bibnamefont
  {Wessel}},\ }\bibfield  {title} {\bibinfo {title} {Critical entropy of
  quantum heisenberg magnets on simple-cubic lattices},\ }\href
  {https://doi.org/10.1103/PhysRevB.81.052405} {\bibfield  {journal} {\bibinfo
  {journal} {Phys. Rev. B}\ }\textbf {\bibinfo {volume} {81}},\ \bibinfo
  {pages} {052405} (\bibinfo {year} {2010})}\BibitemShut {NoStop}%
\bibitem [{\citenamefont {Richardson}(1997)}]{Richardson1997}%
  \BibitemOpen
  \bibfield  {author} {\bibinfo {author} {\bibfnamefont {R.~C.}\ \bibnamefont
  {Richardson}},\ }\bibfield  {title} {\bibinfo {title} {The pomeranchuk
  effect},\ }\href {https://doi.org/10.1103/RevModPhys.69.683} {\bibfield
  {journal} {\bibinfo  {journal} {Rev. Mod. Phys.}\ }\textbf {\bibinfo {volume}
  {69}},\ \bibinfo {pages} {683} (\bibinfo {year} {1997})}\BibitemShut
  {NoStop}%
\bibitem [{\citenamefont {Georges}\ and\ \citenamefont
  {Krauth}(1992)}]{Georges1992}%
  \BibitemOpen
  \bibfield  {author} {\bibinfo {author} {\bibfnamefont {A.}~\bibnamefont
  {Georges}}\ and\ \bibinfo {author} {\bibfnamefont {W.}~\bibnamefont
  {Krauth}},\ }\bibfield  {title} {\bibinfo {title} {Numerical solution of the
  d=\ensuremath{\infty} hubbard model: Evidence for a mott transition},\ }\href
  {https://doi.org/10.1103/PhysRevLett.69.1240} {\bibfield  {journal} {\bibinfo
   {journal} {Phys. Rev. Lett.}\ }\textbf {\bibinfo {volume} {69}},\ \bibinfo
  {pages} {1240} (\bibinfo {year} {1992})}\BibitemShut {NoStop}%
\bibitem [{\citenamefont {Li}\ \emph {et~al.}(2014)\citenamefont {Li},
  \citenamefont {Antipov}, \citenamefont {Rubtsov}, \citenamefont {Kirchner},\
  and\ \citenamefont {Hanke}}]{GangLi2014}%
  \BibitemOpen
  \bibfield  {author} {\bibinfo {author} {\bibfnamefont {G.}~\bibnamefont
  {Li}}, \bibinfo {author} {\bibfnamefont {A.~E.}\ \bibnamefont {Antipov}},
  \bibinfo {author} {\bibfnamefont {A.~N.}\ \bibnamefont {Rubtsov}}, \bibinfo
  {author} {\bibfnamefont {S.}~\bibnamefont {Kirchner}},\ and\ \bibinfo
  {author} {\bibfnamefont {W.}~\bibnamefont {Hanke}},\ }\bibfield  {title}
  {\bibinfo {title} {Competing phases of the hubbard model on a triangular
  lattice: Insights from the entropy},\ }\href
  {https://doi.org/10.1103/PhysRevB.89.161118} {\bibfield  {journal} {\bibinfo
  {journal} {Phys. Rev. B}\ }\textbf {\bibinfo {volume} {89}},\ \bibinfo
  {pages} {161118} (\bibinfo {year} {2014})}\BibitemShut {NoStop}%
\bibitem [{\citenamefont {Wietek}\ \emph {et~al.}(2021)\citenamefont {Wietek},
  \citenamefont {Rossi}, \citenamefont {\ifmmode~\check{S}\else
  \v{S}\fi{}imkovic}, \citenamefont {Klett}, \citenamefont {Hansmann},
  \citenamefont {Ferrero}, \citenamefont {Stoudenmire}, \citenamefont
  {Sch\"afer},\ and\ \citenamefont {Georges}}]{Wietek2021}%
  \BibitemOpen
  \bibfield  {author} {\bibinfo {author} {\bibfnamefont {A.}~\bibnamefont
  {Wietek}}, \bibinfo {author} {\bibfnamefont {R.}~\bibnamefont {Rossi}},
  \bibinfo {author} {\bibfnamefont {F.}~\bibnamefont {\ifmmode~\check{S}\else
  \v{S}\fi{}imkovic}}, \bibinfo {author} {\bibfnamefont {M.}~\bibnamefont
  {Klett}}, \bibinfo {author} {\bibfnamefont {P.}~\bibnamefont {Hansmann}},
  \bibinfo {author} {\bibfnamefont {M.}~\bibnamefont {Ferrero}}, \bibinfo
  {author} {\bibfnamefont {E.~M.}\ \bibnamefont {Stoudenmire}}, \bibinfo
  {author} {\bibfnamefont {T.}~\bibnamefont {Sch\"afer}},\ and\ \bibinfo
  {author} {\bibfnamefont {A.}~\bibnamefont {Georges}},\ }\bibfield  {title}
  {\bibinfo {title} {Mott insulating states with competing orders in the
  triangular lattice hubbard model},\ }\href
  {https://doi.org/10.1103/PhysRevX.11.041013} {\bibfield  {journal} {\bibinfo
  {journal} {Phys. Rev. X}\ }\textbf {\bibinfo {volume} {11}},\ \bibinfo
  {pages} {041013} (\bibinfo {year} {2021})}\BibitemShut {NoStop}%
\bibitem [{\citenamefont {Li}\ \emph {et~al.}(2023)\citenamefont {Li},
  \citenamefont {Gao}, \citenamefont {He}, \citenamefont {Qi}, \citenamefont
  {Chen},\ and\ \citenamefont {Li}}]{Qiaoyi2023}%
  \BibitemOpen
  \bibfield  {author} {\bibinfo {author} {\bibfnamefont {Q.}~\bibnamefont
  {Li}}, \bibinfo {author} {\bibfnamefont {Y.}~\bibnamefont {Gao}}, \bibinfo
  {author} {\bibfnamefont {Y.-Y.}\ \bibnamefont {He}}, \bibinfo {author}
  {\bibfnamefont {Y.}~\bibnamefont {Qi}}, \bibinfo {author} {\bibfnamefont
  {B.-B.}\ \bibnamefont {Chen}},\ and\ \bibinfo {author} {\bibfnamefont
  {W.}~\bibnamefont {Li}},\ }\bibfield  {title} {\bibinfo {title} {Tangent
  space approach for thermal tensor network simulations of the 2d hubbard
  model},\ }\href {https://doi.org/10.1103/PhysRevLett.130.226502} {\bibfield
  {journal} {\bibinfo  {journal} {Phys. Rev. Lett.}\ }\textbf {\bibinfo
  {volume} {130}},\ \bibinfo {pages} {226502} (\bibinfo {year}
  {2023})}\BibitemShut {NoStop}%
\bibitem [{\citenamefont {Rozen}\ \emph {et~al.}(2021)\citenamefont {Rozen},
  \citenamefont {Park}, \citenamefont {Zondiner}, \citenamefont {Cao},
  \citenamefont {Rodan-Legrain}, \citenamefont {Taniguchi}, \citenamefont
  {Watanabe}, \citenamefont {Oreg}, \citenamefont {Stern}, \citenamefont
  {Berg}, \citenamefont {Jarillo-Herrero},\ and\ \citenamefont
  {Ilani}}]{Rozen2021}%
  \BibitemOpen
  \bibfield  {author} {\bibinfo {author} {\bibfnamefont {A.}~\bibnamefont
  {Rozen}}, \bibinfo {author} {\bibfnamefont {J.~M.}\ \bibnamefont {Park}},
  \bibinfo {author} {\bibfnamefont {U.}~\bibnamefont {Zondiner}}, \bibinfo
  {author} {\bibfnamefont {Y.}~\bibnamefont {Cao}}, \bibinfo {author}
  {\bibfnamefont {D.}~\bibnamefont {Rodan-Legrain}}, \bibinfo {author}
  {\bibfnamefont {T.}~\bibnamefont {Taniguchi}}, \bibinfo {author}
  {\bibfnamefont {K.}~\bibnamefont {Watanabe}}, \bibinfo {author}
  {\bibfnamefont {Y.}~\bibnamefont {Oreg}}, \bibinfo {author} {\bibfnamefont
  {A.}~\bibnamefont {Stern}}, \bibinfo {author} {\bibfnamefont
  {E.}~\bibnamefont {Berg}}, \bibinfo {author} {\bibfnamefont {P.}~\bibnamefont
  {Jarillo-Herrero}},\ and\ \bibinfo {author} {\bibfnamefont {S.}~\bibnamefont
  {Ilani}},\ }\bibfield  {title} {\bibinfo {title} {Entropic evidence for a
  pomeranchuk effect in magic-angle graphene},\ }\href
  {https://doi.org/10.1038/s41586-021-03319-3} {\bibfield  {journal} {\bibinfo
  {journal} {Nature}\ }\textbf {\bibinfo {volume} {592}},\ \bibinfo {pages}
  {214} (\bibinfo {year} {2021})}\BibitemShut {NoStop}%
\bibitem [{\citenamefont {Taie}\ \emph {et~al.}(2012)\citenamefont {Taie},
  \citenamefont {Yamazaki}, \citenamefont {Sugawa},\ and\ \citenamefont
  {Takahashi}}]{Taie2012}%
  \BibitemOpen
  \bibfield  {author} {\bibinfo {author} {\bibfnamefont {S.}~\bibnamefont
  {Taie}}, \bibinfo {author} {\bibfnamefont {R.}~\bibnamefont {Yamazaki}},
  \bibinfo {author} {\bibfnamefont {S.}~\bibnamefont {Sugawa}},\ and\ \bibinfo
  {author} {\bibfnamefont {Y.}~\bibnamefont {Takahashi}},\ }\bibfield  {title}
  {\bibinfo {title} {An su(6) mott insulator of an atomic fermi gas realized by
  large-spin pomeranchuk cooling},\ }\href {https://doi.org/10.1038/nphys2430}
  {\bibfield  {journal} {\bibinfo  {journal} {Nature Physics}\ }\textbf
  {\bibinfo {volume} {8}},\ \bibinfo {pages} {825} (\bibinfo {year}
  {2012})}\BibitemShut {NoStop}%
\bibitem [{\citenamefont {Duffy}\ and\ \citenamefont
  {Moreo}(1997)}]{Duffy1997}%
  \BibitemOpen
  \bibfield  {author} {\bibinfo {author} {\bibfnamefont {D.}~\bibnamefont
  {Duffy}}\ and\ \bibinfo {author} {\bibfnamefont {A.}~\bibnamefont {Moreo}},\
  }\bibfield  {title} {\bibinfo {title} {Specific heat of the two-dimensional
  hubbard model},\ }\href {https://doi.org/10.1103/PhysRevB.55.12918}
  {\bibfield  {journal} {\bibinfo  {journal} {Phys. Rev. B}\ }\textbf {\bibinfo
  {volume} {55}},\ \bibinfo {pages} {12918} (\bibinfo {year}
  {1997})}\BibitemShut {NoStop}%
\bibitem [{\citenamefont {Paiva}\ \emph {et~al.}(2001)\citenamefont {Paiva},
  \citenamefont {Scalettar}, \citenamefont {Huscroft},\ and\ \citenamefont
  {McMahan}}]{Paiva2001}%
  \BibitemOpen
  \bibfield  {author} {\bibinfo {author} {\bibfnamefont {T.}~\bibnamefont
  {Paiva}}, \bibinfo {author} {\bibfnamefont {R.~T.}\ \bibnamefont
  {Scalettar}}, \bibinfo {author} {\bibfnamefont {C.}~\bibnamefont
  {Huscroft}},\ and\ \bibinfo {author} {\bibfnamefont {A.~K.}\ \bibnamefont
  {McMahan}},\ }\bibfield  {title} {\bibinfo {title} {Signatures of spin and
  charge energy scales in the local moment and specific heat of the half-filled
  two-dimensional hubbard model},\ }\href
  {https://doi.org/10.1103/PhysRevB.63.125116} {\bibfield  {journal} {\bibinfo
  {journal} {Phys. Rev. B}\ }\textbf {\bibinfo {volume} {63}},\ \bibinfo
  {pages} {125116} (\bibinfo {year} {2001})}\BibitemShut {NoStop}%
\bibitem [{\citenamefont {Huang}\ \emph {et~al.}(2019)\citenamefont {Huang},
  \citenamefont {Sheppard}, \citenamefont {Moritz},\ and\ \citenamefont
  {Devereaux}}]{Edwin2019}%
  \BibitemOpen
  \bibfield  {author} {\bibinfo {author} {\bibfnamefont {E.~W.}\ \bibnamefont
  {Huang}}, \bibinfo {author} {\bibfnamefont {R.}~\bibnamefont {Sheppard}},
  \bibinfo {author} {\bibfnamefont {B.}~\bibnamefont {Moritz}},\ and\ \bibinfo
  {author} {\bibfnamefont {T.~P.}\ \bibnamefont {Devereaux}},\ }\bibfield
  {title} {\bibinfo {title} {Strange metallicity in the doped hubbard model},\
  }\href {https://doi.org/10.1126/science.aau7063} {\bibfield  {journal}
  {\bibinfo  {journal} {Science}\ }\textbf {\bibinfo {volume} {366}},\ \bibinfo
  {pages} {987} (\bibinfo {year} {2019})}\BibitemShut {NoStop}%
\bibitem [{\citenamefont {Brown}\ \emph {et~al.}(2019)\citenamefont {Brown},
  \citenamefont {Mitra}, \citenamefont {Guardado-Sanchez}, \citenamefont
  {Nourafkan}, \citenamefont {Reymbaut}, \citenamefont {Hébert}, \citenamefont
  {Bergeron}, \citenamefont {Tremblay}, \citenamefont {Kokalj}, \citenamefont
  {Huse}, \citenamefont {Schauß},\ and\ \citenamefont {Bakr}}]{Peter2019}%
  \BibitemOpen
  \bibfield  {author} {\bibinfo {author} {\bibfnamefont {P.~T.}\ \bibnamefont
  {Brown}}, \bibinfo {author} {\bibfnamefont {D.}~\bibnamefont {Mitra}},
  \bibinfo {author} {\bibfnamefont {E.}~\bibnamefont {Guardado-Sanchez}},
  \bibinfo {author} {\bibfnamefont {R.}~\bibnamefont {Nourafkan}}, \bibinfo
  {author} {\bibfnamefont {A.}~\bibnamefont {Reymbaut}}, \bibinfo {author}
  {\bibfnamefont {C.-D.}\ \bibnamefont {Hébert}}, \bibinfo {author}
  {\bibfnamefont {S.}~\bibnamefont {Bergeron}}, \bibinfo {author}
  {\bibfnamefont {A.-M.~S.}\ \bibnamefont {Tremblay}}, \bibinfo {author}
  {\bibfnamefont {J.}~\bibnamefont {Kokalj}}, \bibinfo {author} {\bibfnamefont
  {D.~A.}\ \bibnamefont {Huse}}, \bibinfo {author} {\bibfnamefont
  {P.}~\bibnamefont {Schauß}},\ and\ \bibinfo {author} {\bibfnamefont {W.~S.}\
  \bibnamefont {Bakr}},\ }\bibfield  {title} {\bibinfo {title} {Bad metallic
  transport in a cold atom fermi-hubbard system},\ }\href
  {https://doi.org/10.1126/science.aat4134} {\bibfield  {journal} {\bibinfo
  {journal} {Science}\ }\textbf {\bibinfo {volume} {363}},\ \bibinfo {pages}
  {379} (\bibinfo {year} {2019})}\BibitemShut {NoStop}%
\bibitem [{\citenamefont {Rozenberg}\ \emph {et~al.}(1999)\citenamefont
  {Rozenberg}, \citenamefont {Chitra},\ and\ \citenamefont
  {Kotliar}}]{Rozenberg1999}%
  \BibitemOpen
  \bibfield  {author} {\bibinfo {author} {\bibfnamefont {M.~J.}\ \bibnamefont
  {Rozenberg}}, \bibinfo {author} {\bibfnamefont {R.}~\bibnamefont {Chitra}},\
  and\ \bibinfo {author} {\bibfnamefont {G.}~\bibnamefont {Kotliar}},\
  }\bibfield  {title} {\bibinfo {title} {Finite temperature mott transition in
  the hubbard model in infinite dimensions},\ }\href
  {https://doi.org/10.1103/PhysRevLett.83.3498} {\bibfield  {journal} {\bibinfo
   {journal} {Phys. Rev. Lett.}\ }\textbf {\bibinfo {volume} {83}},\ \bibinfo
  {pages} {3498} (\bibinfo {year} {1999})}\BibitemShut {NoStop}%
\bibitem [{\citenamefont {Loh}\ \emph {et~al.}(1990)\citenamefont {Loh},
  \citenamefont {Gubernatis}, \citenamefont {Scalettar}, \citenamefont {White},
  \citenamefont {Scalapino},\ and\ \citenamefont {Sugar}}]{Loh1990}%
  \BibitemOpen
  \bibfield  {author} {\bibinfo {author} {\bibfnamefont {E.~Y.}\ \bibnamefont
  {Loh}}, \bibinfo {author} {\bibfnamefont {J.~E.}\ \bibnamefont {Gubernatis}},
  \bibinfo {author} {\bibfnamefont {R.~T.}\ \bibnamefont {Scalettar}}, \bibinfo
  {author} {\bibfnamefont {S.~R.}\ \bibnamefont {White}}, \bibinfo {author}
  {\bibfnamefont {D.~J.}\ \bibnamefont {Scalapino}},\ and\ \bibinfo {author}
  {\bibfnamefont {R.~L.}\ \bibnamefont {Sugar}},\ }\bibfield  {title} {\bibinfo
  {title} {Sign problem in the numerical simulation of many-electron systems},\
  }\href {https://doi.org/10.1103/PhysRevB.41.9301} {\bibfield  {journal}
  {\bibinfo  {journal} {Phys. Rev. B}\ }\textbf {\bibinfo {volume} {41}},\
  \bibinfo {pages} {9301} (\bibinfo {year} {1990})}\BibitemShut {NoStop}%
\bibitem [{\citenamefont {Zhang}(1999)}]{Zhang1999}%
  \BibitemOpen
  \bibfield  {author} {\bibinfo {author} {\bibfnamefont {S.}~\bibnamefont
  {Zhang}},\ }\bibfield  {title} {\bibinfo {title} {Finite-temperature monte
  carlo calculations for systems with fermions},\ }\href
  {https://doi.org/10.1103/PhysRevLett.83.2777} {\bibfield  {journal} {\bibinfo
   {journal} {Phys. Rev. Lett.}\ }\textbf {\bibinfo {volume} {83}},\ \bibinfo
  {pages} {2777} (\bibinfo {year} {1999})}\BibitemShut {NoStop}%
\end{thebibliography}%
\end{document}